\PassOptionsToPackage{hyphens}{url} % must come first - allowing linebreaks in footnote 2

\documentclass[a4paper,fleqn]{cas-dc}
\renewcommand{\printorcid}{} 

\usepackage[numbers]{natbib}
\usepackage{subfig}
\usepackage{algorithm,algpseudocode}
\usepackage{tabularx}
\usepackage{stfloats}

\usepackage{placeins} % https://chatgpt.com/share/68d2f1fd-26c8-800a-a33c-6c609803bcbc

\usepackage{caption} 

\makeatletter % https://chatgpt.com/share/68bda991-9fc0-800a-bb5b-241f4c964231
\newcommand\subsubsubsection{%
  \@startsection{subsubsubsection}{4}{\z@}%
    {1.0ex \@plus .3ex \@minus .2ex}%  % before-skip
    {.1ex}%                             % after-skip
    {\normalfont\normalsize\bfseries\itshape}* % <-- the star disables toc/mark
}
\makeatother

\usepackage{fix-cm}

\usepackage[strings]{underscore}

\usepackage{esvect}
\usepackage{graphicx}
\usepackage{epstopdf}
\usepackage{rotating}   % for 'sidewaysfigure' environment

\usepackage{csquotes} %for quotes
\usepackage[justification=centering]{caption}

\usepackage{quotmark}
\setquotemarks{{(}{)},{[}{]}}
\newcommand{\paren}[1]{\tqt{#1}}
\usepackage[toc, acronym, nonumberlist, shortcuts=ac, automake]{glossaries-extra}
\usepackage[toc,acronym]{glossaries-prefix} %https://tex.stackexchange.com/questions/391297/how-can-glossaries-handle-the-article-a-or-an-before-abbreviations

\makeglossaries
\setabbreviationstyle[acronym]{long-short} % 
\glssetcategoryattribute{acronym}{nohyperfirst}{true}
\newacronym[prefixfirst={a\ },prefix={an\ }]{RNN}{RNN}{recurrent neural network}
\newacronym{RTRL}{RTRL}{real-time recurrent learning}
\newacronym{CT}{CT}{computed tomography}
\newacronym{1D}{1D}{one-dimensional}
\newacronym{2D}{2D}{two-dimensional}
\newacronym{3D}{3D}{three-dimensional}
\newacronym{4D}{4D}{four-dimensional}
\newacronym[prefixfirst={a\ },prefix={an\ }]{ROI}{ROI}{region of interest}
\newacronym[prefixfirst={a\ },prefix={an\ }]{RMS}{RMS}{root-mean-square}
\newacronym[prefixfirst={a\ },prefix={an\ }]{RMSE}{RMSE}{root-mean-square error}
\newacronym[%
prefix={an\space},prefixfirst={a~}%
]{nRMSE}{nRMSE}{normalized root-mean-square error}
\newacronym{UORO}{UORO}{unbiased online recurrent optimization}
\newacronym{DNI}{DNI}{decoupled neural interfaces}
\newacronym{LMS}{LMS}{least mean squares}
\newacronym{MR}{MR}{magnetic resonance}
\newacronym{MRI}{MRI}{magnetic resonance imaging}
\newacronym{LINAC}{LINAC}{linear accelerator}
\newacronym{PCA}{PCA}{principal component analysis}
\newacronym{kPCA}{kPCA}{kernel principal component analysis}
\newacronym{SnAp-1}{SnAp-1}{sparse one-step approximation}
\newacronym{SnAp-n}{SnAp-$n$}{sparse $n$-step approximation}
\newacronym[prefixfirst={a\ },prefix={an\ }]{SSIM}{SSIM}{structural similarity index measure}
\newacronym{IGRT}{IGRT}{image-guided radiotherapy}
\newacronym{4DCT}{4DCT}{four-dimensional computed tomography}
\newacronym{kV}{kV}{kilovoltage}
\newacronym{DVF}{DVF}{displacement vector field}
\newacronym{DIR}{DIR}{deformable image registration}
\newacronym{ANN}{ANN}{artificial neural network}
\newacronym{4D-CBCT}{4D-CBCT}{four-dimensional cone beam computed tomography}
\newacronym[prefixfirst={a\ },prefix={an\ }]{SHL}{SHL}{signal history length}
\newacronym{AIP}{AIP}{average intensity projection}
\newacronym[prefixfirst={a\ },prefix={an\ }]{MAE}{MAE}{mean average error}
\newacronym{GPU}{GPU}{graphics processing unit}
\newacronym{RPM}{RPM}{Real-Time Position Management}
\newacronym{BPTT}{BPTT}{backpropagation through time}
\newacronym[prefixfirst={a\ },prefix={an\ }]{LSTM}{LSTM}{long short-term memory}
\newacronym[prefixfirst={a\ },prefix={an\ }]{MLP}{MLP}{multilayer perceptron}
\newacronym{LR}{LR}{left--right}
\newacronym{AP}{AP}{antero-posterior}
\newacronym{SI}{SI}{superior--inferior}
\newacronym{GRU}{GRU}{gated recurrent unit}
\newacronym{bi-GRU}{bi-GRU}{bidirectional gated recurrent unit}
\newacronym{bi-LSTM}{bi-LSTM}{bidirectional long short-term memory}
\newacronym[prefixfirst={a\ },prefix={an\ }]{FCL}{FCL}{fully connected layer}
\newacronym[prefixfirst={a\ },prefix={an\ }]{FWHM}{FWHM}{full width at half maximum}
\newacronym{EPID}{EPID}{electronic portal imaging device}
\newacronym{GPR}{GPR}{gamma passing rate}
\newacronym{QA}{QA}{quality assurance}
\newacronym{AI}{AI}{artificial intelligence}
\newacronym{OSTL}{OSTL}{online spatio-temporal learning}
\newacronym{PET}{PET}{positron emission tomography}
\newacronym{IMRT}{IMRT}{intensity modulated radiation therapy}
\newacronym{CNN}{CNN}{convolutional neural network}
\newacronym[prefixfirst={a\ },prefix={an\ }]{FCN}{FCN}{fully convolutional network}
\newacronym{GAN}{GAN}{generative adversarial network}
\newacronym[longplural={organs at risk}]{OAR}{OAR}{organ at risk}
\newacronym{ART}{ART}{adaptive radiotherapy}
\newacronym[prefixfirst={a\ },prefix={an\ }]{SVM}{SVM}{support vector machine}
\newacronym{ITV}{ITV}{internal target volume}
\newacronym{PTV}{PTV}{planning target volume}
\newacronym{GTV}{GTV}{gross tumor volume}
\newacronym[prefixfirst={a\ },prefix={an\ }]{MLC}{MLC}{multileaf collimator}
\newacronym{ARIMA}{ARIMA}{Auto-Regressive Integrated Moving Average}
\newacronym{DTA}{DTA}{distance to agreement}
\newacronym{LS-SVM}{LS-SVM}{least-squares support vector machine}
\newacronym{RF}{RF}{radio-frequency}
\newacronym{MSSA}{MSSA}{multi-channel singular spectrum analysis}
\newacronym{NCAT}{NCAT}{non-uniform rational B-splines cardiac-torso}
\newacronym{VAE}{VAE}{variational autoencoder}
\newacronym{AE}{AE}{autoencoder}
\newacronym{PredNet}{PredNet}{predictive coding network}
\newacronym{CDNA}{CDNA}{Convolutional Dynamic Neural Advection}
\newacronym{IR}{IR}{infra-red}
\newacronym{DCE}{DCE}{dynamic contrast-enhanced}
\newacronym{ADAM}{ADAM}{adaptive momentum}
\newacronym{ANFIS}{ANFIS}{adaptive neuro-fuzzy inference system}
\newacronym{SNR}{SNR}{signal-to-noise ratio}
\newacronym{XCAT}{XCAT}{extended cardiac-torso}
\newacronym{TE}{TE}{tracking error}
\newacronym{NLP}{NLP}{natural language processing}
\newacronym{SVR}{SVR}{support vector regression}
\newacronym{LED}{LED}{light-emitting diode}
\newacronym[prefixfirst={a\ },prefix={an\ }]{STN}{STN}{spatial transformer network}
\newacronym{CBCT}{CBCT}{cone beam computed tomography}
\newacronym{TBPTT}{TBPTT}{truncated backpropagation through time}
\newacronym{FOV}{FOV}{field of view}
\newacronym{OvGU}{OvGU}{Otto-von-Guericke University Magdeburg}
\newacronym{TCN}{TCN}{temporal convolutional network}
\newacronym{NCC}{NCC}{normalized cross-correlation}
\newacronym{LPIPS}{LPIPS}{learned perceptual image patch similarity}
\newacronym{VGG}{VGG}{Visual Geometry Group}
\newacronym{ConvLSTM}{ConvLSTM}{convolutional long short-term memory}
\newacronym{ConvGRU}{ConvGRU}{convolutional gated recurrent unit}
\newacronym{FIR}{FIR}{finite impulse response}
\newacronym{ReLU}{ReLU}{rectified linear unit}
\newacronym{FFN}{FFN}{feed-forward network}
\newacronym{MSE}{MSE}{mean squared error}
\newacronym[prefixfirst={a\ },prefix={an\ }]{EE}{EE}{end-expiration}
\newacronym[prefixfirst={a\ },prefix={an\ }]{EI}{EI}{end-inspiration}
\newacronym{MR-LINAC}{MR-LINAC}{MR-guided linear accelerator}
\newacronym{MR-guided}{MR-guided}{magnetic resonance--guided}
\newacronym{AR}{AR}{autoregressive}
\newacronym{OLS}{OLS}{ordinary least squares}
\newacronym{SGD}{SGD}{stochastic gradient descent}
\newacronym{CI}{CI}{confidence interval}
\newacronym{CKS}{CKS}{CyberKnife Synchrony}
\newacronym{HGRN}{HGRN}{hierarchically gated linear RNN}
\newacronym{VRNN}{VRNN}{variational recurrent neural network}
\newacronym[prefixfirst={a\ },prefix={an\ }]{LG-SSM}{LG-SSM}{linear Gaussian state-space model}

\newcommand{\glsfirstbracket}[1]{%
  \acrlong{#1}~[\acrshort{#1}]%
  \glsunset{#1}%
}

\usepackage{xcolor}

\usepackage{bookmark}

\usepackage{cleveref}

\usepackage{hyperref}
\newcounter{figfn}

\makeatletter
\DeclareRobustCommand{\FNmark}[1]{%
  \begingroup \setcounter{footnote}{#1}\footnotemark[#1]\endgroup}
\newcommand{\FNtext}[2]{%
  \begingroup \setcounter{footnote}{#1}\footnotetext[#1]{#2}\endgroup}
\makeatother

\usepackage[hyphens]{url}  % No option clash because of PassOptionsToPackage earlier
\usepackage{etoolbox}      % For toggling font locally
\newcommand{\datasetid}[1]{%
  \begingroup
    \def\UrlFont{\the\font}% inherit current font fully (shape, series, size)
    \nolinkurl{#1}%
  \endgroup
}
\def\tsc#1{\csdef{#1}{\textsc{\lowercase{#1}}\xspace}}
\tsc{WGM}
\tsc{QE}
\tsc{EP}
\tsc{PMS}
\tsc{BEC}
\tsc{DE}

\begin{document}
\let\WriteBookmarks\relax
\def\floatpagepagefraction{1}
\def\textpagefraction{.001}
\shorttitle{Frame forecasting in cine MRI using the PCA respiratory motion model: RNNs trained online vs. transformers}
\shortauthors{Pohl, Uesaka, Demachi and Chhatkuli}

\title [mode = title]{Frame forecasting in cine MRI using the PCA respiratory motion model: comparing recurrent neural networks trained online and transformers}              

\author[1]{Michel Pohl}
\cormark[1]
\ead{michel.pohl@centrale-marseille.fr}
\address[1]{The University of Tokyo, 113-8654 Tokyo, Japan}

\author[2]{Mitsuru Uesaka}
\address[2]{Japan Atomic Energy Commission, 100-8914 Tokyo, Japan}

\author[1]{Hiroyuki Takahashi}
\author[1]{Kazuyuki Demachi}

\author[3]{Ritu Bhusal Chhatkuli}
\address[3]{National Institutes for Quantum and Radiological Science and Technology, 263-8555 Chiba, Japan}

\cortext[cor1]{Corresponding author}

\begin{abstract}
Respiratory-induced motion complicates accurate irradiation of thoraco-abdominal tumors during radiotherapy, as treatment-system latency entails target-location uncertainties. This work addresses frame forecasting in dynamic chest and liver \acs{MRI} to compensate for such delays. We investigate \acsp{RNN} trained with online learning algorithms, enabling adaptation to changing respiratory patterns via on-the-fly parameter updates, and transformers, increasingly common in time-series forecasting for their ability to capture long-term dependencies.

Experiments were conducted using twelve sagittal thoracic and upper-abdominal cine-\acs{MRI} sequences from ETH Zürich and \gls{OvGU}; the \gls{OvGU} data exhibited higher motion variability, noise, and lower contrast. \acs{PCA} decomposes the Lucas--Kanade optical-flow field into static deformation modes and low-dimensional, time-dependent weights. We compare various methods for forecasting these weights: linear filters, population and sequence-specific encoder-only transformers, and \acsp{RNN} trained with \gls{RTRL}, unbiased online recurrent optimization, decoupled neural interfaces, and \gls{SnAp-1}. Predicted displacements were used to warp the reference frame and generate future images.

Prediction accuracy decreased with the horizon $h$. Linear regression performed best at short horizons (1.3mm geometrical error at $h = 0.32\text{s}$, ETH Zürich dataset), while \gls{RTRL} and \gls{SnAp-1} outperformed the other algorithms at medium-to-long horizons, with geometrical errors below 1.4mm and 2.8mm on the sequences from ETH Zürich and \gls{OvGU}, respectively. The sequence-specific transformer was competitive for low-to-medium horizons, but transformers remained overall limited by data scarcity and domain shift between datasets. Predicted frames visually resembled the ground truth, with notable errors occurring near the diaphragm at end-inspiration and regions affected by out-of-plane motion.

\end{abstract}
\glsresetall % makes the main text treat all acronyms as unused again

%\includegraphics{figs/grabs.pdf}

\begin{highlights}
\item RNNs trained online can forecast chest and liver cine-MRI frames accurately and fast. % 85 characters
\item Online learning helps adapt to unsteady motion and reach high accuracy with few data. % 85 characters
\item Breathing was represented mainly by the PCA motion model's first or second component. % 85 characters
%\item Predicting higher-order PCA weights related with minor deformations boosted accuracy. % 85 characters - removed because 5 bullet points max
\item Data scarcity, narrow input window, and domain shift capped transformer accuracy.
\item Linear filters and RNNs predicted motion well at low and high horizons, respectively. % 85 characters

\end{highlights}


\maketitle
\glsresetall
\glsunset{SnAp-1}
% I don't reuse SnAp-1 in full form in the main text, because of how it is defined:
% \gls{SnAp-n} introduces bias during loss gradient estimation but yields deterministic closed-form updates. The specific case $n=1$ \paren{\gls{SnAp-1}} is equivalent to a diagonal approximation of the sensitivity matrix analogous to that proposed in the original LSTM paper \cite{hochreiter1997long}.
\glsunset{GPU} % No need to write "graphics processing units" in a long paper, and the acronym is well understood

\section{Introduction}
\label{section: introduction}
% Possible changes/improvements:
% - cite other papers in the intro regarding tumor motion amplitude
% - I would possibly need to cite developments in MR-guided liver radiotehrapy

\subsection{Respiratory motion management in MR-guided radiotherapy}\label{section: intro respiratory motion management in MRgRT}
% If I expand the first two paragraphs, I could divide further into:
% - 1.1.1. Motion management in MR-guided radiotherapy
% - 1.1.2. Correspondence models in radiotherapy

% General intro about respiratory motion and the associated challenges in radiotherapy
Machine learning can benefit external beam radiotherapy at several stages of the clinical workflow, ranging from optimal treatment plan selection and dose planning to radiation delivery and assessment of patient response to therapy \cite{huynh2020artificial}. Among these steps, respiratory motion forecasting during treatment is a critical application that can enhance therapeutic irradiation precision and patient outcomes. Indeed, the beam may partially miss the moving target (e.g., a lung or pancreatic tumor) and negatively affect surrounding healthy tissue instead, due to intrinsic treatment-system latency. Specifically, rotations, deformations, and translations of the tumor and surrounding \glspl{OAR} induced by breathing can cause geometrical and dosimetric errors. Thoracic and upper-abdominal tumor motion is primarily periodic, with a range in the cranio-caudal direction sometimes exceeding 5cm \cite{sarudis2017systematic}. Nonetheless, it is influenced by phase shifts and local variations in frequency and amplitude. Amplitude shifts refer to sudden, occasional variations in the average position of the organs, while the term "drift" describes more gradual changes occurring during a single treatment session. For example, baseline intrafractional drifts of 1.65 $\pm$ 5.95 mm, 1.50 $\pm$ 2.54 mm, and 0.45 $\pm$ 2.23 mm (mean $\pm$ standard deviation) have been observed in the \gls{SI}, \gls{AP}, and left--right directions, respectively \cite{takao2016intrafractional}. Moreover, general posture changes due to patient relaxation over time or minor positional adjustments on the treatment bed also contribute to variability in motion records. Each radiotherapy treatment system has a specific characteristic time delay, but "for most radiation treatments, the latency will be more than 100ms, and can be up to two seconds" \cite{verma2010survey}.

% General intro about surrogates in radiotherapy and MR-guided radiotherapy
% I have not added an hyphen in 4D-MRI because it's the first time the acronym is used
During external beam radiotherapy for thoracic and abdominal tumors, the \gls{3D} target cannot be fully imaged in real time. Conventional methods rely on tracking fiducial markers implanted near the target using \gls{kV} fluoroscopic imaging (e.g., the \glsfirstbracket{CKS} system), which requires an invasive surgical procedure, or on recording the positions of external markers on the chest and abdomen (e.g., the Varian \glsfirstbracket{RPM} system), whose trajectories may correlate only weakly with tumor motion, partly due to phase shifts. Recent advancements in \gls{MRI} have enabled real-time observation of soft tissue and organs with high contrast in a fixed imaging plane. Moreover, \gls{MR-guided} radiotherapy systems (e.g., MRIdian) do not involve additional imaging dose, unlike systems relying on \gls{kV} fluoroscopy or \gls{4D-CBCT}. Consequently, they can image the slice of interest during free breathing over relatively long periods, enabling the observation of inter-cycle variations. By contrast, \gls{4D-CBCT} can only capture an average breathing cycle, due to dose and hardware constraints. This improved visualization, combined with online adaptive planning capabilities (i.e., on-table plan re-optimization based on real-time imaging) specific to \gls{MR-guided} radiotherapy, can widen eligibility for high-dose focal liver irradiation and support improved local control in cases where motion and \gls{OAR} dose constraints would otherwise limit treatment due to toxicity \cite{witt2020mri}. Fast \gls{4D} \gls{MRI} is an active area of research, motivated by the relatively low spatial resolution and image quality of current \gls{4D}-\gls{MRI} acquisition techniques. In that context, \acrlong{AI} research for \gls{MR-guided} radiotherapy has focused on \gls{3D} motion estimation from the observed \gls{2D} slices and irradiation delay mitigation---our work addresses the latter challenge.

% Correspondence models in general
Previous work has explored mathematical correspondence models that derive \gls{3D} internal motion, which is not directly observable in real time, from surrogate signals (also referred to as partial observations). Fitting such models requires prior, simultaneous acquisition of image and surrogate signal data. Correspondence and forecasting models may be subject-specific, when fitted to data from a single patient, or population-based, also termed cross-subject, when trained on data from multiple individuals. \citeauthor{wang2021real} inferred liver motion from ultrasound images using the positions of light-emitting diodes placed on the chest as surrogate signals (AccuTrack 250 system), and further forecast these positions to achieve spatiotemporal prediction \cite{wang2021real}. They reported that \gls{LSTM} networks outperformed \gls{SVR} on both tasks and that continuously updating the correlation model improved accuracy.

% PCA in correspondence models: introduction and direct models
\Gls{PCA}, an unsupervised dimensionality-reduction algorithm that can be interpreted as fitting an ellipsoid to the data and relies on eigendecomposition of the data matrix, has been widely used in correspondence models. In radiotherapy, \gls{PCA} has commonly been applied to motion information obtained via \gls{DIR} between a reference phase and other phases in a four-dimensional dataset (Appendix \ref{appendix: PCA respiratory motion model}). This approach implicitly assumes that linear combinations of the leading eigenvectors can approximate relevant organ motion states. Two broad strategies utilizing \gls{PCA} in correspondence modeling are highlighted in \cite{ehrhardt20134d}: direct and indirect modeling. In direct models, \gls{PCA} is applied to internal or surrogate data and a regression method is then fitted using the \gls{PCA} weights, or \gls{PCA} itself is used to fit the correspondence model. The originally proposed \gls{PCA} respiratory motion model belongs to the latter subcategory \cite{zhang2007patient}. In that work, \citeauthor{zhang2007patient} inferred the \gls{DVF} throughout the chest from the diaphragm position in \gls{4DCT}. They argued that two principal components were sufficient to describe breathing motion accurately. Likewise, \citeauthor{chen2018internal} applied \gls{PCA} to fit a correspondence model relating tumor and lung motion to that of the external chest surface in \gls{4DCT}, using particle-based surface meshing and topology-preserving non-rigid point-matching registration \cite{chen2018internal}. Conversely, \citeauthor{li2011pca} performed regression-based direct modeling using \gls{PCA} fitted with internal data, inferring the entire chest motion from a single artificial marker via Bayesian maximum a posteriori estimation \cite{li2011pca}. They recommended selecting the number of components on a per-patient basis and also experimentally found that two components were generally sufficient. 
This finding was supported by the fact that two eigenvectors can exactly represent a cosine breathing model.

% Indirect PCA models
By contrast, in indirect models, the weights associated with principal components derived from internal motion are estimated by maximizing the similarity between partial \gls{2D} observations and the corresponding cross-sections from inferred \gls{3D} data (the warped reference volume). Subject-specific indirect models were used in \cite{stemkens2016image, harris2016technique} to synthesize volumetric \gls{MR} images from \gls{2D} cine \gls{MRI}. 
% In those two studies, iterative optimization of the eigenvector weights was conducted until satisfactory alignment between the observed slices and warped reference volumes was reached. % Is that necessary
The Bayesian information criterion was applied in \cite{stemkens2016image} to select the \gls{PCA}-subspace dimension% in 10-phase respiratory-correlated abdominal \gls{MRI}
; one component was favored as providing the best bias--variance trade-off, but a second component was retained in practice to model respiratory motion hysteresis.
\citeauthor{romaguera2021predictive} mentioned that "results reported for these patient-specific models are often more accurate than for population-based methods. In a clinical scenario, their reliability depends, however, on the degree of patient-specific inter-fraction motion variations" \cite{romaguera2021predictive}.

\subsection{Breathing motion prediction with recurrent and attention-based neural networks}
\label{section: breathing motion pred. with RNNs and attention-based ANNs}

% General facts about RNNs and adaptive retraining - reviewed
Recurrent connections are widespread in recent artificial neural network architectures proposed for respiratory motion forecasting in radiotherapy \cite{wang2018feasibility, lin2019towards, yu2020rapid, lombardo2022offline, samadi2023respiratory}. At the core of all \gls{RNN} variants, these feedback loops act as a form of memory, enabling the storage and retrieval of information over time. This distinctive feature allows them to effectively learn patterns and relationships within sequential data and often surpass conventional \glspl{MLP} and linear models in time-series processing and natural language processing. %\gls{NLP}. 
Studies have shown that recurrent models such as deep \glspl{bi-GRU} or \glspl{LSTM} can outperform simpler algorithms, for instance, adaptive-boosted \glspl{MLP} \cite{wang2018feasibility} and linear filters \cite{lombardo2022offline}, in breathing motion prediction.

% Introduction on attention mechanisms and transformers - reviewed
Respiratory motion forecasting has also been influenced by research on attention-based algorithms, most notably transformers. Originally introduced by \citeauthor{bahdanau2014neural} to improve neural machine translation, attention mechanisms allow models to focus on relevant parts of an input sequence dynamically \cite{bahdanau2014neural}. This helps reduce the information bottleneck resulting from encoding a long history into a single recurrent state and mitigate long-range dependency issues often associated with vanishing gradients during \gls{RNN} training. While early forms of attention operated sequentially alongside \glspl{RNN}, transformer self-attention helps avoid the strictly sequential update of recurrent models by enabling parallel computation over sequence positions \cite{vaswani2017attention}. This can accelerate training, while allowing each token to attend to all others simultaneously. 

% General ML research on transformers for time-series forecasting - reviewed
Recent research on general time-series forecasting with transformers has explored improvements regarding complexity and sequential data modeling. For instance, \citeauthor{li2019enhancing} proposed the LogSparse transformer, making each cell attend to previous cells at exponentially increasing intervals and using causal convolutions in self-attention to better capture local context \cite{li2019enhancing}. Also leveraging sparsity to reduce complexity, \citeauthor{zhou2021informer} focused on long sequence forecasting and introduced the Informer \cite{zhou2021informer}. The latter relies on ProbSparse self-attention, which operates on only a selected subset of dominant queries based on a sparsity measure derived from the Kullback--Leibler divergence. This architecture is characterized by shrinking encoder representations at each layer via max-pooling and a generative-style decoder that performs prediction in one forward operation. In the Autoformer model, proposed for long-term forecasting, self-attention is replaced with auto-correlation, helping discover periodicity-based similarities, and the input is decomposed into seasonality and trend, whose representations are progressively refined \cite{wu2021autoformer}.

% Respiratory motion forecasting with transformers only - reviewed
Regarding respiratory motion forecasting, \citeauthor{jeong2022clinical} compared a full encoder--decoder transformer 
% a transformer (comprising a 6-layer encoder and a 6-layer decoder) 
with \pgls{LSTM} and a \gls{bi-LSTM} network to predict the distance from a laser source to the chest surface of cancer patients (Anzai respiration-gating system). The transformer outperformed the other two models at horizons $h \geq 0.50\text{s}$ (the time interval in the future for which the prediction is made, also called response time or look-ahead time). Moreover, \citeauthor{shi2022respiratory} predicted respiratory signals using an architecture combining squeeze-and-excitation attention with a \gls{TCN} component \cite{shi2022respiratory}. It consistently yielded lower \glsentrylong{MAE} and \gls{RMSE} than the \gls{bi-LSTM} model examined in \cite{wang2018feasibility} and \gls{CNN}--\acs{TCN} and \acs{CNN}--\gls{bi-LSTM} baselines, for $h \geq 0.30\text{s}$.
% MAE spelled out fully because I only use MAE once in this paper.
% If I have time I can cite "Lightweight attention TCN based on multi-scale feature fusion for respiratory prediction in radiotherapy" which was published in 2024 (from Shi et al.)

% Practical transformer limitations
Despite recent progress enabled by transformers in time-series processing, models in this family remain constrained by computational and memory bottlenecks. In particular, the complexity of the canonical transformer architecture grows quadratically with the input length. Notably, \citeauthor{romaguera2023conditional} reported that their transformer-based approach for predicting future volumes from \gls{2D} dynamic liver \gls{MRI} was approximately three times slower than baselines relying on a \gls{ConvLSTM} or \gls{ConvGRU} module \cite{romaguera2023conditional}. These costs practically limit usable context length, which in turn can hinder long-range dependency modeling.
% This can also limit their ability to model long-range dependencies in practice. [These computational constraints / These costs]
In addition, standard transformers lack built-in notions of order or locality, unlike \glspl{RNN}. The latter have stronger inductive biases, as their core definition, involving a hidden state evolving over time, implies assumptions regarding continuity and causality. Accordingly, recent evidence suggests that although transformers trained on large datasets tend to perform well, \glspl{RNN} can be more sample-efficient and generalize better in low-data regimes, sometimes with equivalent parameter counts.
% (e.g., see \cite{haller2024babyhgrnexploring} for a comparative analysis in low-resource language modeling scenarios with controlled parameter budget).
\citeauthor{haller2024babyhgrnexploring} provided such a comparative analysis in the context of low-resource language modeling using datasets of 10M--100M words \cite{haller2024babyhgrnexploring}. They showed that recurrent architectures, including hierarchically gated linear \glspl{RNN}, can surpass decoder-only transformers on multiple benchmarks under a controlled parameter budget (approximately 300M--360M weights). In biomedical imaging and signal processing, datasets are often relatively small, partly due to privacy and data-governance constraints. This has been noted to pose "challenges for training and validating transformer models effectively" \cite{islam2024comprehensive}. Importantly, there is an ongoing discussion regarding the practical applicability of transformers in radiotherapy. For example, \citeauthor{wimmert2024benchmarking} reported that \pgls{LSTM} outperformed a limited-history transformer for predicting Varian \gls{RPM} signals despite a large cohort (2,502 traces, 416 patients, mean duration of 130s) \cite{wimmert2024benchmarking}. However, their full-history encoder-only transformer (without window limitation) surpassed the \gls{LSTM} in accuracy. Attention maps revealed that its third and fourth layers attended to similar patterns relatively far in the past (up to approximately 80s). Nonetheless, its inference time was roughly 100 times higher than that of the \gls{LSTM}, potentially constraining clinical feasibility.

% Motivations regarding RNN research
In this context, research on \glspl{RNN} has evolved alongside that on transformers, with recent works proposing hybrid models that integrate elements of both architectural families. For instance, \citeauthor{tan2022lstformer} proposed the LSTformer, which combines transformer-encoder and \gls{LSTM}-based layers, to predict respiratory motion traces using a dataset of 304 CyberKnife records (26Hz sampling, 71 min average duration) \cite{tan2022lstformer}.
This model consistently outperformed the \gls{LSTM}, \gls{bi-LSTM}, and \gls{bi-GRU} baselines in \cite{lin2019towards, wang2018feasibility, yu2020rapid} across various metrics for horizons between 0.20s and 0.60s, although its inference time was nearly 1.5 times higher than that of the \gls{LSTM}. Moreover, \citeauthor{zhang2023lgeanet} introduced LGEANet, a network for multivariate time-series forecasting that combines \gls{LSTM} layers, temporal convolutional and linear \gls{AR} modules, and external attention (small, trainable, and shared memory units accounting for relationships between all samples) \cite{zhang2023lgeanet}. When trained and evaluated on the same CyberKnife dataset, LGEANet consistently surpassed several attention-based and recurrent baselines over the four test sequences and horizons ranging from 0.07s to 0.92s. These observations underscore the continued relevance of approaches involving recurrent architectures in breathing motion prediction.

\subsection{Online learning algorithms for RNNs applied to respiratory motion forecasting in radiotherapy}\label{section: online RNN algorithms applied to radiotherapy}

% Introduction to adaptive learning: sliding-window retraining and TBPTT - paragraph reviewed
A common strategy in sequence processing is to adjust model parameters as new samples arrive, improving robustness to patterns unseen during training. To achieve that, one can fine-tune the model using the information contained within a sliding time window, beyond which past data is discarded. For instance, this approach was used to personalize and update over time a cross-subject \gls{MLP} predicting lung-tumor \gls{SI} positions recorded by the \gls{CKS} system \cite{teo2018feasibility}. Similarly, \citeauthor{lombardo2022offline} adapted a population \gls{LSTM} forecasting \gls{SI} tumor motion in \gls{2D} cine \gls{MRI} (acquired on MRIdian) to a specific patient via sliding-window updates \cite{lombardo2022offline}. They reported \gls{RMSE} decreases from 2.02mm to 1.77mm and from 1.59mm to 1.34mm on both test sets at $h=0.75\text{s}$ compared to offline learning (the training procedure without continuous adaptation). This strategy can improve performance but induces forgetting, as earlier examples outside the current window no longer contribute to parameter updates.
% This strategy can enhance performance but has a significant drawback: with each new retraining cycle, the algorithm progressively forgets the characteristics of earlier data outside the most recent windows.
A related and widely used approach for dynamic training of \glspl{RNN} is \gls{TBPTT}, which limits gradient propagation to a fixed history length \cite{jaeger2002tutorial}. %In \gls{TBPTT}, past data influence persists softly as forward recursion from non-detached hidden states is preserved, unlike sliding window retraining with \gls{BPTT}, subject to representation drift. 
In \gls{TBPTT}, past data influence persists implicitly via forward recursion using non-detached states. By contrast, sliding-window retraining with \gls{BPTT}, the standard offline training procedure for \glspl{RNN}, is prone to representation drift due to repeated reinitialization. 
While \gls{TBPTT} remains relatively resource-efficient, with a time complexity of $\mathcal{O}(T d^2)$, where $d$ and $T$ denote the number of hidden units and truncation length, respectively, it still biases learning towards more recent dependencies.

% Online learning and RTRL
Unlike sliding-window retraining or \gls{TBPTT}, truly online training methods retain knowledge over a longer effective time span. Indeed, parameter update equations do not directly reference a past time point beyond which information is lost. \Gls{RTRL}, a fundamental online learning algorithm for \glspl{RNN}, recursively updates the sensitivity matrix---the derivative of the hidden state with respect to the parameters---also referred to as the influence matrix, at each time step \cite{williams1989learning}. This algorithm has been applied to forecast the locations of fiducial markers implanted in the lungs (SyncTraX system) \cite{jiang2019prediction}, abdominal and thoracic cancer lesions recorded by the \gls{CKS} system \cite{mafi2020real}, points within the chest tracked with \gls{DIR} in \gls{4D-CBCT} images \cite{pohl2021prediction}, and external markers on the chest and abdomen of volunteers (NDI Polaris system) \cite{pohl2022prediction, pohl2025real}. Nonetheless, \gls{RTRL} is limited by its high computational complexity, which scales as $\mathcal{O}(d^4)$, rendering inference impractical even for moderately sized networks. 

\begin{table}[htb!]
\setlength{\tabcolsep}{1.3pt}
\begin{center}
\begin{tabular}{lll}
\hline
Algorithm                                                       & \multicolumn{2}{l}{Complexity}  \\
                                                                & Memory & Time \\
\hline
Real-time recurrent learning \cite{williams1989learning}        & $\mathcal{O}(d^3)$  & $\mathcal{O}(d^4)$      \\
Truncated BPTT \cite{williams1990efficient}                     & $\mathcal{O}(T d)$  & $\mathcal{O}(T d^2)$    \\
Unbiased online recurrent optimization \cite{tallec2018unbiased}& $\mathcal{O}(d^2)$  & $\mathcal{O}(d^2)$      \\
Kronecker-factored RTRL \cite{mujika2018approximating}          & $\mathcal{O}(d^2)$  & $\mathcal{O}(d^3)$      \\
Kernel RNN learning \cite{roth2018kernel}                       & $\mathcal{O}(d^2)$  & $\mathcal{O}(d^2)$      \\
$r$-optimal Kronecker-sum approximation \cite{benzing2019optimal} & $\mathcal{O}(rd^2)$ & $\mathcal{O}(rd^3)$     \\
Random-feedback online learning \cite{murray2019local}          & $\mathcal{O}(d^2)$  & $\mathcal{O}(d^2)$      \\
Sparse $n$-step approximation \cite{menick2020practical}        & $\mathcal{O}(d^2)$  & $\mathcal{O}(d^2)$      \\
Reverse Kronecker-factored RTRL \cite{marschall2020unified}     & $\mathcal{O}(d^2)$  & $\mathcal{O}(d^3)$      \\
Efficient BPTT \cite{marschall2020unified}                      & $\mathcal{O}(T d)$  & $\mathcal{O}(d^2)$      \\
Future-facing BPTT \cite{marschall2020unified}                  & $\mathcal{O}(T d)$  & $\mathcal{O}(T d^2)$    \\
Decoupled neural interfaces \cite{jaderberg2017decoupled}       & $\mathcal{O}(d^2)$  & $\mathcal{O}(d^2)$      \\
\hline
\end{tabular}
\caption{Memory and time complexity of several online learning algorithms for RNNs. Here, $d$ and $T$ designate the number of hidden units and the truncation length, respectively.\protect\footnotemark}
\label{table:RNN online learning comparison}
\end{center}
\end{table}

\footnotetext{Adapted from \citep{marschall2020unified}, first published in modified form in \cite{pohl2025real}, Copyright JMLR 2020.}

% Recent research on online learning algorithms for RNNs
Several online-training algorithms for \glspl{RNN} have been proposed to lower computational requirements compared to \gls{RTRL}, while avoiding the truncation bias incurred by \gls{TBPTT}, thereby better preserving learning signals from distant past time steps (Table \ref{table:RNN online learning comparison}) \cite{marschall2020unified}. 
% preserve long-range credit assignment % (variant)
In \gls{UORO}, the sensitivity matrix is approximated by an unbiased, random rank-one estimator updated recursively %in closed form 
\cite{tallec2018unbiased}. This, in turn, produces unbiased loss-gradient estimates, but introduces additional stochasticity. Other approaches leverage sparsity to reduce time complexity; for instance, \gls{SnAp-n} tracks only parameter-to-state dependencies that affect hidden units within $n$ recurrent steps \cite{menick2020practical}. \gls{SnAp-n} introduces bias during loss-gradient estimation but yields deterministic closed-form updates. %Although \gls{SnAp-n} introduces a bias in the loss gradient computation, it maintains a non-stochastic closed-form update. 
The specific case $n=1$ \paren{\gls{SnAp-1}} %corresponds to a diagonal approximation of the state-to-state Jacobian% corresponds to block-diagonal approximation of the sensitivity matrix
corresponds to a diagonal-like approximation of the sensitivity matrix, similar in spirit to early forward-mode training formulations for \glspl{LSTM} \cite{hochreiter1997long}. Unlike \gls{UORO} and \gls{SnAp-1}, which compress past information encoded in the sensitivity matrix, \gls{DNI} is future-facing, as it seeks to predict the error signal, i.e., the derivative of the future accumulated loss with respect to the hidden states \cite{marschall2020unified, jaderberg2017decoupled}. Originally introduced as a framework relevant to both recurrent and non-recurrent networks, \gls{DNI} aims to eliminate the dependency of modules on the completion of backward or forward computations by other modules before updating their own weights. This is achieved by learning a "synthetic gradient," an independent prediction of the error signal, at each layer. The updates in \gls{DNI} are biased, deterministic, and numerical, as the synthetic gradient is obtained via auxiliary optimization rather than closed-form recursion. Notably, \Gls{UORO}, \gls{SnAp-1}, and \gls{DNI} benefit from a relatively low time complexity of $\mathcal{O}(d^2)$; efficient implementations for standard (i.e., vanilla) \glspl{RNN} were proposed in \cite{pohl2022prediction, pohl2025real}. Specifically, closed-form simplifications for quantities involved in the loss-gradient calculation in \gls{UORO} were derived in \cite{pohl2022prediction}. Regarding \gls{SnAp-1}, compression of sparse matrices (including the influence matrix) into a compact, dense form was introduced in \cite{pohl2025real} to reduce practical memory and computational requirements. The same work also improved the linear-operator update used to fit the synthetic gradient compared with the formulation in \cite{marschall2020unified}, thereby yielding higher empirical accuracy. These three algorithms were first applied to external marker position forecasting for radiotherapy in \cite{pohl2022prediction, pohl2025real} and generally outperformed \gls{RTRL} under roughly comparable computational budgets (e.g., smaller value of $d$ for \gls{RTRL}).

\subsection{Future frame prediction in natural videos and medical imaging}

\subsubsection{Video prediction in general computer vision}

% General introduction about video prediction and related model architectures
Closely related to time-series forecasting, video prediction---the self-supervised task of estimating future frames given a sequence of past frames---has attracted considerable interest in the computer vision community, with applications in robotic control \cite{minderer2019unsupervised, gupta2022maskvit}, autonomous driving \cite{jin2017predicting, lotter2017deep, luc2018predicting}, and precipitation nowcasting \cite{shi2015convolutional}. Natural videos exhibit complex dynamics across scales and may involve occlusions, camera motion, and illumination changes, which makes forecasting challenging. Furthermore, the future is multimodal, since multiple outcomes may be consistent with past frames. As a result, predictions may become blurred, especially at long horizons $h$. Overall, prediction accuracy typically degrades as $h$ increases. Recurrent models have long been a common backbone for temporal modeling in video processing. Early \gls{RNN}-based video predictors include \gls{ConvLSTM}, in which fully connected layers are replaced with convolutional layers to better capture spatial structure \cite{shi2015convolutional}, and extensions of the \gls{LSTM} encoder--decoder model originally developed for machine translation \cite{sutskever2014sequence, srivastava2015unsupervised}. Notably, \citeauthor{jin2017predicting} reported that sliding-window fine-tuning at test time with \gls{BPTT} improved multi-step predictions relative to purely autoregressive rollouts \cite{jin2017predicting}. More recently, transformer-based approaches have been proposed for video prediction, with the task framed as the generation of a fixed block of future frames conditioned on a short context. For example, MaskViT uses a masked visual modeling objective (i.e., it masks a subset of tokens and predicts the remaining ones) together with iterative decoding, i.e., progressively filling in masked tokens based on the most confident predictions \cite{gupta2022maskvit}.
%More recently, transformer-based approaches have been explored for video prediction; for instance, MaskViT leverages masked visual modeling as a pretraining task and computes self-attention in windows (a spatial window and spatiotemporal window) to increase computational efficiency \cite{gupta2022maskvit}.

% Literature about video prediction with natural images
\citeauthor{oprea2020review} comprehensively surveyed deep learning methods for video prediction and proposed a taxonomy of approaches \cite{oprea2020review}. Early works attempted to forecast raw pixel values directly by implicitly capturing fine details and scene dynamics \cite{shi2015convolutional, srivastava2015unsupervised, lotter2017deep}. Nevertheless, learning a stable and meaningful representation from raw frames proved difficult due to the high dimensionality and variability of the pixel space, motivating efforts to structure and simplify internal representations.
%That challenge led to developments aimed at reducing the dimensionality of internal representations and supervision complexity. 
One line of work has focused on separating sources of variability from visual content, which \citeauthor{oprea2020review} refer to as "factorizing the prediction space." One such strategy is to disentangle motion and appearance using a two-stream processing approach \cite{villegas2017decomposing}. In another subcategory of methods, change %variability factors are modeled
is represented as explicit transformations between subsequent frames to leverage the latter's high level of similarity, using kernel- or vector-based resampling \cite{finn2016unsupervised, liu2017video}. Architectures relying on vector-based resampling predict a dense motion field between an observed frame and a future frame, then warp the former to synthesize the latter. This resampling approach often relies on a %\gls{STN}
spatial transformer-style differentiable warping module \cite{jaderberg2015spatial}. 
%The latter learns parameterized geometrical transformations, which are then applied to the input image or feature map.
The original \glsentrylong{STN} comprises a localization network that predicts the parameters of geometrical transformations and a sampling layer that applies them to the input image or feature map.
% It generally relies on the \gls{STN}, a module that learns to perform geometrical transformations on an input image or feature map \cite{jaderberg2015spatial}. 
It has been argued that such approaches "can avoid [the] blurring problem by copying coherent regions of pixels from existing frames" and "lead to more realistic and sharper results than techniques that hallucinate pixels from scratch" \cite{liu2017video}. By contrast, \citeauthor{oprea2020review} group methods that "narrow the prediction space" either by "conditioning the predictions on extra variables" (e.g., the state or actions of a robot \cite{finn2016unsupervised, babaeizadeh2017stochastic, gupta2022maskvit}) or by reframing the forecasting task in a higher-level space (e.g., %semantic or instance
segmentation maps \cite{jin2017predicting, luc2018predicting} or keypoint coordinates \cite{minderer2019unsupervised}). In parallel, the challenge posed by future multimodality has % been addressed via 
% motivated 
spurred the development of probabilistic approaches based on latent-variable models, including \glspl{VAE} and \glspl{VRNN} \cite{babaeizadeh2017stochastic, denton2018stochastic, minderer2019unsupervised}. A \gls{VAE} learns a distribution over a low-dimensional representation $z$ from which inputs can be reconstructed; 
%A \gls{VAE} compresses and recovers the input $x$ to capture a probability distribution over a low-dimensional representation $z$ that encapsulates the most significant factors of variability in $x$
\glspl{VRNN} extend this framework to recurrent architectures. In computer vision, such models can generate images 
% When applied to image generation, \glspl{VAE} seek to produce new images 
by sampling from a prior over the encoding $z$, thus introducing a probabilistic element to the otherwise deterministic latent space of traditional \glspl{AE}.

\subsubsection{Future frame estimation in thoraco-abdominal and cardiac image sequences}
\label{section:intro future frame forecasting in chest and liver image sequences}

% Video forecasting within the context of respiratory motion management for radiotherapy
Video prediction in dynamic chest and liver imaging is valuable for radiotherapy, as it helps characterize the future motion of anatomical structures of interest. However, many studies on tumor position estimation have attempted to predict only its
% center of mass, whose sole trajectory is an incomplete representation of motion. 
centroid trajectory, which does not capture spatially varying deformations.
Several studies on motion compensation in radiotherapy have also addressed organ and tumor contour estimation or interpolation between subsequent \gls{4DCT} frames to better estimate the target volume during treatment planning.
%(cf. for instance \cite{GONG2024102385}, which used dynamic mode decomposition---another eigenvalue decomposition method---for motion modeling). 
For instance, dynamic mode decomposition---an eigenvalue decomposition method that identifies a linear evolution operator capturing temporal dynamics---was applied in \cite{GONG2024102385} to tumor contour samples for motion modeling.
By contrast, our contribution focuses on video forecasting in cine \gls{MRI} %using past information only
% conditioned only on past frames, 
from past frames alone, without requiring or inferring contours. % , as \gls{MR} scans can be acquired in real time during the treatment. 
This task is particularly challenging due to the limited temporal resolution, potentially low contrast and signal-to-noise ratio, and specific degradations (e.g., susceptibility, flow, chemical shift, and magnetic field inhomogeneity artifacts) associated with this imaging modality. Future multimodality, arising from respiratory irregularities, cardiac beating, or out-of-plane motion (Section \ref{section: intro respiratory motion management in MRgRT}), as well as limited cohort sizes and sequence lengths, can further complicate prediction. Nonetheless, cine \gls{MRI}-forecasting is a critical step towards enabling real-time motion tracking in \gls{MR-guided} radiotherapy and connects to broader video prediction research. % contributes more broadly to the field of video prediction.

% Literature about video prediction using classical ML/PCA for chest imaging
\citeauthor{chhatkuli2015dynamic} applied \gls{PCA} to raw pixel intensities and predicted the associated time-dependent weights with \gls{MSSA} to forecast chest image sequences \cite{chhatkuli2015dynamic}. They evaluated this method using \gls{kV} fluoroscopic phantom images and coronal \gls{4DCT} cross-sections.
% ... to forecast \gls{kV} fluoroscopic chest phantom images and coronal cross-sections in thoracic \gls{4DCT}.
Likewise, \citeauthor{pham2019predicting} employed \gls{PCA} to model the \gls{DVF} between the reference and incoming frames in \gls{3D} chest cine-\gls{MRI} sequences obtained from the \gls{XCAT} phantom and a liver cancer patient \cite{pham2019predicting}. They forecast the corresponding coefficients using adaptive-boosted \glspl{MLP} to generate the next frames. A similar approach was used in \cite{liu2016prediction} to predict chest surfaces reconstructed from point clouds captured by a \gls{3D} photogrammetry system. In that work, \gls{kPCA} was applied to the time-varying surface height map, and a linear \gls{AR} model was used to forecast the low-dimensional motion representation in the kernel feature space. The \gls{PCA} models in \cite{chhatkuli2015dynamic, pham2019predicting, liu2016prediction} were subject-specific; an oscillatory pattern following the breathing motion characterized the first-order weights.

% Literature about video prediction using deep learning for chest and liver imaging
Regarding video prediction with deep learning, \citeauthor{nabavi2020respiratory} applied the \gls{PredNet}---a direct-pixel-synthesis \gls{ConvLSTM}-based architecture rooted in the neuroscience concept of predictive coding---to forecast chest \gls{4DCT} cross-sections \cite{nabavi2020respiratory, lotter2017deep}. By contrast, \citeauthor{romaguera2020prediction} adopted a warping (vector-based resampling) strategy. They designed a neural network inspired by VoxelMorph---an encoder--decoder model for self-supervised \gls{DIR}---to forecast \gls{2D} liver images from multiple modalities \paren{\gls{MRI}, \gls{CT}, and ultrasound} \cite{romaguera2020prediction, balakrishnan2019voxelmorph}. This architecture included multi-scale residual blocks and a \gls{ConvLSTM} module to model temporal dynamics. More recent works have focused on the generation of future volumetric \gls{MR} images from \gls{2D} cine \gls{MRI} to simultaneously address the intrinsic delays and real-time \gls{3D} imaging limitations of \gls{MR-LINAC} systems (Section \ref{section: intro respiratory motion management in MRgRT}). Notably, \citeauthor{romaguera2021predictive} proposed an encoder--decoder network (similar to that in \cite{romaguera2020prediction}) that estimates the latent encodings of future \gls{3D} deformations from the incoming \gls{2D} frames \cite{romaguera2021predictive}. They later reformulated that architecture using a conditional \gls{VAE} backbone and achieved lower vessel-\gls{TE} \cite{romaguera2021probabilistic}. Replacing the \gls{ConvLSTM} component in the latter work with a transformer that uses prior-based conditioning and learnable queries for temporal prediction yielded further accuracy improvements \cite{romaguera2023conditional}. In these studies, predictions were generally less accurate near \gls{EI}, especially around the diaphragm edge, as that phase exhibits pronounced inter-cycle variability. It has also been reported that blood vessels entering or leaving the imaged slice due to out-of-plane motion were challenging to predict. Also using a latent-variable probabilistic \gls{AE} to learn low-dimensional motion representations of intra-interventional image sequences, \citeauthor{gunnarsson2024online} modeled temporal dynamics in cardiac imaging with \pgls{LG-SSM} \cite{gunnarsson2024online}. They introduced a patient-specific online adaptation of the \gls{LG-SSM} by maximizing, via Kalman filtering, the marginal log-likelihood of recently inferred latent states. This improved ultrasound forecasting performance compared with a fixed pretrained model. They also reported competitive results for offline cine-\gls{MRI} interpolation relative to conventional registration baselines.

\subsection{Contributions and scope of this study}

Most works on breathing motion forecasting for radiotherapy have focused on low-dimensional respiratory surrogate signals, such as external motion traces acquired by optical tracking systems \cite{lin2019towards, lombardo2022offline, pohl2025real}. 
% low-dimensional respiratory signals, such as the position of external markers on the chest surface \cite{pohl2025real}, \gls{RPM} system acquisitions \cite{lin2019towards}, or the tumor center of mass in cine \gls{MRI} \cite{lombardo2022offline},
By contrast, studies on future frame prediction in thoraco-abdominal imaging have been scarce \cite{romaguera2020prediction}.
% Could be even shorter if needed, e.g. "Frame forecasting remains less common than low-dimensional signal prediction [citation]"
To our knowledge, this study is the first to apply online learning algorithms for \glspl{RNN} to chest and liver cine-\gls{MR} image prediction and to compare them with transformers, which have underpinned many recent developments in time-series forecasting. Regarding \gls{RNN} algorithms, we selected \gls{RTRL} as a standard, computationally exact baseline, along with more efficient, recent approaches---\gls{UORO}, \gls{SnAp-1}, and \gls{DNI}---due to their more favorable time and memory complexity (Table \ref{table:RNN online learning comparison}). We chose a simple, standard \gls{RNN} structure given the small size of our datasets and the scope of this study---a proof of concept focused on comparing broad architectural and training paradigms. Such vanilla \glspl{RNN} can suffice in low-data regimes. In contrast, architectures based on \gls{LSTM} or \gls{GRU} cells, whose greater gating capacity can help capture longer-range dependencies, might benefit from more training data. Although transformers typically require large amounts of data, we selected a lightweight encoder-only model as a modern, representative attention-based forecasting method. Notably, \citeauthor{romaguera2023conditional} proposed an end-to-end transformer-based algorithm to forecast frames in respiratory cine-\gls{MRI} \cite{romaguera2023conditional}; we explore transformers from a different experimental standpoint. Indeed, our modular approach remains suitable for smaller datasets. We also discuss paradigms such as population-based versus subject-specific training and online versus offline learning. In addition, we highlight domain-adaptation challenges when scanning protocols and image characteristics differ and assess robustness to varying levels of noise and contrast. Such insights are particularly needed as "the application of attention mechanisms for deformation prediction is still underexplored in the literature" \cite{romaguera2023conditional}.

Prior \gls{PCA}-based approaches to respiratory motion prediction relied on simple surrogates, forecasting algorithms, or pixel-space representations. For instance, the baseline forecasting pipeline in \cite{romaguera2020prediction} combined a \gls{PCA} motion model driven by a one-dimensional surrogate signal with temporal prediction via an adaptive linear filter. By contrast, we derive a sequence-specific internal motion model by applying \gls{PCA} to dense deformation fields and predict temporal dynamics with modern algorithms. Specifically, we use the Lucas--Kanade optical-flow algorithm to estimate deformations between the incoming and reference frames. This \gls{DIR} technique has been commonly employed in chest imaging \cite{xu2008lung, akino2014evaluation, dhont2019multi}, and here we analyze the optimization of its parameters using chest cine-\gls{MRI} data. % for the first time, to our knowledge. 
\citeauthor{li2011pca} selected the \gls{PCA}-subspace dimension by minimizing the validation error between the ground-truth and reference \glspl{DVF} in \gls{4DCT} imaging \cite{li2011pca}%, but their 8-phase \gls{4DCT} data could not represent all possible motion states \cite{li2011pca}
; we extend this general validation-based model selection approach to cine-\gls{MRI} forecasting. We forecast the time-dependent \gls{PCA} weights to obtain the future \glspl{DVF}, which are then used to warp the reference frame and estimate subsequent frames.
%; the next frames are estimated as the result of warping the reference frame with those predicted \glspl{DVF}. 
As such, our approach belongs to the vector-based resampling category in the classification of video forecasting models proposed by \citeauthor{oprea2020review} \cite{oprea2020review}. 
We comprehensively evaluate performance across multiple accuracy metrics and a wide range of horizons (up to 2.2s).
% We evaluate the performance of forecasting algorithms for various accuracy metrics and horizons up to 2.2s; the horizon range considered is the most extensive within the literature in frame forecasting in chest and liver imaging. 
Finally, we highlight the interpretability of our approach by analyzing deformation modes and linking \gls{PCA}-weight prediction errors to mismatches between predicted and ground-truth frames.

\section{Materials and methods}
% Possible improvement: I can compute the extent of the motion and its velocity (as computed by the Lucas Kanade method)

\subsection{Chest and liver cine-MRI data}

This study uses two publicly available datasets. The first one, referred to here as the ETH Zürich dataset, comprises two volumetric chest \gls{MR} image sequences \cite{ETHdataset}. The original \acs{4D}-\acs{MRI} data were acquired using a technique based on \enquote{stacking of dynamic \gls{2D} images using internal image-based sorting} \cite{von20074d, boye2013population}. We resampled the original volumes to an isotropic resolution of 1mm$^3$ via bicubic spline interpolation. We then selected two sagittal planes for each subject, yielding a total of four \gls{2D} image sequences. These slices have an isotropic in-plane resolution of 1mm$^2$ and are encoded in 8 bits. Lastly, we shifted the frames in each sequence so that the first frame corresponds to mid-expiration. The resulting sagittal \gls{MRI} sequences each comprise 200 frames, as in the original \gls{4D} data. Sequences 1 and 4 correspond to the left hemithorax and feature cardiac motion, whereas sequences 2 and 3 correspond to the right hemithorax. % Fig. \ref{fig:next frame pred all sq SnAp-1 h=6} displays the mean image associated with each \gls{2D}$+t$ sequence. % Cite only at the end otherwise that is completely off

\begin{figure}[pos=htbp,align=\centering] %[pos=htbp,width=10cm,align=\centering]
    \captionsetup[subfigure]{labelformat=empty}
    \centering
    \subfloat[Sequence 1]{\includegraphics[width=.23\columnwidth]{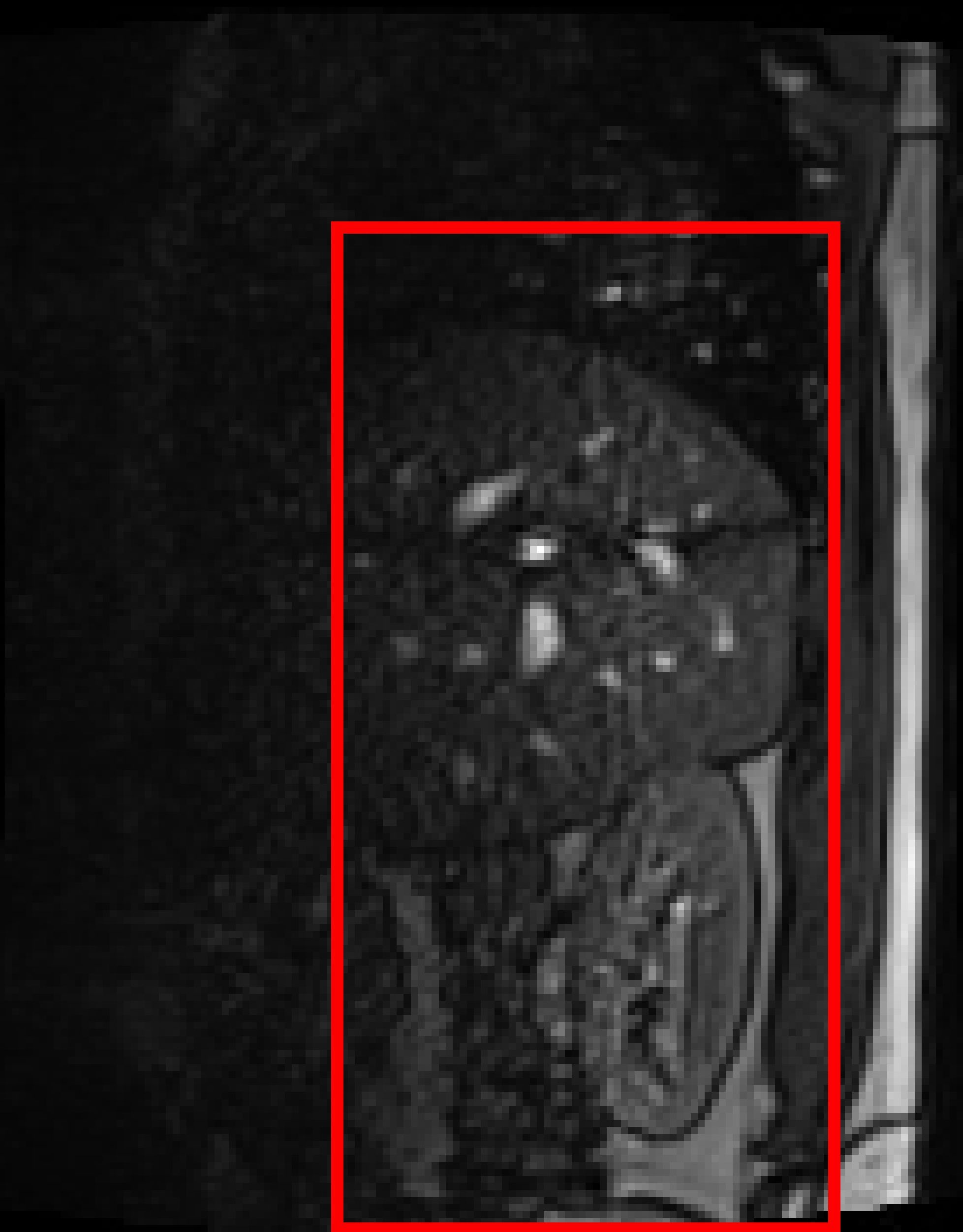}}% 
    \,
    \subfloat[Sequence 2]{\includegraphics[width=.23\columnwidth]{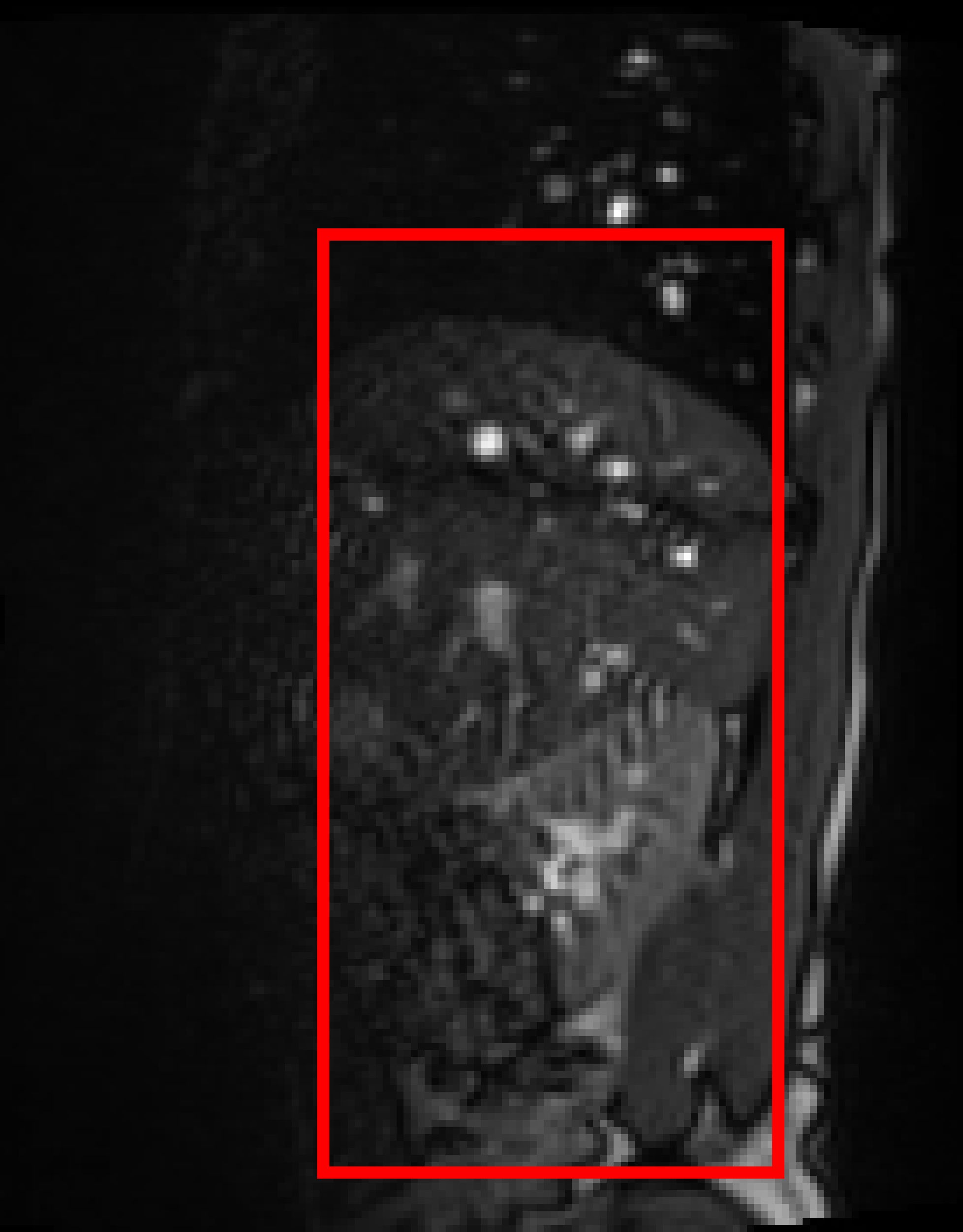}}% 
    \,
    \subfloat[Sequence 3]{\includegraphics[width=.23\columnwidth]{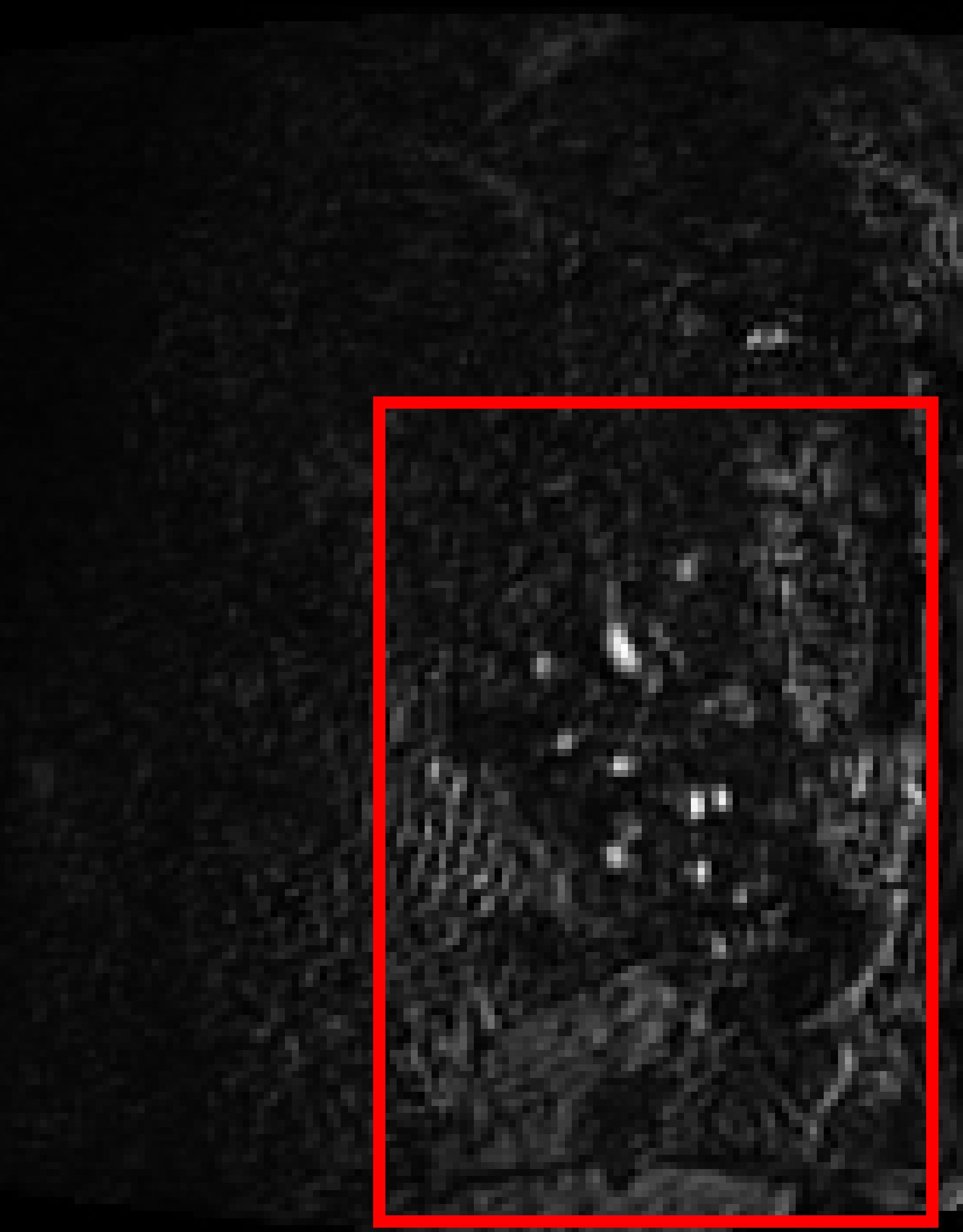}}% 
    \,
    \subfloat[Sequence 4]{\includegraphics[width=.23\columnwidth]{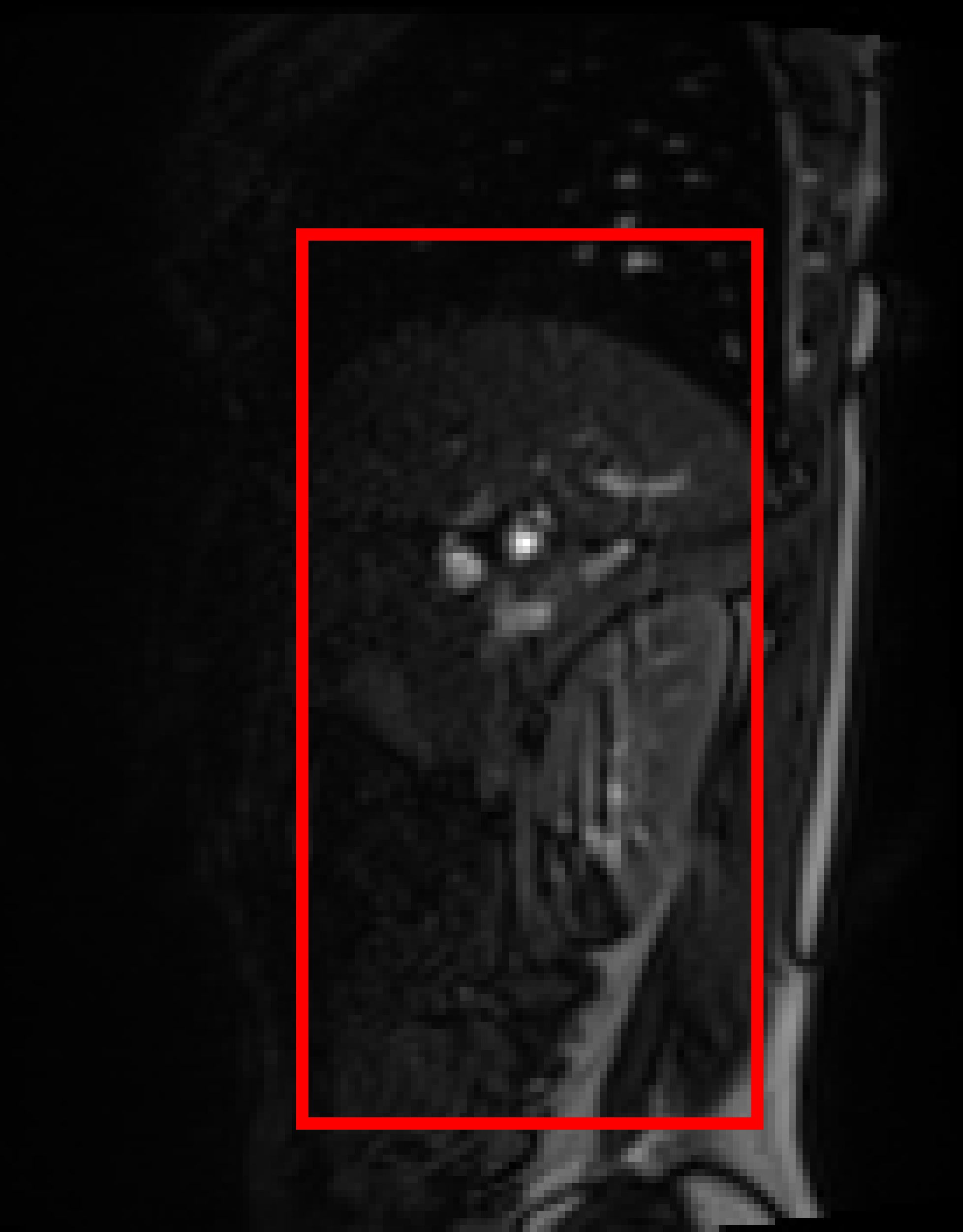}}%  
    \\
    \subfloat[Sequence 5]{\includegraphics[width=.23\columnwidth]{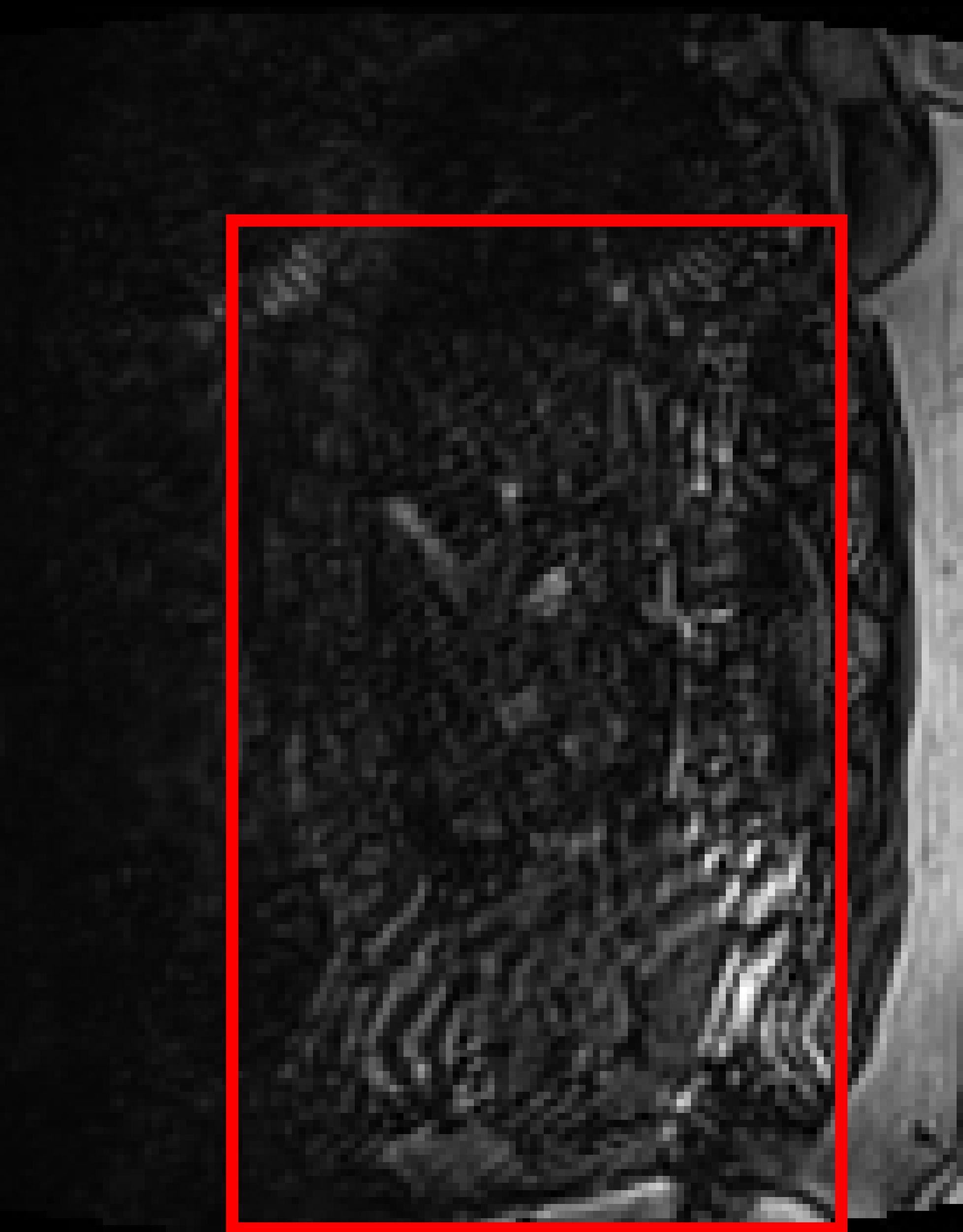}}% 
    \,
    \subfloat[Sequence 6]{\includegraphics[width=.23\columnwidth]{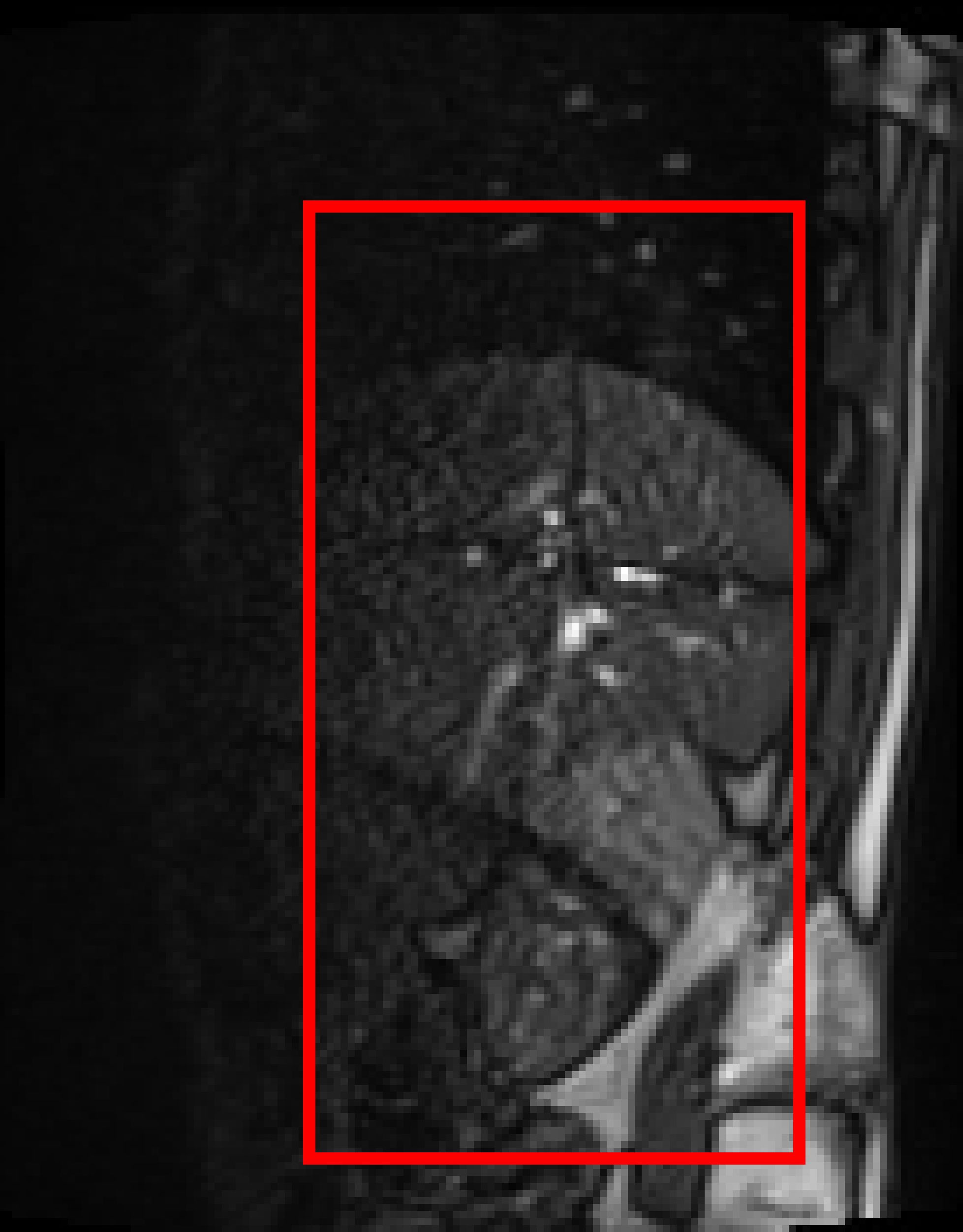}}% 
    \,
    \subfloat[Sequence 7]{\includegraphics[width=.23\columnwidth]{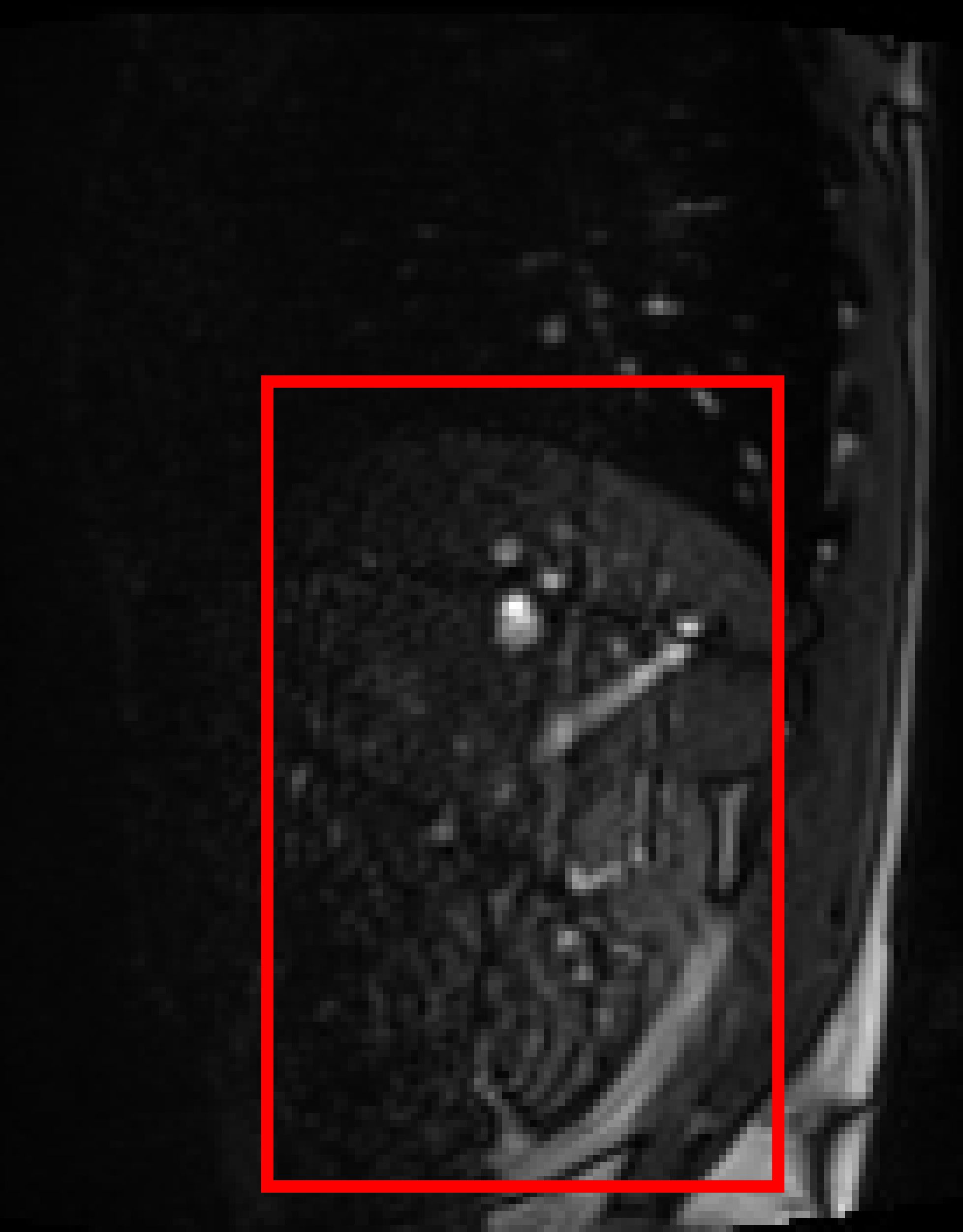}}% 
    \,
    \subfloat[Sequence 8]{\includegraphics[width=.23\columnwidth]{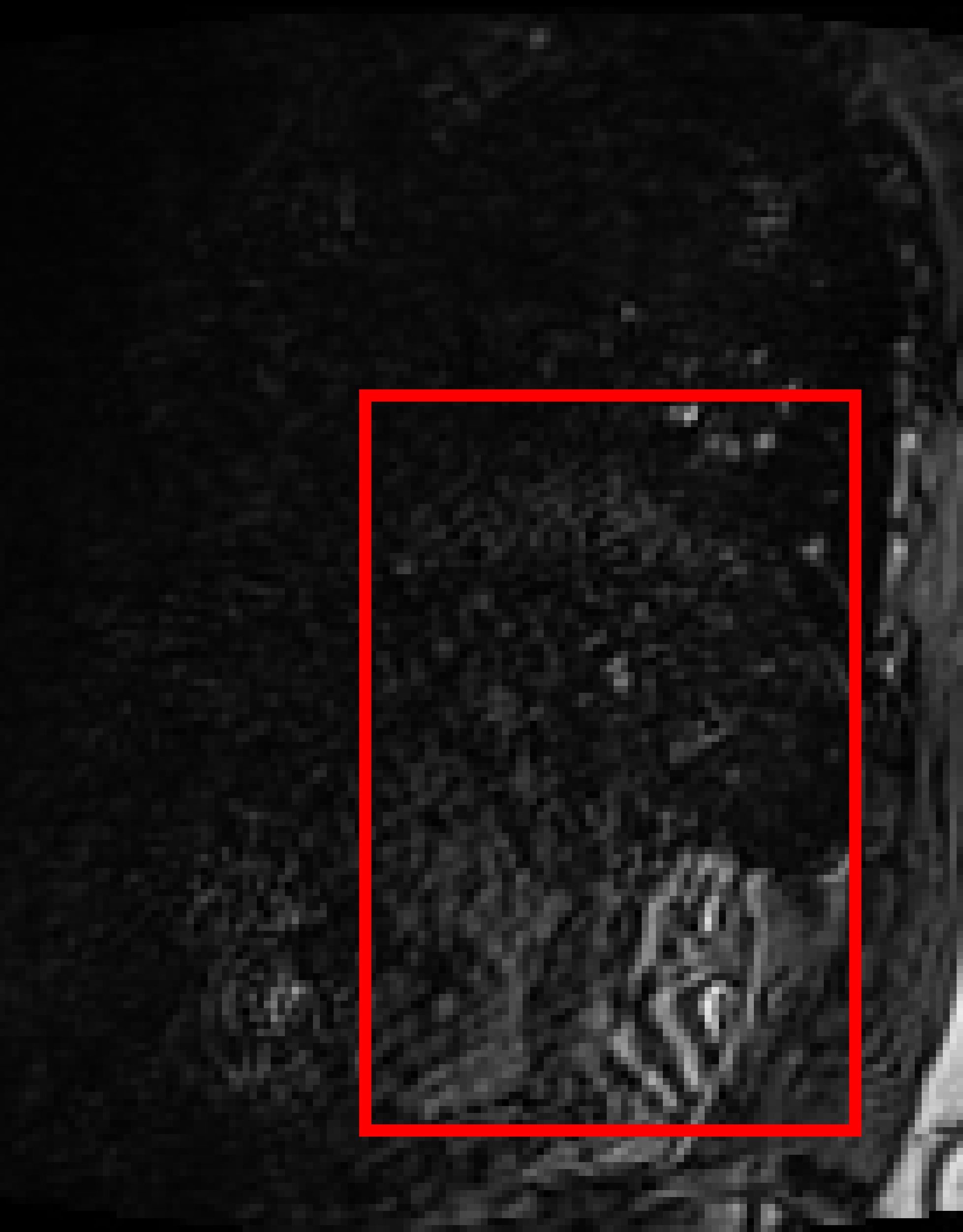}}%\\           
	\caption{Initial frame from each sequence in the \acs{OvGU} dataset ($t = t_1$), with the high-contrast \acsp{ROI} for evaluation displayed as red rectangles.}  
	\label{fig:Original Magdeburg MR images}
\end{figure}

% Description of the Magdeburg dataset
We also use eight sagittal cine-\acs{MRI} sequences, each from a single individual, from another public dataset released by \gls{OvGU}, which cover mainly the liver and surrounding organs, including the kidneys \cite{gulamhussene__2d_2021, gulamhussene2022predicting}.
Images were acquired on a MAGNETOM Skyra \gls{MRI} scanner (Siemens Medical Solutions, Erlangen, Germany) using a TRUFI sequence and just enough contrast to detect respiratory motion. No body-array coil was utilized; acquisitions relied on the fixed receiver coil of the bore. Two reference \gls{2D} \gls{MRI} sequences were available for each subject in that dataset. These are also referred to as navigator sequences in the context of \acs{4D}-\gls{MRI} reconstruction, as each corresponds to a static plane representative of breathing motion. We selected one per patient based on visual assessment of image quality. Specifically, we chose the acquisition whose frames were the least noisy and exhibited the least vessel flickering due to blood circulation. The \gls{OvGU} navigator slices are characterized by lower contrast, higher noise, and greater breathing variability than the ETH Zürich frames. In particular, sequences 3, 5, and 8 feature high noise, contrast variations over time are more pronounced in sequence 6, and sequence 4 exhibits particularly irregular breathing combining deep and shallow inspirations. We did not perform resampling and retained the original in-plane resolution of 1.82 $\times$ 1.82 mm (through-plane resolution: 4.0mm). However, we discarded the first 15 images from each original acquisition, as those were subject to higher contrast variability; the resulting sequences comprise 498 frames. Finally, for each subject, we selected \pgls{ROI} with higher contrast for algorithm evaluation (Fig. \ref{fig:Original Magdeburg MR images}). We avoided including the spine in the \glspl{ROI}, as it is largely immobile during respiration and may introduce spurious registration errors due to out-of-plane motion or imaging artifacts.

Table \ref{table:MRI general characteristics} in Appendix \ref{appendix: dataset characteristics} summarizes the characteristics of both datasets. Notably, the ETH Zürich sequences have a lower sampling rate (approximately 3.18Hz) than the \gls{OvGU} sequences (6.0Hz). In addition, the former sequences have a shorter duration (63s) than the latter (83s).

\subsection{Breathing motion modeling with PCA}
\label{Breathing motion modeling with PCA}

For each \gls{MR} image sequence, \gls{DIR} between the initial frame (at time $t_1$) and the frame at time $t$ is performed using the pyramidal, iterative Lucas--Kanade optical-flow algorithm \cite{lucas1981iterative, bouguet2001pyramidal, fleet2006optical}, following the implementation in \cite{pohl2022prediction} (with a straightforward adaptation from \gls{3D} to \gls{2D}). In the following, $\vec{u}(\vec{x}, t)$ denotes the push-forward \gls{2D} deformation vector at pixel $\vec{x}$ and time $t$, satisfying the local brightness constancy assumption, by definition:%
\begin{equation} \label{eq:DVF_def}
 I(\vec{x}, t_1) \approx I(\vec{x} + \vec{u}(\vec{x}, t_k), t_k)
\end{equation}%
We optimize the optical-flow parameters for each sequence via grid search, minimizing the registration error averaged over its first 28.3s (Appendix \ref{appendix: chest MR image registration optimization}). 
% not sure about fleet2006optical citation, I could remove it if necessary (e.g., if too many references). No I should keep it because it is clearly written and it has the kernel implementation with the Gaussian kernel weighting the moment matrix

The vector obtained by concatenating and centering the deformations $\vec{u}(\vec{x}, t)$ at time $t$, $X_{\text{c}}(t)$, lies in a $2|I|$-dimensional space, where $|I|$ designates the number of pixels (Eq. \ref{eq:instantaneous centered DVF} in Appendix \ref{appendix: PCA respiratory motion model}). Yet, these complex spatiotemporal variations are driven by relatively simple underlying phenomena that can be described with few degrees of freedom. We use \gls{PCA} to project the centered motion vector $X_{\text{c}}(t)$ onto a linear subspace of dimension $n_{\text{cp}}$, with $n_{\text{cp}} \ll 2|I|$. For a sequence of $M_{\text{max}}$ frames, we compute, for each $j \in \{1, \dots, n_{\text{cp}}\}$, the $j^{\text{th}}$ principal component $\big(\vec{u_j}(\vec{x})\big)_{\vec{x} \in I}$ and its associated coefficients $\big( w_j(t_k) \big)_{k \in \{ 1, \dots, M_{\text{max}}\}}$, approximately satisfying the following relationship for all pixels $\vec{x}$ at any time $t$:%
\begin{equation} \label{eq:PCA respiratory model}
\vec{u}(\vec{x}, t) \approx \vec{\mu}(\vec{x}) + \sum_{j=1}^{n_{\text{cp}}} w_j(t) \vec{u_j}(\vec{x})
\end{equation}%
This formula expresses the high-dimensional, time-dependent \gls{DVF}, $\vec{u}(\vec{x}, t)$, as a linear combination of a few static vector fields, $\vec{u_j}(\vec{x})$, weighted by the time-dependent \gls{PCA} coefficients, $w_j(t)$. The latter are also referred to as the \gls{PCA} weights or \gls{PCA} scores in this work. $\vec{\mu}(\vec{x})$ represents the temporal mean of $\vec{u}(\vec{x}, t_k)$ over $k \in \{1, \dots, M_{\text{train}}\}$. The principal components $\vec{u_j}(\vec{x})$, also called here the principal \glspl{DVF}, are likewise computed using the first $M_{\text{train}}$ frames of each sequence. The selection of $M_{\text{train}}$ is detailed in Section \ref{section: PCA weight cross-validation}. To estimate the weights at time $t_k$ for $k \geq M_{\text{train}}$, we project $X_{\text{c}}(t_k)$ onto the principal components:%
\begin{equation} \label{eq:PCA weight calculation}
w_j(t) = \sum_{\vec{x} \in I} \big\langle \vec{u}(\vec{x}, t) - \vec{\mu}(\vec{x}) , \vec{u_j}(\vec{x}) \big\rangle    
\end{equation}%
In this equation, $\langle\cdot,\cdot\rangle$ designates the Euclidean inner product, and the sum is over all pixels $\vec{x}$ in the image at time $t$. The principal components stay constant throughout each sequence, as the motion model is not updated as time elapses. Indeed, breathing dynamics are assumed to remain relatively stable across all sequences, as the latter are relatively short (approximately one minute) and show no significant out-of-plane motion or changes in motion patterns. Eq. \ref{eq:PCA weight calculation} follows directly from Eq. \ref{eq:PCA respiratory model} using the orthonormality of the $2|I|$-dimensional principal components:%
\begin{equation} \label{eq:principal components orthonormality}
\sum_{\vec{x} \in I} \big\langle \vec{u_i}(\vec{x}) | \vec{u_j}(\vec{x}) \big\rangle =
\begin{cases}
      1 & \text{if}\ i=j \\
      0 & \text{otherwise}
\end{cases}
\end{equation}%
Appendix \ref{appendix: PCA respiratory motion model} provides the mathematical derivation of Eqs. \ref{eq:PCA respiratory model} and \ref{eq:PCA weight calculation}.

\subsection{Prediction of the time-dependent PCA weights}
\label{section:methods PCA time-dependant weights forecasting}

To predict future frames in a given \gls{MRI} sequence, we first forecast the \gls{PCA} weights $w_j(t)$. To this end, we compare several adaptive methods, namely standard \glspl{RNN} trained online and \gls{LMS}, with offline predictors, namely transformers trained with backpropagation and an \gls{OLS} linear \gls{AR} model. For brevity, we refer to the latter as linear regression in this article.

\subsubsection{Input and target definition}
\label{section: methods - input and target definition}

The input to all forecasting algorithms consists of the concatenation of the \gls{PCA} coefficients $w_j(t_n),\allowbreak \dots,\allowbreak w_j(t_{n+L-1})$ for each component index $j \in \{1, \dots, n_{\text{cp}}\}$, where $L$ designates the \gls{SHL}---the input-window width---expressed in number of time steps. The scores $w_1(t), \dots, w_{n_{\text{cp}}}(t)$ are predicted simultaneously to exploit correlation information. A unit bias term is added to the input vector, providing more flexibility to the model. The input and output vectors, denoted by $x_n$ and $y_{n+1}$, respectively, are defined as follows, where $h$ refers to the horizon expressed in number of time steps:%
\begin{equation} \label{eq:RNN_in_out_def_wth_horizon chap 4}
x_n
=
\begin{pmatrix}
1 \\
w_1(t_n)\\
w_2(t_n)\\
\dots \\
w_{n_{\text{cp}}}(t_n)\\
w_1(t_{n+1})\\
\dots\\
w_{n_{\text{cp}}}(t_{n+L-1})\\
\end{pmatrix}
\, ,
\enspace
y_{n+1}
=
\begin{pmatrix}
w_1(t_{n+L+h-1})\\
\dots\\
w_{n_{\text{cp}}}(t_{n+L+h-1})\\
\end{pmatrix}
\end{equation}

\subsubsection{Model architectures and learning algorithms}

\subsubsubsection{Standard RNNs trained dynamically}

We first provide a brief description of the \gls{RNN} models used in this study. To prevent overfitting given the short length of each sequence, we adopt a minimal architecture: a standard \gls{RNN} with a single hidden layer of limited size, $d \leq 110$. Its hidden state vector $s_n \in \mathbb{R}^d$ encodes past information and serves as its internal memory. At each time step $n$, the state is updated by applying a non-linear activation function, $\Phi$, to a linear combination of the previous state, $s_n$, and current input, $x_n$. The output, $y_{n+1}$, is obtained via a linear transformation of the updated hidden state:%
\begin{equation} \label{eq:RNN_equations}
s_{n+1} = \Phi(A_n x_n + B_n s_n) \, , 
\enspace
y_{n+1} = C_n s_{n+1}
\end{equation}
In these equations, $A_n$, $B_n$, and $C_n$ denote the input-to-hidden, hidden-to-hidden, and hidden-to-output learnable weight matrices, respectively. Their shapes are $d \times (n_{\text{cp}} L +1)$, $d \times d$, and $n_{\text{cp}} \times d$. These matrices depend on $n$, as online learning algorithms update them continually. $\Phi$ is set as the coordinate-wise hyperbolic tangent function. The training algorithms for \glspl{RNN} compared in this study are \gls{RTRL}, \gls{UORO}, \gls{SnAp-1}, and \gls{DNI}; they minimize the instantaneous squared error between the latest prediction and incoming sample at each time step. We use the efficient vanilla-\gls{RNN} implementations of \gls{UORO}, \gls{SnAp-1}, and \gls{DNI} described in \cite{pohl2022prediction, pohl2025real} (see also Section \ref{section: online RNN algorithms applied to radiotherapy}). Gradient-norm clipping is applied to \gls{LMS} and \glspl{RNN} to enhance numerical stability \cite{pascanu2013difficulty}. We also consider \pgls{RNN} baseline whose hidden parameter matrices, $A$ and $B$, are randomly initialized and then frozen during inference (they do not depend on $n$), but whose output-layer weights, $C_n$, are updated online. \Glspl{RNN} are fully implemented in MATLAB \cite{pohl2024MRforecastingcode}.

\begin{table*}[htb!] %[pos=htbp,width=0.9\textwidth,align=\centering]
%\normalsize %\small
\setlength{\tabcolsep}{1.2pt}
\begin{center}
\begin{tabular}{llll}
\hline
Prediction              &  Mathematical model                                   & Training/validation split  & Range of hyperparameters \\
method \vspace{0.15cm}  &                                                       &                        & in the tuning grid     \\
\hline
\acs{RTRL}, \acs{UORO}, & $s_{n+1} = \Phi(A_n x_n + B_n s_n)$                   & 28.3s/28.3s            & $\eta \in \{0.005, 0.01, 0.015, 0.02\}$ \rule{0pt}{2.6ex}\\
\acs{SnAp-1}, \acs{DNI} & $y_{n+1} = C_n s_{n+1}$                               &                        & $L \in \{ 1.9\text{s}, 3.8\text{s}, 5.7\text{s}, 7.6\text{s}, 9.5\text{s} \}$ \\
\vspace{0.15cm}         &                                                       &                        & $d \in \{10, 30, 50, 70, 90, 110\}$  \\ 
\acs{LMS}               & $y_{n+1} = A_n x_n$                                   & 28.3s/28.3s            & $\eta \in \{ 0.02, 0.05, 0.1, 0.2 \}$ \\
\vspace{0.15cm}         &                                                       &                        & $L \in \{ 1.9\text{s}, 3.8\text{s}, 5.7\text{s}, 7.6\text{s}, 9.5\text{s} \}$\\
Linear                  & $y_{n+1} = A x_n$                                     & 50.4s/6.2s             & $L \in \{ 1.9\text{s}, 3.8\text{s}, 5.7\text{s}, 7.6\text{s}, 9.5\text{s} \}$\\
regression \vspace{0.15cm} &                                                    &                        & \\
Transformer & $y_{n+1} = \mathrm{FFN}_{\mathrm{out}}\Big(\!\operatorname{vec}\!\big(E^{(n_{\text{layer}})} \circ\cdots\circ E^{(1)}(A_{\mathrm{in}} x_n +\mathrm{PE})\big)\!\Big)$ & Sequence-specific: 50.4s/6.2s& $\eta \in \{0.0001, 0.0005\}$\\ 
encoder     & with $E^{(l)}=\mathrm{FFN}^{(l)} \circ \mathrm{MHA}^{(l)}$        & Population: first 80\%/last 20\%           & $L \in \{1.9\text{s}, 3.8\text{s}, 5.7\text{s}, 7.6\text{s}, 9.5\text{s}\}$\\ 
                        &                                                       & of each sequence (samples from             & $n_{\text{layer}} \in \{1, 2\}$  \\ 
\vspace{0.15cm}         &                                                       & all sequences are concatenated)                 & $d_{\text{emb}} \in \{8, 16\}$  \\ 
\acs{RNN} with a        & $s_{n+1} = \Phi(A x_n + B s_n)$                       & 28.3s/28.3s            & $\eta \in \{ 0.02, 0.05, 0.1, 0.2 \}$ \\
frozen hidden           & $y_{n+1} = C_n s_{n+1}$                               &                        & $L \in \{ 1.9\text{s}, 3.8\text{s}, 5.7\text{s}, 7.6\text{s}, 9.5\text{s} \}$ \\
layer                   &                                                       &                        & $d \in \{10, 30, 50, 70, 90, 110\}$  \\ 
\hline
\end{tabular}
\end{center}
\caption{Overview of the forecasting models and validation schemes considered. The second column specifies the relationship between the input $x_n$, containing the past \acs{PCA} weights, and the output $y_{n+1}$, corresponding to the predicted weights (both defined in Eq. \ref{eq:RNN_in_out_def_wth_horizon chap 4}).\protect\footnotemark~The fourth column lists the hyperparameter values evaluated during grid search. $\eta$, $L$, $d$, $n_{\text{layer}}$, and $d_{\text{emb}}$ designate the learning rate of neural networks and \gls{LMS}, the \gls{SHL},\protect\footnotemark~the hidden-state dimension of the standard \gls{RNN}, and the number of layers and embedding dimension of the encoder-only transformer, respectively.} 
\label{table:models comparison}
\end{table*}
% https://tex.stackexchange.com/questions/536265/using-two-footnote-in-a-figures-caption
\addtocounter{footnote}{-1}
\footnotetext[\thefootnote]{For linear regression and \gls{LMS}, the coefficient matrices $A$ and $A_n$, respectively, both have size $n_{\text{cp}} \times (n_{\text{cp}} L +1)$. $A_n$ depends on the time-step index $n$, since it is updated whenever a new sample arrives.}
\addtocounter{footnote}{1}
\footnotetext[\thefootnote]{Due to the difference in sampling rates, $L$, expressed in number of time steps, is selected from $\{ 6, 12, 18, 24, 30 \}$ or $\{ 11, 23, 34, 45, 57\}$ when testing is conducted on the ETH Zürich dataset or the \gls{OvGU} dataset, respectively. For instance, for a population transformer whose training examples are drawn from the former dataset and resampled to 6Hz to match the \gls{OvGU} test sequences, the range of values for $L$ in the search grid is $\{ 11, 23, 34, 45, 57\}$. For sequence-specific models, the validation and test sets are extracted from the same sequence, so resampling is not needed.}

\subsubsubsection{Transformers}

In this work, \glspl{RNN} are compared with transformers, whose self-attention mechanism supports long-range dependency modeling. Specifically, we adopt an encoder-only architecture as a natural choice for deterministic time-series forecasting. Indeed, we do not seek to generate sequences autoregressively using a decoder layer, but instead focus on fixed-horizon regression.
%, as a deterministic regression task. Indeed, unlike \gls{NLP} applications, where decoder layers generate tokens autoregressively from a probability distribution, we do not seek to generate a new sequence, but rather focus on simplicity and fair and direct comparison with \glspl{RNN}, and avoid complexity that would practically hamper model performance in a low-data setting. 
In our study, transformers are trained with backpropagation.

By abuse of notation, the transformer input, $x_n$, is reshaped as a matrix of size $(n_{\text{cp}}+1) \times L$, where a bias term is added to each column. It is first linearly mapped to an embedding space of dimension $d_{\text{emb}}$ by multiplication with the matrix $A_{\text{in}}$. Sinusoidal positional encodings, denoted by "$\mathrm{PE}$", are then added to preserve temporal order. Subsequently, the sequence is processed by a succession of $n_{\text{layer}}$ transformer-encoder layers. The resulting representation is flattened across both temporal and embedding dimensions ("$\operatorname{vec}$" operator below) and passed through the output-head \gls{FFN}, denoted as $\mathrm{FFN}_{\mathrm{out}}$. %\paren{Linear--\gls{ReLU}--Linear}. 
% The latter consists of one hidden layer, sized as the geometric mean of its input and output dimensions, followed by a \gls{ReLU} activation:%
The latter stacks two linear layers separated by a \gls{ReLU} activation:%
\begin{equation}
y_{n+1} = \mathrm{FFN}_{\mathrm{out}}\Big(\!\operatorname{vec}\!\big(E^{(n_{\text{layer}})} \circ\cdots\circ E^{(1)}(A_{\mathrm{in}} x_n +\mathrm{PE})\big)\Big)
\end{equation}%
Each encoder layer, $E^{(l)}$, consists of multi-head self-attention (with two heads), $\mathrm{MHA}^{(l)}$, followed by a position-wise \gls{FFN}, denoted as $\mathrm{FFN}^{(l)}$:%
\begin{equation}
E^{(l)}=\mathrm{FFN}^{(l)} \circ \mathrm{MHA}^{(l)}
\end{equation}

Unlike online-trained \gls{RNN}, which only encounter each training example once, as they perform one-shot sequence forecasting and learn parameters during that process, transformers are shown the same examples multiple times during training, as we minimize the \gls{MSE} loss on the training set across several epochs. The optimization algorithm selected for transformers is \gls{ADAM}. Shuffling is applied at each epoch to mitigate data-ordering bias and improve convergence stability. As for \glspl{RNN}, transformer capacity is kept relatively low ($d_{\text{emb}} \leq 16$, $n_{\text{layer}} \leq 2$, and position-wise \gls{FFN} dimension below 32) to speed up computations and avoid overfitting, given the limited amount of data available, while preserving expressivity. In addition, dropout is applied to the encoder layers (with probability $p = 0.5$) for further regularization. Nonetheless, weight decay is not applied. Transformers are implemented using the PyTorch library in Python \cite{pohl2024MRforecastingcode}. The \gls{RNN} and transformer configurations are summarized in Table \ref{table:models comparison} and in Table \ref{table:RNNs_configuration} in Appendix \ref{appendix: RNN experimental setup}.

\subsubsection{Training and validation schemes}
\label{section: PCA weight cross-validation}

\subsubsubsection{Overview}

We compare two types of models: sequence-specific and population models. For sequence-specific models, the train/validation/test split is defined within the same sequence, with offline predictors (linear regression and transformers) using a relatively longer training window. By contrast, for the population transformer, training and hyperparameter tuning are conducted on one dataset, while testing is performed using the end of each sequence from the other dataset, ensuring consistent performance evaluation across both model categories. We recall that registration and \gls{PCA}-based motion modeling are sequence-specific, regardless of the forecasting method employed (Section \ref{Breathing motion modeling with PCA}). When discussing results on the \gls{OvGU} data, we use the terms "sequence-specific" and "subject-specific" interchangeably, as each subject corresponds to exactly one image sequence (this does not hold for the ETH Zürich data). During grid-search--based validation, we optimize $L$ for all algorithms, the learning rate of \gls{LMS} and neural networks (\glspl{RNN} and transformers), denoted by $\eta$, the \gls{RNN} hidden layer size, and the number of layers and the embedding dimension of the transformer (Table \ref{table:models comparison}).

In our experiments, we examine the influence of the horizon $h \in \{1, \dots, h_{\text{max}}\}$, with $h_{\text{max}}$ corresponding to 2.2s, on prediction accuracy. In other words, we set $h_{\text{max}} = 7$ for the ETH Zürich dataset and $h_{\text{max}} = 13$ for the \gls{OvGU} dataset due to differences in sampling rates. Each model is specific to a given horizon $h$, that is, we perform training and validation separately for each value of $h$ considered. This choice---avoiding recursive forecasting---prevents error accumulation over time. %experimentally.
% Not sure that the two sentences of the first paragraph are at the best place here

\subsubsubsection{Sequence-specific models}

For sequence-specific models, each time series is split into training and validation sets that together span its first 56.6s, and a test set spanning the final 6.3s and 26.3s for the ETH Zürich and \gls{OvGU} datasets, respectively. The training set comprises the data from 0s to $t_{M_{\text{train}}} = 28.3 \text{s}$, except for offline methods, whose training period ends at $t_{M_{\text{train}}} = 50.4 \text{s}$, as using more training data may improve their performance. % As a result, the principal components and \gls{PCA} scores slightly differ between offline and adaptive algorithms, the latter comprising \gls{LMS} and \glspl{RNN} trained online (Section \ref{section: PCA breathing motion modeling results}). %This is illustrated in Fig. \ref{fig:1st cpt prediction (SnAp-1 vs vs UORO vs transformer) Magdeburg}. % figure order!
$M_{\text{train}}$, denoting the last time index of the training set, also depends on the sampling rate and therefore on the dataset (Table \ref{table:general experimental setup} in Appendix \ref{appendix: RNN experimental setup}). Before forming the inputs $x_n$ via Eq. \ref{eq:RNN_in_out_def_wth_horizon chap 4}, each \gls{PCA}-score vector $w(t) = [w_1(t), \dots, w_{n_{\text{cp}}}(t)]$ is standardized using the mean and standard deviation over the training set, which accelerates the convergence of \gls{SGD}. Predictions on the validation and test sets are then mapped back to the original scale.

The validation segment for sequence-specific models corresponds to the time indices $k \in \{M_{\text{train}}+1, \dots, M_{\text{val}}\}$, with $t_{M_{\text{val}}}=56.6\text{s}$ (Table \ref{table:general experimental setup}). We select the hyperparameters in the grid that minimize the \gls{nRMSE} on the validation set, defined by the following formula (with $k_{\text{min}} = M_{\text{train}}+1$ and $k_{\text{max}} = M_{\text{val}}$):%
\begin{equation} \label{eq:predicted weights nRMSE}
\text{nRMSE} = \sqrt{\frac{\sum_{k = k_{\text{min}}}^{k_{\text{max}}} \sum_{j =1}^{n_{\text{cp}}}
 						\big( \widehat{w_j}(t_k) - w_j^{\text{true}}(t_k) \big)^2}
 				  {\sum_{k = k_{\text{min}}}^{k_{\text{max}}} \sum_{j =1}^{n_{\text{cp}}}
 				 		\big( \overline{w_j^{\text{true}}} - w_j^{\text{true}}(t_k) \big)^2}}
\end{equation}
In this equation, for each component order $j$, $w_j^{\text{true}}(t_k)$, $\widehat{w_j}(t_k)$, and $\overline{w_j^{\text{true}}}$ designate the reference (i.e., ground-truth) \gls{PCA} score at time $t_k$, computed by projecting the deformations between the initial and incoming images onto the \gls{PCA} linear subspace (Eq. \ref{eq:PCA weight calculation}), the predicted score at time $t_k$, and the mean of $w_j^{\text{true}}(t_k)$ over $k \in \{k_{\text{min}}, \dots, k_{\text{max}}\}$, respectively. For sequence-specific models, the hyperparameters selected for testing depend both on $h$ and the sequence considered.

For neural networks, $\widehat{w_j}(t)$, and by extension the validation \gls{nRMSE}, depend on the run index due to stochasticity (e.g., random parameter initialization). Consequently, for each combination of hyperparameters in the tuning grid (Table \ref{table:models comparison}), we average this error over $n_{\text{val}}$ successive runs (each with a different random seed). This averaged error is minimized during validation to determine neural-network hyperparameters. We set $n_{\text{val}} = 250$ for \gls{RNN} algorithms, except for \gls{RTRL}, for which we set $n_{\text{val}} = 10$, as the latter was more computationally intensive but yielded error measurements with lower uncertainty. This is likely because \gls{RTRL} computes the exact loss gradient, while the other \gls{RNN} algorithms considered only approximate it. \Gls{UORO} and \gls{DNI} were associated with larger \glspl{CI} for sequence-level average accuracy metrics,\footnotemark~possibly due to gradient-update stochasticity in \gls{UORO} and random initialization of the coefficients involved in \gls{DNI}'s linear error-signal estimation. We also set $n_{\text{val}} = 10$ for the sequence-specific transformer, as it requires more computational resources for training.

\footnotetext{This is not reflected in Tables \ref{table:signal pred nRMSE 3 PCA cpts avg over horizon} and \ref{table:frame pred perf}, where sequence-level performance, averaged over $n_{\text{test}}^{\text{PCA}}$ runs and $n_{\text{test}}$ runs, respectively, is treated as one sample.}

\subsubsubsection{The population transformer}

% General experimental setup
Besides considering sequence-specific models, we also train a population transformer on multiple sequences outside the test sequences. Specifically, when evaluating performance on the ETH Zürich dataset, this model is trained on the \gls{OvGU} data, and vice versa, ensuring subject-wise and scanner-wise splitting. The training set is formed using batches from the first 80\% of each training sequence, while the last 20\% is used for validation. We do not use the whole separate dataset for testing, but only the end of each sequence (starting from 56.6s), for consistency with sequence-specific model evaluation. Since sampling frequencies differ between datasets, we resample the \gls{PCA} weights from the combined training and validation sets to match the test-set sampling rate, using bandlimited finite-impulse-response (FIR) interpolation via polyphase filtering (SciPy library). In the downsampling case (evaluation on the ETH Zürich sequences), polyphase filtering applies an anti-aliasing filter before reducing the sampling rate, which helps limit high-frequency artifacts. Training and validation data are standardized using the mean and standard deviation of all concatenated samples in the training set. During test-set performance evaluation, \gls{PCA} weights are first standardized using the portion corresponding to the first 50.4s of the evaluation sequence, before the inputs $x_n$ are fed to the population model, as in inference with sequence-specific offline models. The transformer outputs are then scaled back to evaluate accuracy. Notably, for the population transformer, standardization during inference does not involve training-set data, but only data from the sequence used for evaluation. To mitigate stochasticity, we train $n_{\text{val}} = 5$ different models per horizon. During grid search, we select the hyperparameters that minimize the validation \gls{RMSE}---the numerator on the right-hand side of Eq. \ref{eq:predicted weights nRMSE}---averaged over $n_{\text{val}}$ trained models. The hyperparameter grid for the population transformer is the same as that for sequence-specific transformers (Table \ref{table:models comparison}). The Optuna library in Python is used to tune the population transformers \cite{optuna_2019}. Population-transformer hyperparameters only depend on the specified value of $h$, not on the evaluation sequence, unlike those of sequence-specific models.

% Specific remarks regarding regularization
The number of epochs for cross-subject transformers (300) is set higher than for sequence-specific transformers (50), as training time constrains feasibility less for the former, while a lower epoch count for the latter acts as lightweight regularization, helping counteract more pronounced data scarcity. After hyperparameter selection, the final population model is trained using early stopping, which terminates training if the validation loss does not improve for 30 consecutive epochs, to potentially improve generalization. Indeed, \gls{PCA} semantics between the ETH Zürich and \gls{OvGU} acquisitions may misalign, and breathing patterns differ between the two datasets. To further prevent overfitting with cross-subject transformers, we apply stochastic data augmentation to each training-set batch, operating per feature and mirroring all relevant transformations on the targets to preserve input--output consistency. Specifically, we (i) scale amplitudes by a factor drawn from the range $[0.8,1.2]$ with probability $p=0.8$, (ii) permute feature channels ($p=0.5$), (iii) add a constant bias (offset) proportional to the amplitude of the input-window signal ($p=0.3$, multiplicative factor $\leq$ 0.2), and (iv) inject a linear drift (time-dependent slope) with a magnitude scaled to the feature amplitudes ($p=0.3$, slope $\leq$ 0.05). The amplitude of the input \gls{PCA} coefficients within the current window is estimated robustly as the 5$^{\text{th}}$--95$^{\text{th}}$ percentile range to normalize the strength of bias and drift. Random noise is not injected during data augmentation, given the noise already present in the \gls{PCA} coefficients.

\subsection{Image prediction}
\label{section:methods: frame prediction from PCA weights}

\begin{figure}%[thb!]
\centering
\includegraphics[width=\columnwidth]{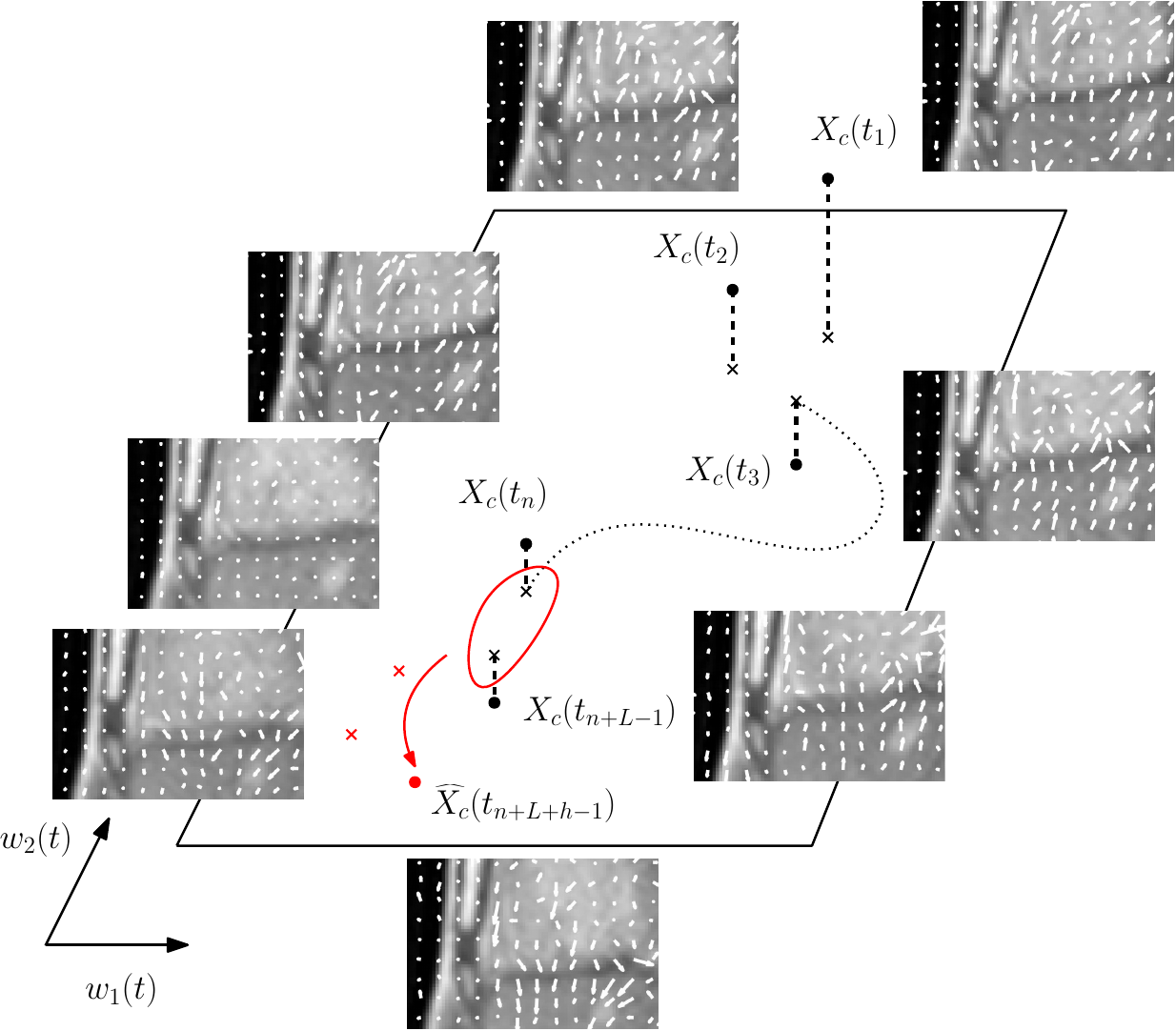}
\caption{Geometrical interpretation of motion-field prediction. The centered flattened \gls{DVF} at time $t$, $X_{\text{c}}(t)$, lies in a $2 |I|$-dimensional space, with $|I|$ denoting the number of pixels. It is projected onto the $n_{\text{cp}}$-dimensional linear subspace spanned by the principal components. The coordinates of this projection are the \gls{PCA} weights, $w_j(t)$ (Eq. \ref{eq:PCA weight calculation}). The \acs{DVF} predicted at horizon $h$ from the past time steps $n, \dots, n+L-1$, $\widehat{X_{\text{c}}}(t_{n+L+h-1})$, also lies in this subspace. The case illustrated corresponds to $h=3$, $L=2$, and $n_{\text{cp}} = 2$. The diaphragm region in sequence 1 of the ETH Zürich dataset (frames at times $t_{36}$--$t_{43}$) and the associated Lucas--Kanade optical-flow vectors are shown for illustration. Only a subset of the computed dense motion field (a grid of vectors separated by 6 pixels from one another) is displayed for readability.} 
\label{fig:Geometrical viewpoint}
\end{figure}

% The counter stuff is so that the footnote does not skip to footnote N+2 after footnote N
\refstepcounter{footnote}\setcounter{figfn}{\value{footnote}}   % reserve N
\begin{figure*}[pos=htbp,width=\textwidth,align=\centering]
    \centering
    \includegraphics[width=\textwidth]{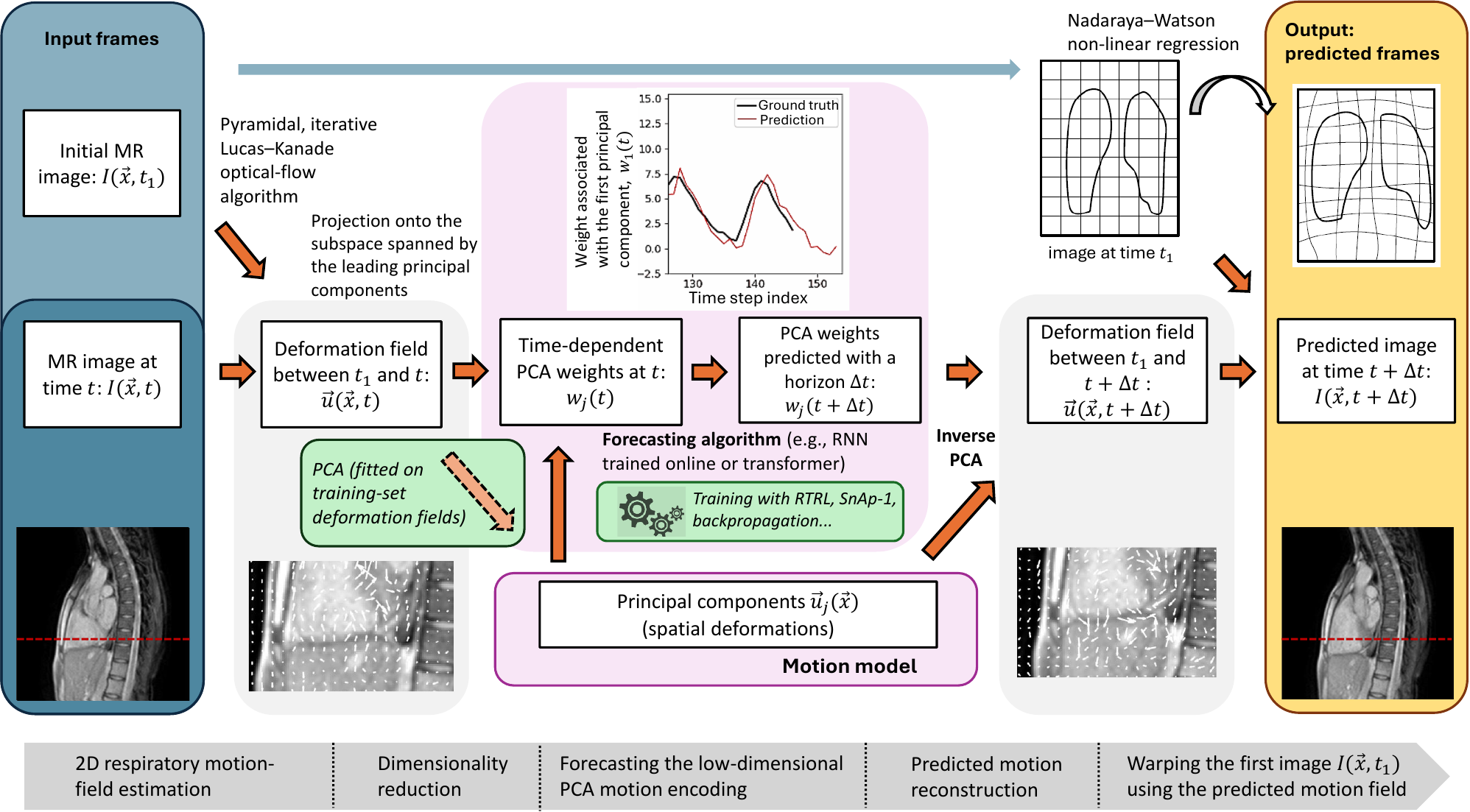}
    \caption{Overview of the proposed frame-forecasting pipeline. The inputs to the algorithm are the \gls{MR} frames at times $t_1$ and $t$ (blue boxes). First, the optical-flow algorithm computes the deformation field between them. During training, \gls{PCA} is fitted using these deformations, which produces a small set of (static) principal components and associated time-dependent weights, which a model learns to forecast (green boxes). During inference, current weights are estimated by projecting the optical-flow field at time $t$ onto the \gls{PCA} subspace and then forecast by the trained model. The reference image is warped by the resulting reconstructed \gls{DVF}, yielding the predicted frame at $t + \Delta t$ (yellow box).\FNmark{\value{figfn}}}
%Outline of the algorithm pipeline, showcasing the articulation between \gls{PCA} and time-series forecasting. The algorithm inputs, consisting of the \acs{MR} images at time $t_1$ and the current time step $t$, and the principal \glspl{DVF}, are highlighted in blue. During the initial learning step, \gls{PCA} is used to compute those principal components, and the forecasting algorithm is trained to predict the corresponding time-dependent weights, as shown in the green boxes. The algorithm outputs during inference are the forecast images, computed in real time by warping the initial image according to the predicted \gls{DVF} (yellow box)
    \label{fig:algorithm pipeline}%
\end{figure*}
\FNtext{\value{figfn}}{Includes content from \cite{pohl2021prediction}, Copyright 2024, with permission from Elsevier.}

\begin{figure}%[htb!]
	\centering
		\includegraphics[width=0.8\columnwidth]{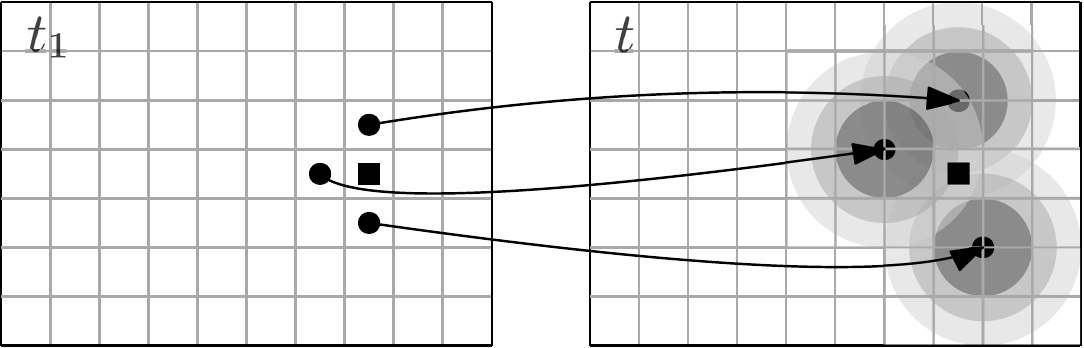}
	\caption{Warping the first image (at time $t_1$) using Nadaraya--Watson regression with a Gaussian kernel. The closer a point in the source image lands to the square point in the target image (at time $t$), the more it influences the intensity of that target pixel at time $t$.\FNmark{\value{figfn}}}% \protect\footnotemark.}
	\label{fig:warp_vectors_principle}
\end{figure}

We forecast the \glspl{DVF} by plugging the predicted \gls{PCA} weights, $\widehat{w_j}(t_{k+h})$, where $h$ denotes the horizon, into Eq. \ref{eq:PCA respiratory model}:%
\begin{equation} \label{eq:PCA weight forecasting}
\vec{u}(\vec{x}, t_{k+h}) \approx \vec{\mu}(\vec{x}) + \sum_{j=1}^{n_{\text{cp}}} \widehat{w_j}(t_{k+h}) \vec{u_j}(\vec{x})
\end{equation}
By doing so, we assume that the predicted motion lies in the $n_{\text{cp}}$-dimensional linear subspace spanned by the principal \glspl{DVF} estimated from the first $M_{\text{train}}$ images (Fig. \ref{fig:Geometrical viewpoint}).
%computed using the first $M_{\text{train}}$ images: is that necessary
The future frames are synthesized by warping the initial image according to the predicted \glspl{DVF} (Fig. \ref{fig:algorithm pipeline}). This operation maps each pixel in the source frame to a new, generally non-integer location in the target frame---the tip of its associated displacement vector. Since intensity values are then required at integer pixel coordinates on the regular grid, they must be interpolated. To achieve this, we perform Nadaraya--Watson regression, whose implementation in this work follows that in \cite{pohl2021prediction} (Fig. \ref{fig:warp_vectors_principle}). This forecasting strategy assumes that the motion field is reasonably smooth (i.e., mostly invertible) and that out-of-plane motion, artifacts, and brightness variations corresponding to the same tissue patch throughout the video are minimal.
% We select $\sigma_w = 0.5$ and a window size $h_w = 3$ in Eq. \ref{eq:Nadaraya_Watson}, based on visual assessment of the quality of the resulting images (Appendix \ref{appendix: determination of image warping parameters}).

% The counter stuff is so that the footnote does not skip to footnote N+2 after footnote N
\refstepcounter{footnote}\setcounter{figfn}{\value{footnote}}   % reserve N
\FNtext{\value{figfn}}{Reprinted from \cite{pohl2021prediction}, Copyright 2024, with permission from Elsevier.}

To evaluate the frame-forecasting performance of neural networks while accounting for randomness in the initialization and updates of their parameters, we perform $n_{\text{test}}$ runs. Each run corresponds to a different random seed and therefore different predicted \gls{PCA} coefficients, $\widehat{w_j}(t_{k+h})$. The latter yield $n_{\text{test}}$ versions of the predicted \gls{DVF}, each of which is used to warp the initial image. Subsequently, the test-set accuracy metrics are averaged over the $n_{\text{test}}$ runs. $n_{\text{test}}$ is set such that $n_{\text{test}} \leq n_{\text{val}}$, reflecting the computational cost of the warping process (Table \ref{table:general experimental setup}). In the following, when clear from the context, we write $w_j(t)$ instead of $\widehat{w_j}(t)$ for simplicity.

\subsection{Optimization of the PCA-subspace dimension}\label{section: methods - optimization of n_cp}

For the population transformer, we set the number of principal components (the \gls{PCA}-subspace dimension) to $n_{\text{cp}} = 3$, following prior work on respiratory motion modeling \cite{chhatkuli2015dynamic, harris2016technique}. For sequence-specific predictors, $n_{\text{cp}}$ is selected as described below. First, for each value of $n_{\text{cp}} \in \{1,2,3,4\}$, the hyperparameters that minimize the validation \gls{nRMSE} are determined via grid search (Section \ref{section: PCA weight cross-validation}; Eq. \ref{eq:predicted weights nRMSE}). To select $n_{\text{cp}}$, we minimize the registration error $E_{\text{pred}}(n_{\text{cp}})$, which characterizes the \glspl{DVF} predicted using these hyperparameters and the first $n_{\text{cp}}$ components (definition below in Eq. \ref{eq:val registration error}).%, $\vec{u}^{(i)}(n_{\text{cp}})$, where $i$ designates the run index. 

We first define $\delta(\vec{u}, \vec{x}, t)$, the absolute registration error at pixel $\vec{x}$ and time $t$, for a vector field $\vec{u}$ defined over a \gls{3D} (\gls{2D} + time) space:%
\begin{equation}\label{eq:instant registration error}
\delta(\vec{u}, \vec{x}, t_k) = |I(\vec{x}+\vec{u}(\vec{x}, t_k), t_k) - I(\vec{x}, t_1)|
\end{equation}%
Using the latter quantity, one can calculate $\epsilon(\vec{u}, t_k)$, the instantaneous normalized \gls{RMS} registration error at time $t_k$ using the vector field $\vec{u}$:%
\begin{equation}
\epsilon(\vec{u}, t_k) = 
\sqrt{\frac{\sum_{\vec{x}} \delta(\vec{u}, \vec{x}, t_k)^2}
{\sum_{\vec{x}} \big( \overline{I}(t_1) - I(\vec{x}, t_1) \big)^2}}
\end{equation}%
In this expression, $\overline{I}(t_1)$ denotes the mean intensity of the initial image. This error, evaluated using $\vec{u}^{(i)}(n_{\text{cp}})$---the \gls{DVF} predicted at run index $i$ using the hyperparameters selected for the leading $n_{\text{cp}}$ components---is averaged over the validation time steps and $n_{\text{val}}^{\text{dim}}$ runs to account for neural-network stochasticity:%
\begin{equation}\label{eq:val registration error} 
E_{\text{pred}}(n_{\text{cp}}) \!= \!\frac{1}{n_{\text{val}}^{\text{dim}}(M_{\text{val}}\!-\!M_{\text{train}})} \!\sum_{i=1}^{n_{\text{val}}^{\text{dim}}} \!\sum_{k= M_{\text{train}} + 1}^{M_{\text{val}}} \!\epsilon(\vec{u}^{(i)}(n_{\text{cp}}), t_k)
\end{equation}%
% \! helps reduce the spacing so that the equation fits in one line
We refer to $E_{\text{pred}}(n_{\text{cp}})$ as the (mean) normalized \gls{RMS} registration error and select the value of $n_{\text{cp}}$ that minimizes it. Computing this error is relatively affordable, as it does not require warping the initial image. We set $n_{\text{val}}^{\text{dim}} = n_{\text{test}}$, the number of runs used to evaluate image-prediction accuracy (Section \ref{section:methods: frame prediction from PCA weights}); $n_{\text{val}}^{\text{dim}} = 1$ for deterministic algorithms. Fig. \ref{fig:algorithm pipeline} and Table \ref{table:general experimental setup} outline the overall experimental setting and related parameters.

\begin{figure*}[pos=htbp,align=\centering]
    %\centering
    \includegraphics[width=0.9\textwidth]{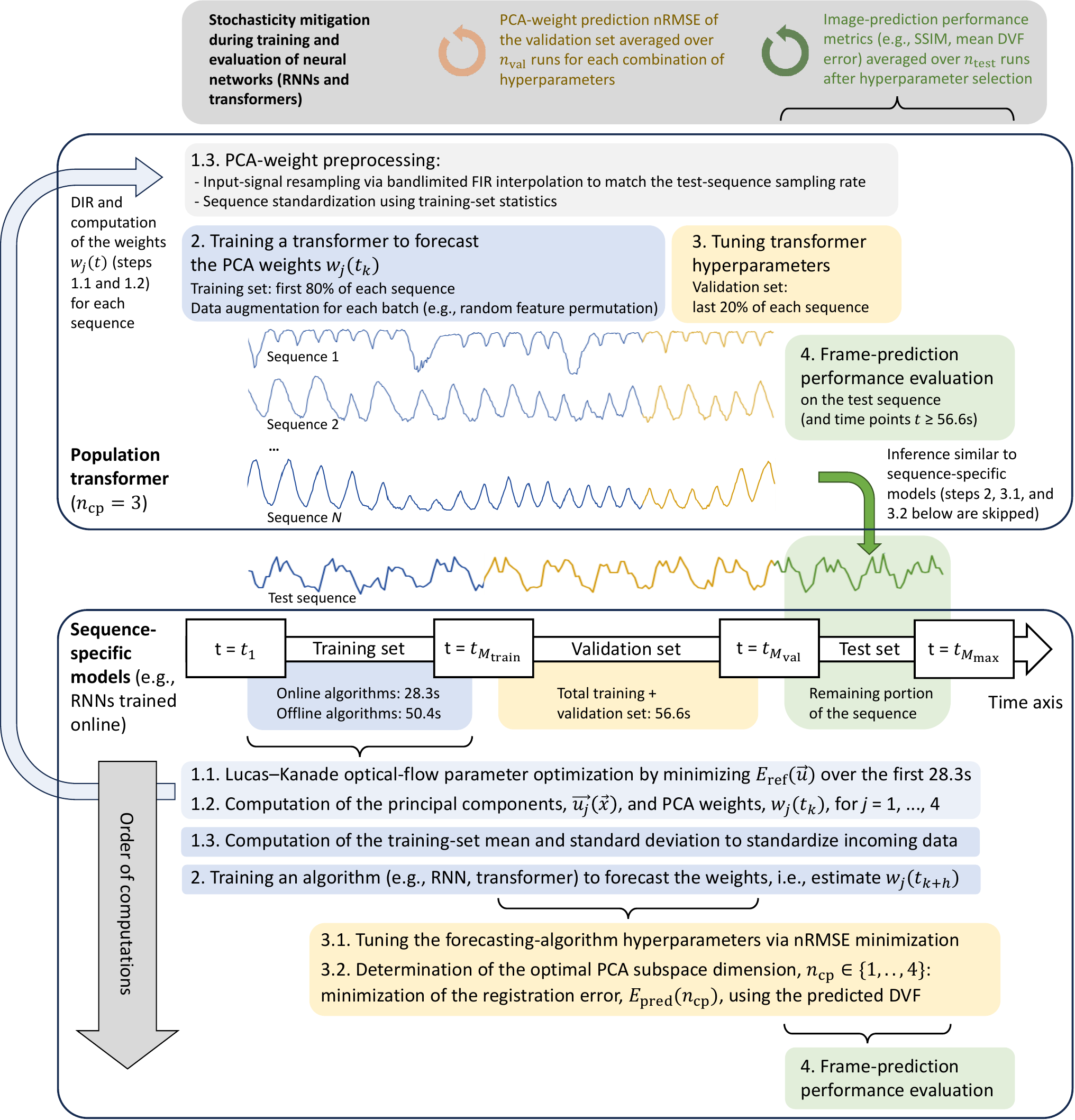}
	%\captionsetup{width=\linewidth}    
    \caption{Overall experimental setting. Only the first-order weight is shown for each sequence for clarity. We visually aligned the end of the training sequences in the population-transformer workflow with the start of the test-set period to emphasize that all algorithms are evaluated on the same test segment for consistency. This alignment is schematic, as the durations of the sequences vary in practice.}
    \label{fig:overall experimental setting}%
\end{figure*}

\section{Results}

\subsection{Breathing motion modeling with PCA}
\label{section: PCA breathing motion modeling results}

\begin{figure*}[pos=htbp, align=\centering]
    \centering
    \subfloat[Optical flow between $t_1$ and $t_{28}$, $\vec{u}(\vec{x}, t_{28})$.\\ Background: frame at $t=t_{28}$.]{\includegraphics[width=.325\textwidth]{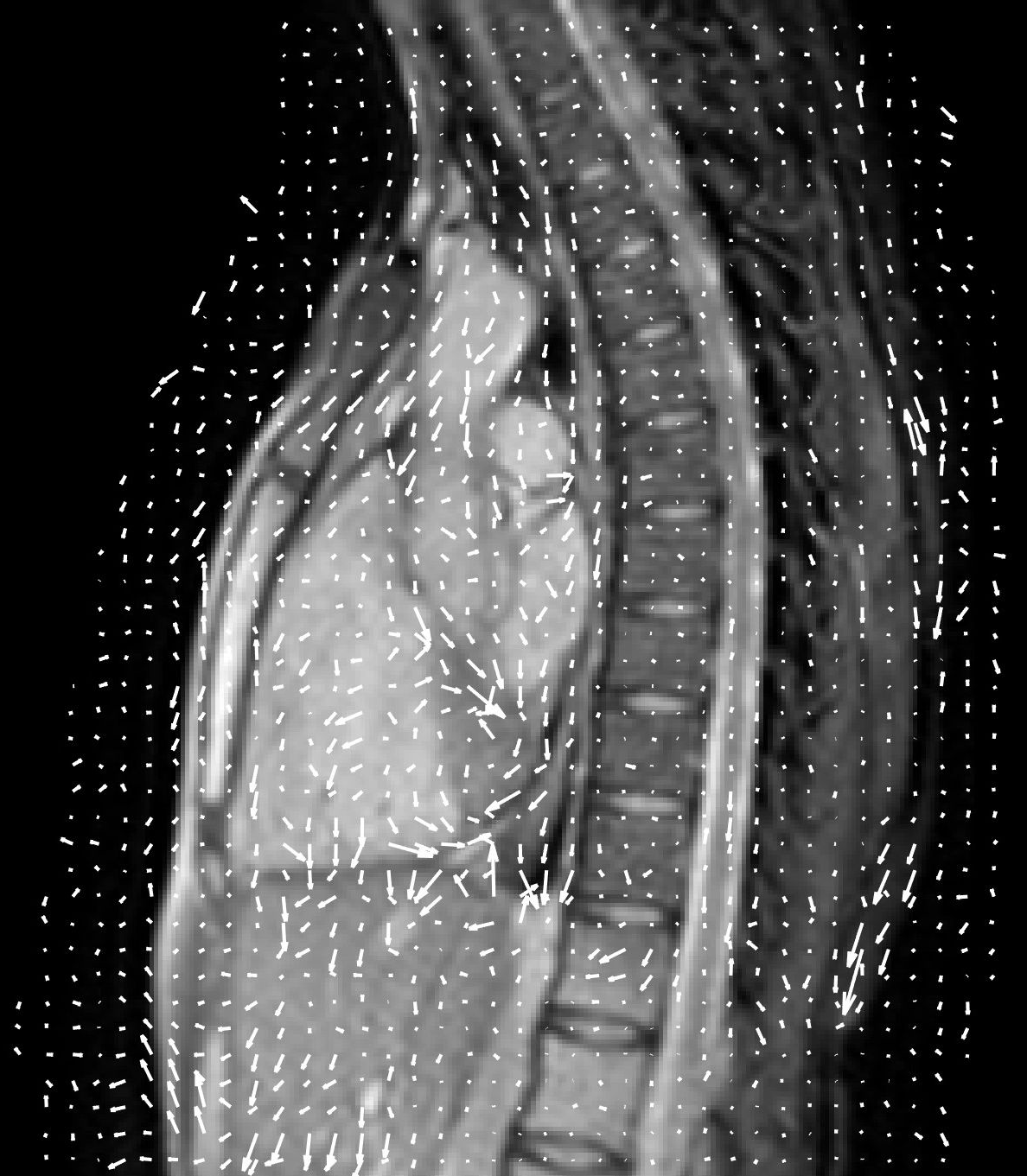}\label{fig:DVF sq 1 ETH inspiration}}% 
    \,
    \subfloat[Optical flow between $t_1$ and $t_{32}$, $\vec{u}(\vec{x}, t_{32})$. \\ Background: frame at $t=t_{32}$.]{\includegraphics[width=.325\textwidth]{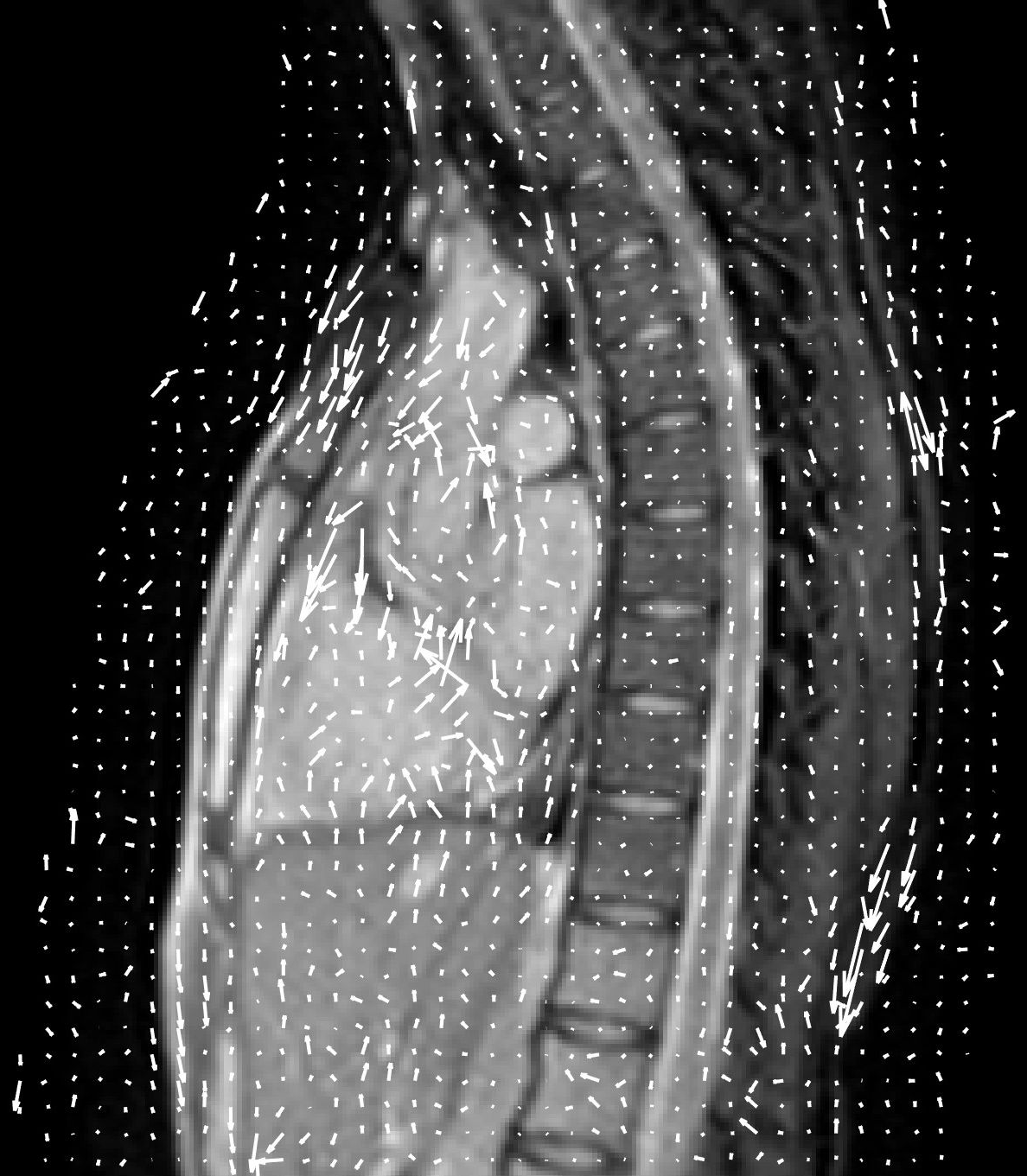}\label{fig:DVF sq 1 ETH expiration}}% 
    \,
    \subfloat[Optical-flow mean over $t \leq t_{90}$, $\vec{\mu}(\vec{x})$. \\ Background: frame at $t=t_{1}$.]{\includegraphics[width=.325\textwidth]{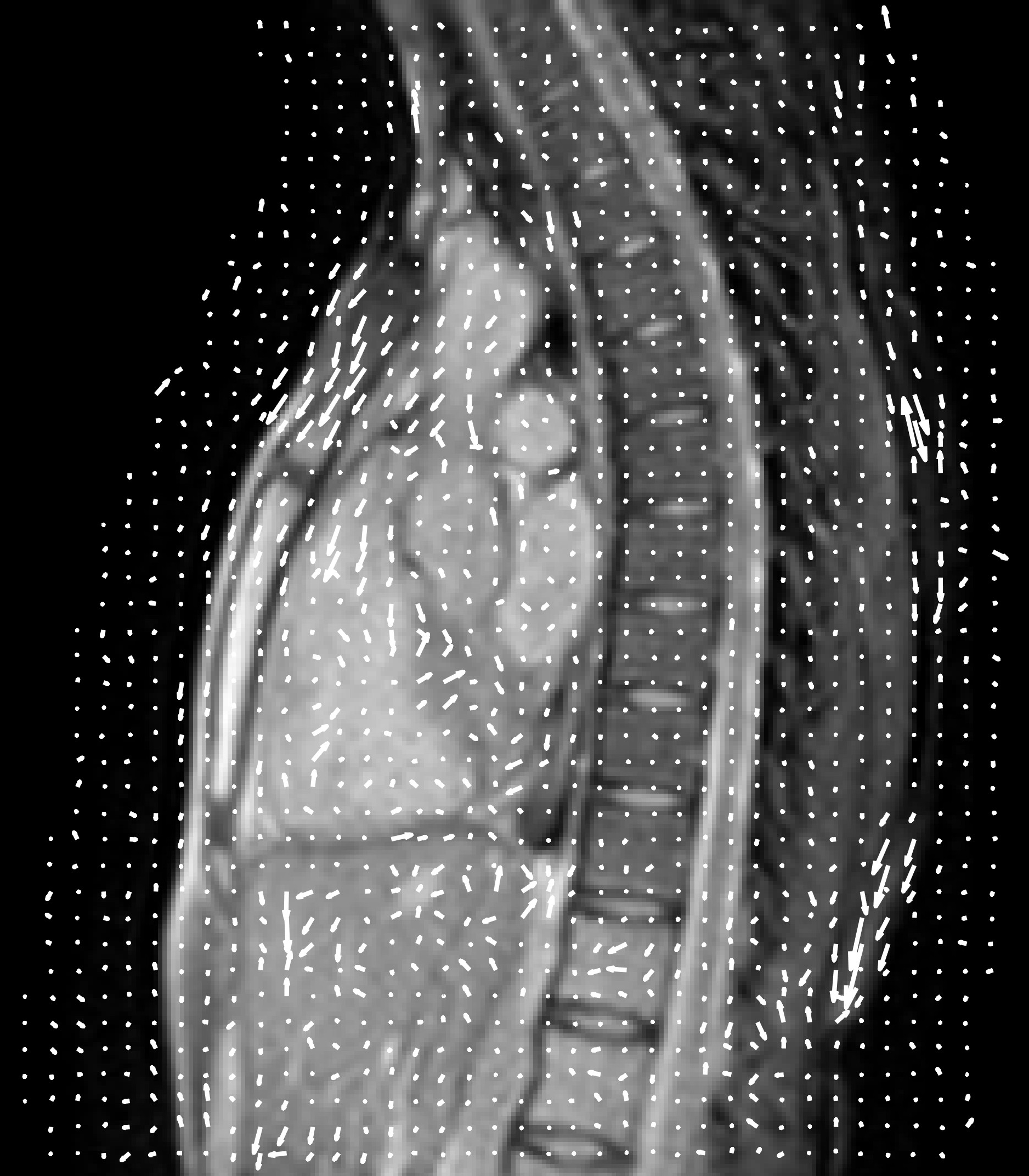}\label{fig:temporal average DVF sq 1 ETH}}% 
    \\
    \subfloat[1$^{\text{st}}$ principal component, $\vec{u_1}(\vec{x})$. \\ Background: frame at $t=t_{1}$.]{\includegraphics[width=.325\textwidth]{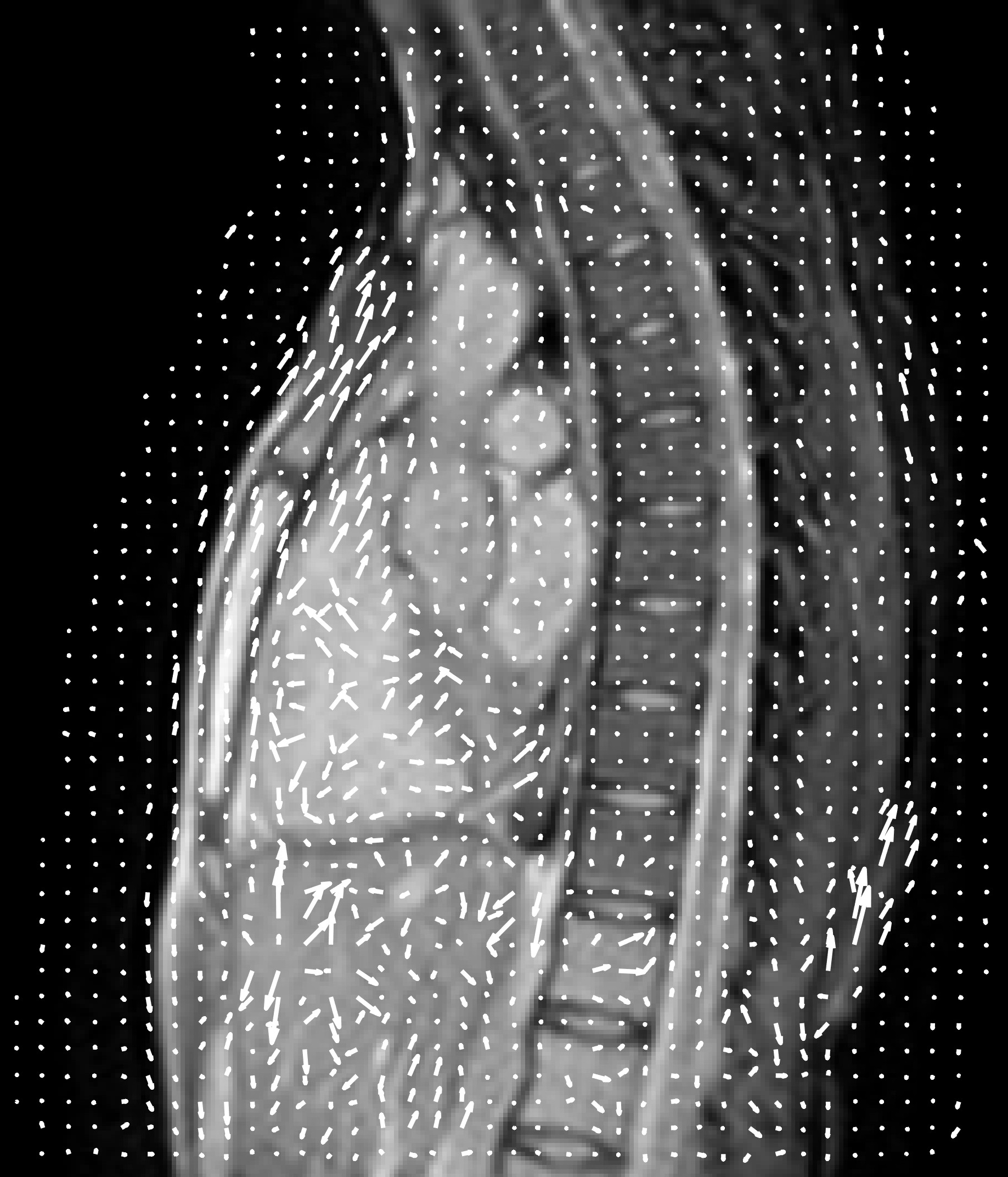}\label{fig:first principal component sq 1 ETH}}%
    \,
    \subfloat[2$^{\text{nd}}$ principal component, $\vec{u_2}(\vec{x})$. \\ Background: frame at $t=t_{1}$.]{\includegraphics[width=.325\textwidth]{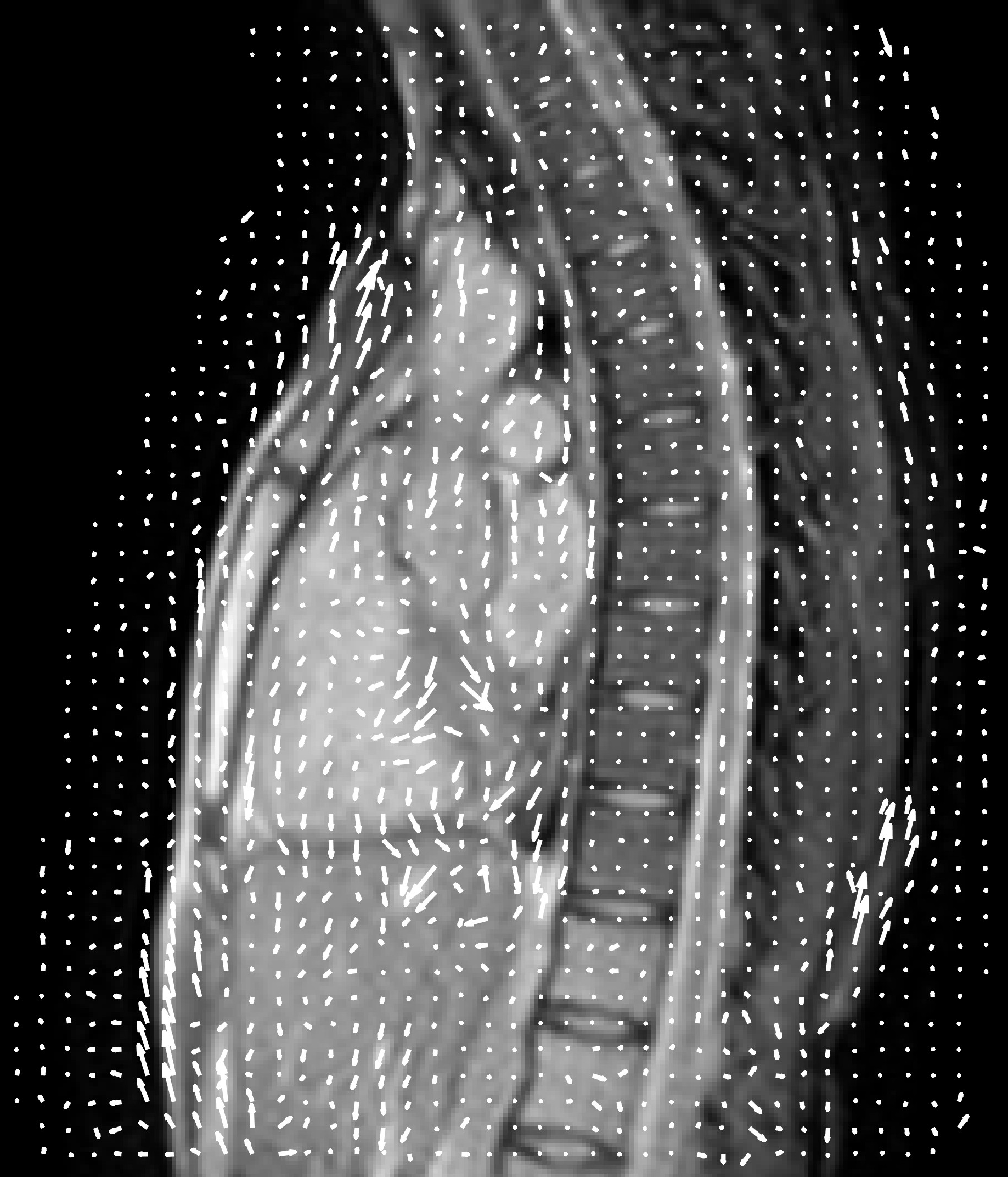}\label{fig:second principal component sq 1 ETH}}%  
    \,  
    \subfloat[3$^{\text{rd}}$ principal component, $\vec{u_3}(\vec{x})$. \\ Background: frame at $t=t_{1}$.]{\includegraphics[width=.325\textwidth]{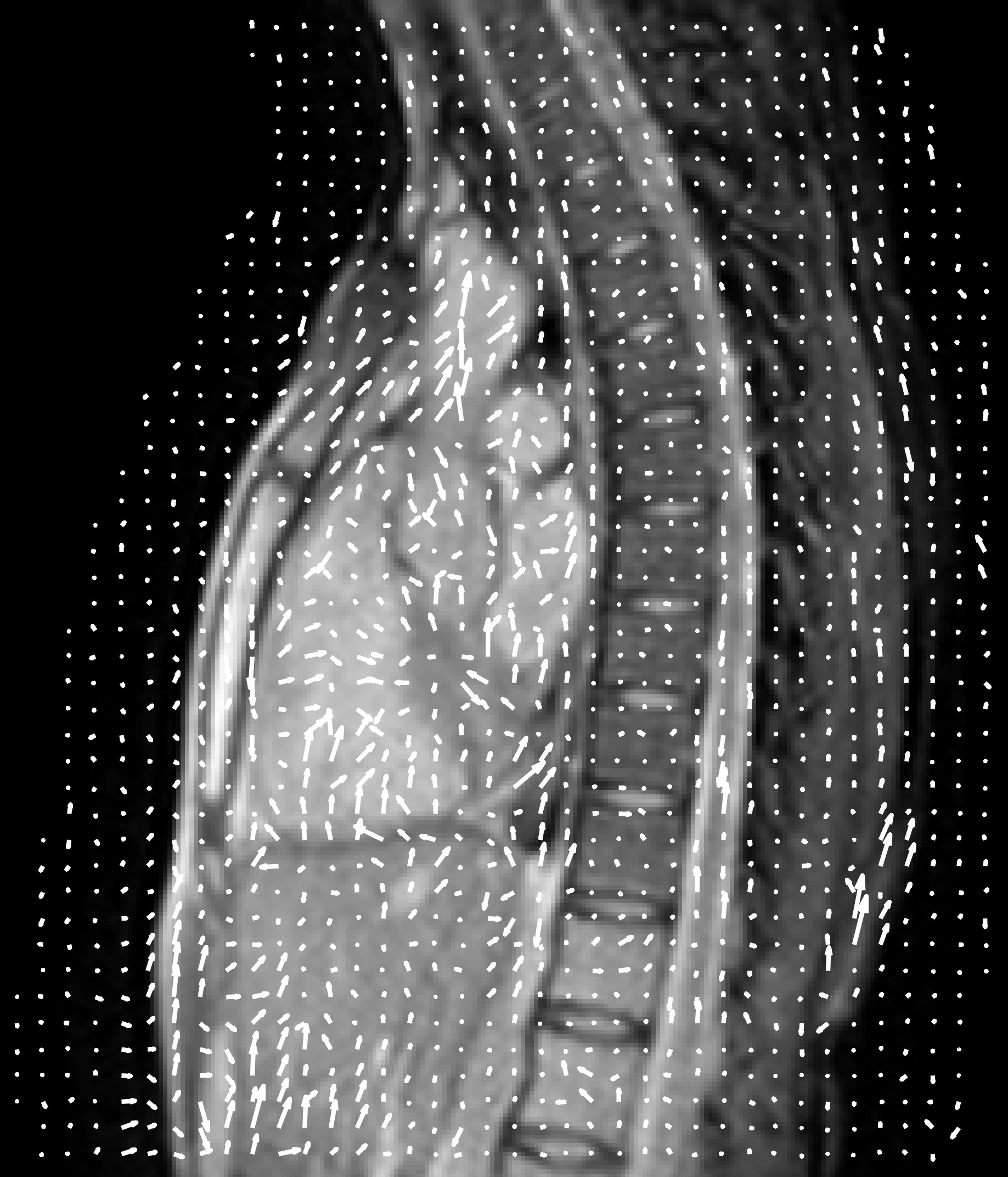}\label{fig:third principal component sq 1 ETH}}%     
    \caption{Top row: Lucas--Kanade optical-flow field in sequence 1 of the ETH Zürich dataset at inspiration (left, $t=t_{28}$) and expiration (middle, $t=t_{32}$), and its temporal average between $t_1$ and $t_{90}$. Bottom row: leading three principal components, computed using the first 90 frames of the \gls{MRI} sequence (online learning setting; Table \ref{table:general experimental setup}). Deformable registration parameters were optimized on the same first 90 images (Appendix \ref{appendix: chest MR image registration optimization}). For readability, the displacement vectors displayed are sampled on a grid with 6-pixel spacing. In addition, the principal components are scaled by a factor of 500, and the left and right black borders containing null vectors are removed in each subfigure. Best viewed with zoom on a digital display.}
    \label{fig:DIR and principal components sequence 1 ETH}
\end{figure*}

% Lucas Kanade optical flow
The optical-flow vectors around the liver and diaphragm mainly followed the \gls{SI} tissue motion associated with breathing. In sequence 1 of the ETH Zürich dataset, they mostly pointed downwards during inspiration and upwards during expiration (Figs. \ref{fig:Geometrical viewpoint}, \ref{fig:DVF sq 1 ETH inspiration}, and \ref{fig:DVF sq 1 ETH expiration}), as the initial frame corresponds to a middle phase in the cycle (background in Fig. \ref{fig:temporal average DVF sq 1 ETH}). Their average over $t \leq t_{90}$, $\vec{\mu}(\vec{x})$, did not capture a noticeable vertical displacement trend in those areas, which reflects the absence of respiratory drift (Fig. \ref{fig:temporal average DVF sq 1 ETH}). Artifacts and transverse motion causing sudden brightness variations led to relatively strong fluctuations in the norms of the deformation vectors, for example, near the sternum and lower back. In general, the optical-flow vectors did not lie vertically in a homogeneous manner, as organs move and deform in specific ways. Appendix \ref{appendix: chest MR image registration optimization} contains results regarding \gls{DIR} parameter optimization.

\begin{figure*}[pos=htbp, align=\centering]
    \centering
    \subfloat[First principal component, $\vec{u_1}(\vec{x})$. \\ Background: frame at $t=t_{1}$.]{\includegraphics[width=.30\textwidth]{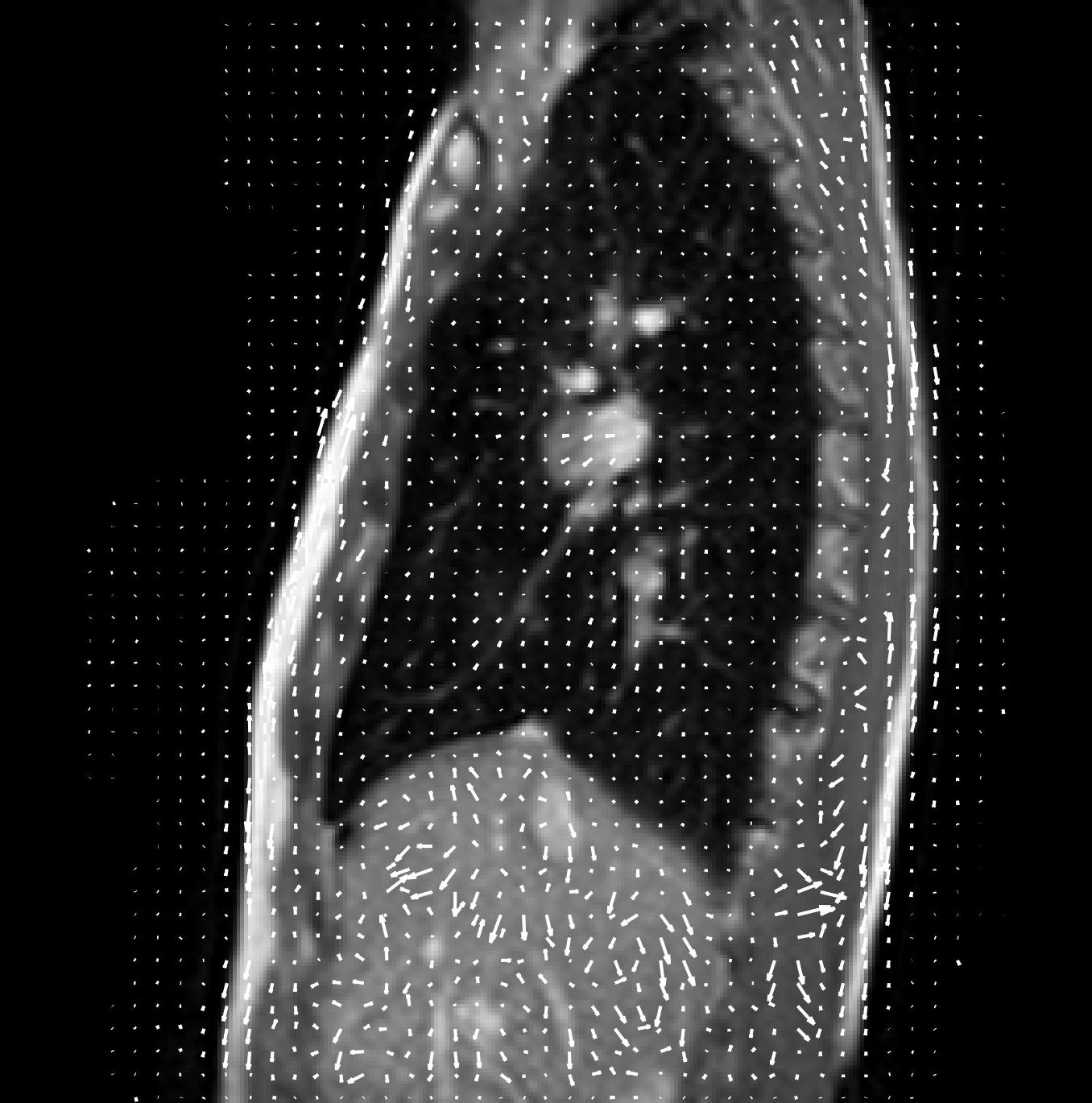}\label{fig:1st DVF principal component sq 4}}% 
    \subfloat[First-order \acs{PCA} weight, $w_1(t_k)$.]{\includegraphics[width=.34\textwidth]{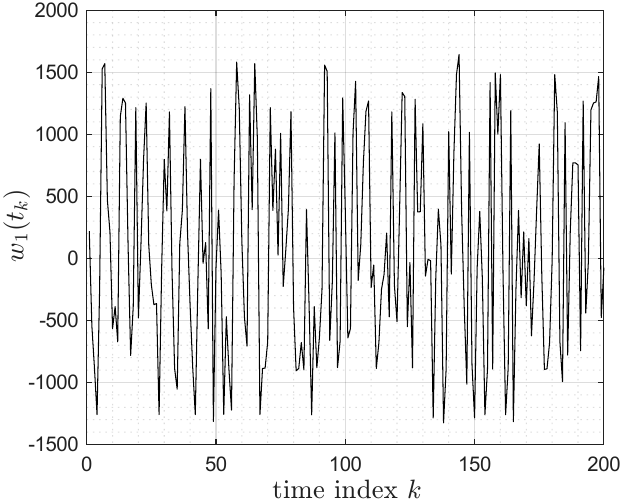} }%
    \subfloat[Third-order PCA weight, $w_3(t_k)$.]{\includegraphics[width=.34\textwidth]{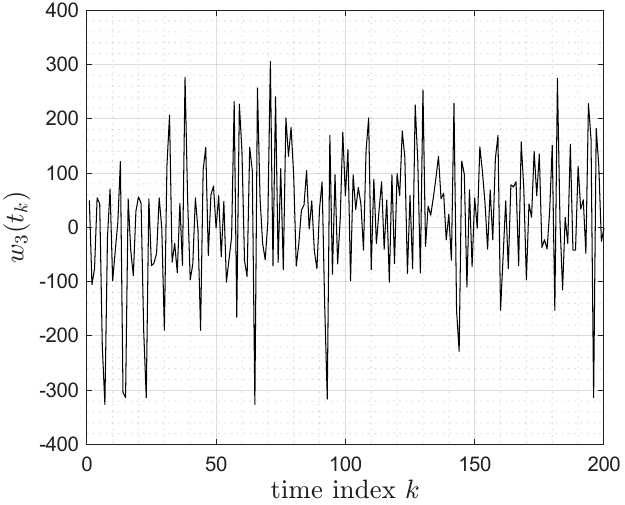} }%    
    \,
    \subfloat[Second principal component, $\vec{u_2}(\vec{x})$. \\ Background: frame at $t=t_{1}$.]{\includegraphics[width=.30\textwidth]{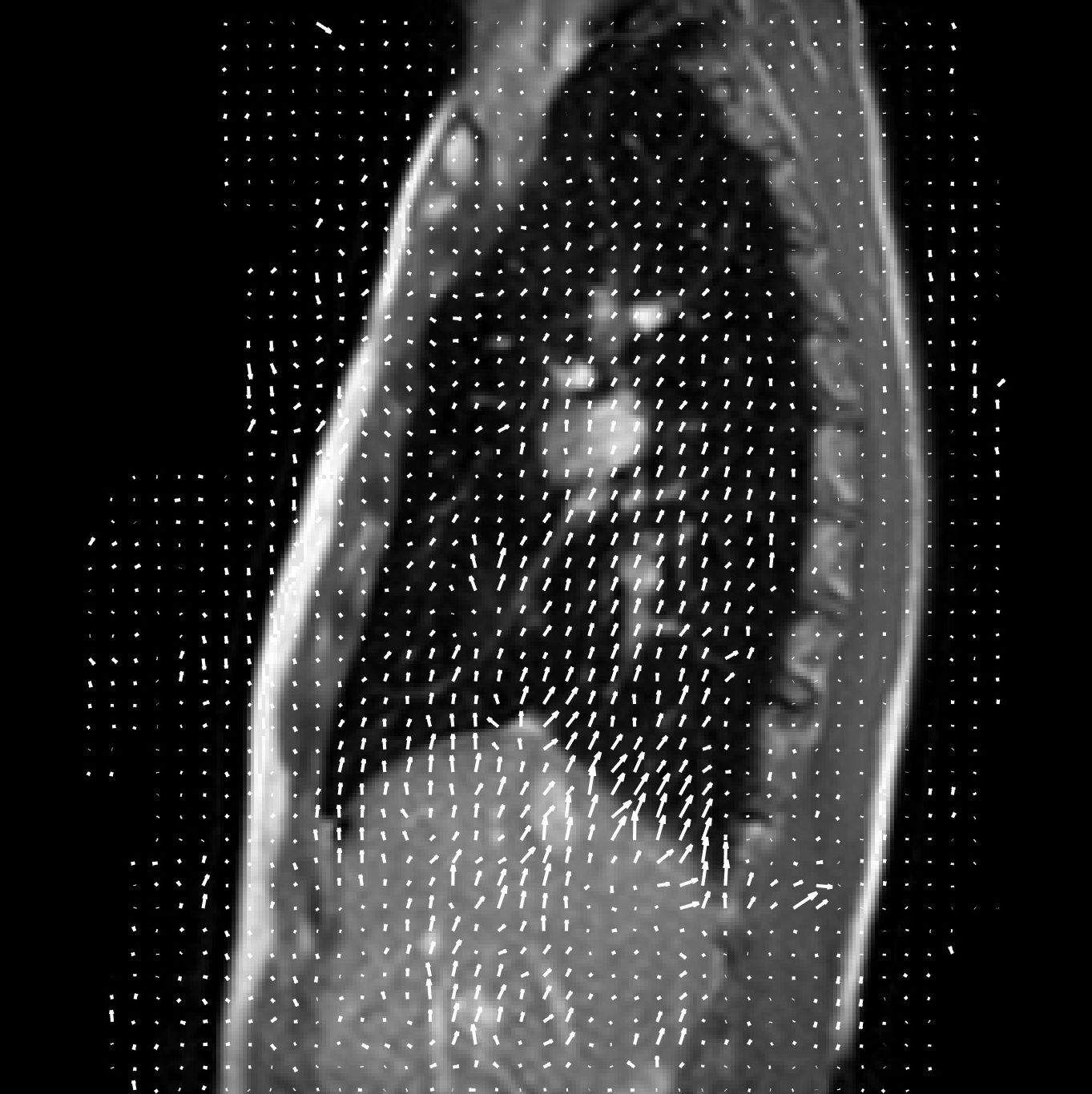}\label{fig:2nd DVF principal component sq 4}}% 
    \subfloat[Second-order PCA weight, $w_2(t_k)$.]{\includegraphics[width=.34\textwidth]{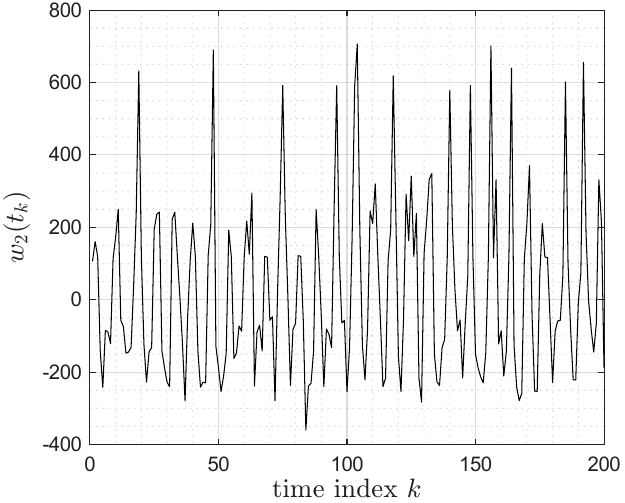} \label{fig:2nd PCA weight sq 4}}%
    \subfloat[Fourth-order PCA weight, $w_4(t_k)$.]{\includegraphics[width=.34\textwidth]{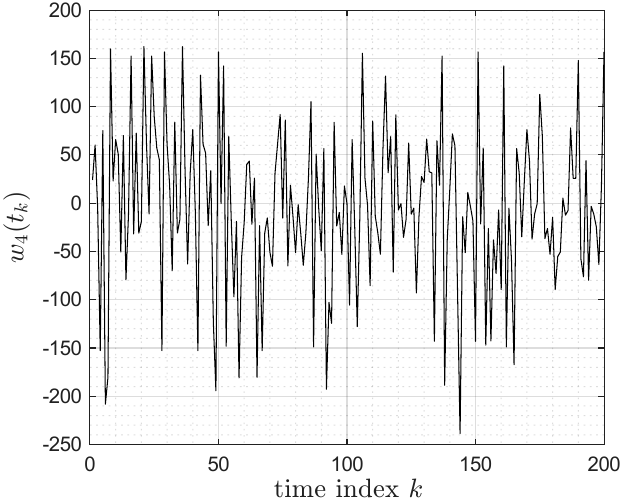} }%            
    \caption{Leading two principal components and first four time-dependent \gls{PCA} coefficients associated with sequence 4 of the ETH Zürich dataset. The principal components were computed using the first 90 frames (online learning setting). Same remarks as in Fig. \ref{fig:DIR and principal components sequence 1 ETH} regarding vector spacing, scaling, border removal, and recommended viewing conditions.}
    \label{fig:DVF principal components and weights sequence 4}
\end{figure*}

\begin{figure*}[pos=htbp, align=\centering]
    \centering
    \subfloat[1$^{\text{st}}$ principal component $\vec{u_1}(\vec{x})$. \\* Sequence 1, background: frame at $t_1$.]{\includegraphics[width=.265\textwidth]{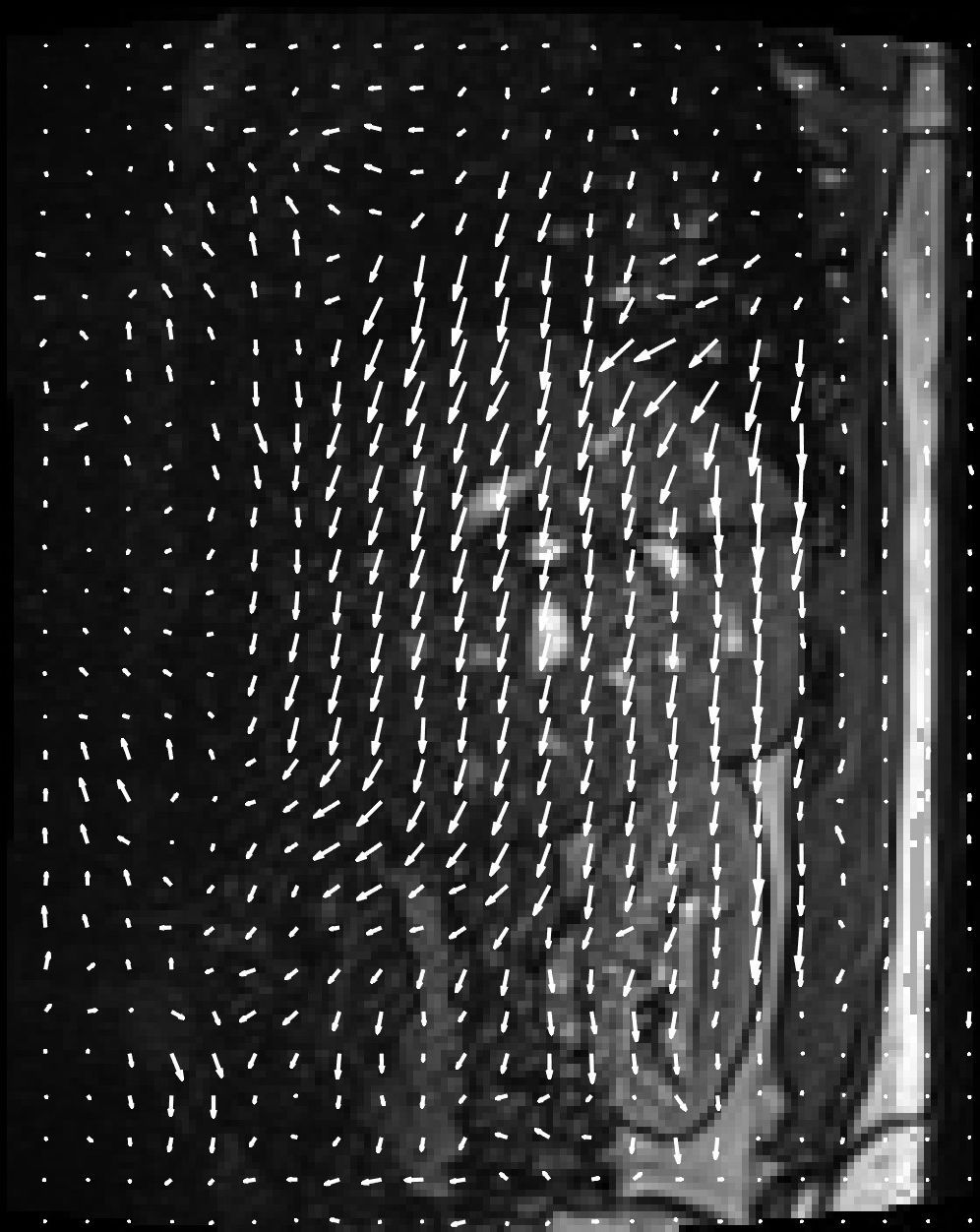}\label{fig:1st PCA component Sq 1 OvGU dataset}}
    \, 
    \subfloat[Prediction of the first-order \acs{PCA} coefficient in sequence 1 ($h=1.66\text{s}$).]{\includegraphics[width=.72\textwidth]{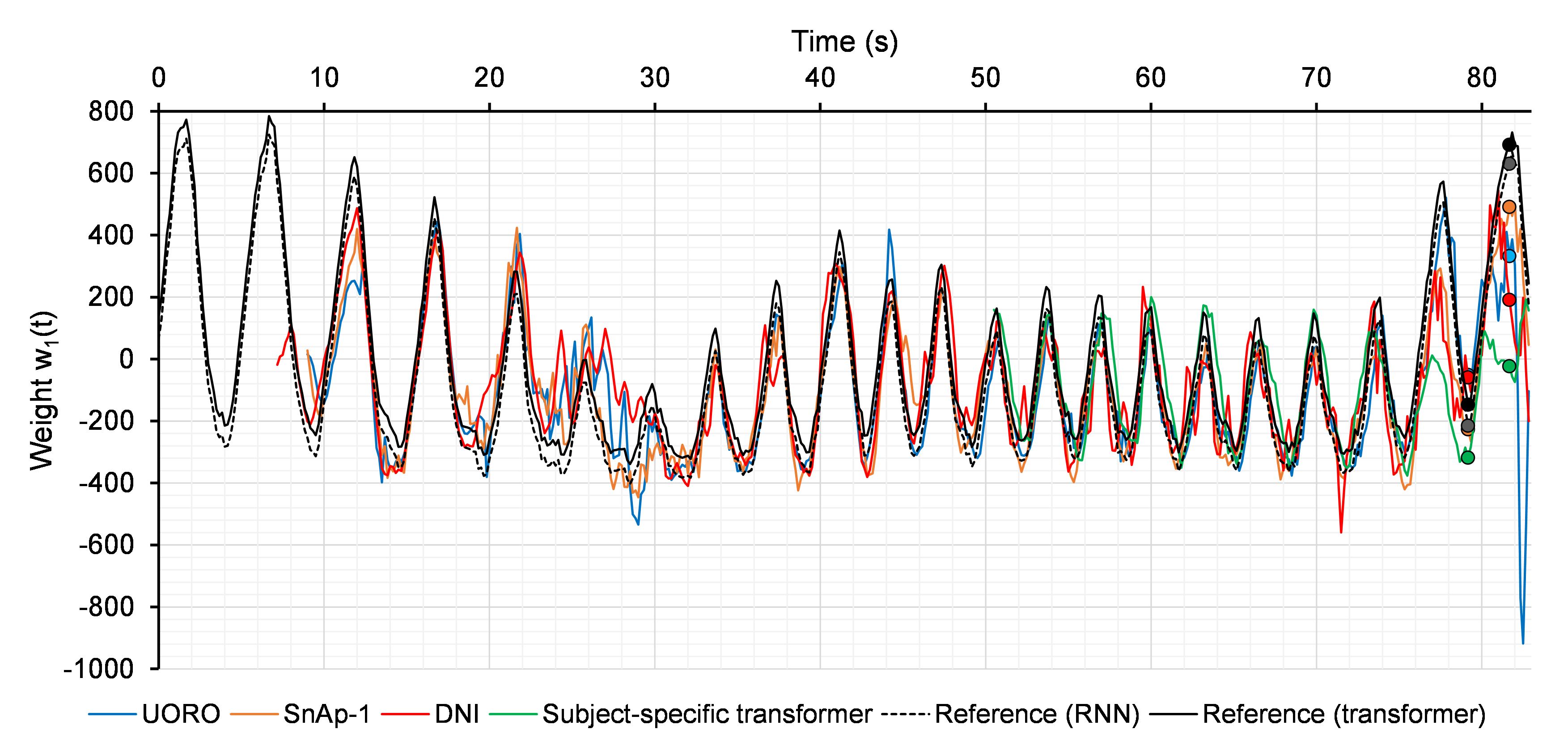}\label{fig:1st PCA component prediction in Sq 1 of OgVU dataset}}% 
    \hfill
    \subfloat[1$^{\text{st}}$ principal component $\vec{u_1}(\vec{x})$. \\* Sequence 4, background: frame at $t_1$.]{\includegraphics[width=.265\textwidth]{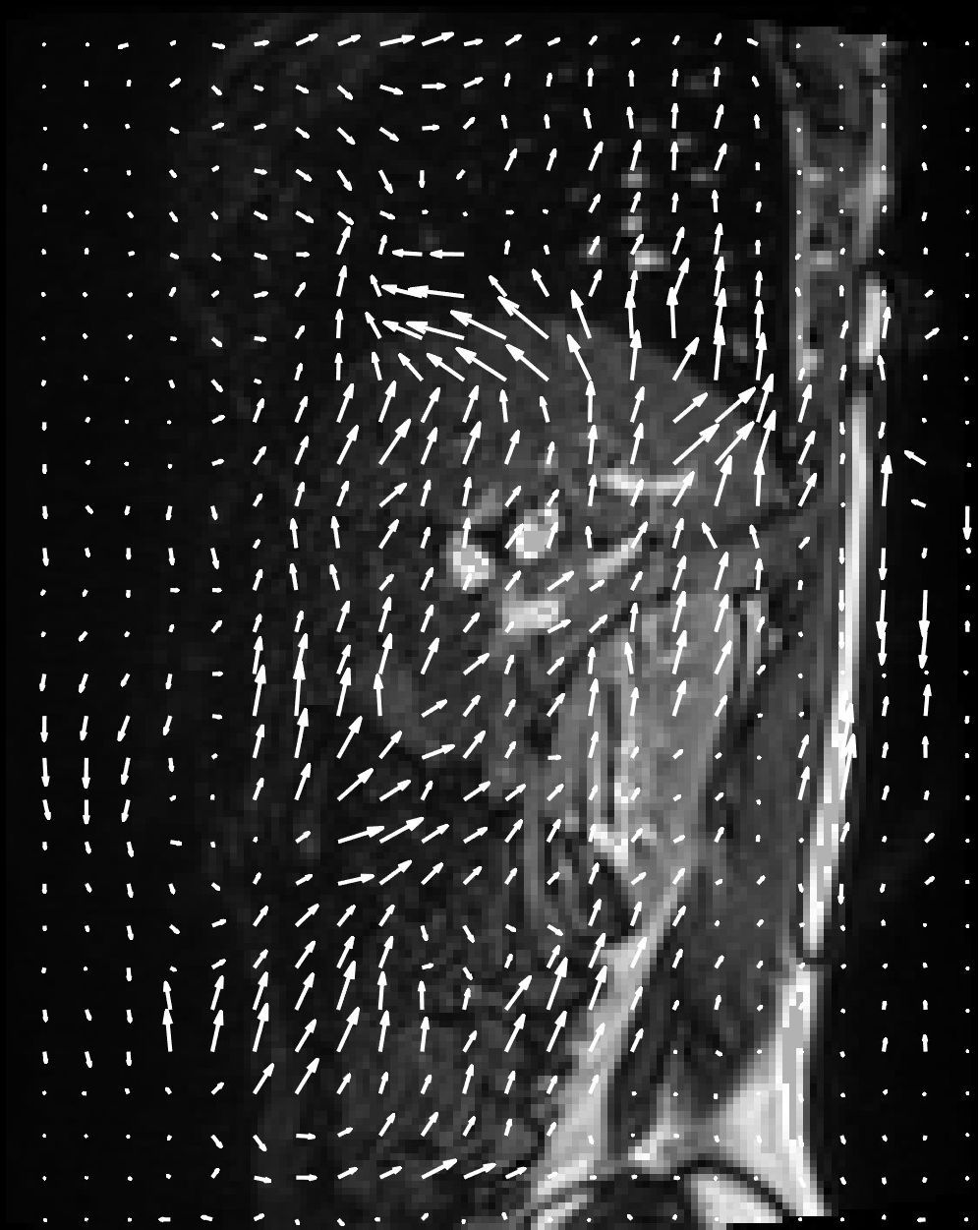}\label{fig:1st PCA component Sq 4 OvGU dataset}}
    \, 
    \subfloat[Prediction of the first-order \acs{PCA} coefficient in sequence 4 ($h=1.66\text{s}$).]{\includegraphics[width=.72\textwidth]{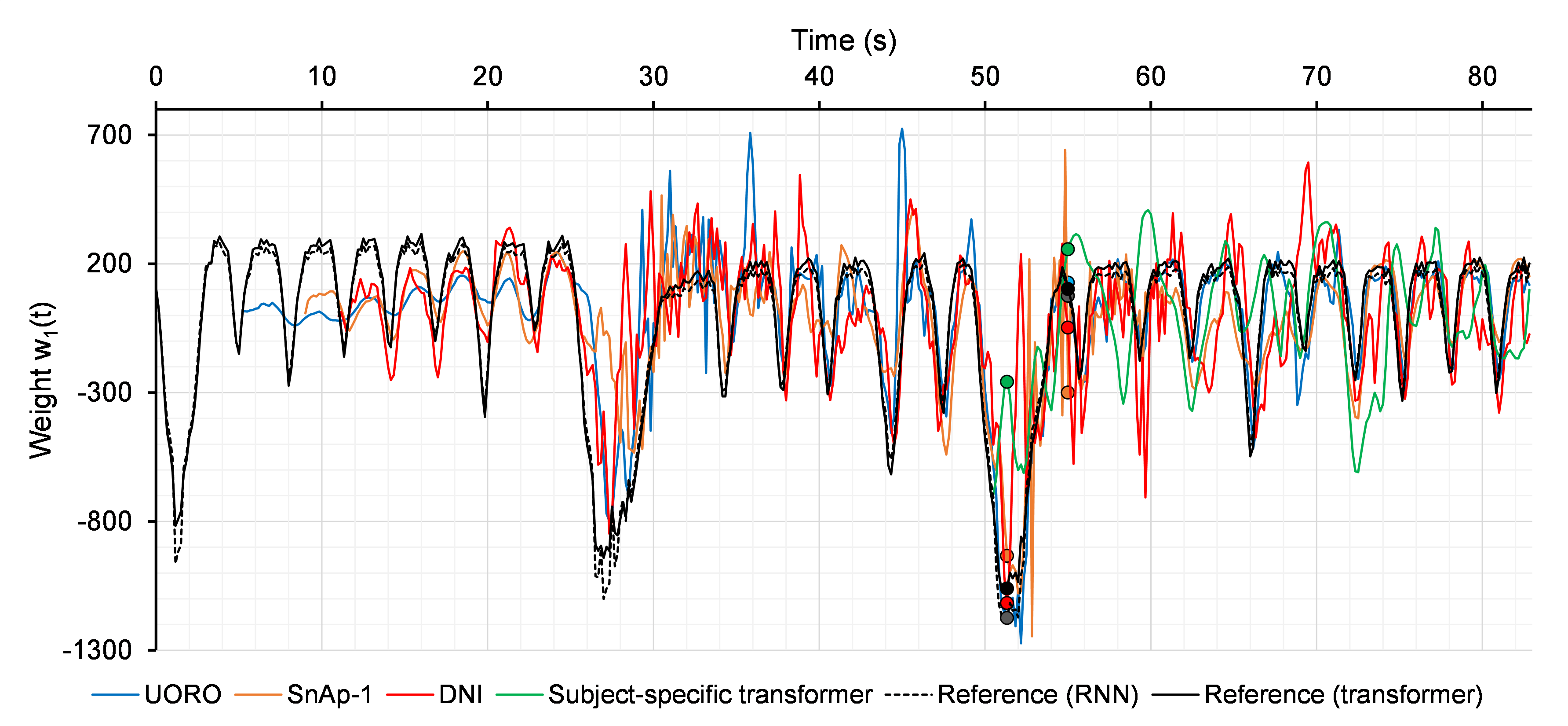}\label{fig:1st PCA component prediction in Sq 4 of OgVU dataset}}% 
    \hfill
    \subfloat[1$^{\text{st}}$ principal component $\vec{u_1}(\vec{x})$. \\* Sequence 6, background: frame at $t_1$.]{\includegraphics[width=.265\textwidth]{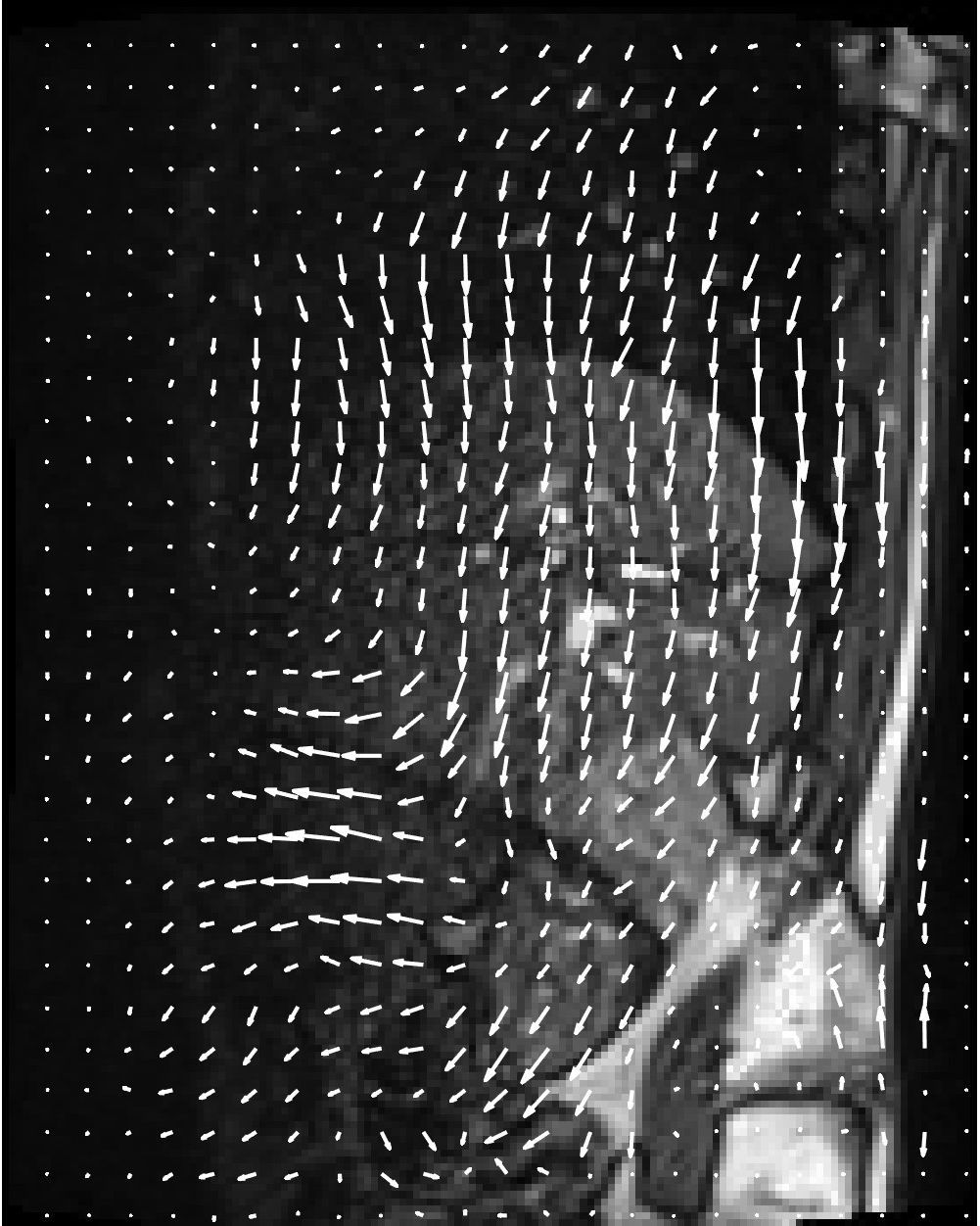}\label{fig:1st PCA component Sq 6 OvGU dataset}}
    \,
    \subfloat[Prediction of the first-order \acs{PCA} coefficient in sequence 6 ($h=1.66\text{s}$).]{\includegraphics[width=.72\textwidth]{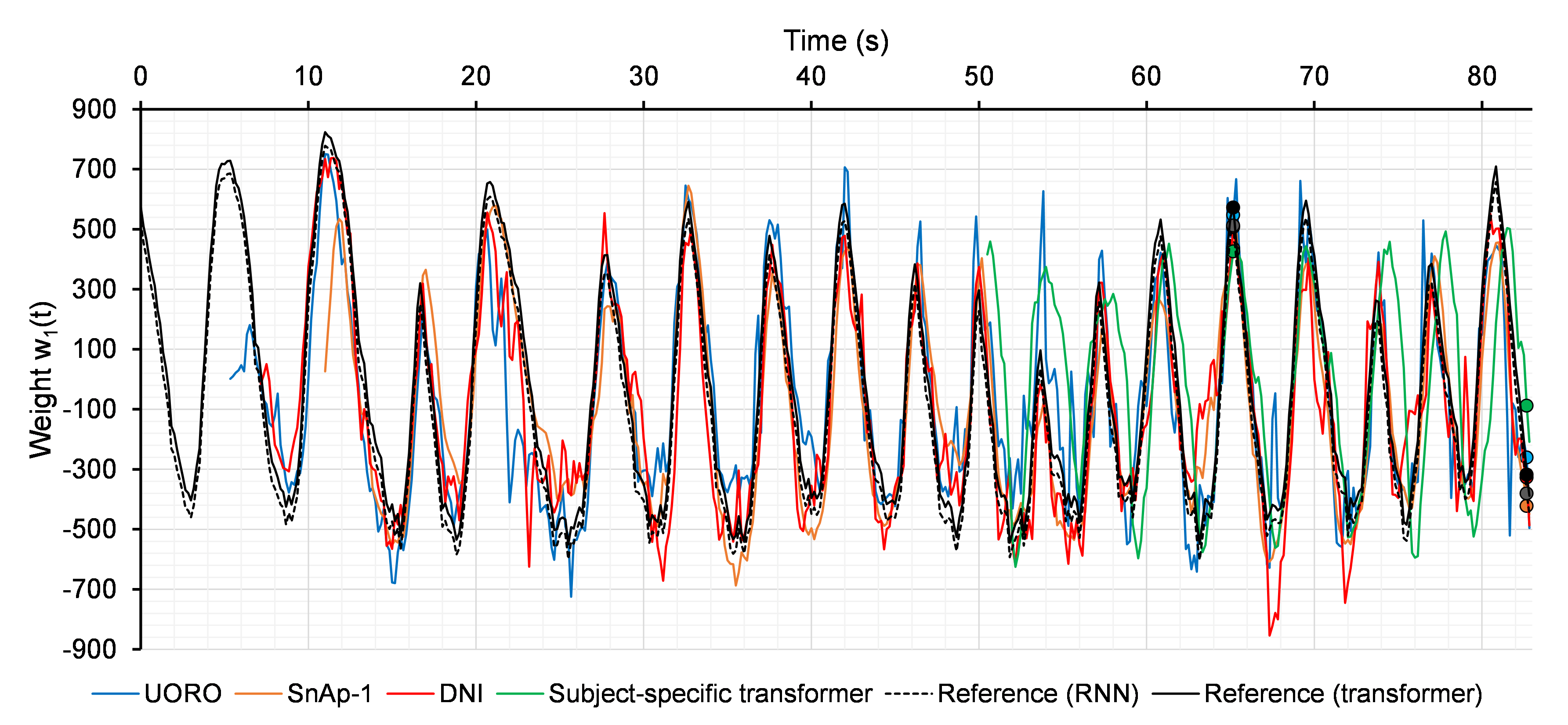}\label{fig:1st PCA component prediction in Sq 6 of OgVU dataset}}%         
    \caption{First-order principal components and associated time-dependent \acs{PCA} weights, along with their forecasts using \pgls{RNN} trained with \gls{UORO}, \acs{SnAp-1}, and \gls{DNI} and a subject-specific transformer at $h=1.66\text{s}$, for sequences 1, 4, and 6 of the \acs{OvGU} dataset. The principal \glspl{DVF} plotted here were estimated from the first $t_{M_{\text{train}}}=50.4\text{s}$ of each sequence; this corresponds to the transformer setting (black solid lines in the plots on the right side), as opposed to the \acs{RNN} setting ($t_{M_{\text{train}}}=28.3\text{s}$, black dotted lines). Hyperparameters, including $n_{\text{cp}}$, were optimized using the validation segment for each sequence individually. The round markers in the forecasting graphs correspond to the predicted images shown in Fig. \ref{fig:visual comparison of ROI prediction with SnAp-1 and transformer at h=10 on OvGU dataset} (both depict the same runs). Same remarks as in Fig. \ref{fig:DIR and principal components sequence 1 ETH} regarding vector spacing, scaling, and recommended viewing conditions.}
    \label{fig:1st cpt prediction (SnAp-1 vs vs UORO vs transformer) Magdeburg}
\end{figure*}

% Principal components
The first-order principal \gls{DVF} in sequence 1 of the ETH Zürich dataset was primarily associated with expansion and contraction of the right cardiac ventricle and internal deformations of the liver (Fig. \ref{fig:first principal component sq 1 ETH}). By contrast, the second \gls{PCA} component mainly reflected respiratory motion, as most of the corresponding deformation vectors leaned downwards uniformly within the thoracic cavity  (Fig. \ref{fig:second principal component sq 1 ETH}). The third component combined heart deformations and respiratory elements, with motion vectors pointing superiorly or posteriorly (Fig. \ref{fig:third principal component sq 1 ETH}). These three components also captured periodic intensity variations around the sternum and lumbar regions, largely due to transversal displacements. Likewise, in the fourth ETH Zürich sequence, the first-order principal \gls{DVF} corresponded to liver deformations and out-of-plane motion artifacts near the back of the body (Fig. \ref{fig:1st DVF principal component sq 4}). The \gls{2D} vectors forming the second component aligned upwards relatively homogeneously, reflecting breathing motion (Fig. \ref{fig:2nd DVF principal component sq 4}). By contrast, for each \gls{OvGU} sequence, the first principal component primarily represented the dominant \gls{SI} constituent of respiratory motion (Fig. \ref{fig:1st cpt prediction (SnAp-1 vs vs UORO vs transformer) Magdeburg}). The \gls{AP} displacements of the upper abdominal organs, particularly those below the liver and near the anterior abdominal wall, were also reflected in the first component, notably in sequence 6. In all sequences, the norm of the components $\vec{u}_j(\vec{x})$ in air/background regions tended to increase with $j$, consistent with \gls{PCA} progressively capturing variance attributable to noise (not displayed for brevity).

\begin{figure*}[thb!]
    %\captionsetup[subfigure]{labelformat=empty}
    %\captionsetup[subfloat]{farskip=2pt} %{farskip=2pt,captionskip=1pt}
    \centering
    \subfloat[Prediction of $w_1(t)$ with \acs{RTRL}]{\includegraphics[width=.49\textwidth, height=.28\textwidth]{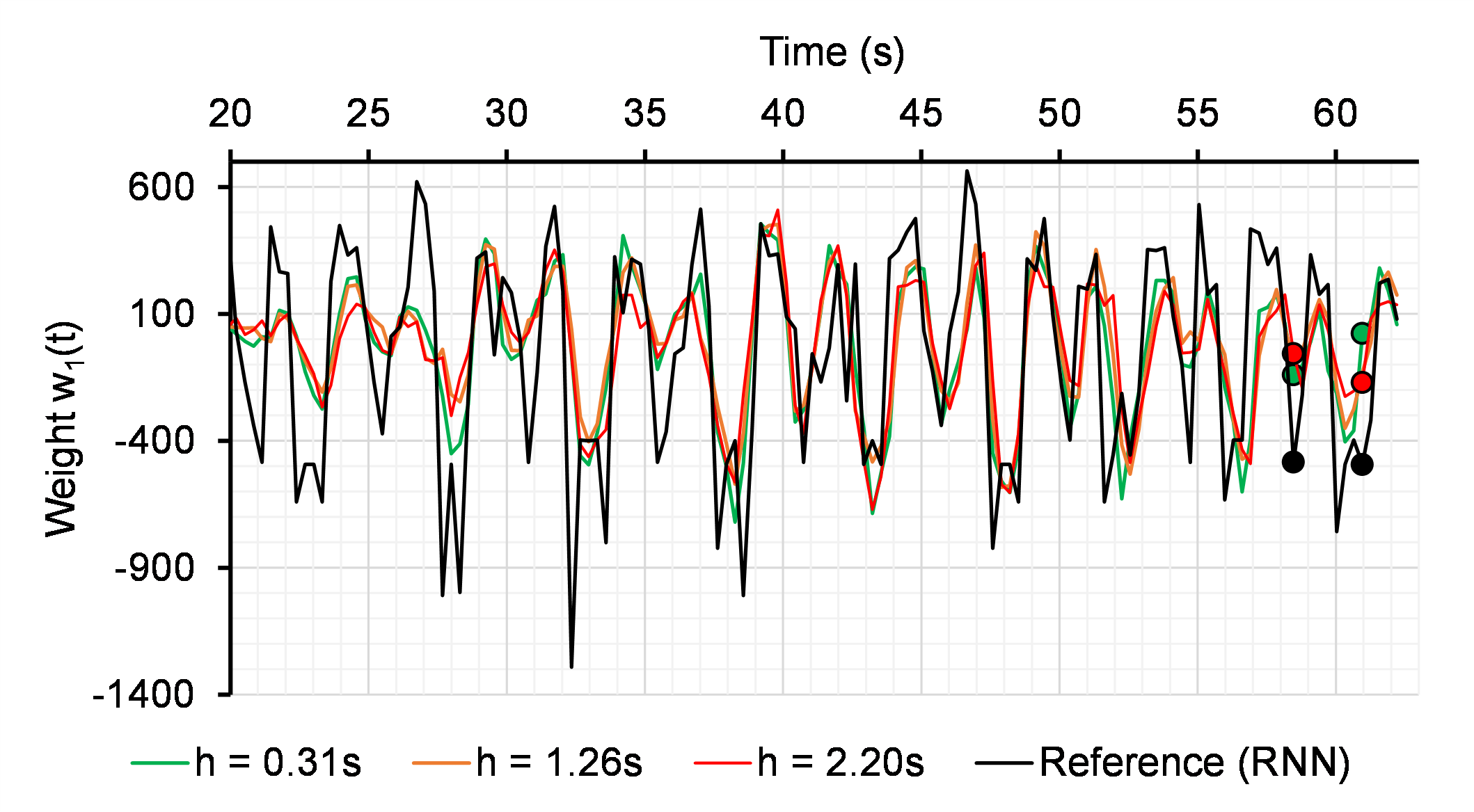}}% 
    \,
    \subfloat[Prediction of $w_1(t)$ with a population transformer]{\includegraphics[width=.49\textwidth, height=.28\textwidth]{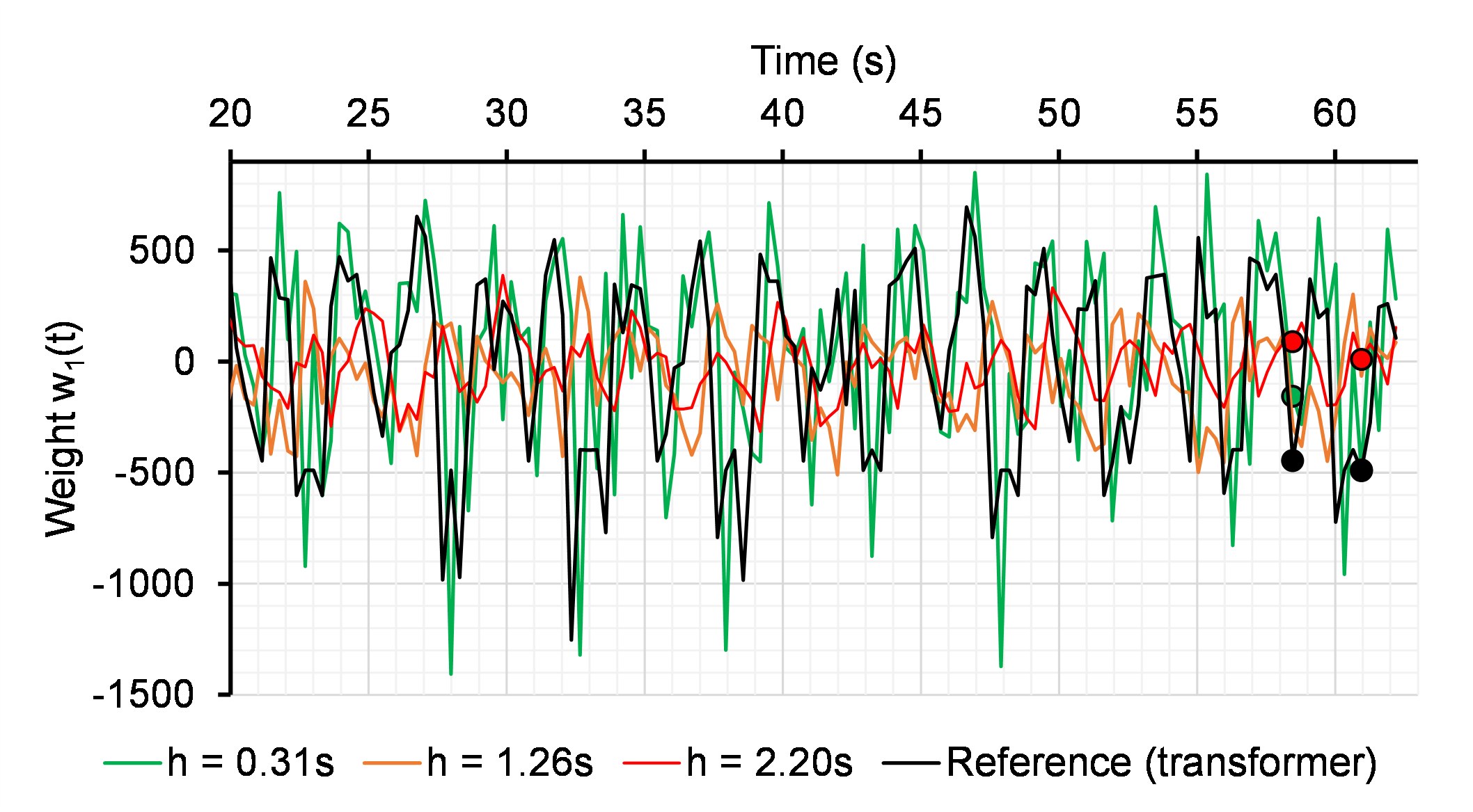}}%
    \\
    \subfloat[Prediction of $w_2(t)$ with \acs{RTRL}]{\includegraphics[width=.49\textwidth, height=.28\textwidth]{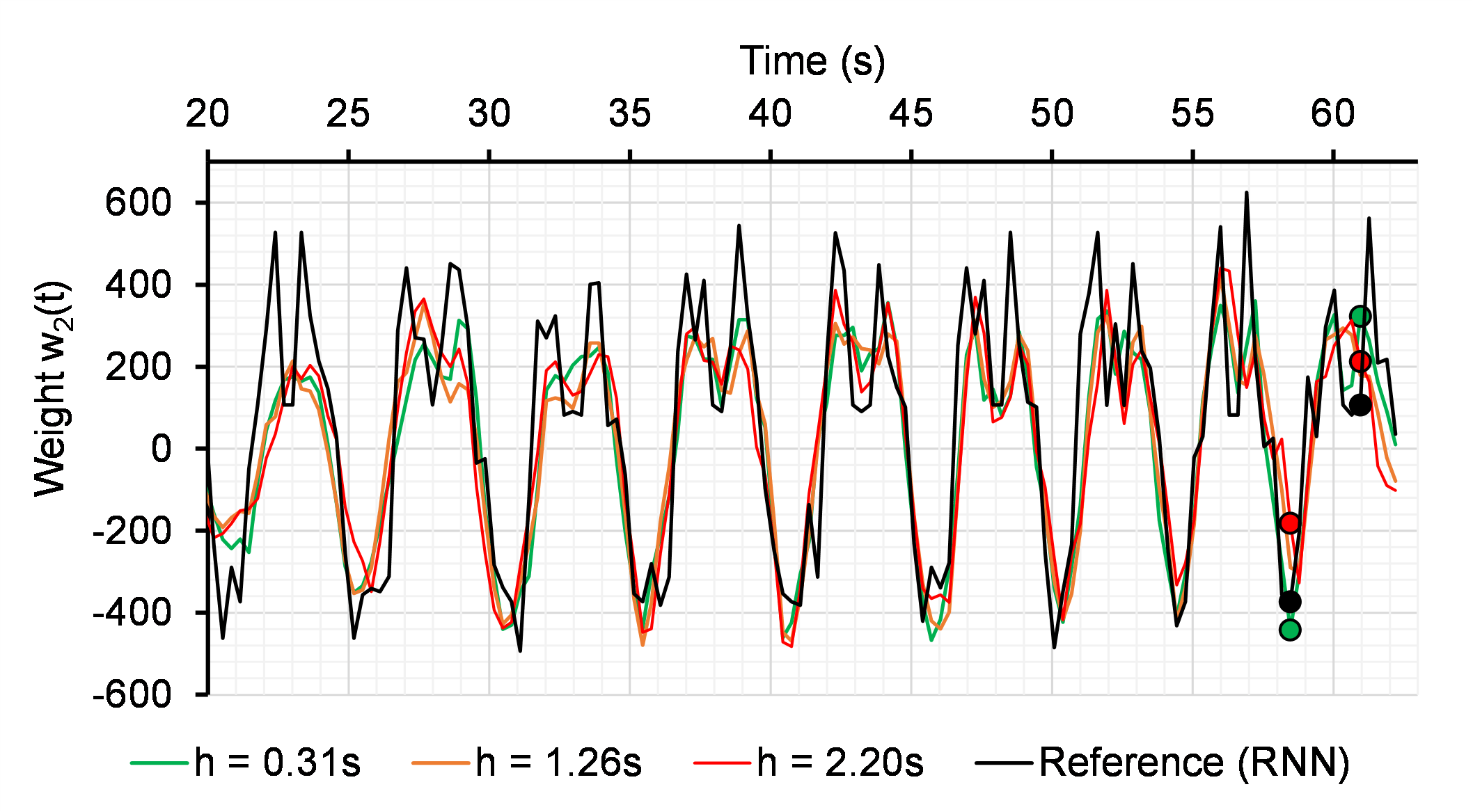}\label{fig:2nd PCA weight sq 1 RTRL prediction}}%
    \,
    \subfloat[Prediction of $w_2(t)$ with a population transformer]{\includegraphics[width=.49\textwidth, height=.28\textwidth]{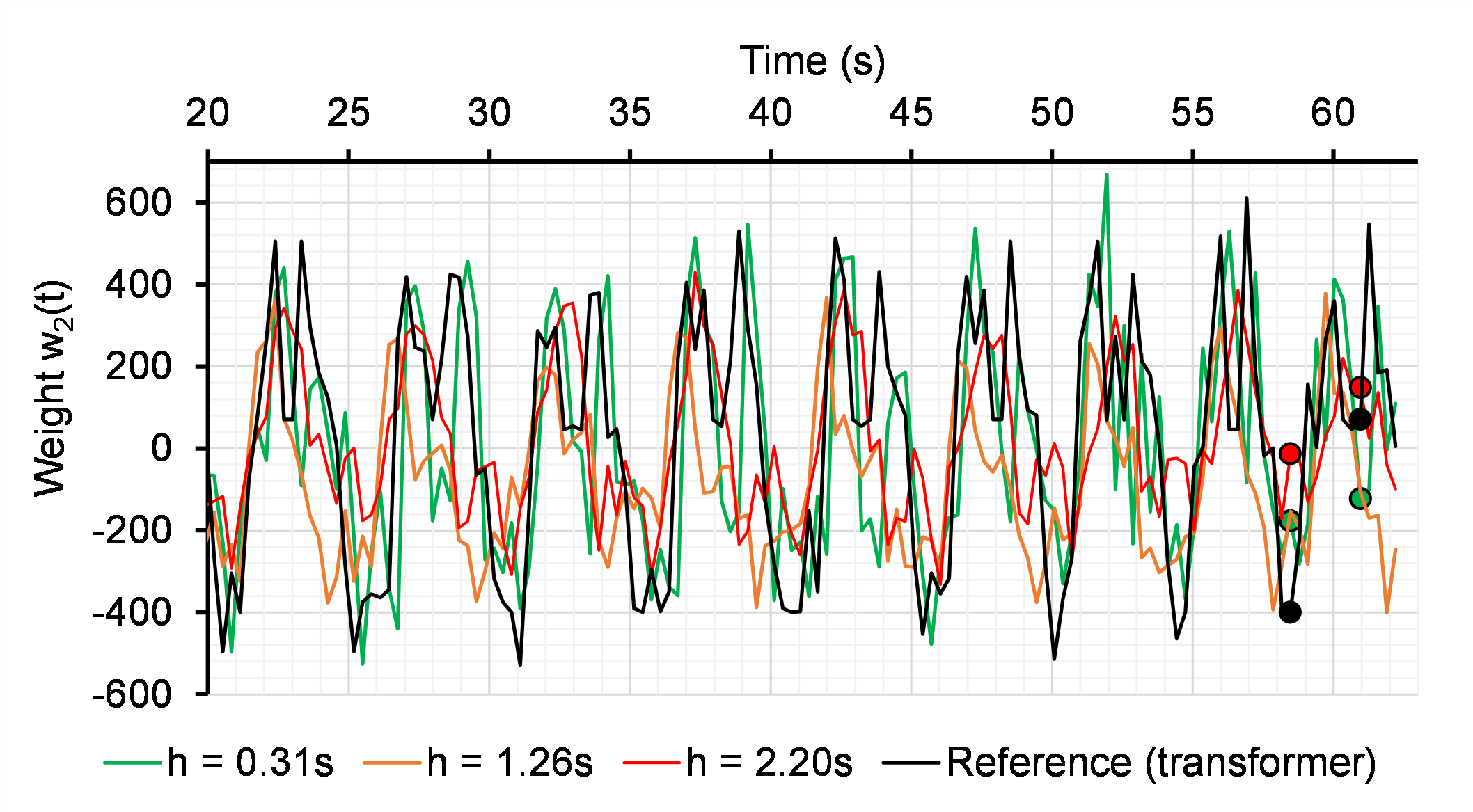}\label{fig:2nd PCA weight sq 1 pop transformer prediction}}%
    \\
    \subfloat[Prediction of $w_3(t)$ with \acs{RTRL}]{\includegraphics[width=.49\textwidth, height=.28\textwidth]{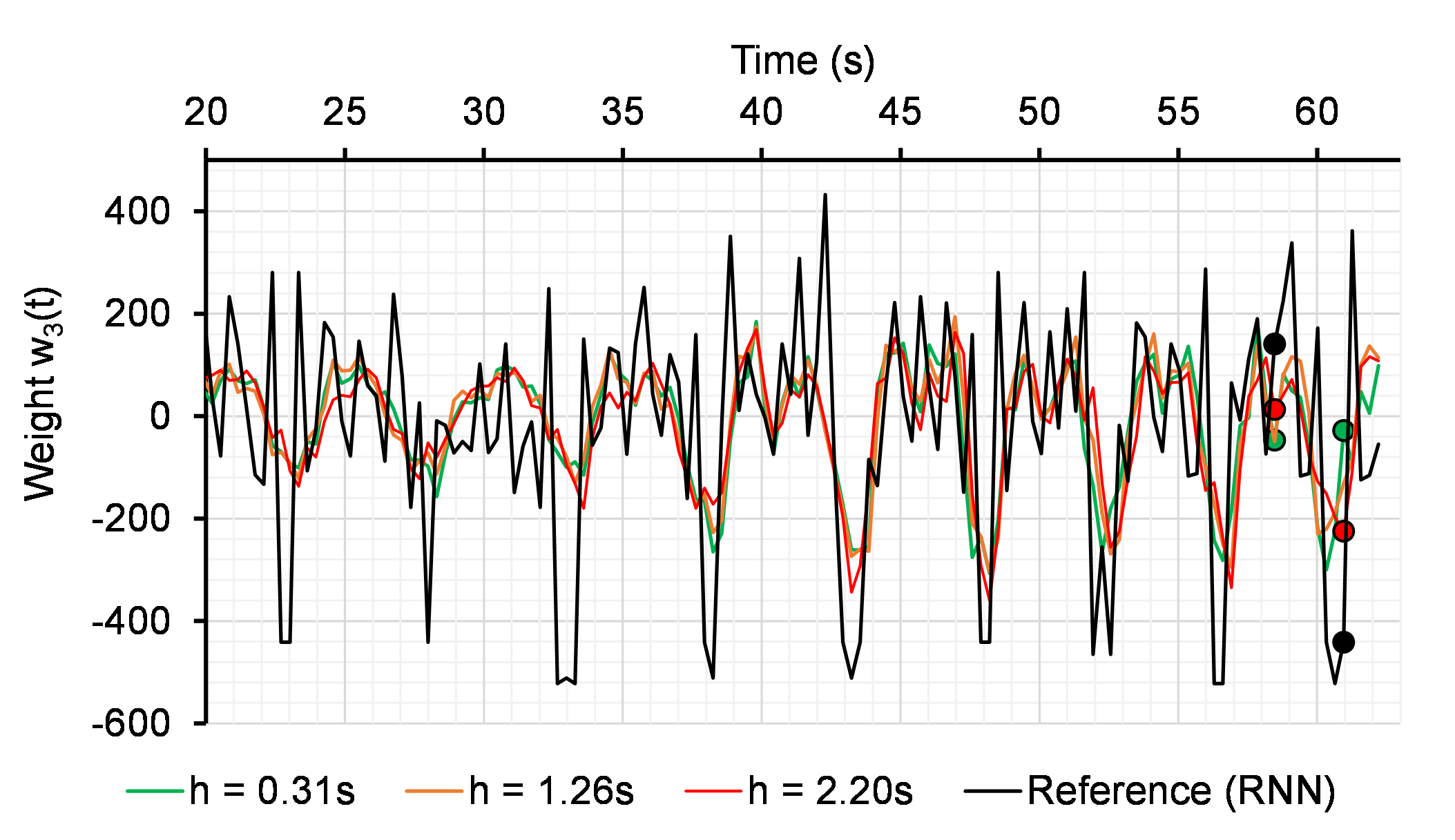}\label{fig:3rd PCA weight sq 1 RTRL prediction}}%
    \,
    \subfloat[Prediction of $w_3(t)$ with a population transformer]{\includegraphics[width=.49\textwidth, height=.28\textwidth]{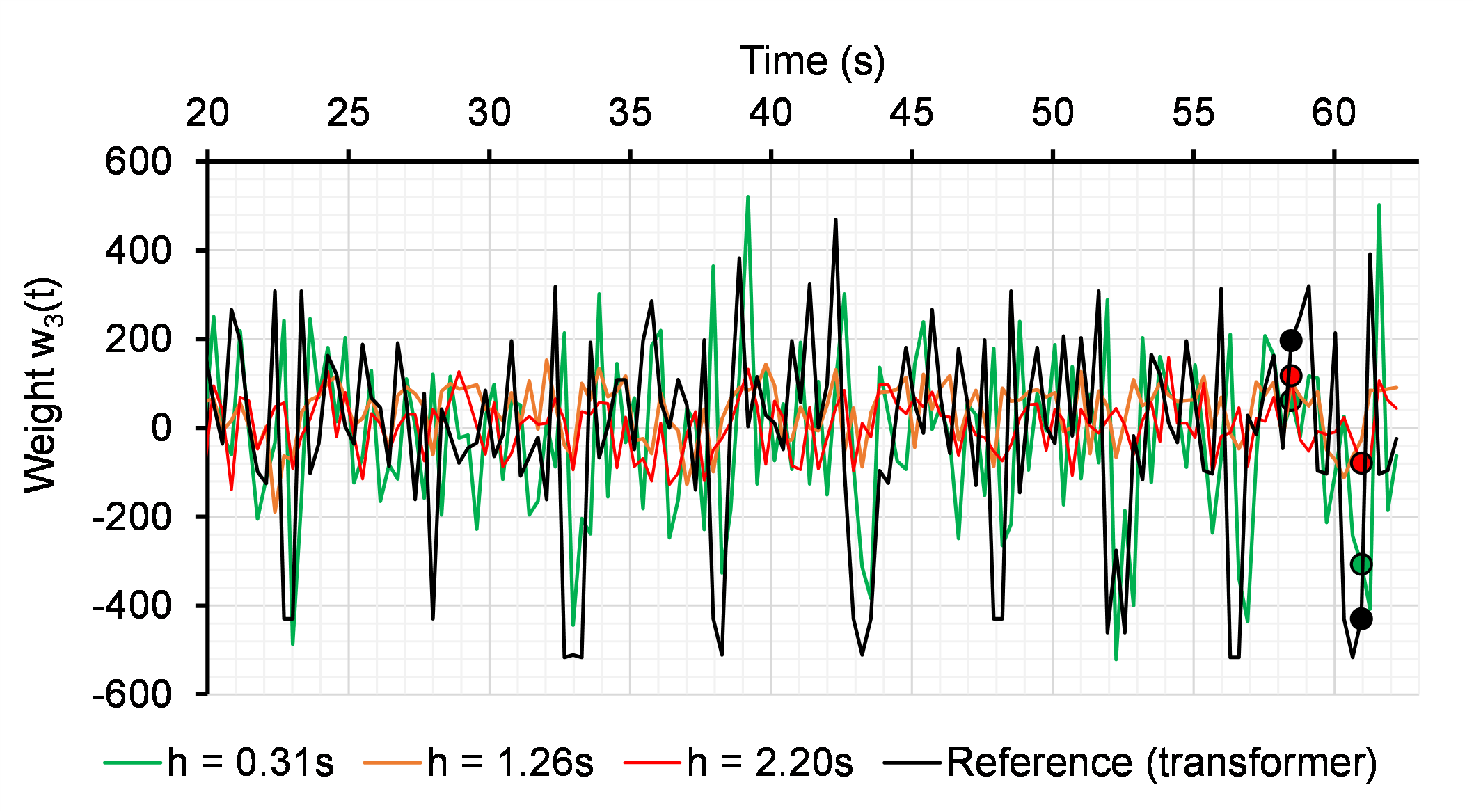}\label{fig:3rd PCA weight sq 1 pop transformer prediction}}% 
    \caption{Reference \acs{PCA} weights for sequence 1 of the ETH Zürich dataset, along with their predictions for several horizons $h$ using \pgls{RNN} trained with \gls{RTRL} and a population transformer trained on the \gls{OvGU} data. Hyperparameters, including $n_{\text{cp}}$ for \gls{RTRL}, were selected by grid search on the validation set for each value of $h$ (Sections \ref{section: PCA weight cross-validation} and \ref{section: methods - optimization of n_cp}). Accordingly, both algorithms were trained to forecast the leading three \gls{PCA} coefficients, except for \gls{RTRL} at $h=2.20\text{s}$, where $n_{\text{cp}} = 4$ yielded higher validation accuracy ($w_4(t)$ is not shown due to space constraints). The reference weights were computed using \gls{PCA} fitted to motion data from the first 28.3s and 50.4s of the sequence for \gls{RTRL} and the population transformer, respectively. Test-set evaluation only involves data beyond 56.6s, but earlier predictions are displayed to better assess prediction quality visually. Round markers at $t=58.4\text{s}$ and $t=60.9\text{s}$ correspond to the predicted images shown in Fig. \ref{fig:next frame pred sq 1 RTRL vs pop transformer}, as both figures depict the same runs.}
    \label{fig:PCA weights pred RTRL vs pop transformer}
\end{figure*}

% PCA weights
Correspondingly, the time-varying \acs{PCA} weights became noisier and their amplitude decreased as the component order $j$ increased (Fig. \ref{fig:DVF principal components and weights sequence 4}). Temporal noise and instability were more pronounced in the ETH Zürich sequences, owing to their lower acquisition rate and cardiac motion, which has a higher characteristic frequency. Respiratory motion was associated with the most regular and cyclical \gls{PCA} coefficients. In the ETH Zürich dataset, breathing variability in frequency and amplitude was higher in sequence 4 than in sequence 1, which was reflected in the oscillations of the second-order weight, primarily capturing respiratory motion in both sequences (Figs. \ref{fig:2nd PCA weight sq 4} and \ref{fig:2nd PCA weight sq 1 RTRL prediction}). Nonetheless, the peaks of most coefficients exhibited some degree of synchronization, suggesting that \acs{PCA} may not have fully isolated respiratory motion into a single deformation mode. Incidentally, the first-order coefficient in sequence 1 was very similar to the second-order coefficient in sequence 2 (not shown here for brevity), which was also related to liver deformations. This occurred even though the cross-sections differed visually, as both were extracted from the same \acs{4D}-\acs{MRI} acquisition. % Likewise, the first two PCA scores from sequence 3 resembled those from sequence 4. % true but removed to decrease text length because it adds no new conceptual information and the sequences aren’t displayed.
In each \gls{OvGU} sequence, the first-order weight trajectory followed the respiratory motion apparent in the cine-\gls{MR} images despite their high noise and low contrast, and captured irregularities in amplitude and frequency. Across both datasets, the coefficients (and components) corresponding to online and offline learning remained similar, even though $M_{\text{train}}$ was almost halved in the online setting (Figs. \ref{fig:PCA weights pred RTRL vs pop transformer}--\ref{fig:1st cpt prediction (SnAp-1 vs vs UORO vs transformer) Magdeburg}). This mainly stems from the inherent redundancy in the quasi-periodic nature of respiratory motion.

\subsection{Prediction of the time-dependent PCA weights}
\label{section:results weights forecasting}

\subsubsection{Quantitative evaluation on the ETH Zürich dataset}\label{section: prediction of weights on ETH Zurich data}

In this section, we report the performance of \gls{PCA}-weight predictors on the ETH Zürich acquisitions. To compare all algorithms using identical time series, \gls{PCA} is fitted separately to the first 28.3s of each sequence with $n_{\text{cp}}=3$.\footnote{This differs from the end-to-end frame-prediction pipeline (Sections \ref{section: PCA weight cross-validation} and \ref{section: methods - optimization of n_cp}), where offline models use weights obtained from \gls{PCA} fitted on the first 50.4s of each sequence and $n_{\text{cp}}$ is selected via a validation-based procedure.}~Sequence-specific offline algorithms are still trained on the first 50.4s of each sequence.

% \subsubsubsection{PCA weight prediction accuracy of the test set}%\label{section:PCA weights prediction accuracy}

\begin{table}[htb!]
\setlength{\tabcolsep}{1.2pt}
\begin{tabular}{ll}
\hline
Prediction method                                  & Test \acs{nRMSE} \\
\hline 
\acs{UORO}                                         & 0.895 $\pm$ 0.081 \\
\acs{SnAp-1}                                       & \textbf{0.857 $\pm$ 0.100} \\
\acs{DNI}                                          & 0.887 $\pm$ 0.079 \\
\acs{RTRL}                                         & \textbf{0.859 $\pm$ 0.098} \\
\acs{LMS}                                          & 0.898 $\pm$ 0.134 \\
Linear regression                                  & 0.928 $\pm$ 0.063 \\
Sequence-specific transformer                      & 0.879 $\pm$ 0.048 \\
Population transformer (training: \acs{OvGU} data)  & 1.171 $\pm$ 0.084 \\                
\acs{RNN} with a frozen hidden layer               & 0.977 $\pm$ 0.019 \\   
Latest \gls{PCA} weight as prediction (persistence)& 1.458 $\pm$ 0.010 \\                 
\hline
\end{tabular}
\caption{Test-set \acs{nRMSE} associated with the prediction of the first three time-dependent \acs{PCA} weights in the ETH Zürich dataset for each algorithm (Eq. \ref{eq:predicted weights nRMSE} with $k_{\text{min}} = M_{\text{val}} + 1 = 181$ and $k_{\text{max}} = 200$). Error values are averaged over horizons between 0.32s and 2.20s, the four sequences, and $n_{\text{test}}^{\text{PCA}}$ runs to account for neural-network stochasticity (Table \ref{table:general experimental setup}). Each error reported in the second column corresponds to the mean of the horizon-wise errors shown in Fig. \ref{fig:signal pred error 3 PCA cpts vs horizon}. Hyperparameters were selected by grid search on the validation set for each sequence and horizon individually, except the population transformer for which selection was horizon-wise only (Section \ref{section: PCA weight cross-validation}). The 70\% \glspl{CI} using the Student t-distribution with 3 degrees of freedom are also reported. The two lowest \glspl{nRMSE} are bolded for readability.}
\label{table:signal pred nRMSE 3 PCA cpts avg over horizon}
\end{table}

\begin{figure}%[thb!]
    \centering
	\includegraphics[width=0.8\columnwidth]{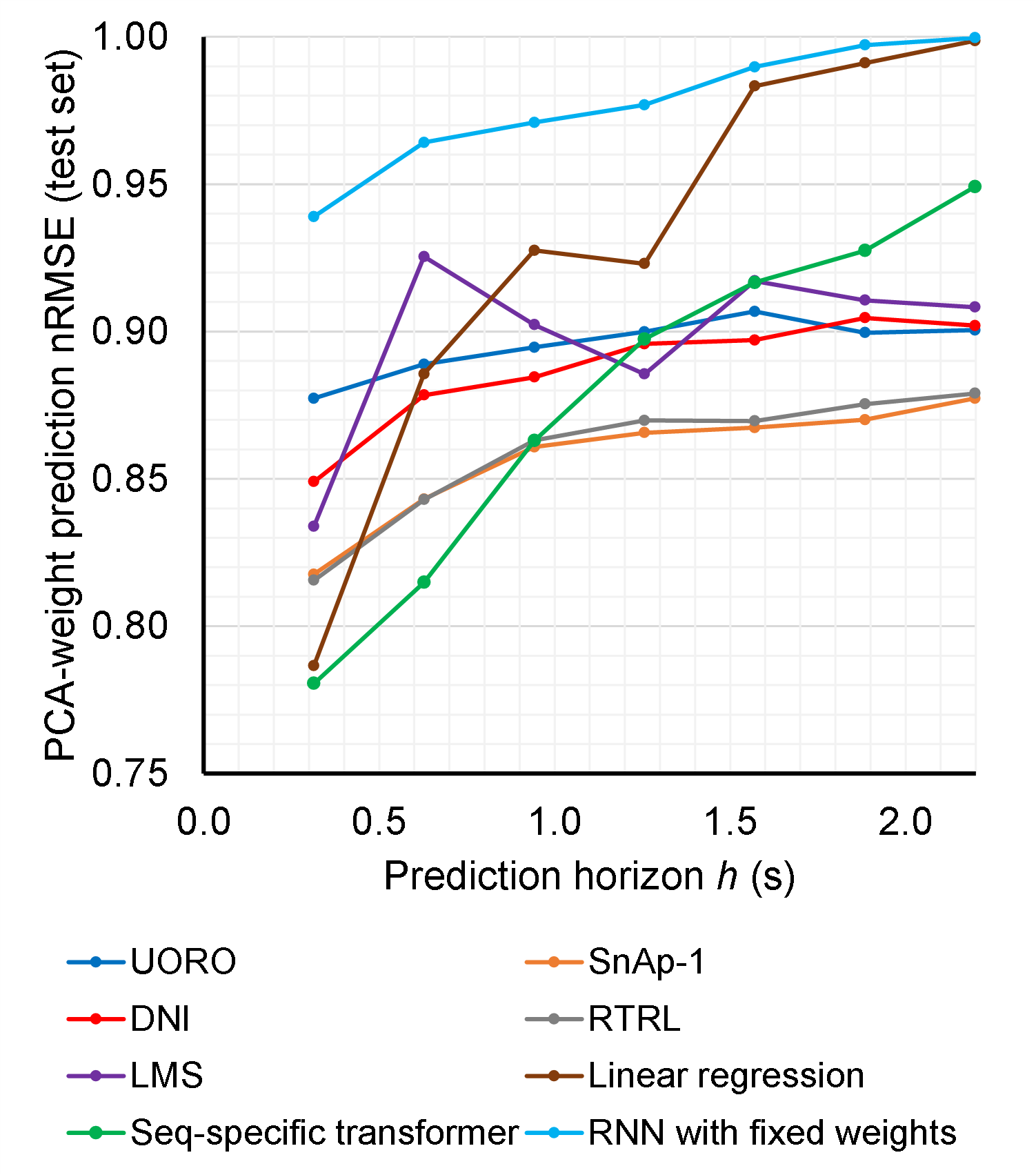} %          
    \caption{Test-set \acs{nRMSE} between the first three reference time-dependent \acs{PCA} weights relative to the ETH Zürich dataset and their prediction with several algorithms as a function of the horizon $h$ (Eq. \ref{eq:predicted weights nRMSE}). Each point represents the error for a given value of $h$, averaged over the four sequences and $n_{\text{test}}^{\text{PCA}}$ runs to account for neural-network stochasticity (Table \ref{table:general experimental setup}). The population transformer and the baseline using the latest incoming weight as the prediction were not shown, as the corresponding errors were relatively high. Hyperparameters were tuned by grid search on the validation set for each value of $h$.}
    \label{fig:signal pred error 3 PCA cpts vs horizon}
\end{figure}

Performance is quantified using the test-set \gls{nRMSE} between predicted and ground-truth time-dependent \gls{PCA} weights (Eq. \ref{eq:predicted weights nRMSE} with $k_{\text{min}} = M_{\text{val}}+1 = 181$ and $k_{\text{max}} = 200$). For neural networks, metrics are averaged over $n_{\text{test}}^{\text{PCA}}$ evaluation runs; for simplicity, we set $n_{\text{test}}^{\text{PCA}}$ equal to $n_{\text{val}}$. \Gls{SnAp-1} and \gls{RTRL} achieved the lowest \glspl{nRMSE} (averaged over the four sequences and horizon values examined, $h \leq  2.20\text{s}$), followed by the sequence-specific transformer, \gls{DNI}, and then \gls{UORO} (Table \ref{table:signal pred nRMSE 3 PCA cpts avg over horizon}). The naive predictor using the latest weight as the prediction (i.e., the \acs{PCA}-score persistence model) yielded the worst accuracy, followed by the population transformer. The population transformer significantly underperformed relative to the sequence-specific models, with non-overlapping 70\% \glspl{CI}. However, the \glspl{CI} associated with the latter models intersected, likely due to the limited number of sequences ($N=4$). Errors generally increased with $h$ (Fig. \ref{fig:signal pred error 3 PCA cpts vs horizon}). The sequence-specific transformer yielded the lowest sequence-mean \glspl{nRMSE} at $h=0.31\text{s}$ and $h=0.62\text{s}$ (0.781 and 0.815, respectively); that of the linear \gls{AR} model at $h=0.31\text{s}$ (0.787) was comparable. \gls{SnAp-1} outperformed all other algorithms for $h \geq 0.94\text{s}$; its \gls{nRMSE} remained below 0.879. It was nearly matched by \gls{RTRL}, which suggests that the diagonal approximation of the influence matrix was accurate. Linear regression had lower accuracy than all \gls{RNN} predictors for $h \geq 0.94\text{s}$, except the baseline with a frozen hidden layer.

% \subsubsubsection{Hyperparameter influence on accuracy}

\begin{figure*}[pos=htbp, align=\centering]%
    \captionsetup[subfloat]{farskip=2pt,captionskip=1pt} % {farskip=2pt,captionskip=1pt}
    \centering
    \subfloat[Influence of $\eta$ on UORO accuracy]{\includegraphics[width=.31\textwidth]{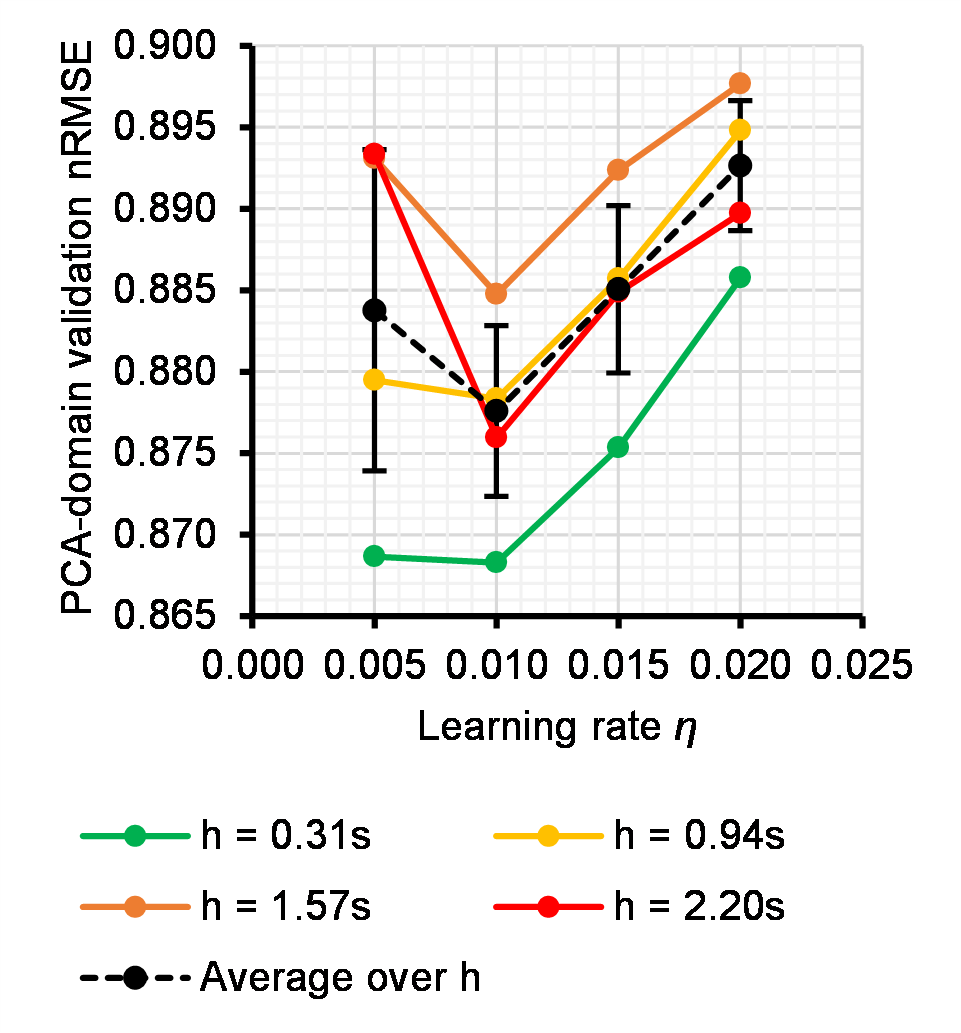}}
    \quad
    \subfloat[Influence of $\eta$ on SnAp-1 accuracy]{\includegraphics[width=.31\textwidth]{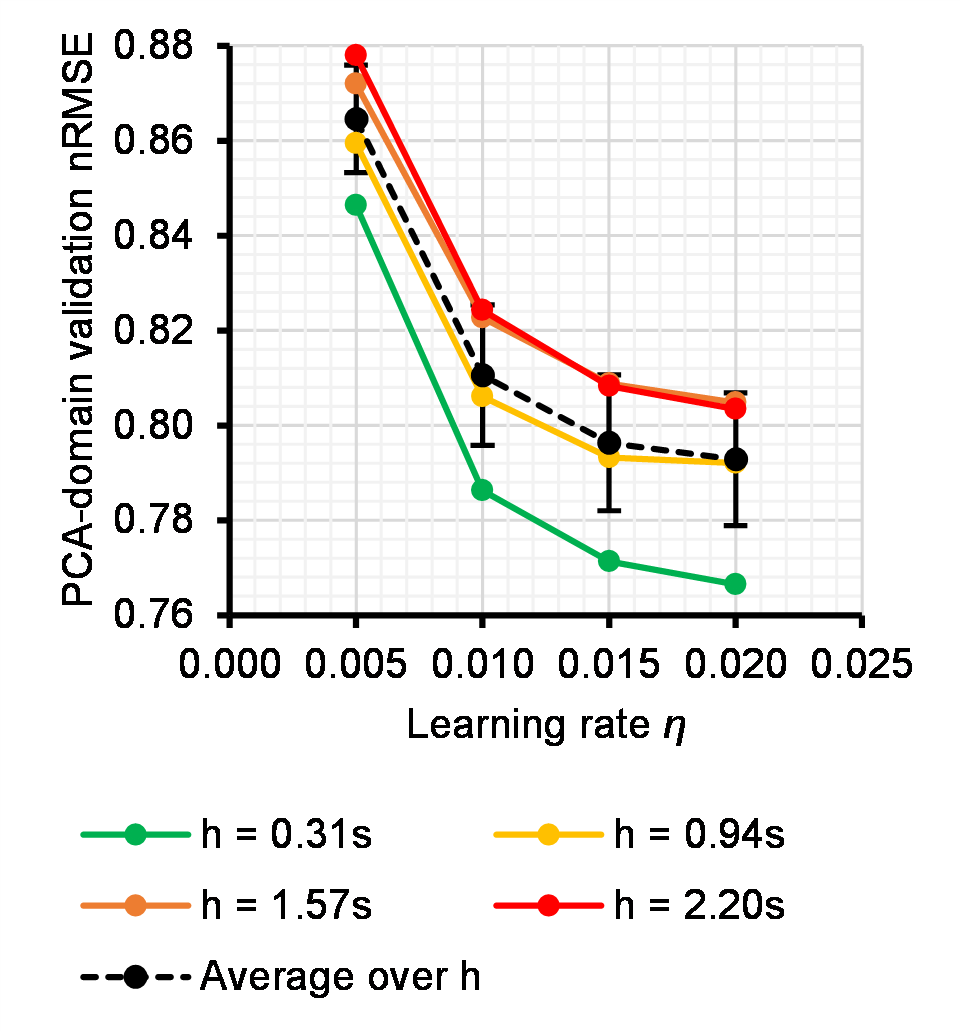}}
    \quad
    \subfloat[Influence of $\eta$ on DNI accuracy]{\includegraphics[width=.31\textwidth]{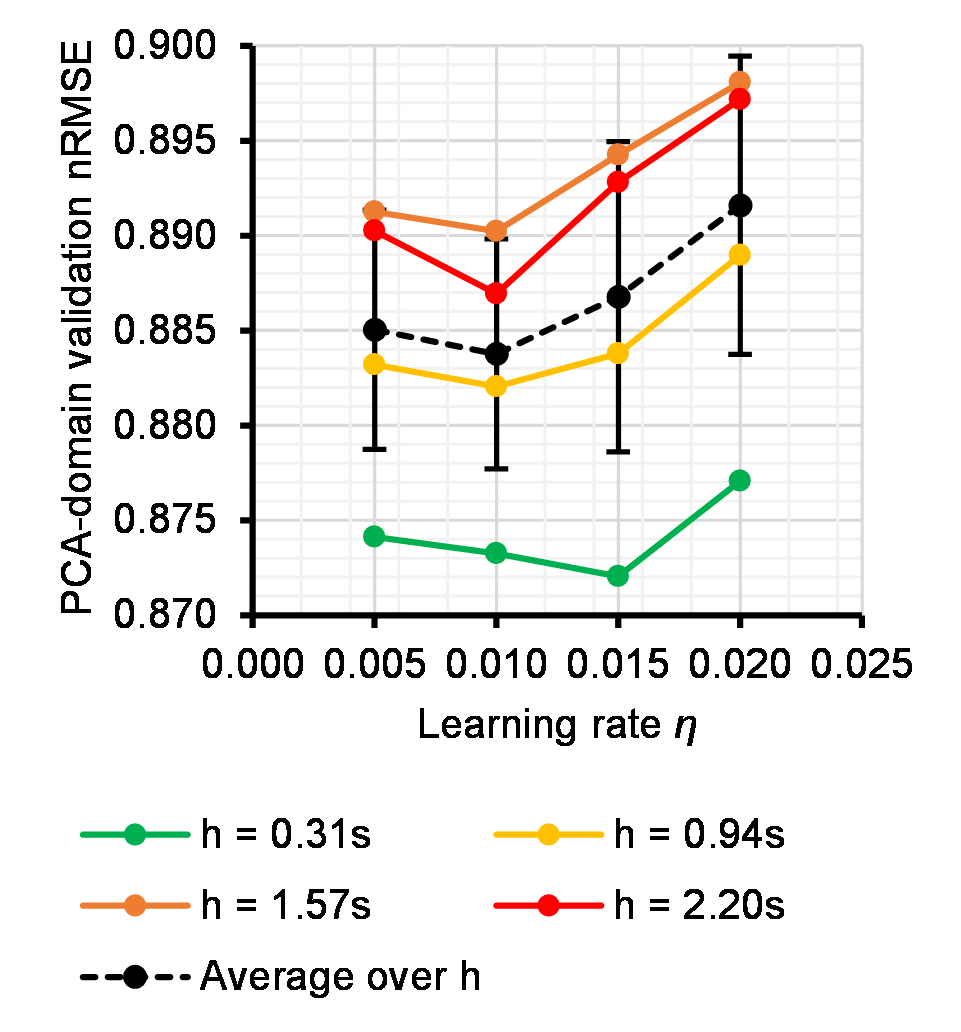}}
    \quad
    \subfloat[Influence of $d$ on UORO accuracy]{\includegraphics[width=.31\textwidth]{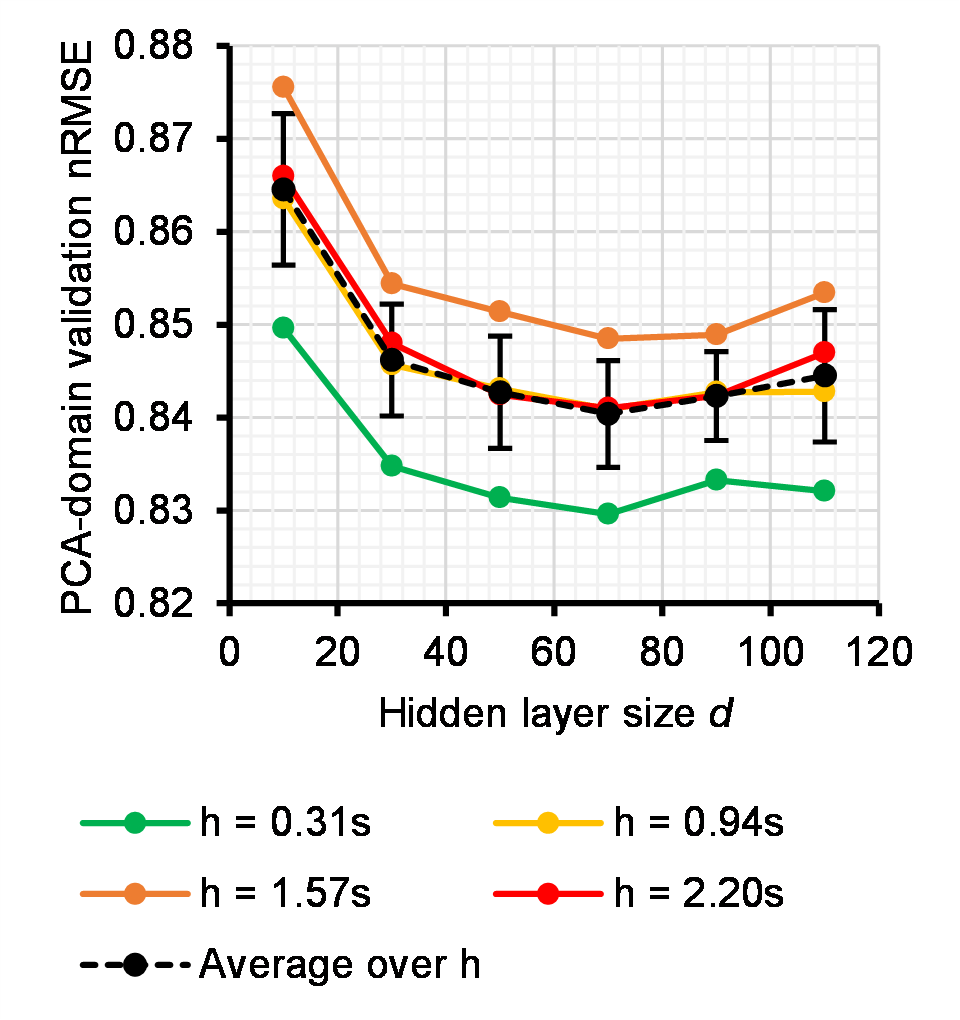}}
    \quad
    \subfloat[Influence of $d$ on SnAp-1 accuracy]{\includegraphics[width=.31\textwidth]{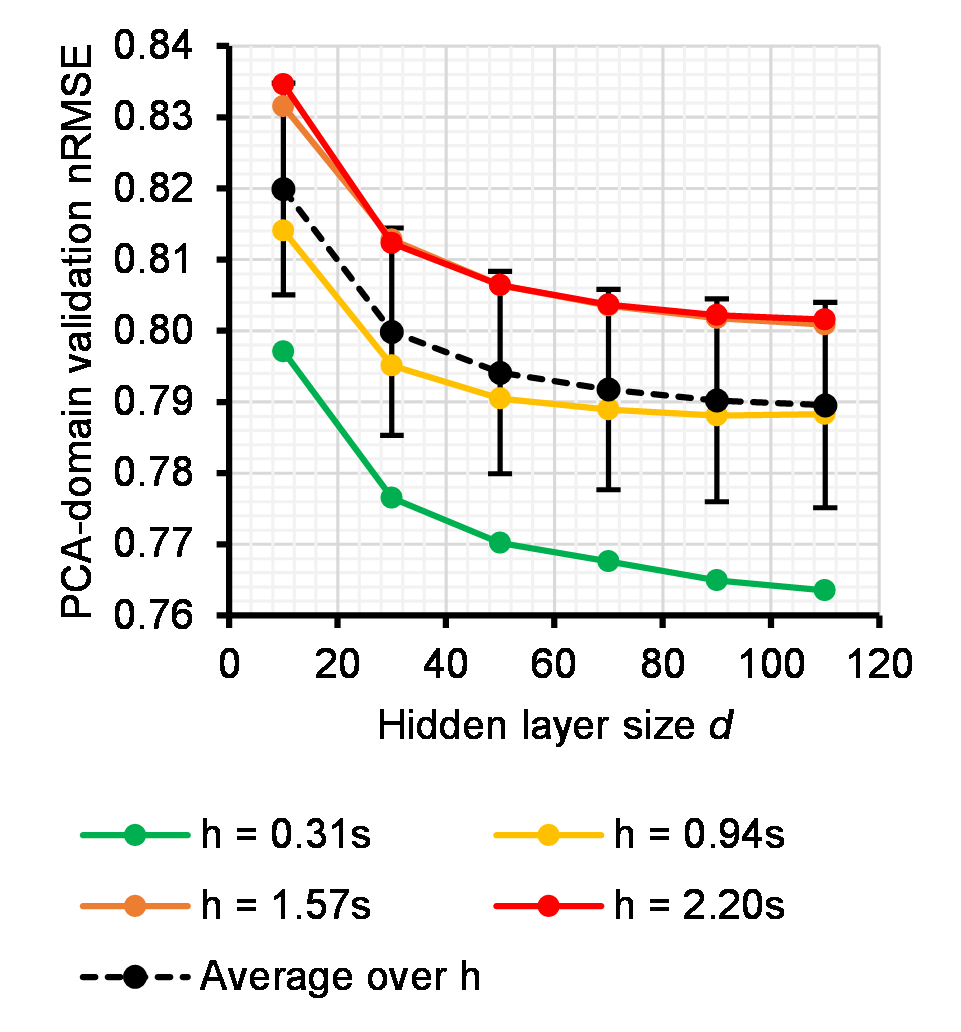}}
    \quad
    \subfloat[Influence of $d$ on DNI accuracy]{\includegraphics[width=.31\textwidth]{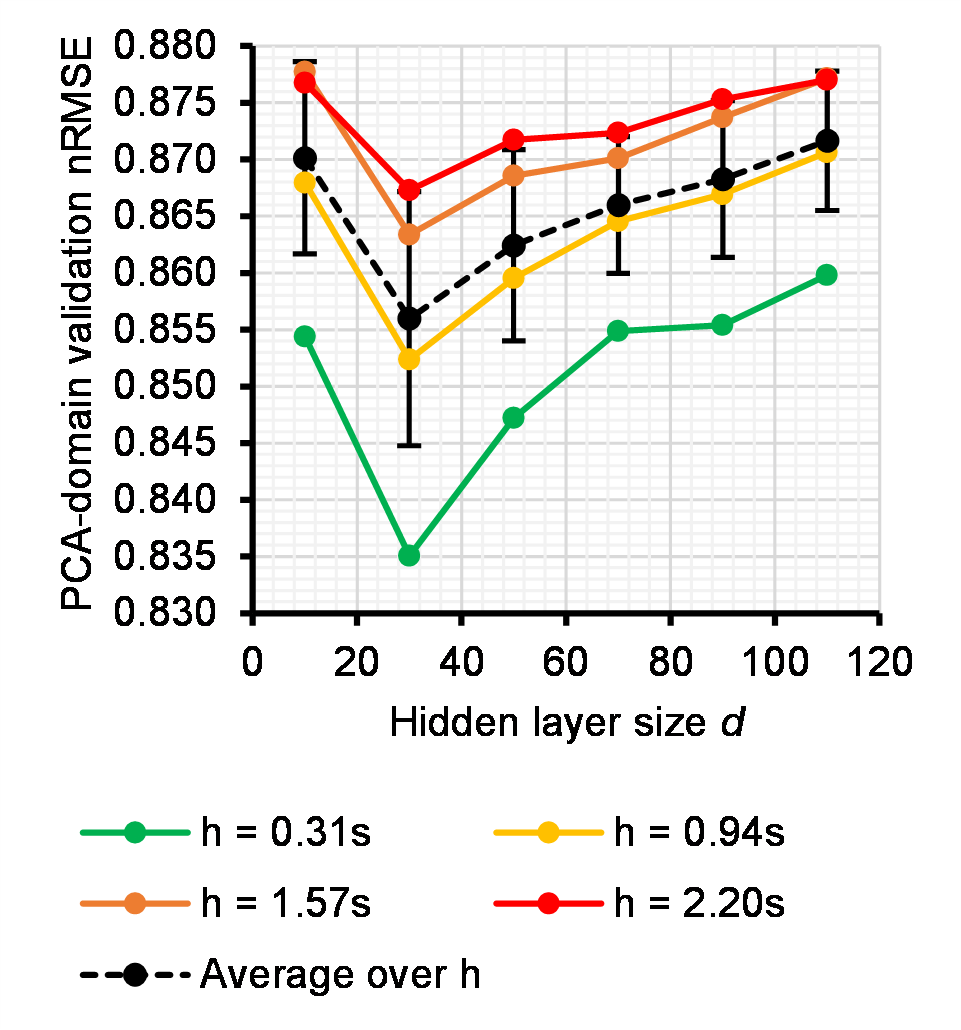}}
    \quad
    \subfloat[Influence of $L$ on UORO accuracy]{\includegraphics[width=.31\textwidth]{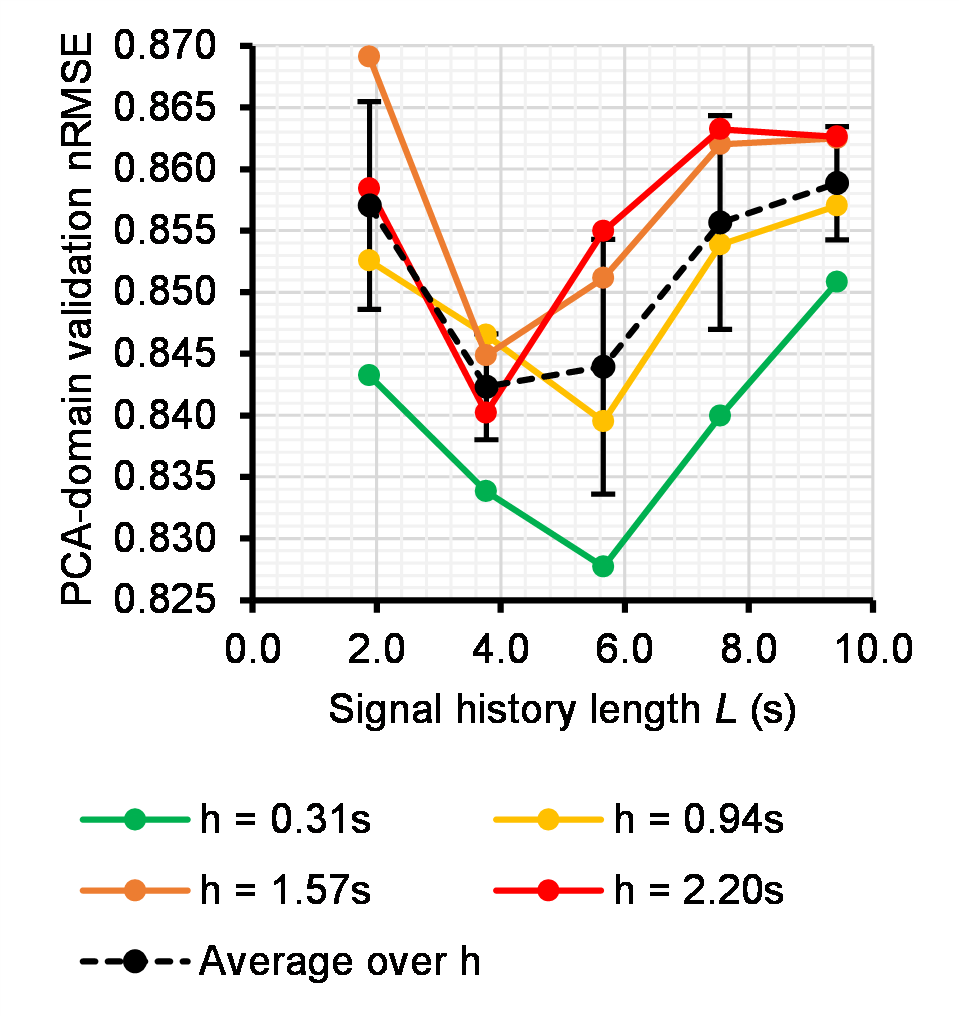}}
    \quad
    \subfloat[Influence of $L$ on SnAp-1 accuracy]{\includegraphics[width=.31\textwidth]{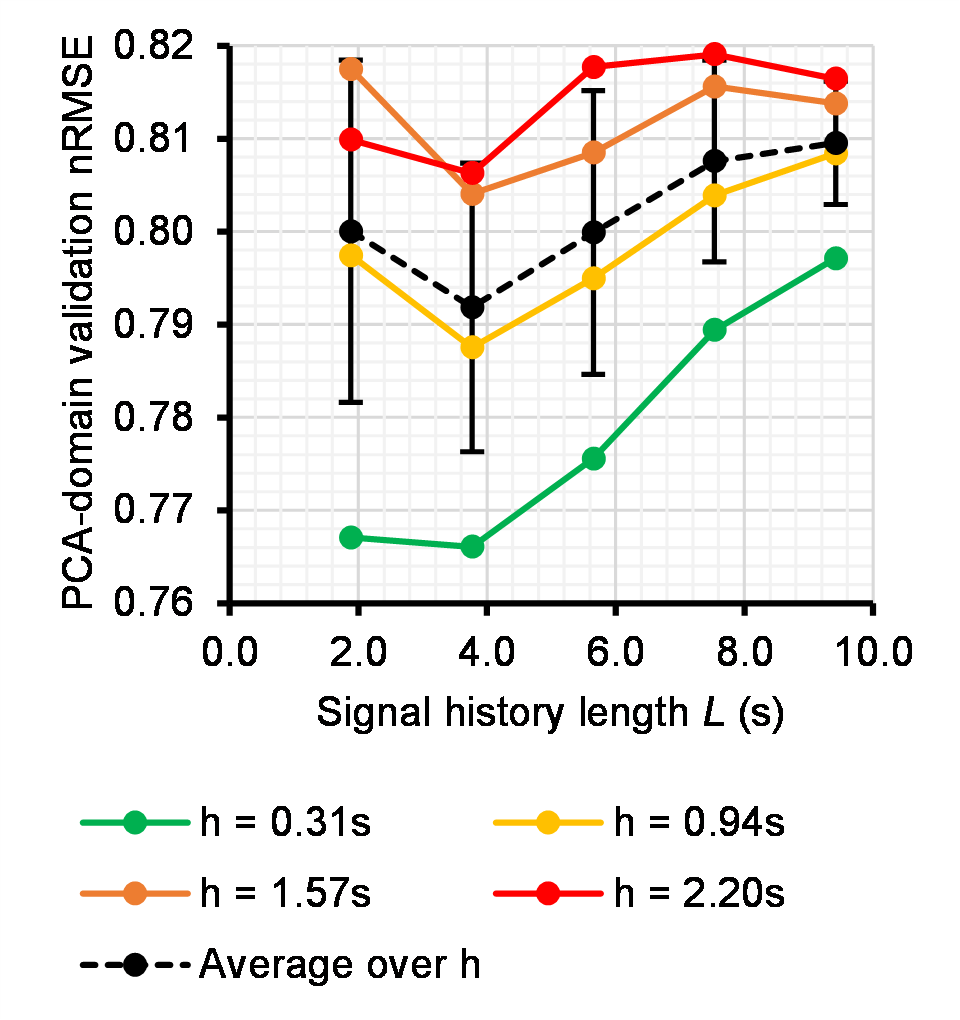}}
    \quad
    \subfloat[Influence of $L$ on DNI accuracy]{\includegraphics[width=.31\textwidth]{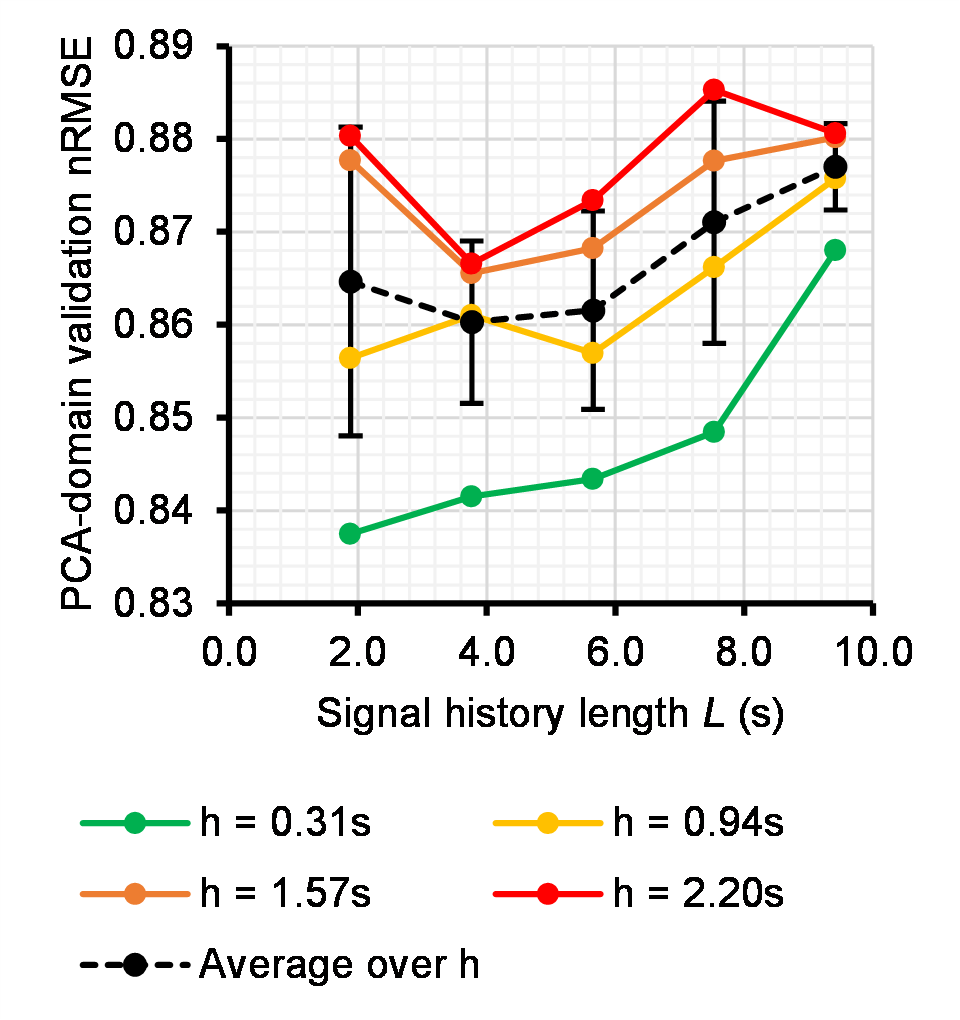}}
    \quad
    \caption{Validation \acs{nRMSE} between the first three ground-truth time-dependent \acs{PCA} weights relative to the ETH Zürich dataset and their prediction using \pgls{RNN} trained with \gls{UORO}, \gls{SnAp-1}, and \gls{DNI} (Eq. \ref{eq:predicted weights nRMSE}). For a given hyperparameter, each colored point in the corresponding graph represents the minimum, over all combinations of the other hyperparameters in the search grid, of the \acs{nRMSE} first averaged across the four sequences and $n_{\text{val}} = 250$ runs, at a specific horizon $h$. The black dotted curves and associated error bars represent the average and standard deviation, respectively, of these error minima averaged over $h$ between 0.31s and 2.20s.}
    \label{fig:hyperpar influence on PCA cpts prediction}
\end{figure*}

%Fig. \ref{fig:hyperpar influence on PCA cpts prediction} describes the influence of hyperparameters on the cross-validation \gls{nRMSE} for \gls{UORO}, \gls{SnAp-1}, and \gls{DNI}.
The validation \glspl{nRMSE} for \gls{UORO}, \gls{SnAp-1}, and \gls{DNI} (Eq. \ref{eq:predicted weights nRMSE} with $k_{\text{min}} = M_{\text{train}}+1 = 161$ and $k_{\text{max}} = M_{\text{val}} = 180$) also tended to increase with $h$ (Fig. \ref{fig:hyperpar influence on PCA cpts prediction}). These three algorithms attained the minimum of the sequence-mean validation \gls{nRMSE} (averaged over $h \leq 2.20\text{s}$) at $L=12$ time steps, corresponding to a history window of 3.77s. Furthermore, the value of $\eta$ that minimized this error was 0.01 for both \gls{UORO} and \gls{DNI} and 0.02 for \gls{SnAp-1}. The validation error of \gls{SnAp-1} decreased with both $\eta$ and $d$, irrespective of $h$. % Optimizing $\eta$ led to the most significant average \gls{nRMSE} decrease for \gls{SnAp-1} (a 8.3\% relative decrease), from 0.865 at $h=0.31\text{s}$ to 0.793 at $h=2.20\text{s}$. % maybe last sentence not so useful 

\subsubsection{Qualitative evaluation}\label{section: PCA weight forecasting qualitative eval}

In what follows, $n_{\text{cp}}$ is selected using the validation set, and the number of frames used to fit \gls{PCA} depends on the type of training algorithm (online or offline), as detailed in Sections \ref{section: PCA weight cross-validation} and \ref{section: methods - optimization of n_cp}. In sequence 1 of the ETH Zürich dataset, the predictions from \acs{RTRL} appeared visually correct overall. However, their extrema did not perfectly match those of the ground truth, especially for the third-order weight, possibly due to the short acquisition time and low sampling rate (Fig. \ref{fig:PCA weights pred RTRL vs pop transformer}). 
% This may lead to a lower breathing extent in the corresponding predicted \gls{2D} frames.  % it looks like this sentence about predicted images should appear later and not at this stage.
By comparison, the predictions of the population transformer seemed less accurate. At $h=0.31\text{s}$, they captured the general trends but were overly oscillatory, often overshooting near the extrema. At medium-to-long horizons, forecasts for the second-order weight remained moderately faithful overall, yet with suboptimal extrema, whereas those corresponding to the other weights were low-amplitude and inaccurate.

On the \gls{OvGU} dataset, the subject-specific transformer was qualitatively less accurate than \gls{UORO}, \gls{SnAp-1}, and \gls{DNI} at $h=1.66\text{s}$ (Fig. \ref{fig:1st cpt prediction (SnAp-1 vs vs UORO vs transformer) Magdeburg}). This relatively long horizon and the stronger breathing irregularities in this dataset likely amplified the negative impact of training-data scarcity on offline algorithms compared with online algorithms.
% By contrast, online learning algorithms can learn on-the-fly, hence they have access to more data over time and adapt to the most recent trends. % discussion-ish and tautological
Motion variability was generally higher near \gls{EI}, making prediction more challenging at that phase.
% during inspiration, making the \gls{EI} phase harder to predict. 
Forecasts appeared less accurate in sequence 4, as it featured more pronounced amplitude fluctuations, including deep inspirations around $t=28\text{s}$ and $t=52\text{s}$. Although the three \gls{RNN} algorithms above performed reasonably well across all breathing traces, they also exhibited substantial high-frequency oscillations, for instance, near the large-amplitude irregularities in sequence 4. Early \acs{RNN} forecasts occasionally had lower amplitudes, which reflects the adaptation (warm-up) period typical of online training from scratch before convergence stabilization (cf. \gls{RTRL} and \gls{UORO} predictions in Figs. \ref{fig:PCA weights pred RTRL vs pop transformer} and \ref{fig:1st PCA component prediction in Sq 4 of OgVU dataset}, respectively).  %Nonetheless, accurate prediction of respiratory motion with limited training data and $h=1.66\text{s}$, which is high relative to the acquisition rate (6Hz), is already a non-trivial problem.

\subsection{Optimization of the PCA-subspace dimension}\label{section: optimization of nb of PCA cpts}

\begin{figure*}[pos=htbp,width=\textwidth,align=\centering]%
	% \captionsetup[subfloat]{justification=raggedright,singlelinecheck=false}    
    \centering
    \subfloat[Prediction with \acs{UORO}]{\includegraphics[width=.25\textwidth]{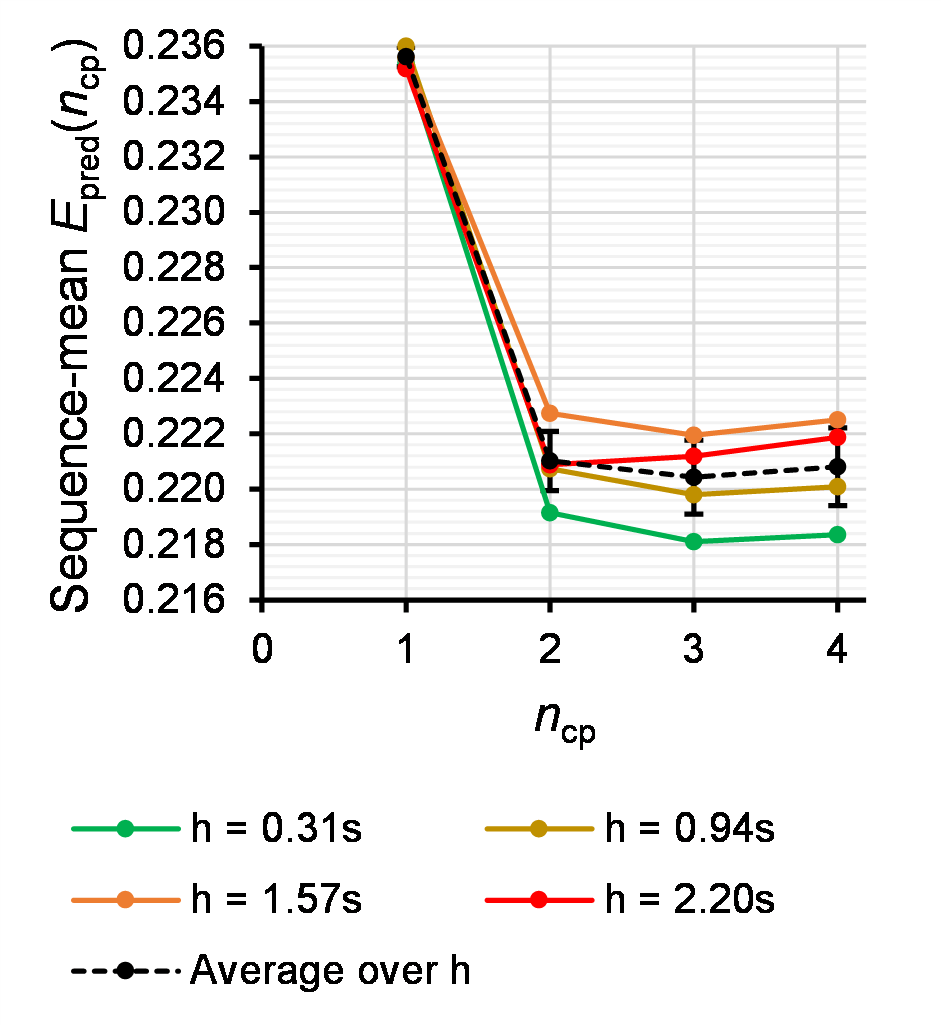}}%
    %\quad
    \subfloat[Prediction with \acs{SnAp-1}]{\includegraphics[width=.25\textwidth]{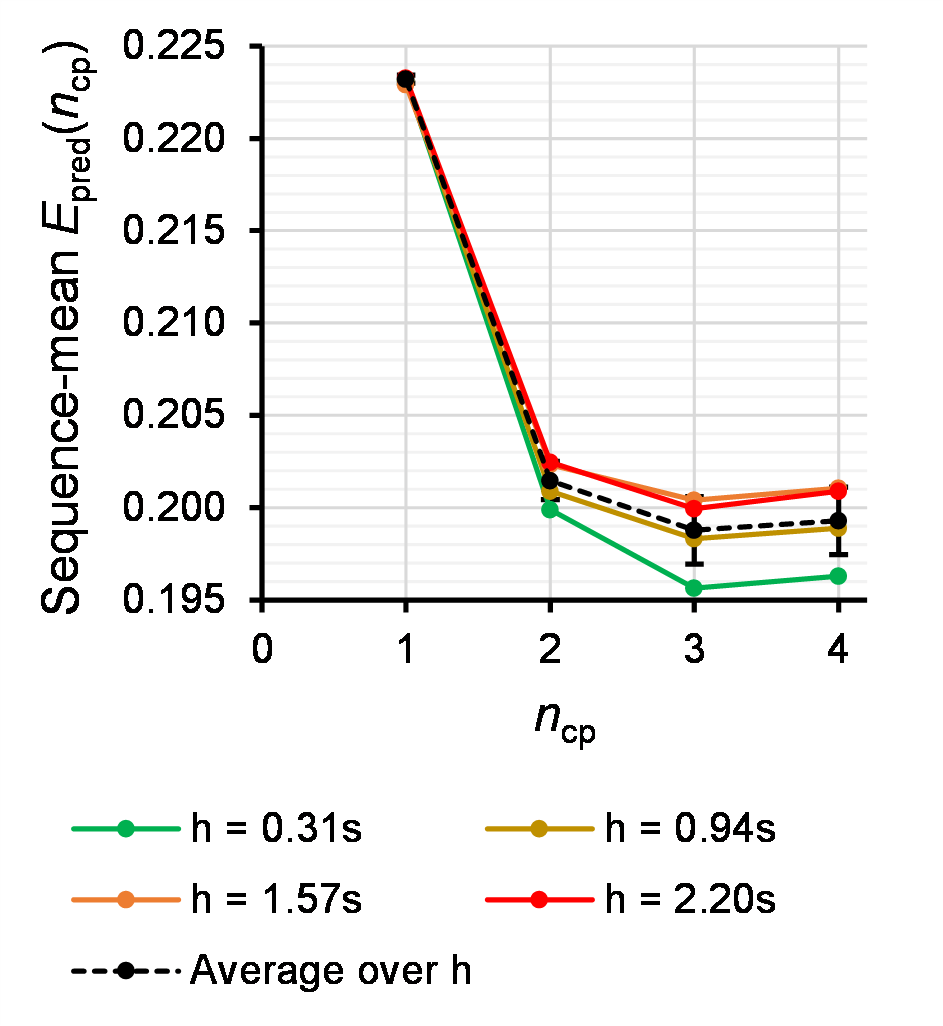}}%
    %\quad
    \subfloat[Prediction with \acs{DNI}]{\includegraphics[width=.25\textwidth]{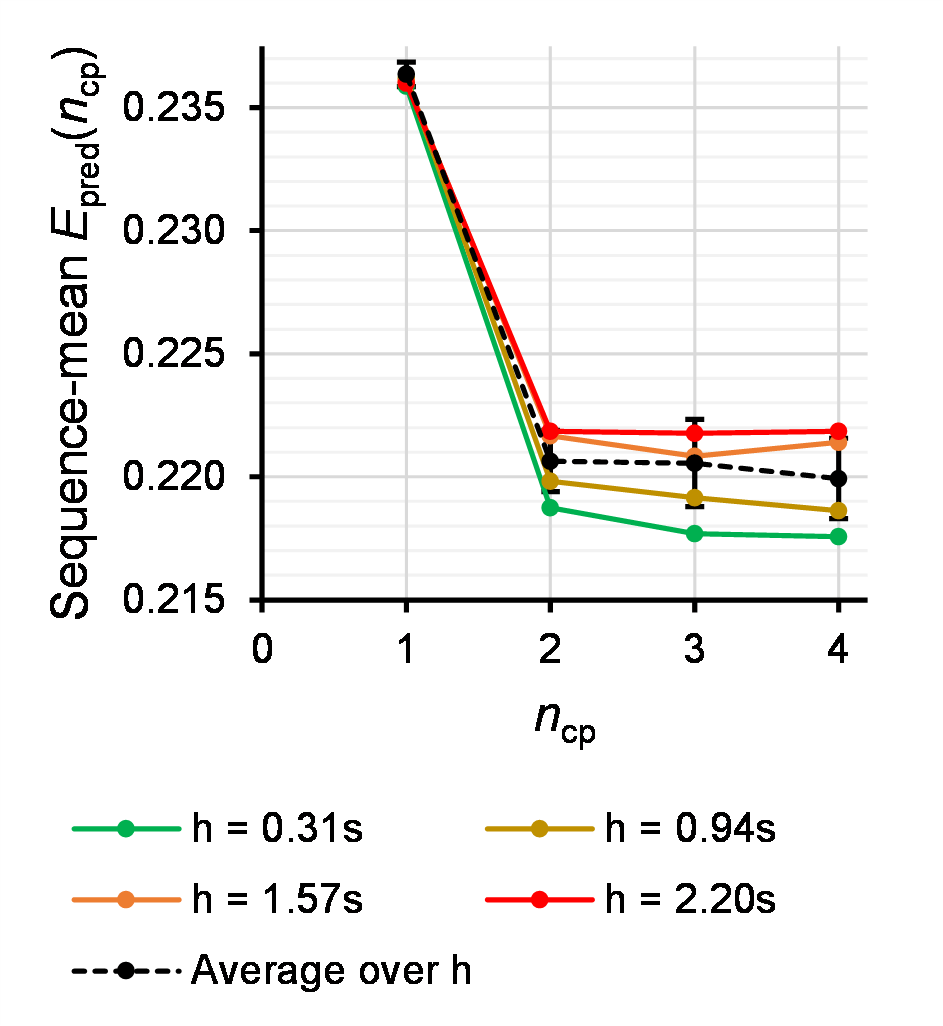}}%
    %\quad
    \subfloat[Prediction with \acs{RTRL}]{\includegraphics[width=.25\textwidth]{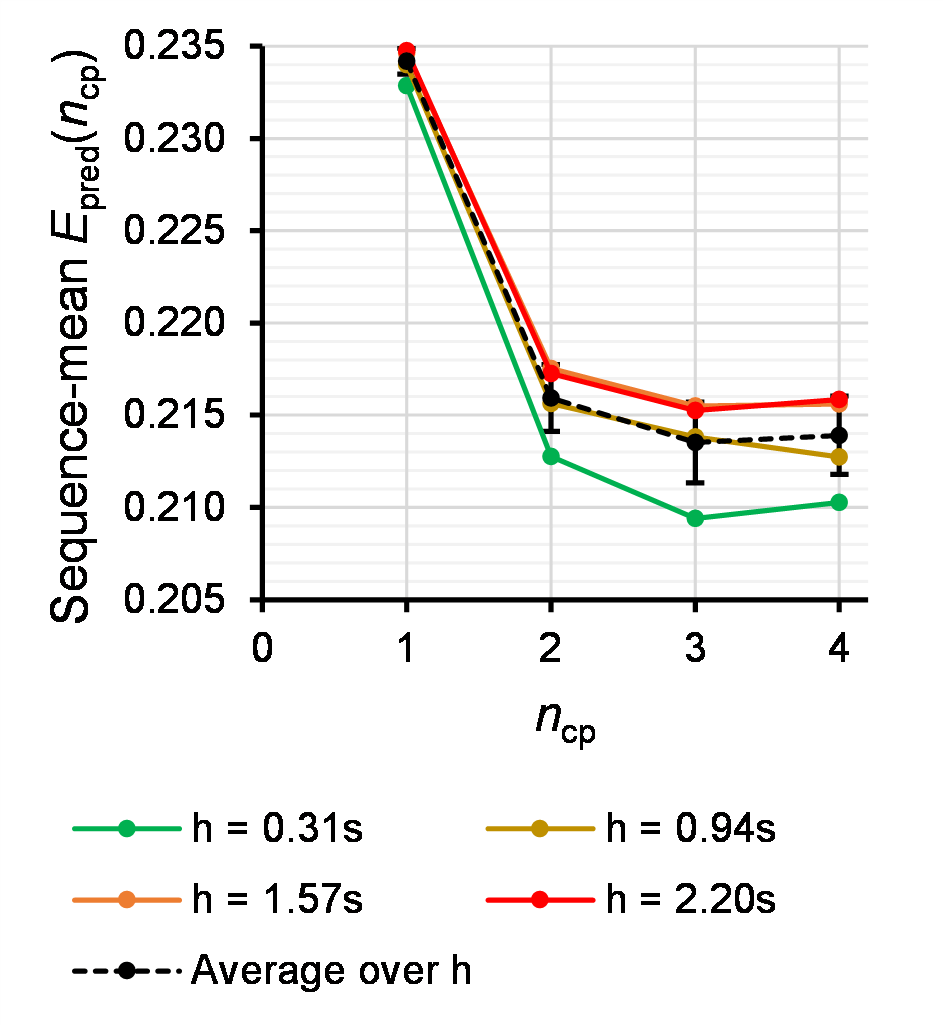}}\\
    %\quad
    \subfloat[Prediction with \acs{LMS}]{\includegraphics[width=.25\textwidth]{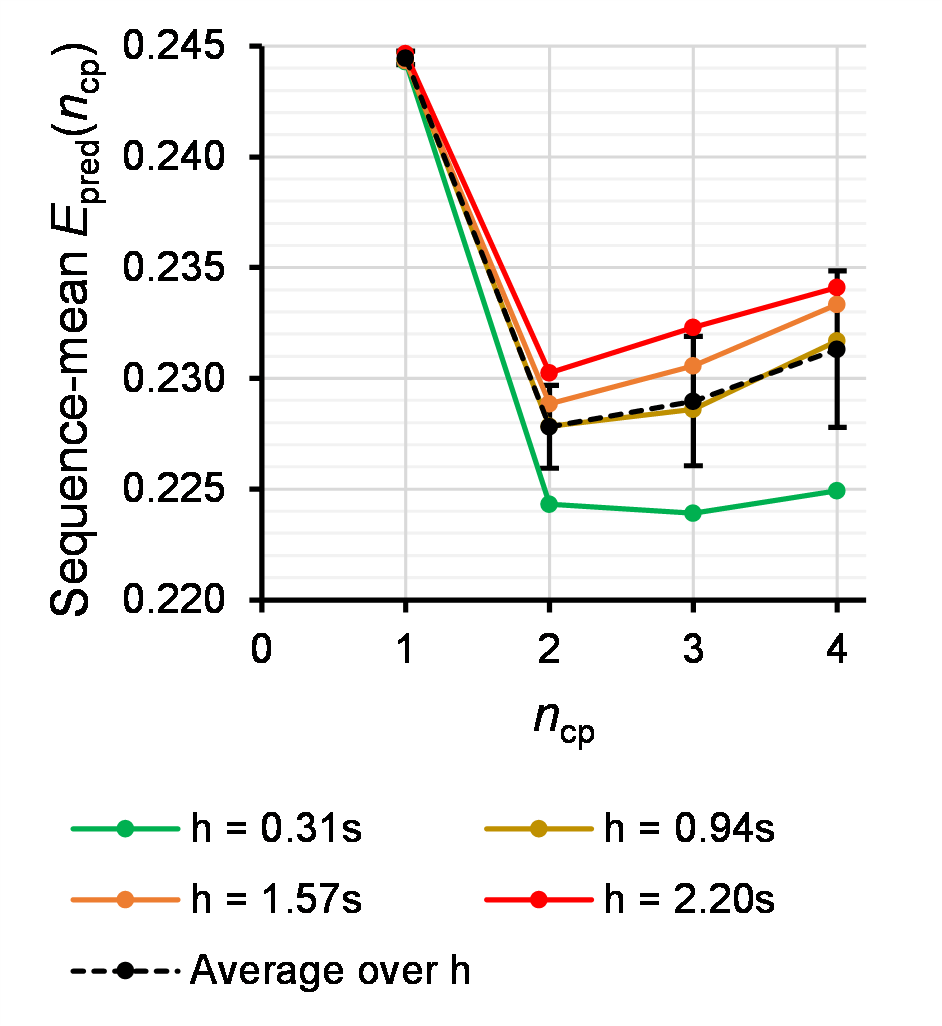}}%
    %\quad
    \subfloat[Prediction with \\ linear regression]{\includegraphics[width=.25\textwidth]{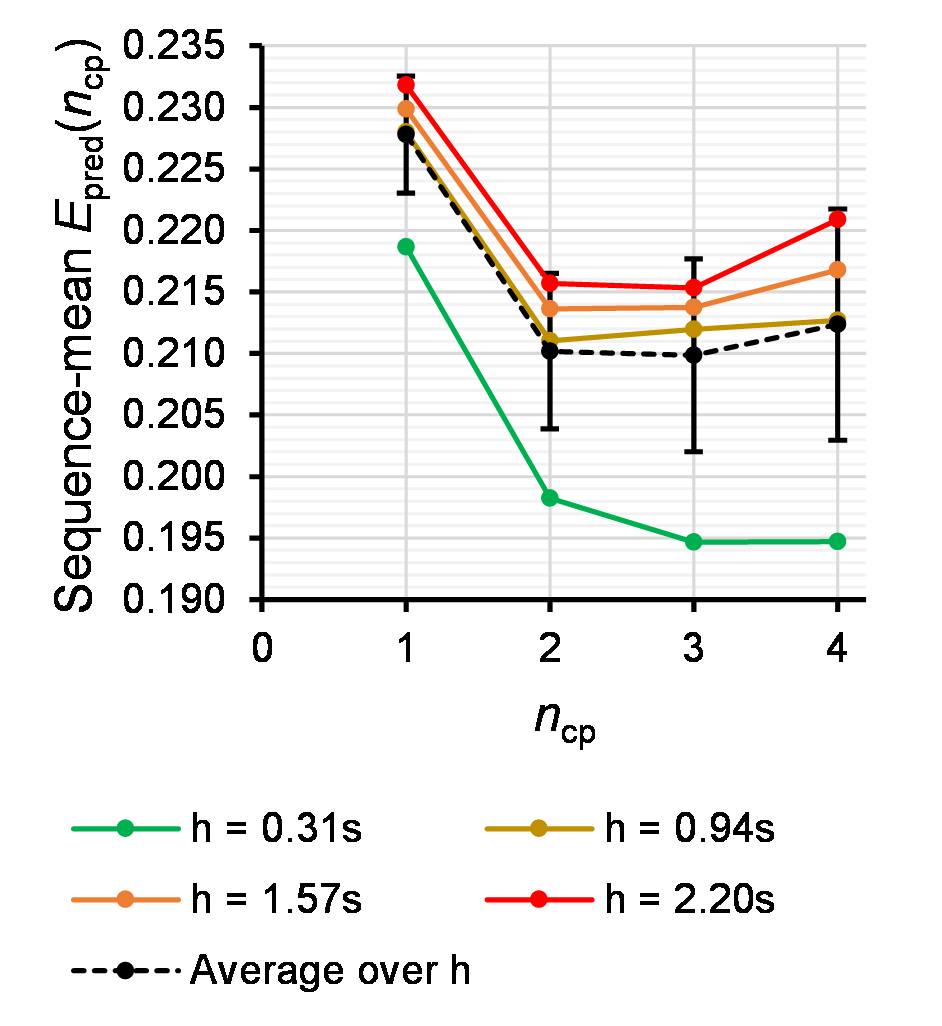}}%
    %\quad
    \subfloat[Prediction with \\a sequence-specific transformer]{\includegraphics[width=.25\textwidth]{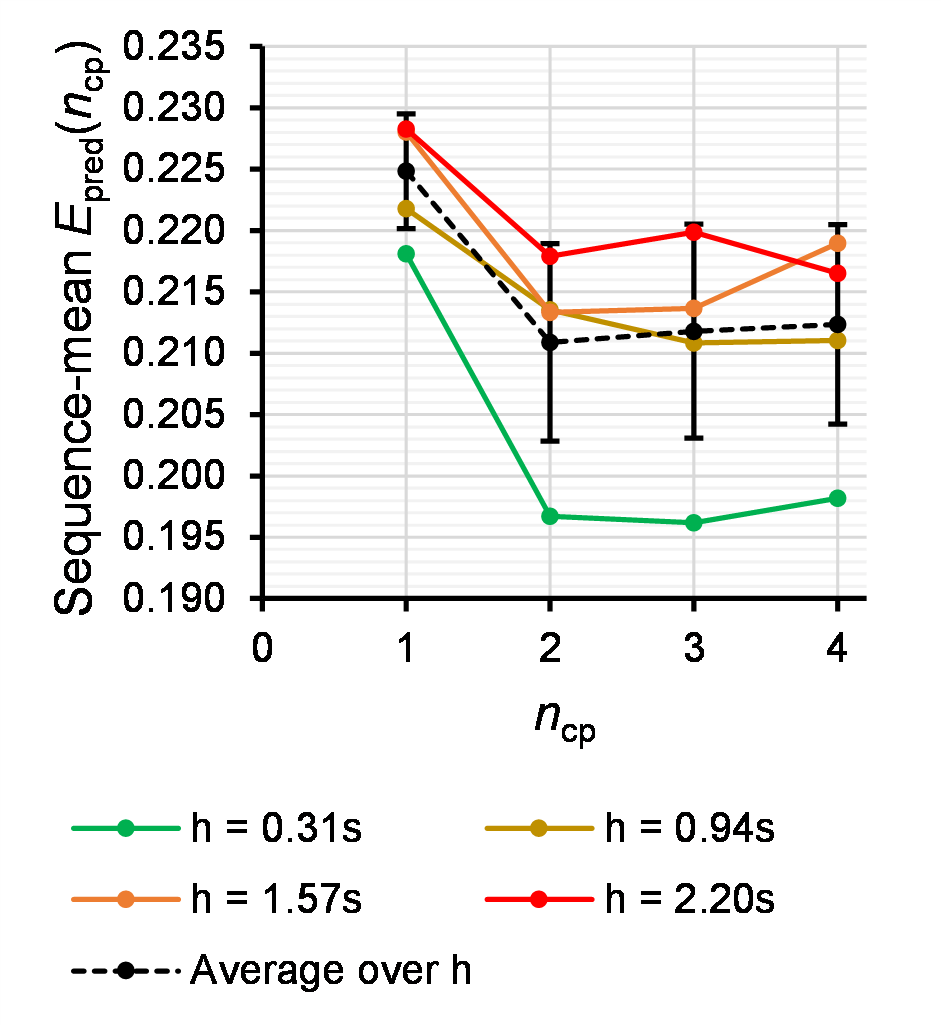}}%
    %\quad        
    \caption{Validation-set registration error, $E_{\text{pred}}(n_{\text{cp}})$, computed using the \gls{DVF} predicted with $n_{\text{cp}}$ principal components, averaged over the four sequences of the ETH Zürich dataset, as a function of $n_{\text{cp}}$, for all sequence-specific algorithms and several horizons $h$ (Eq. \ref{eq:val registration error}). Hyperparameters were optimized via grid search for each sequence and each combination of $n_{\text{cp}}$ and $h$ (Section \ref{section: PCA weight cross-validation}). The black dashed curves and associated error bars represent the average and standard deviation, respectively, of the sequence-mean $E_{\text{pred}}(n_{\text{cp}})$ over $h$ between 0.31s and 2.20s.}
    \label{fig:nb PCA cpts optim}
\end{figure*}

In this section, we examine the optimization of $n_{\text{cp}}$ for sequence-specific predictors on the ETH Zürich dataset, following the procedure described in Section \ref{section: methods - optimization of n_cp} (step 3.2 in Fig. \ref{fig:overall experimental setting}). The same selection strategy was applied to the \gls{OvGU} acquisitions to ensure comparability across video-prediction experiments. The validation-set registration error using the predicted \gls{DVF}, $E_{\text{pred}}(n_{\text{cp}})$, defined in Eq. \ref{eq:val registration error}, generally increased with $h$ (Fig. \ref{fig:nb PCA cpts optim}), with more pronounced differences between horizon extrema for offline algorithms. This aligns with previous observations in Figs. \ref{fig:PCA weights pred RTRL vs pop transformer}, \ref{fig:signal pred error 3 PCA cpts vs horizon}, and \ref{fig:hyperpar influence on PCA cpts prediction} (Section \ref{section:results weights forecasting}). For all algorithms and horizons, the mean of $E_{\text{pred}}(n_{\text{cp}})$ over the four sequences attained its maximum at $n_{\text{cp}}= 1$. Indeed, the second-order principal component captured the main \gls{SI} respiratory pattern in three of the four ETH Zürich acquisitions (Section \ref{section: PCA breathing motion modeling results}). The variations of $E_{\text{pred}}(n_{\text{cp}})$ between $n_{\text{cp}}=2$ and $n_{\text{cp}}=4$ were moderate, compared with its sharp drop when increasing $n_{\text{cp}}$ from 1 to 2. $n_{\text{cp}} = 2$ was generally optimal for \gls{LMS}, as $E_{\text{pred}}(n_{\text{cp}})$ clearly increased across most horizons when $n_{\text{cp}}$ rose from 2 to 4, and for the sequence-specific transformer when considering that error averaged over $h$ (black dotted curves in Fig. \ref{fig:nb PCA cpts optim}). By contrast, the sequence- and horizon-averaged error was minimized at $n_{\text{cp}} = 3$ for \gls{UORO}, \gls{SnAp-1}, \gls{RTRL}, and linear regression, and at $n_{\text{cp}} = 4$ for \gls{DNI}. The higher-order \gls{PCA} coefficients, which were generally noisier, carried complementary information regarding minor deformation modes and exhibited mild synchronization with lower-order weights, potentially helping predict the latter. Algorithms with stronger predictive capabilities (e.g., \gls{SnAp-1} and \gls{RTRL}) were better able to exploit cues related to these higher-order modes, which may explain their higher performance for larger values of 
$n_{\text{cp}}$.

Overall, predictions at low horizons were more accurate as $n_{\text{cp}}$ increased. For \gls{UORO}, the sequence-mean $E_{\text{pred}}(n_{\text{cp}})$ was minimized at $n_{\text{cp}} = 2$ for $h \geq 1.88\text{s}$ and at $n_{\text{cp}} = 3$ for $h \leq 1.57\text{s}$. Similarly, for \gls{LMS}, $n_{\text{cp}} = 3$ and $n_{\text{cp}} = 4$ minimized that error for $h \geq 0.94\text{s}$ and $h \leq 0.63\text{s}$, respectively. At short horizons, the higher-order, noisier \gls{PCA} weights could still be predicted reasonably well. This helped improve \gls{DVF} estimation, as the corresponding principal deformations captured finer deformation patterns within the chest and liver. At longer horizons, however, their prediction became unreliable, and discarding them (i.e., using a lower value of $n_{\text{cp}}$) was often preferable. In other words, the information added by increasingly noisier \gls{PCA} weights was worth considering only if they could be predicted with a relatively high degree of confidence, which was more often the case when $h$ was small.

\subsection{Image prediction}\label{section:image prediction results}

\subsubsection{Quantitative evaluation}\label{section:image prediction accuracy}

% General methodology 
In this section, we evaluate image-prediction performance on the test set using pixel-intensity--based metrics, namely the Pearson correlation coefficient $r$, the \gls{SSIM} \citep{wang2004image}, and the \gls{nRMSE} between the ground-truth and predicted frames. Formally, the image-domain \gls{nRMSE} is obtained by replacing $\widehat{w_j}$ and $w_j^{\text{true}}$ in Eq. \ref{eq:predicted weights nRMSE} with the predicted and ground-truth intensities at pixel locations indexed by $j \in \{1, \dots, |I|\}$, respectively. In our implementation, the \gls{SSIM} is computed over local windows and then averaged across the image. Alongside evaluation in the image space, we quantify motion-forecasting performance via the mean and maximum Euclidean endpoint errors, also referred to as geometrical errors, between the predicted and reference \gls{DVF}. The latter, denoting the Lucas--Kanade optical-flow field, serves as a reasonable benchmark for respiratory motion estimation \cite{xu2008lung, akino2014evaluation, dhont2019multi}, although the true underlying local tissue displacements are not directly observable. Even though mean deformation errors are reported to two decimals for consistency and comparability, differences below approximately 0.1mm should not be over-interpreted, given the underlying image resolution and potential inaccuracies in deformable registration. Besides evaluating \glspl{RNN}, transformers, and linear filters, we assess the performance of two baseline naive predictors: a \gls{PCA}-score--domain persistence model and an image-domain persistence model, using the latest frame and \gls{PCA} weight at time $t_k$ as the predicted frame and weight at $t_{k+h}$, respectively. Finally, we provide the accuracy of an "oracle" that has access to the reference optical-flow field and estimates the frame at $t_{k+h}$ as the initial frame at $t_1$ warped by that field. For the \gls{OvGU} sequences, evaluation was performed both on the entire images and within the manually defined higher-contrast \glspl{ROI}.

\subsubsubsection{Overall accuracy and stability}

% Ranking of methods in terms of average performance
\Gls{RTRL} and \gls{SnAp-1} consistently yielded either the best or second-best test-set performance, averaged over the sequences and horizons considered ($h \leq 2.2\text{s}$), across all metrics and both datasets (including both evaluation regions for the OvGU sequences), with mean geometrical errors of $1.4\pm0.1$mm and $2.7\pm0.3$mm on the ETH Zürich (full-frame) and \gls{OvGU} (\acs{ROI}) data, respectively (Table \ref{table:frame pred perf}). \Gls{LMS} generally ranked second or third in terms of intensity-based metrics, and, on the ETH Zürich dataset, achieved the highest $r$ and \gls{SSIM} jointly with \gls{RTRL} and \gls{SnAp-1}. \Gls{UORO} and \gls{DNI} invariably ranked either just below or on par with \gls{LMS} regarding the three intensity-based measures. However, the \gls{DVF} errors associated with \gls{LMS} were greater than or equal to those of the four online \gls{RNN} algorithms. One exception was the \gls{ROI}-based maximum error on the \gls{OvGU} sequences, marginally lower for \gls{LMS} than for \gls{UORO}. Linear regression always matched or surpassed the performance of the sequence-specific transformer. Both models ranked immediately below the adaptive algorithms regarding intensity-based metrics and mean geometrical error, except on the ETH Zürich acquisitions, where they attained a slightly lower mean endpoint error than \gls{LMS}. The population transformer was invariably outperformed by all non-baseline predictors (corresponding to the first seven rows after the header in Table \ref{table:frame pred perf}). However, it consistently surpassed both last-value naive models across all metrics and both datasets. Noticeably, the \acs{PCA}-score persistence model generally achieved higher accuracy than the image-domain persistence model. This suggests that discarding the higher-order \gls{PCA} components may have a significant regularizing effect on the initial \gls{DVF} and may thereby reduce noise in the predicted motion. The \gls{RNN} with frozen hidden-layer parameters yielded lower accuracy than dynamically trained \glspl{RNN}, which confirms the benefit of online training in this setting. The oracle attained the highest performance across all metrics and both datasets (e.g., full-frame \glspl{SSIM} of $0.92\pm0.01$ and $0.75\pm0.03$ on the ETH Zürich and \gls{OvGU} datasets, respectively). The relatively high oracle scores, representing an upper bound on image-prediction accuracy intrinsic to the \gls{DIR} method, provide evidence of the effectiveness of our warping-based approach. %and suggests that further enhancing \gls{PCA} weight prediction would likely translate into better video prediction. % too much discussion-ish

% about vertical spacing: https://tex.stackexchange.com/questions/65127/extra-vertical-space-after-hline-causes-a-gap-in-the-right-border-of-an-array 
\captionsetup[table*]{width=.9\textwidth}
\begin{table*}[width=\textwidth, thb!]
%\scriptsize
%\setlength{\tabcolsep}{3.5pt}
\footnotesize
\setlength{\tabcolsep}{1.7pt}
\begin{tabular}{llllllllllllllll}
\hline
                  &\multicolumn{3}{l}{Correlation coefficient $r$}                   & \multicolumn{3}{l}{\acs{nRMSE}}                                 & \multicolumn{3}{l}{\acs{SSIM}}                                  & \multicolumn{3}{l}{Mean \acs{DVF} error (mm)}                & \multicolumn{3}{l}{Max \acs{DVF} error (mm)}               \\[0.1cm]
Dataset           & ETH                 & \acs{OvGU}          & \acs{OvGU}          & ETH                 & \acs{OvGU}          & \acs{OvGU}          & ETH                 & \acs{OvGU}          & \acs{OvGU}          & ETH                & \acs{OvGU}         & \acs{OvGU}         & ETH               & \acs{OvGU}        & \acs{OvGU}         \\
                  & Zürich              &                     &                     & Zürich              &                     &                     & Zürich              &                     &                     & Zürich             &                    &                    & Zürich            &                   &                    \\[0.1cm]
Evaluation        & Whole               & Whole               & \acs{ROI}           & Whole               & Whole               & \acs{ROI}           & Whole               & Whole               & \acs{ROI}           & Whole              & Whole              & \acs{ROI}          & Whole             & Whole             & \acs{ROI}          \\
region            & image               & image               &                     & image               & image               &                     & image               & image               &                     & image              & image              &                    & image             & image             &                   \\[0.05cm]
\hline
\acs{UORO}        & 0.986               & 0.939               & 0.832               & 0.214               & 0.340               & 0.548               & 0.897               & 0.655               & 0.468               & 1.45               & 2.82               & 2.83               & 21.8              & 23.7              & 15.6 \rule{0pt}{2.6ex}\\
                  & $\pm$ 0.005         & $\pm$ 0.004         & $\pm$ 0.023         & $\pm$ 0.016         & $\pm$ 0.012         & $\pm$ 0.034         & $\pm$ 0.022         & $\pm$ 0.025         & $\pm$ 0.033         & $\pm$ 0.14         & $\pm$ 0.21         & $\pm$ 0.24         & $\pm$ 3.2         & $\pm$ 3.6         & $\pm$ 0.9         \\[0.05cm]
\acs{SnAp-1}      & \textbf{0.987}      & \textbf{0.942}      & \textbf{0.839}      & 0.212               & \textbf{0.333}      & \textbf{0.536}      & \textbf{0.899}      & \textbf{0.660}      & 0.476               & \textbf{1.41}      & \textbf{2.73}      & 2.68               & \textbf{21.3}     & \textbf{22.2}     & \textbf{14.8}     \\
                  & \textbf{$\pm$ 0.005}& \textbf{$\pm$ 0.004}& \textbf{$\pm$ 0.024}& $\pm$ 0.016         & \textbf{$\pm$ 0.013}& \textbf{$\pm$ 0.036}& \textbf{$\pm$ 0.022}& \textbf{$\pm$ 0.026}& $\pm$ 0.036         & \textbf{$\pm$ 0.14}& \textbf{$\pm$ 0.22}& $\pm$ 0.25         & \textbf{$\pm$ 3.3}& \textbf{$\pm$ 3.1}& \textbf{$\pm$ 0.9}\\[0.05cm]
\acs{DNI}         & 0.986               & 0.940               & 0.831               & 0.215               & 0.339               & 0.548               & 0.898               & 0.654               & 0.465               & 1.46               & 2.83               & 2.84               & 21.7              & 23.0              & 15.6              \\
                  & $\pm$ 0.006         & $\pm$ 0.004         & $\pm$ 0.023         & $\pm$ 0.018         & $\pm$ 0.012         & $\pm$ 0.035         & $\pm$ 0.023         & $\pm$ 0.025         & $\pm$ 0.034         & $\pm$ 0.16         & $\pm$ 0.21         & $\pm$ 0.25         & $\pm$ 3.2         & $\pm$ 3.1         & $\pm$ 0.9         \\[0.05cm]
\acs{RTRL}        & \textbf{0.987}      & \textbf{0.942}      & 0.838               & \textbf{0.211}      & \textbf{0.333}      & \textbf{0.536}      & \textbf{0.899}      & \textbf{0.660}      & \textbf{0.477}      & \textbf{1.41}      & \textbf{2.73}      & \textbf{2.66}      & \textbf{21.3}     & 22.5              & \textbf{14.8}     \\
                  & \textbf{$\pm$ 0.005}& \textbf{$\pm$ 0.005}& $\pm$ 0.024         & \textbf{$\pm$ 0.016}& \textbf{$\pm$ 0.013}& \textbf{$\pm$ 0.037}& \textbf{$\pm$ 0.022}& \textbf{$\pm$ 0.027}& \textbf{$\pm$ 0.037}& \textbf{$\pm$ 0.14}& \textbf{$\pm$ 0.22}& \textbf{$\pm$ 0.24}& \textbf{$\pm$ 3.3}& $\pm$ 3.2         & \textbf{$\pm$ 0.9}\\[0.05cm]
\acs{LMS}         & \textbf{0.987}      & 0.940               & 0.833               & 0.214               & 0.337               & 0.544               & \textbf{0.899}      & 0.656               & 0.470               & 1.48               & 2.83               & 2.85               & 22.0              & 24.0              & 15.4              \\
                  & \textbf{$\pm$ 0.005}& $\pm$ 0.006         & $\pm$ 0.026         & $\pm$ 0.018         & $\pm$ 0.015         & $\pm$ 0.039         & \textbf{$\pm$ 0.021}& $\pm$ 0.027         & $\pm$ 0.037         & $\pm$ 0.19         & $\pm$ 0.23         & $\pm$ 0.29         & $\pm$ 3.4         & $\pm$ 4.0         & $\pm$ 0.9         \\[0.05cm]
Linear            & 0.985               & 0.938               & 0.822               & 0.218               & 0.342               & 0.560               & 0.894               & 0.651               & 0.453               & 1.47               & 2.86               & 3.05               & 21.7              & 23.2              & 15.3              \\
regression        & $\pm$ 0.006         & $\pm$ 0.004         & $\pm$ 0.023         & $\pm$ 0.017         & $\pm$ 0.010         & $\pm$ 0.034         & $\pm$ 0.025         & $\pm$ 0.026         & $\pm$ 0.034         & $\pm$ 0.13         & $\pm$ 0.20         & $\pm$ 0.29         & $\pm$ 2.8         & $\pm$ 3.7         & $\pm$ 0.9         \\[0.05cm]
Seq-specific      & 0.985               & 0.936               & 0.815               & 0.218               & 0.346               & 0.567               & 0.894               & 0.649               & 0.448               & 1.46               & 2.94               & 3.17               & 22.9              & 23.1              & 15.7              \\
transformer       & $\pm$ 0.007         & $\pm$ 0.004         & $\pm$ 0.025         & $\pm$ 0.018         & $\pm$ 0.012         & $\pm$ 0.035         & $\pm$ 0.026         & $\pm$ 0.025         & $\pm$ 0.032         & $\pm$ 0.14         & $\pm$ 0.23         & $\pm$ 0.32         & $\pm$ 2.7         & $\pm$ 3.2         & $\pm$ 0.9         \\[0.05cm]
Population        & 0.978               & 0.922               & 0.769               & 0.255               & 0.382               & 0.635               & 0.874               & 0.620               & 0.386               & 1.95               & 3.37               & 3.98               & 27.2              & 26.8              & 18.1              \\
transformer       & $\pm$ 0.008         & $\pm$ 0.006         & $\pm$ 0.026         & $\pm$ 0.026         & $\pm$ 0.013         & $\pm$ 0.034         & $\pm$ 0.023         & $\pm$ 0.022         & $\pm$ 0.027         & $\pm$ 0.25         & $\pm$ 0.22         & $\pm$ 0.39         & $\pm$ 3.2         & $\pm$ 3.8         & $\pm$ 0.7         \\[0.05cm]
\acs{RNN} with    & 0.981               & 0.935               & 0.806               & 0.234               & 0.350               & 0.585               & 0.887               & 0.646               & 0.423               & 1.61               & 2.92               & 3.09               & 23.3              & 24.4              & 15.5              \\
fixed weights     & $\pm$ 0.010         & $\pm$ 0.004         & $\pm$ 0.026         & $\pm$ 0.024         & $\pm$ 0.011         & $\pm$ 0.037         & $\pm$ 0.030         & $\pm$ 0.024         & $\pm$ 0.039         & $\pm$ 0.16         & $\pm$ 0.22         & $\pm$ 0.28         & $\pm$ 3.0         & $\pm$ 4.1         & $\pm$ 0.8         \\[0.05cm]
Latest            & 0.972               & 0.914               & 0.748               & 0.276               & 0.399               & 0.662               & 0.860               & 0.611               & 0.369               & 2.13               & 3.63               & 4.41               & 27.9              & 29.2              & 18.9              \\
\acs{PCA} score   & $\pm$ 0.014         & $\pm$ 0.005         & $\pm$ 0.030         & $\pm$ 0.034         & $\pm$ 0.011         & $\pm$ 0.036         & $\pm$ 0.028         & $\pm$ 0.023         & $\pm$ 0.029         & $\pm$ 0.26         & $\pm$ 0.28         & $\pm$ 0.49         & $\pm$ 3.2         & $\pm$ 5.2         & $\pm$ 0.9         \\[0.05cm]
Latest            & 0.965               & 0.910               & 0.733               & 0.318               & 0.408               & 0.698               & 0.848               & 0.609               & 0.385               & n/a                & n/a                & n/a                & n/a               & n/a               & n/a                \\
image             & $\pm$ 0.015         & $\pm$ 0.006         & $\pm$ 0.031         & $\pm$ 0.031         & $\pm$ 0.016         & $\pm$ 0.043         & $\pm$ 0.028         & $\pm$ 0.029         & $\pm$ 0.034         &                    &                    &                    &                   &                   &                    \\
\hline               
Reference         & 0.991               & 0.966               & 0.911               & 0.159               & 0.253               & 0.397               & 0.918               & 0.753               & 0.632               & n/a                & n/a                & n/a                & n/a               & n/a               & n/a \rule{0pt}{2.6ex}\\                        
\acs{DVF} (oracle)& $\pm$ 0.002         & $\pm$ 0.004         & $\pm$ 0.018         & $\pm$ 0.013         & $\pm$ 0.014         & $\pm$ 0.040         & $\pm$ 0.012         & $\pm$ 0.025         & $\pm$ 0.035         &                    &                    &                    &                   &                   &                     \\
\hline
\end{tabular}
\caption{Test-set video-prediction performance for each algorithm. The first-line value in each cell represents a metric averaged over all \acs{MRI} sequences in either the ETH Zürich or \gls{OvGU} dataset, the horizons considered ($h \leq 2.2 \text{s}$), and $n_{\text{test}}$ runs to account for neural-network stochasticity (Section \ref{section:methods: frame prediction from PCA weights}). Hyperparameters, including $n_{\text{cp}}$ for sequence-specific models, were optimized for each sequence and horizon individually, except the population transformer, for which selection was horizon-wise only (Sections \ref{section: PCA weight cross-validation} and \ref{section: methods - optimization of n_cp}). We report 70\% \glspl{CI}, computed using the Student's t-distribution with $N-1$ degrees of freedom, where $N$ denotes the number of sequences in the dataset considered. The rows \enquote{latest \acs{PCA} score} and \enquote{latest image} correspond to using the \gls{PCA} score $w_j(t_k)$ and frame at time $t_k$, respectively, as direct estimates of their future values at time $t_{k+h}$. The last row corresponds to an oracle estimating the frame at time $t_{k+h}$ as the initial frame at $t=t_1$ warped by the Lucas--Kanade optical-flow field between $t_1$ and $t_{k+h}$. For each column, the values associated with all best-performing non-oracle methods are bolded (ties included).}
\label{table:frame pred perf}
\end{table*}
% Concerning the case with no prediction (previous image used) I used the results recorded in the following file: D:\OneDrive\Research in Uesaka Lab\3.22. Im pred results after PhD\2022 - 11 - 19 ter perf comparison ETH Zurich.xlsx 

% version with less rows to minimize page count tentatively
\begin{table*}[htb]
%\footnotesize
\setlength{\tabcolsep}{2.8pt}
\centering
\begin{tabular}{llllll}
\hline
                           && \multicolumn{2}{l}{\textbf{Cross-dataset stability}}                  & \multicolumn{2}{l}{\textbf{Intra-dataset stability}} \\
                           && \multicolumn{2}{l}{ETH Zürich $\rightarrow$ \gls{OvGU} (entire image)}& \multicolumn{2}{l}{entire image $\rightarrow$ \acs{ROI} (\gls{OvGU} data)}\\[0.05cm]                       
\cline{3-6}
                                                                   && Test \acs{SSIM}               & Mean test \acs{DVF}     & Test \acs{SSIM}        & Mean test \acs{DVF} \rule{0pt}{2.6ex}\\
                                                                   && decrease (\%)                 & error increase (\%)     & decrease (\%)          & error increase (\%) \\                           
\hline
\textbf{Prediction}& \acs{UORO}                             & 26.9                          & 94.7                    & 28.6                   & 0.2\rule{0pt}{2.6ex}\\
\textbf{methods}                           & \acs{SnAp-1}                           & \textbf{26.6}                 & 94.2                    & \textbf{27.8}          & \textbf{-1.8} \\
                           & \acs{DNI}                              & 27.1                          & 93.4                    & 28.9                   & 0.6  \\
                           & \acs{RTRL}                             & \textbf{26.5}                 & 94.1                    & \textbf{27.7}          & \textbf{-2.3} \\
                           & \acs{LMS}                              & 27.0                          & \textbf{91.2}           & 28.4                   & 0.8  \\
                           & Linear regression                      & 27.2                          & 94.6                    & 30.4                   & 6.5  \\
                           & Sequence-specific transformer          & 27.5                          & 101.2                   & 30.9                   & 8.1  \\
                           & Population transformer                 & 29.1                          & 72.8                    & 37.8                   & 18.1 \\[0.1cm]
\textbf{Baselines}         & \acs{RNN} with a frozen hidden layer   & 27.2                          & 81.4                    & 34.5                   & 6.0  \\
                           & Latest \acs{PCA} weight used as prediction  & 29.0                          & 70.3                    & 39.7                   & 21.5 \\
                           & Latest image used as prediction             & 28.2                          & n/a                     & 36.7                   & n/a  \\[0.1cm]
\textbf{Oracle}            & Warping with Lucas--Kanade optical flow        & 18.0                          & n/a                     & 16.1                   & n/a  \\[0.1cm]
\hline
\multicolumn{2}{l}{\textbf{Average for non-baseline and non-oracle methods}}& 27.2                          & 92.0                    & 30.1                   & 3.8 \rule{0pt}{2.6ex}\\
\hline
\end{tabular}
\caption{Cross-dataset and intra-dataset stability for each prediction method, measured as the test-set performance drop when moving from entire-image evaluation on the ETH Zürich sequences to entire-image evaluation on the \gls{OvGU} sequences, and from entire-image to \acs{ROI}-based evaluation on the \gls{OvGU} sequences, respectively. The metrics used to compute those relative differences, whose values are provided in Table \ref{table:frame pred perf}, are the \gls{SSIM} and mean \gls{DVF} error averaged over the sequences in each dataset, the horizons $h \leq 2.2 \text{s}$, and $n_{\text{test}}$ runs. A negative value indicates improved performance. In each column, the values corresponding to the two most stable predictors are bolded, except for cross-dataset stability evaluation using the \gls{DVF} error. In the latter setting, only the second most robust method is highlighted, since the population transformer---though most stable---produced much higher errors than the other algorithms.}
\label{table:stability performance}
\end{table*}

% Stability assessment
% For instance, it increased from 1.4mm to 2.7mm for \gls{SnAp-1} - repetition with previous paragraph
For all methods, the mean geometrical error was substantially higher on the ETH Zürich acquisitions than on the \gls{OvGU} ones (by about 92\% on average; Table \ref{table:stability performance}), mainly reflecting the latter's higher noise, lower contrast, and stronger motion variability. On the \gls{OvGU} dataset, this error was slightly higher when computed in the \glspl{ROI} than over the entire frames (by 3.8\% on average). \Gls{SnAp-1} and \gls{RTRL} were the only exceptions: their mean \gls{DVF} errors in the \glspl{ROI} were slightly lower than those in the full frames, although the corresponding \glspl{CI} overlapped (e.g., $2.73\pm0.22$mm vs. $2.68\pm0.25$mm for \gls{SnAp-1}). This suggests that both models could robustly forecast the more structured but higher-amplitude \gls{ROI} motion, consistent with their overall top performance among the algorithms considered (Table \ref{table:frame pred perf}). By contrast, the maximum \gls{DVF} error on the \gls{OvGU} data dropped when evaluation was restricted to the \glspl{ROI}, by about 34\% on average for non-baseline methods (e.g., from $22.2\pm3.1$mm to $14.8\pm0.9$mm for \gls{SnAp-1}). Indeed, motion prediction was more challenging in the lower-contrast, noisier peripheral areas, where \gls{DIR} appeared less reliable. Full-frame intensity-based accuracy metrics were also higher on the ETH Zürich sequences than on the \gls{OvGU} sequences, with, for instance, an average 27.2\% decrease in \gls{SSIM} across \glspl{RNN} trained online, transformers, and linear filters. For instance, the full-frame \gls{SSIM} of \gls{SnAp-1} decreased from $0.90\pm0.02$ to $0.66\pm0.03$ when swapping datasets. Likewise, intensity-based performance on the \gls{OvGU} acquisitions was consistently lower when measured in the \glspl{ROI} rather than over the whole images, with a mean 30.1\% decrease in \gls{SSIM} for non-baseline methods (e.g., the \gls{ROI}-based \gls{SSIM} of \gls{SnAp-1} was $0.48\pm0.04$). This contrasts with the high intra-dataset stability observed for the mean geometrical error: even when deformations were accurately predicted within the \glspl{ROI}, image warping could still cause blurring or misalignments in these high-contrast areas, which degraded the \gls{SSIM}. Notably, \gls{SnAp-1} and \gls{RTRL} exhibited the greatest stability for the latter metric across datasets and image regions and the highest intra-dataset stability for the mean \gls{DVF} error.

% Short comment on confidence intervals
Due to the relatively small number of sequences, the 70\% \glspl{CI} often overlapped between methods evaluated on the same dataset and image region (Table \ref{table:stability performance}). Notably, uncertainties for \gls{ROI}-based metrics were consistently larger, reflecting higher motion variability and structural changes in that region, except for the maximum geometrical error, due to higher noise in peripheral areas. When comparing the same algorithm across datasets or image regions, \glspl{CI} generally did not overlap, except for the mean \gls{DVF} error across datasets and the maximum \gls{DVF} error across evaluation regions, respectively. These patterns support the robustness analysis outlined above.

\subsubsubsection{Statistical analysis of accuracy} % Cohen's d -> d (when referring to that the second time!)

\captionsetup[table*]{width=.9\textwidth}
\begin{table*}[width=\textwidth, htb]
\scriptsize %\footnotesize
% \footnotesize
\setlength{\tabcolsep}{1.4pt} %{1.8pt}
\begin{tabular}{llllllllllllllllllllll}
\hline
Prediction     & Evaluation                & \multicolumn{2}{l}{\acs{SnAp-1}}& \multicolumn{2}{l}{\acs{DNI}} & \multicolumn{2}{l}{\acs{RTRL}}& \multicolumn{2}{l}{\acs{LMS}}& \multicolumn{2}{l}{Linear}    & \multicolumn{2}{l}{Subj-specific}& \multicolumn{2}{l}{Population} & \multicolumn{2}{l}{\acs{RNN} with}& \multicolumn{2}{l}{Latest}         & \multicolumn{2}{l}{Latest}   \\
method         & metric and                &          &                      &               &               &               &               &               &              & \multicolumn{2}{l}{regression}& \multicolumn{2}{l}{transformer}  & \multicolumn{2}{l}{transformer}& \multicolumn{2}{l}{fixed weights} & \multicolumn{2}{l}{\acs{PCA} score}& \multicolumn{2}{l}{image}   \\
               & region                    & $p$-val. & $d$                  & $p$-val.      & $d$           & $p$-val.      & $d$           & $p$-val.      & $d$          & $p$-val.      & $d$           & $p$-val. & $d$                   & $p$-val.       & $d$           & $p$-val. & $d$                    & $p$-val. & $d$                     & $p$-val. & $d$   \\[0.1cm]
\hline
\acs{UORO}     & \acs{SSIM}, full frame    & 0.148    & -0.52                & 0.945         & 0.16$^\dag$   & 0.109         & -0.65         & 0.844         & -0.08$^\dag$ & 0.312         & 0.30          & \textbf{0.078} & \textbf{1.03}   & \textbf{0.008} & \textbf{1.74} & 0.008    & 1.68                   & 0.008    & 3.00                    & 0.016    & 1.75 \rule{0pt}{2.6ex}\\
               & \acs{DVF} error, \acs{ROI}& 0.109    & \textbf{0.85}        & 0.945         & -0.07$^\dag$  & \textbf{0.023}& \textbf{1.10} & 0.641         & -0.08$^\dag$ & 0.148         & -0.60         & \textbf{0.016} & \textbf{-1.29}  & \textbf{0.008} & \textbf{-1.75}& 0.008    & -1.58                  & 0.008    & -1.85                   & -        & -    \\[0.05cm]
\acs{SnAp-1}   & \acs{SSIM}, full frame    &          &                      & 0.109         & 0.63          & 0.547         & -0.35         & 0.383         & 0.38         & 0.109         & 0.47          & \textbf{0.039} & \textbf{0.95}   & \textbf{0.008} & \textbf{1.43} & 0.008    & 2.25                   & 0.008    & 2.41                    & 0.016    & 1.67 \\
               & \acs{DVF} error, \acs{ROI}& -        &                      & \textbf{0.039}& \textbf{-0.83}& 0.742         & 0.20          & \textbf{0.055}& -0.74        & \textbf{0.078}& -0.76         & \textbf{0.008} & \textbf{-1.29}  & \textbf{0.008} & \textbf{-1.60}& 0.008    & -1.76                  & 0.008    & -1.78                   & -        & -    \\[0.05cm]
\acs{DNI}      & \acs{SSIM}, full frame    &          &                      &               &               & 0.312         & -0.72         & 0.461         & -0.30        & 0.312         & 0.18          & 0.250          & 0.49            & \textbf{0.008} & \textbf{1.43} & 0.008    & 1.11                   & 0.008    & 2.34                    & 0.016    & 1.49 \\
               & \acs{DVF} error, \acs{ROI}& -        &                      & -             &               & 0.109         & \textbf{0.80} & 0.844         & -0.04$^\dag$ & 0.195         & -0.45         & \textbf{0.016} & \textbf{-0.98}  & \textbf{0.008} & \textbf{-1.58}& 0.016    & -1.30                  & 0.008    & -1.62                   & -        & -    \\[0.05cm]
\acs{RTRL}     & \acs{SSIM}, full frame    &          &                      &               &               &               &               & 0.250         & 0.47         & 0.109         & 0.53          & \textbf{0.008} & \textbf{1.05}   & \textbf{0.008} & \textbf{1.47} & 0.008    & 2.15                   & 0.008    & 2.45                    & 0.016    & 1.73 \\
               & \acs{DVF} error, \acs{ROI}& -        &                      & -             &               & -             &               & \textbf{0.055}& -0.75        & \textbf{0.055}& \textbf{-0.85}& \textbf{0.008} & \textbf{-1.38}  & \textbf{0.008} & \textbf{-1.65}& 0.008    & -1.84                  & 0.008    & -1.86                   & -        & -    \\[0.05cm]
\acs{LMS}      & \acs{SSIM}, full frame    &          &                      &               &               &               &               &               &              & 0.250         & 0.23          & 0.195          & 0.49            & \textbf{0.008} & \textbf{1.23}	& 0.055    & 0.94                   & 0.008    & 1.85                    & 0.016    & 1.35 \\
               & \acs{DVF} error, \acs{ROI}& -        &                      & -             &               & -             &               & -             &              & 0.383         & -0.36         & \textbf{0.055} & -0.78           & \textbf{0.008} & \textbf{-1.35}& 0.078    & -0.84                  & 0.008    & -1.51                   & -        & -    \\[0.05cm]
Linear         & \acs{SSIM}, full frame    &          &                      &               &               &               &               &               &              &               &               & 0.742          & 0.20            & \textbf{0.008} & \textbf{1.62}	& 0.461    & 0.32                   & 0.008    & 2.86                    & 0.008    & 2.07 \\
regression     & \acs{DVF} error, \acs{ROI}& -        &                      & -             &               & -             &               & -             &              & -             &               & 0.461          & -0.38           & \textbf{0.008} & \textbf{-1.76}& 0.945    & -0.14                  & 0.008    & -2.10                   & -        & -    \\[0.05cm]
Subj-specific  & \acs{SSIM}, full frame    &          &                      &               &               &               &               &               &              &               &               &                &                 & \textbf{0.008} & \textbf{1.59} & 0.547    & 0.36                   & 0.008    & 2.81                    & 0.016    & 1.54 \\
transformer    & \acs{DVF} error, \acs{ROI}& -        &                      & -             &               & -             &               & -             &              & -             &               & -              &                 & \textbf{0.008} & \textbf{-1.59}& 0.461    & 0.37                   & 0.008    & -1.84                   & -        & -    \\[0.05cm]
Population     & \acs{SSIM}, full frame    &          &                      &               &               &               &               &               &              &               &               &                &                 &                &               & 0.008    & -1.13                  & 0.078    & 0.76                    & 0.250    & 0.47 \\
transformer    & \acs{DVF} error, \acs{ROI}& -        &                      & -             &               & -             &               & -             &              & -             &               & -              &                 & -              &               & 0.008    & 1.39                   & 0.109    & -0.74                   & -        & -    \\[0.05cm]
\acs{RNN} with & \acs{SSIM}, full frame    &          &                      &               &               &               &               &               &              &               &               &                &                 &                &               &          &                        & 0.008    & 2.24                    & 0.016    & 1.27 \\
fixed weights  & \acs{DVF} error, \acs{ROI}& -        &                      & -             &               & -             &               & -             &              & -             &               & -              &                 & -              &               & -        &                        & 0.008    & -1.64                   & -        & -    \\[0.05cm]
Latest         & \acs{SSIM}, full frame    &          &                      &               &               &               &               &               &              &               &               &                &                 &                &               &          &                        &          &                         & 0.742    & 0.12 \\
\acs{PCA} score    & \acs{DVF} error, \acs{ROI}& -        &                      & -             &               & -             &               & -             &              & -             &               & -              &                 & -              &               & -        &                        & -        &                         & -        & -    \\[0.05cm]
\hline
\end{tabular}
\caption{Statistical tests for image-prediction performance on the \gls{OvGU} dataset. Each value corresponds to the two-tailed $p$-value from the Wilcoxon signed-rank test or Cohen's $d$ for each pair of methods, computed from the sequence-level differences between methods A and B in full-frame \glspl{SSIM} or \gls{ROI}-based mean \gls{DVF} errors (averaged over the test set, all horizons $h \leq 2.2 \text{s}$, and $n_{\text{test}}$ runs). A positive Cohen's $d$ indicates that method A (rows) tends to yield a higher metric value than method B (columns). A dagger symbol is added near the $d$-value when the median of the corresponding sequence-level metric differences across all sequences has a different sign than the mean difference, indicating that the direction of $d$ may not be meaningful in such instances. $p$-values lower than 0.08 and values of $d$ greater than 0.8, reflecting (at least marginal) statistical significance and large effect sizes, respectively, are bolded for non-baseline predictors. This $p$-value threshold, set higher than 0.05, helps highlight trends that are difficult to detect under limited statistical power, given the small dataset size ($N=8$ sequences).}
\label{table:statistical tests}
\end{table*}

% Statistical tests for the RNNs and linear filters
To complement the analysis above, we conducted two-sided Wilcoxon signed-rank tests and computed Cohen's $d$ values using the \gls{OvGU} acquisitions\footnotemark ~(here, $d$ does not refer to the \gls{RNN} hidden-layer dimension). We considered two complementary perspectives, using horizon-averaged full-frame \glspl{SSIM} and mean deformation errors over the \glspl{ROI} for global intensity-based and local geometry-based evaluation, respectively. Statistical evaluation of \glspl{RNN} trained online and linear filters may lack sensitivity due to the small number of sequences ($N=8$). This is reflected, for instance, by the low-to-medium $p$-values but high $d$-values (indicating practical, meaningful differences) obtained when comparing \gls{UORO} with \gls{SnAp-1} ($p=0.109$,  $d=0.85$) and \gls{RTRL} with \gls{DNI} ($p=0.109$, $d=0.80$), using \gls{ROI}-based mean \gls{DVF} errors (Table~\ref{table:statistical tests}). Notably, only local deformation-based analysis yielded pairwise comparisons within the set of adaptive filters and linear regression with marginal $p$-values ($p \leq 0.08$) and large effect sizes ($|d| \geq 0.80$).
%Noticeably, only the local deformation--based analysis produced marginally significant comparisons ($p \leq 0.08$) and large effect sizes ($|d| \geq 0.8$) among adaptive filters and linear regression. 
The high values of $d$ for \gls{RTRL} and \gls{SnAp-1} compared with \gls{UORO} and \gls{DNI} (four pairwise comparisons in total) indicate a practically robust performance gain that is consistent across sequences. These were observed even though mean error differences between algorithms across those two groups were generally modest; the largest average gap (0.18mm) was between \gls{DNI} and \gls{RTRL} (Table \ref{table:frame pred perf}). By contrast, \gls{ROI}-based mean deformation-error comparisons between online-trained \glspl{RNN}, on the one hand, and linear regression, on the other hand, were characterized by medium-to-large effect sizes (with $|d|$ between 0.45 and 0.85) and low-to-moderate $p$-values (below 0.195), pointing to slightly less robust trends despite larger raw error differences (up to 0.39mm, between linear regression and \acs{RTRL}). Comparisons of mean \gls{DVF} errors in the \glspl{ROI} between \gls{RTRL} and \gls{LMS}, as well as between \gls{SnAp-1} and \gls{LMS}, were marginally significant ($p= 0.055$), with moderate effect sizes ($d \approx 0.75$).
% The lower mean \gls{DVF} errors in the \glspl{ROI} for \gls{RTRL} and \gls{SnAp-1} compared with \gls{LMS} were marginally statistically significant ($p= 0.055$), with moderate effect sizes ($d \approx 0.75$).

\footnotetext{Such statistical analysis was not performed on the ETH Zürich data, as $N=4$ sequences would yield underpowered tests (the smallest attainable Wilcoxon signed-rank $p$-value would be 0.0625).}

% Statistical tests for the transformers and baselines
Statistical analysis confirmed the generally lower performance of transformers relative to the non-baseline algorithms (e.g, $p = 0.008$ and $d \geq 1.23$ for all pairwise comparisons with the cross-subject transformer). Nonetheless, there were exceptions for the subject-specific transformer: its comparisons with linear regression were not statistically significant for either metric ($p \geq 0.461$), and those with both \gls{DNI} and \gls{LMS} yielded only moderate $p$-values and effect sizes for the \gls{SSIM} ($p \geq 0.195$, $d = 0.49$). The relatively lower performance of the \gls{RNN} with a frozen hidden layer relative to \glspl{RNN} trained online, and that of the persistence models relative to the other algorithms, were all statistically significant, except for comparisons with the population transformer, which produced low-to-moderate $p$-values ($p \leq 0.250$).

\subsubsubsection{Performance evolution with the horizon}

\begin{figure*}[pos=htbp,width=\textwidth,align=\centering]
    \centering
    \includegraphics[width=.33\textwidth]{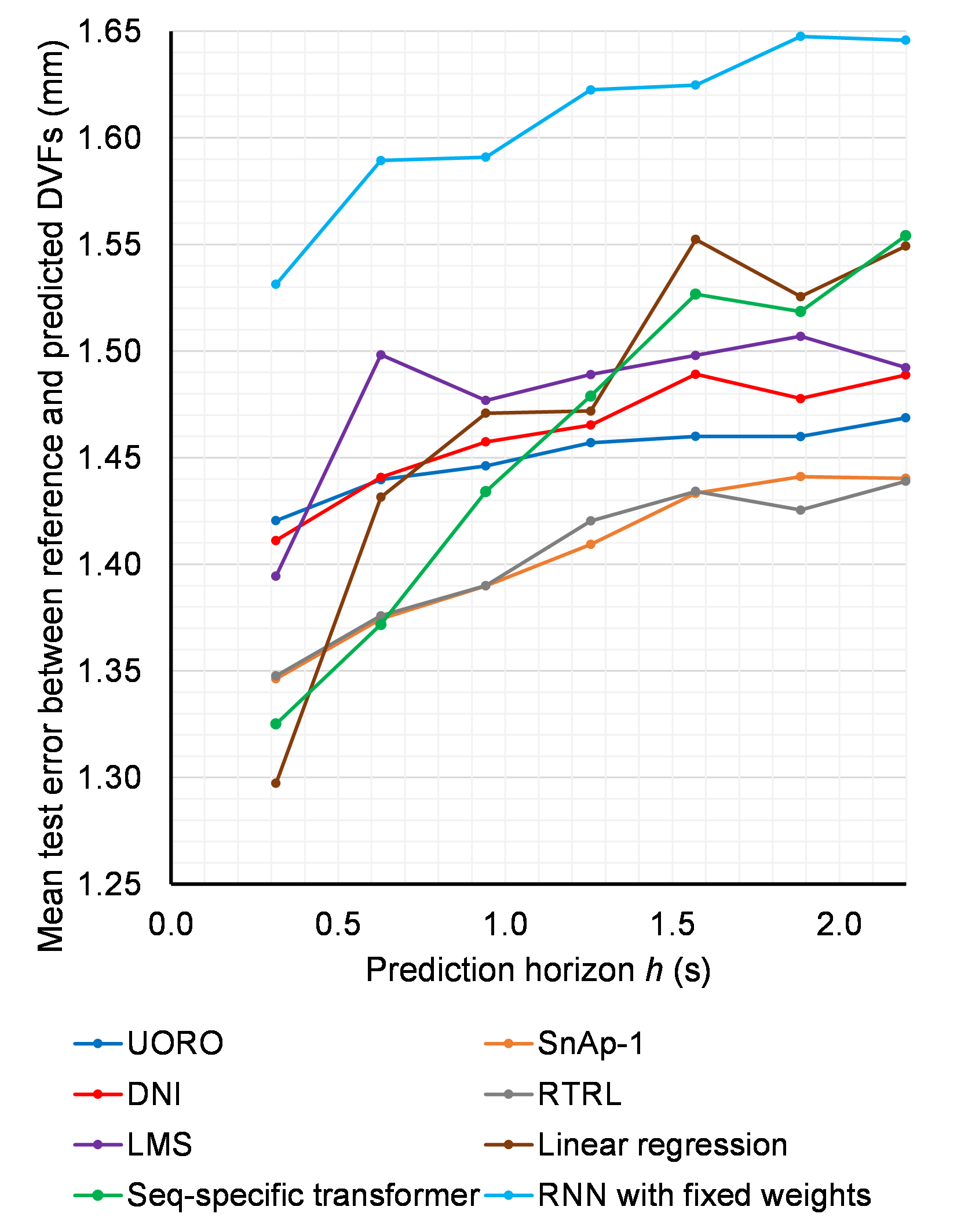}% 
    \includegraphics[width=.33\textwidth]{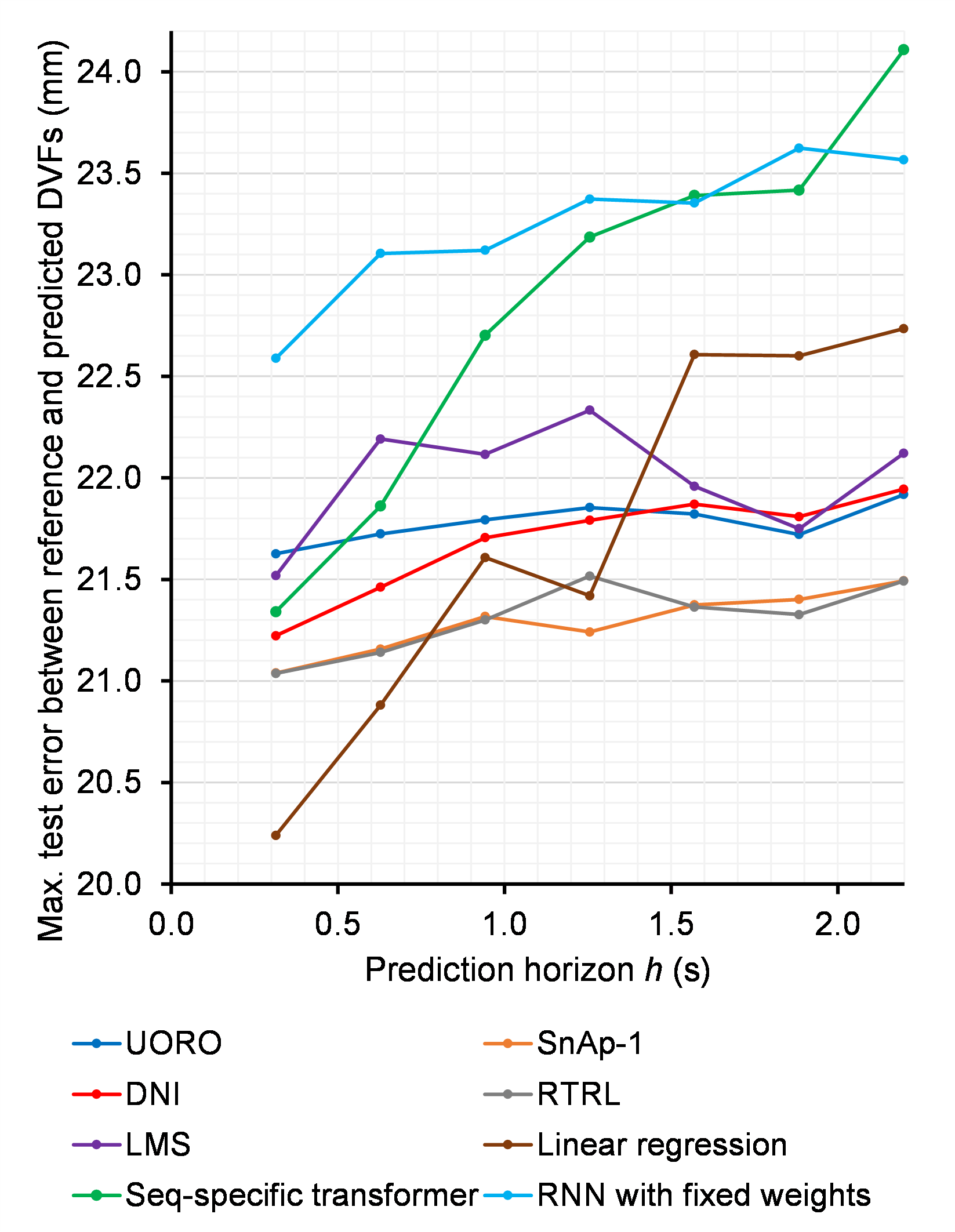} %      
    \includegraphics[width=.33\textwidth]{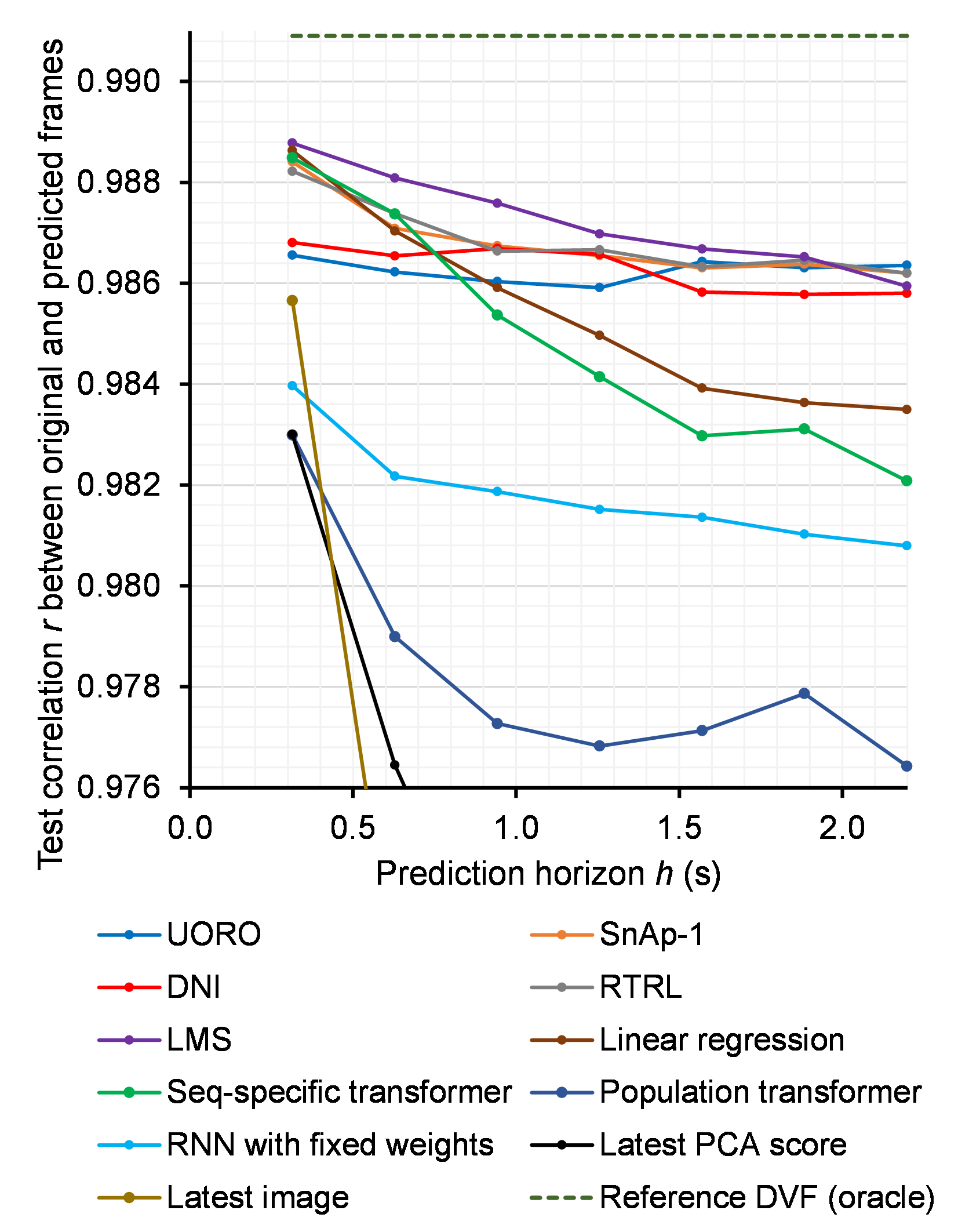}%
    \includegraphics[width=.33\textwidth]{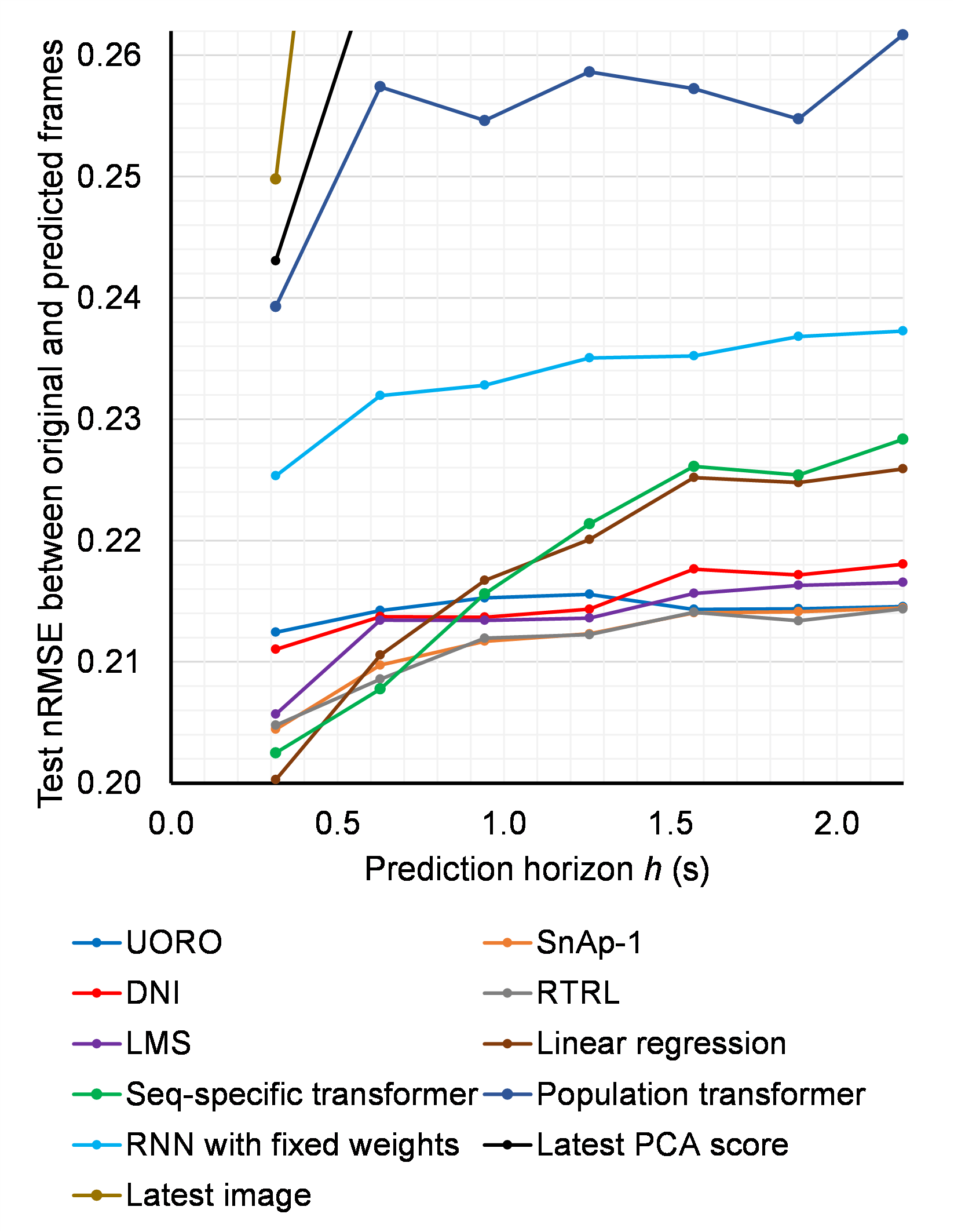}%        
    \includegraphics[width=.33\textwidth]{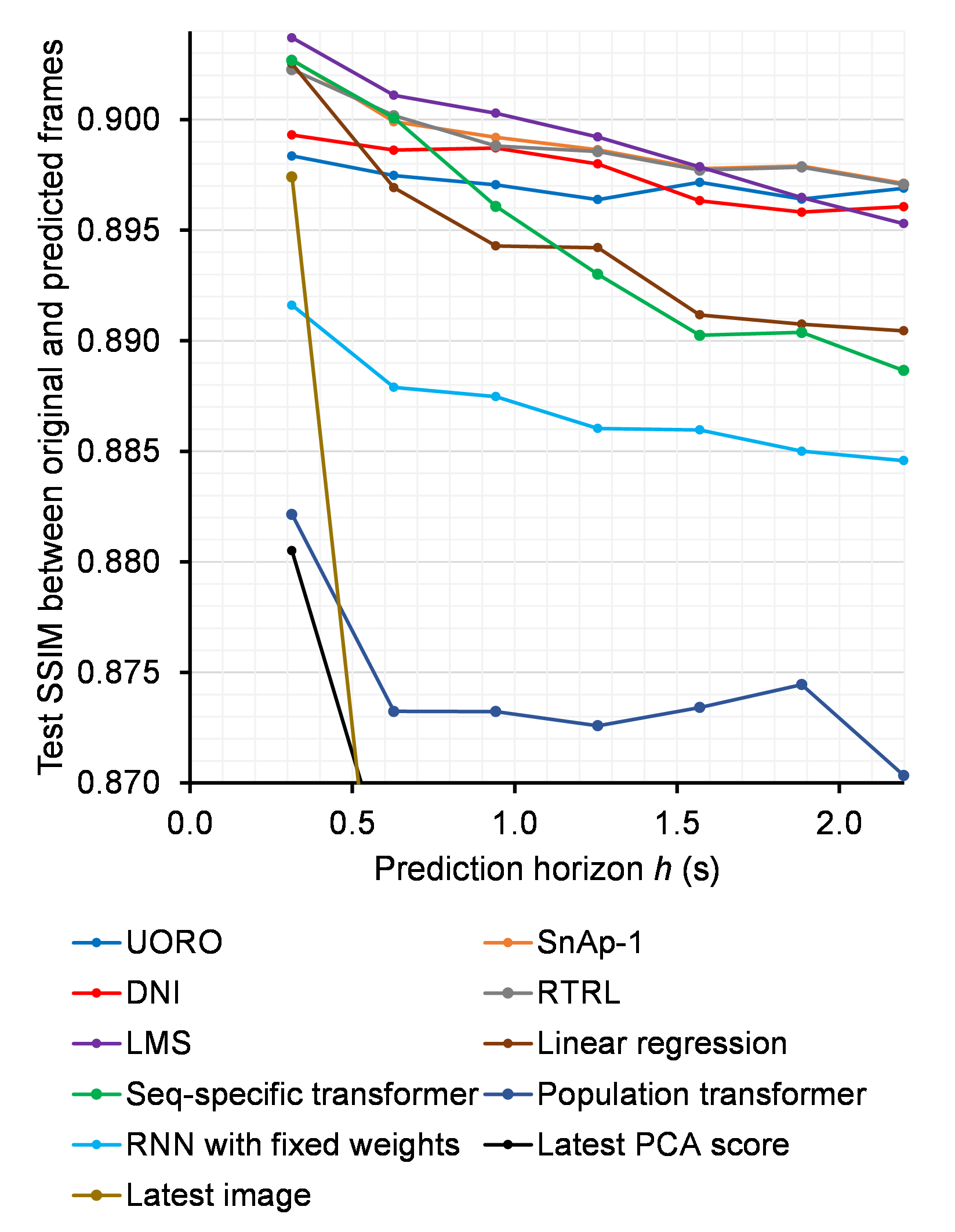} %             
    \caption{Test-set frame-forecasting performance for each algorithm as a function of the horizon $h$, for the ETH Zürich dataset. Each point represents the average of a given metric over the four image sequences and $n_{\text{test}}$ runs (Section \ref{section:methods: frame prediction from PCA weights}). Hyperparameters, including $n_{\text{cp}}$ for sequence-specific models, were optimized for each sequence (except the population transformer) and each value of $h$ (Sections \ref{section: PCA weight cross-validation} and \ref{section: methods - optimization of n_cp}). Baseline and oracle definitions are identical to those in Table \ref{table:frame pred perf}. The mean of each curve (i.e., the performance averaged over $h$ for each method) is reported in the \enquote{ETH Zürich} columns of that table. Curves for algorithms with relatively poor accuracy (\gls{DVF} errors for the population transformer and the naive \gls{PCA}-weight predictor) and for oracle metrics far below or above the other metrics (\gls{nRMSE} and \gls{SSIM}, respectively) are omitted.}
    \label{fig:next frame pred perf vs hrz on ETH}
\end{figure*}

\begin{figure*}[pos=htbp,width=\textwidth,align=\centering]
    \centering
    \includegraphics[width=.33\textwidth]{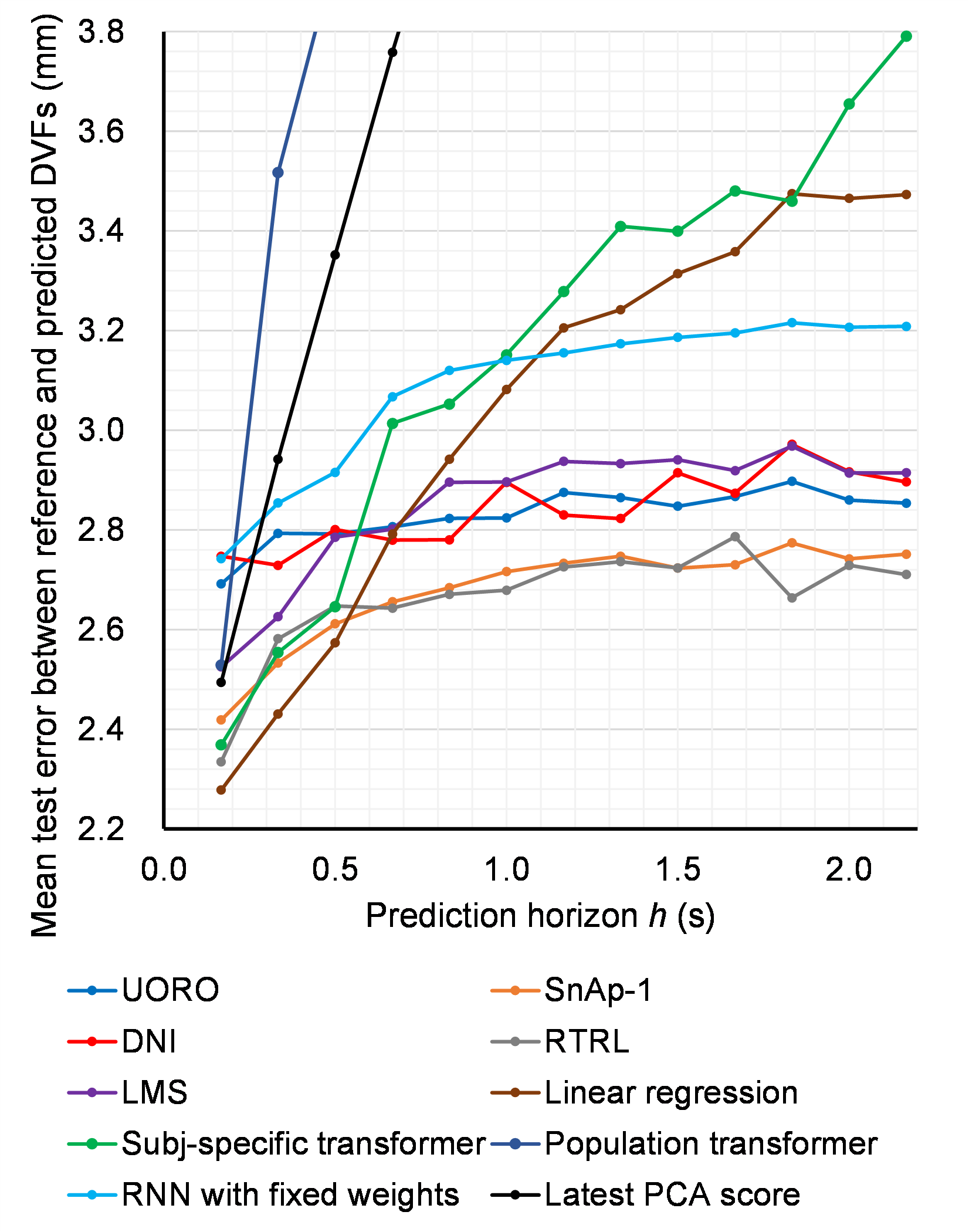}% 
    \includegraphics[width=.33\textwidth]{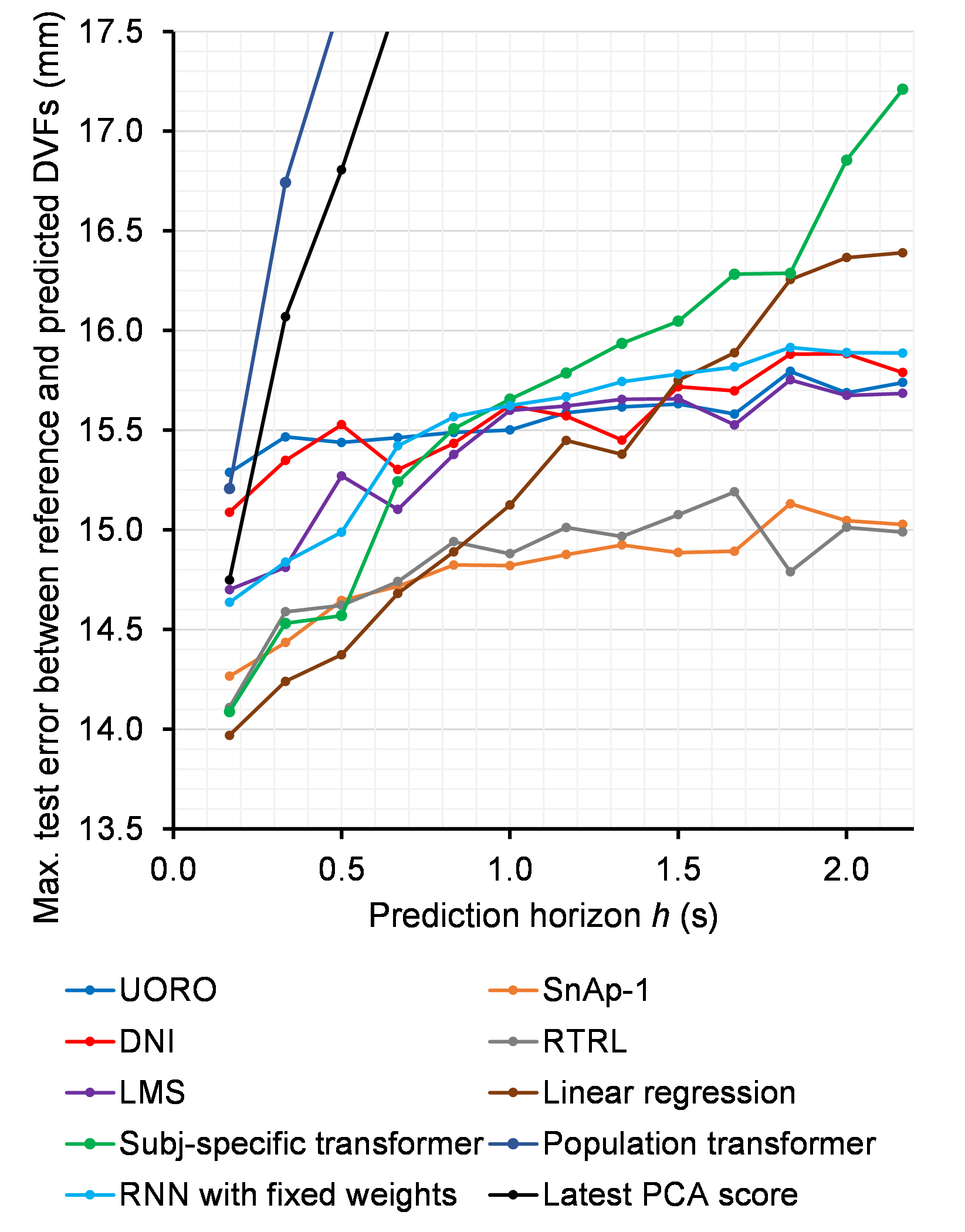} %      
    \includegraphics[width=.33\textwidth]{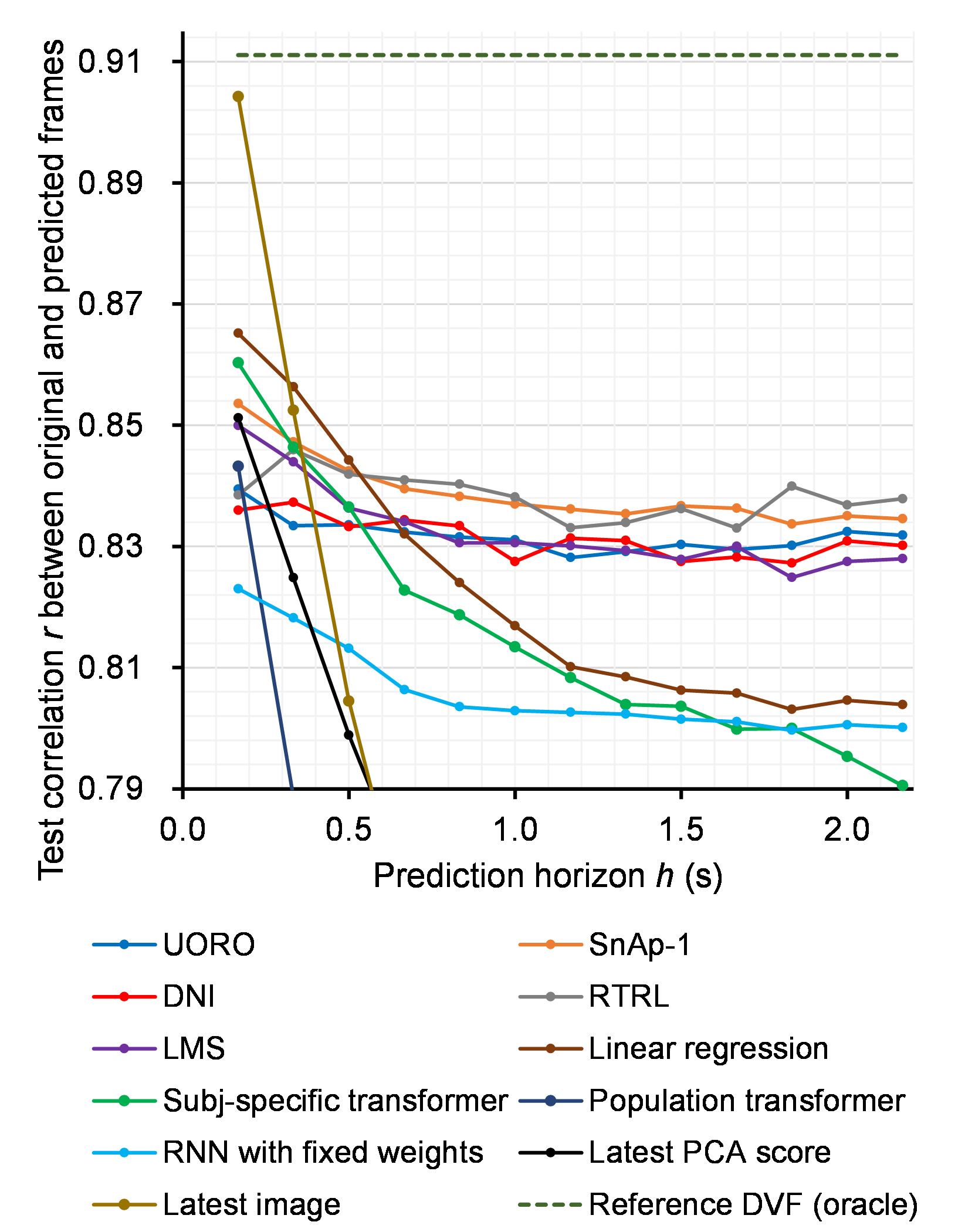}% 
    \includegraphics[width=.33\textwidth]{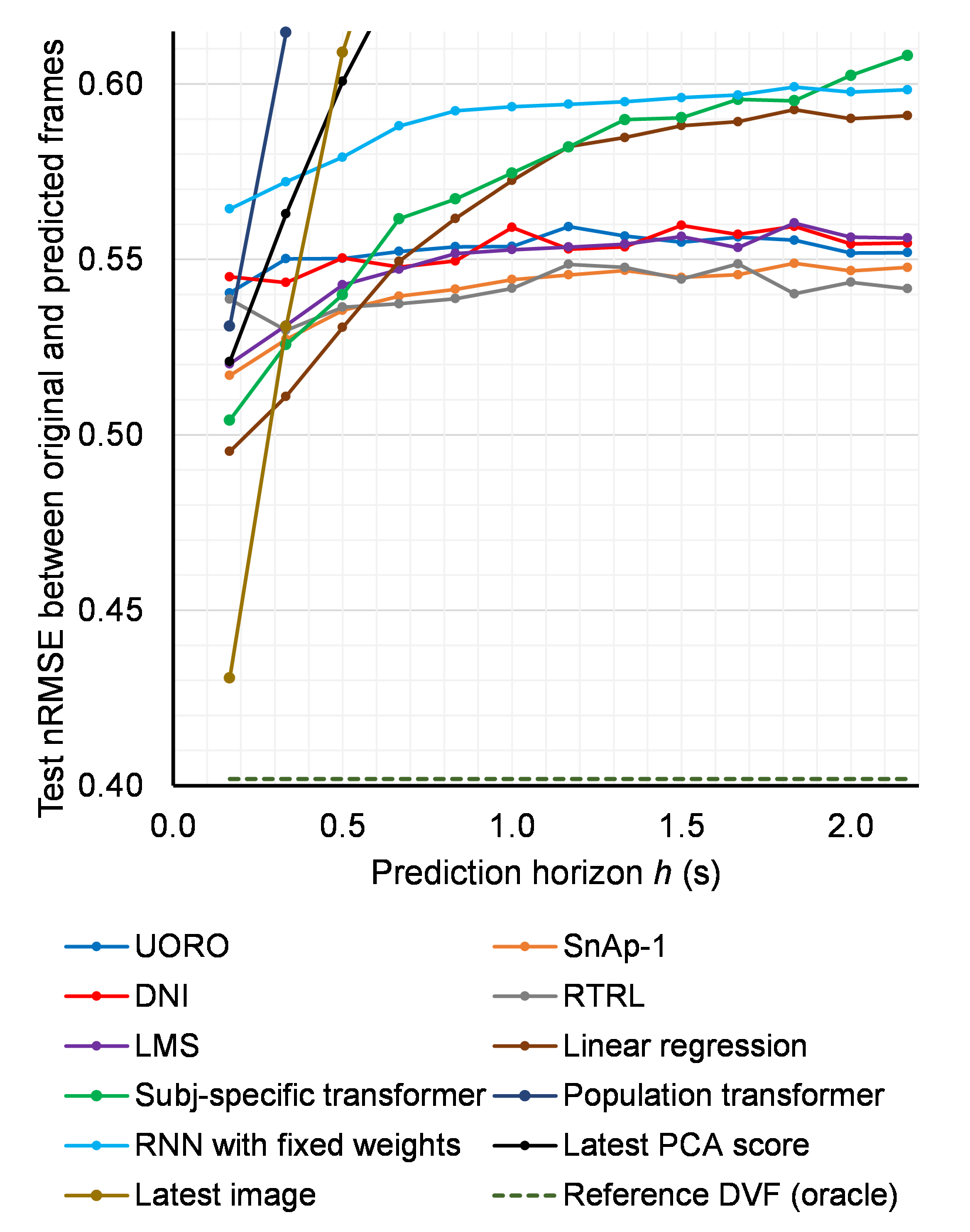}%  
    \includegraphics[width=.33\textwidth]{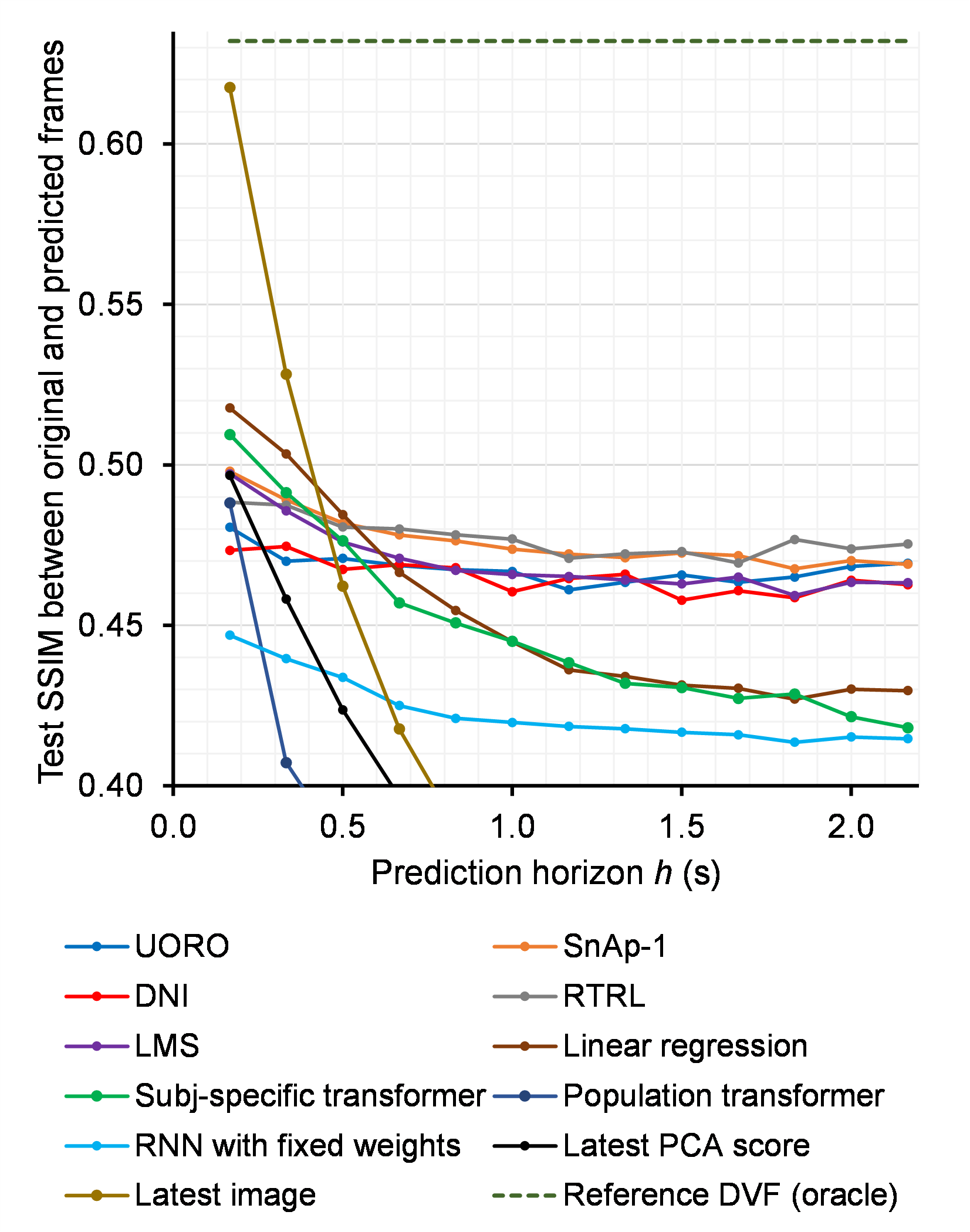} %                  
    \caption{Test-set \gls{ROI}-based frame-forecasting performance for each algorithm as a function of the horizon $h$, for the \gls{OvGU} dataset. Each point represents the average of a given metric over the eight image sequences and $n_{\text{test}}$ runs. Hyperparameters were optimized for each sequence (except the population transformer) and each value of $h$ via grid search on the validation set. The mean of each curve (i.e., the performance averaged over $h$ for each method) is reported in the \enquote{\gls{OvGU} (\gls{ROI})} columns of Table \ref{table:frame pred perf}.}
    \label{fig:next frame pred perf vs hrz on Magdeburg ROI}
\end{figure*}

% General trends
Across all algorithms, accuracy metrics ($r$ and \gls{SSIM}) tended to decrease with the horizon $h$, whereas error metrics (image-domain \gls{nRMSE} and \gls{DVF} errors) generally increased with $h$ (Figs. \ref{fig:next frame pred perf vs hrz on ETH}--\ref{fig:next frame pred perf vs hrz on Magdeburg ROI} and Fig. \ref{fig:next frame pred perf vs hrz on Magdeburg full image} in Appendix \ref{appendix: performance variation with h on Magdeburg}). Nonetheless, performance evolution with $h$ exhibited some instability, due to small dataset sizes ($N \leq 8$ sequences for each dataset) and per-horizon hyperparameter tuning. That phenomenon was most pronounced for the full-frame maximum \gls{DVF} error on the \gls{OvGU} acquisitions, possibly due to less accurate deformable registration near the low-contrast, noise-dominated image borders (Fig. \ref{fig:next frame pred perf vs hrz on Magdeburg full image}).

% Rk: the parameters / brackets after {figure*} have influence on the caption position, so I could modify that to have the correct layout if a layout problem occur (e.g., caption right aligned) https://tex.stackexchange.com/questions/533409/caption-placement-for-double-column-figure-in-elsevier-document

% Rk: same as in previous figure

% General performance and behaviour of RNNs and transformers
\Gls{SnAp-1} and \gls{RTRL}, the strongest predictors for intermediate-to-high values of $h$, exhibited nearly identical performance regardless of $h$. \Gls{UORO}, \gls{DNI}, and \gls{LMS} formed a second group with relatively close performance on the \gls{OvGU} data and similar \gls{DVF} errors and \glspl{nRMSE} on the ETH Zürich data. The gap between the two groups remained small and largely stable as $h$ increased. This pattern mirrors the behavior observed for \gls{PCA}-weight forecasting (Fig. \ref{fig:signal pred error 3 PCA cpts vs horizon}) and aligns with the medium-to-large effect sizes reported in Table \ref{table:statistical tests}. The sequence-specific transformer achieved competitive accuracy at low horizons, performing similarly to \gls{RTRL} and \gls{SnAp-1} for $h \leq 0.63\text{s}$ on the ETH Zürich data (e.g., with a mean \gls{DVF} error below 1.37mm) and $h \leq 0.50\text{s}$ on the \gls{OvGU} sequences (e.g., with a \pgls{ROI}-based mean \gls{DVF} error below 2.65mm). However, its accuracy degraded rapidly as $h$ increased, unlike online-trained \glspl{RNN}. The population transformer remained the least accurate non-baseline predictor across all horizons, except at $h=0.17\text{s}$ on the \gls{OvGU} data (e.g., with \pgls{ROI}-based mean \gls{DVF} error of 2.53 $\pm$ 0.17mm\footnotemark), where it outperformed \gls{UORO} and \gls{DNI} for both evaluation regions and most metrics.

\footnotetext{We report here the 70\% \glspl{CI} computed from the Student's t-distribution with $N-1$ degrees of freedom, where $N$ is the number of sequences in each dataset. Horizon-wise \glspl{CI} were not plotted in Figs. \ref{fig:next frame pred perf vs hrz on ETH}, \ref{fig:next frame pred perf vs hrz on Magdeburg ROI}, and \ref{fig:next frame pred perf vs hrz on Magdeburg full image} to avoid visual clutter and better emphasize trends with $h$.}

% Performance at low horizons
At low horizons, excluding the oracle, the image-domain persistence model yielded the best intensity-based metrics on the \gls{OvGU} data in the \glspl{ROI} at $h=0.17\text{s}$ and over the full images for $h \leq 0.33\text{s}$, respectively. Its \gls{ROI}-based \gls{SSIM} was also the highest among non-oracle predictors at $h=0.33\text{s}$. By contrast, it consistently underperformed the non-baseline algorithms on the ETH Zürich data, except the cross-subject transformer, whose $r$ and \gls{SSIM} were lower at $h=0.31\text{s}$. This discrepancy likely reflects differences in acquisition properties: the ETH Zürich sequences feature lower noise, higher contrast, and less motion variability, which makes prediction easier for non-naive models at short response times. Conversely, matching local noise patterns in the \gls{OvGU} frames can inflate performance metrics at low look-ahead times. For both datasets, linear regression performed strongly at small horizons and reached the lowest mean and maximum \gls{DVF} errors at $h=0.31\text{s}$ on the ETH Zürich sequences (e.g., mean \gls{DVF} error of 1.30 $\pm$ 0.18mm) and for all $h \leq 0.50\text{s}$ on the \gls{OvGU} data in both evaluation regions (e.g., 2.43 $\pm$ 0.23mm \gls{ROI}-based mean \gls{DVF} error at $h=0.33\text{s}$). The lower crossover point at which it began to outperform image copying in terms of \gls{ROI}-based $r$ and \gls{nRMSE} on the \gls{OvGU} data ($h=0.33\text{s}$ instead of $h=0.50\text{s}$ for full-frame evaluation) may reflect the lower noise and more structured motion in that area, which facilitate \gls{DIR} and short-term motion forecasting. However, even for the smallest horizons, the oracle predictor almost systematically surpassed the strongest predictors, including linear regression and the last-image model.

% Performance at medium-to-high horizon for both datasets
% I could remoe the exception for the max DVF error using "generally" or by removing the following sentence to make the text shorter if necessary.
Regarding medium-to-long horizons, \gls{SnAp-1} and \gls{RTRL} were more accurate than linear regression on the \gls{OvGU} dataset when $h \geq 0.66\text{s}$, except for the maximum deformation error. Both \gls{RNN} algorithms attained the lowest maximum \gls{DVF} error among all algorithms over the \glspl{ROI} for $h \geq 1.0\text{s}$ and over the full frames for most horizons $h \geq 0.50\text{s}$, with a few exceptions due to instability in whole-image evaluation. On the ETH Zürich data, they also generally outperformed the other algorithms, including linear regression, in terms of geometrical errors and \glspl{nRMSE} for $h \geq 0.94\text{s}$. Notably, \gls{LMS} reached the highest $r$ and \gls{SSIM} values (among non-oracle methods) for $h \leq 1.88\text{s}$ and $h \leq 1.57\text{s}$, respectively. Performance degradation for \gls{SnAp-1} and \gls{RTRL} beyond $h = 0.66\text{s}$ was less pronounced on the \gls{OvGU} dataset. For instance, the \gls{ROI}-based \gls{DVF} error of \gls{SnAp-1} rose from 2.42 $\pm$ 0.19mm at $h = 0.17\text{s}$ to 2.66 $\pm$ 0.25mm at $h = 0.66\text{s}$ and only reached 2.75 $\pm$ 0.26mm at $h = 2.17\text{s}$. 

% Comments on stability with the horizon
The relative decrease in \gls{SSIM} and relative increase in mean \gls{DVF} error between $h=0.3\text{s}$ and $h=2.2\text{s}$ were lower for online-trained \glspl{RNN} than for all other non-baseline predictors (Table \ref{table:stability_horizon} in Appendix \ref{appendix: performance variation with h on Magdeburg}). \Gls{LMS} consistently ranked immediately below \glspl{RNN} among non-baseline methods regarding stability across horizons. On the \gls{OvGU} dataset, stability decreased from whole-image evaluation to \gls{ROI} evaluation, as relative metric variations between the smallest and largest horizons generally increased by a factor of approximately 2--3. This reflects the presence of static peripheral background regions in the full frames and the more complex, high-amplitude deformations of the diaphragm, liver, and surrounding structures within the \gls{ROI}. By contrast, full-frame stability with $h$ compared across datasets showed no consistent trend: on average over non-baseline algorithms, the relative decrease in \gls{SSIM} between $h=0.3\text{s}$ and $h=2.2\text{s}$ was roughly three times larger for the ETH Zürich acquisitions than for the \gls{OvGU} ones, whereas the relative \gls{DVF}-error increase remained similar.

\subsubsection{Qualitative evaluation}\label{section: frame forecasting qualitative eval}

\subsubsubsection{Visual assessment on the ETH Zürich dataset}

\begin{figure*}[pos=htbp, align=\centering]
    \captionsetup[subfigure]{labelformat=empty} % https://stackoverflow.com/questions/48595559/how-to-refer-sub-figures-without-using-captions
    \centering
% First row: prediction at t = 188 (58.4s) with h=1 (0.3s)
    \subfloat[Ground-truth \\ image, $t=58.4\text{s}$ \\ (expiration)]{\includegraphics[width=.125\textwidth]{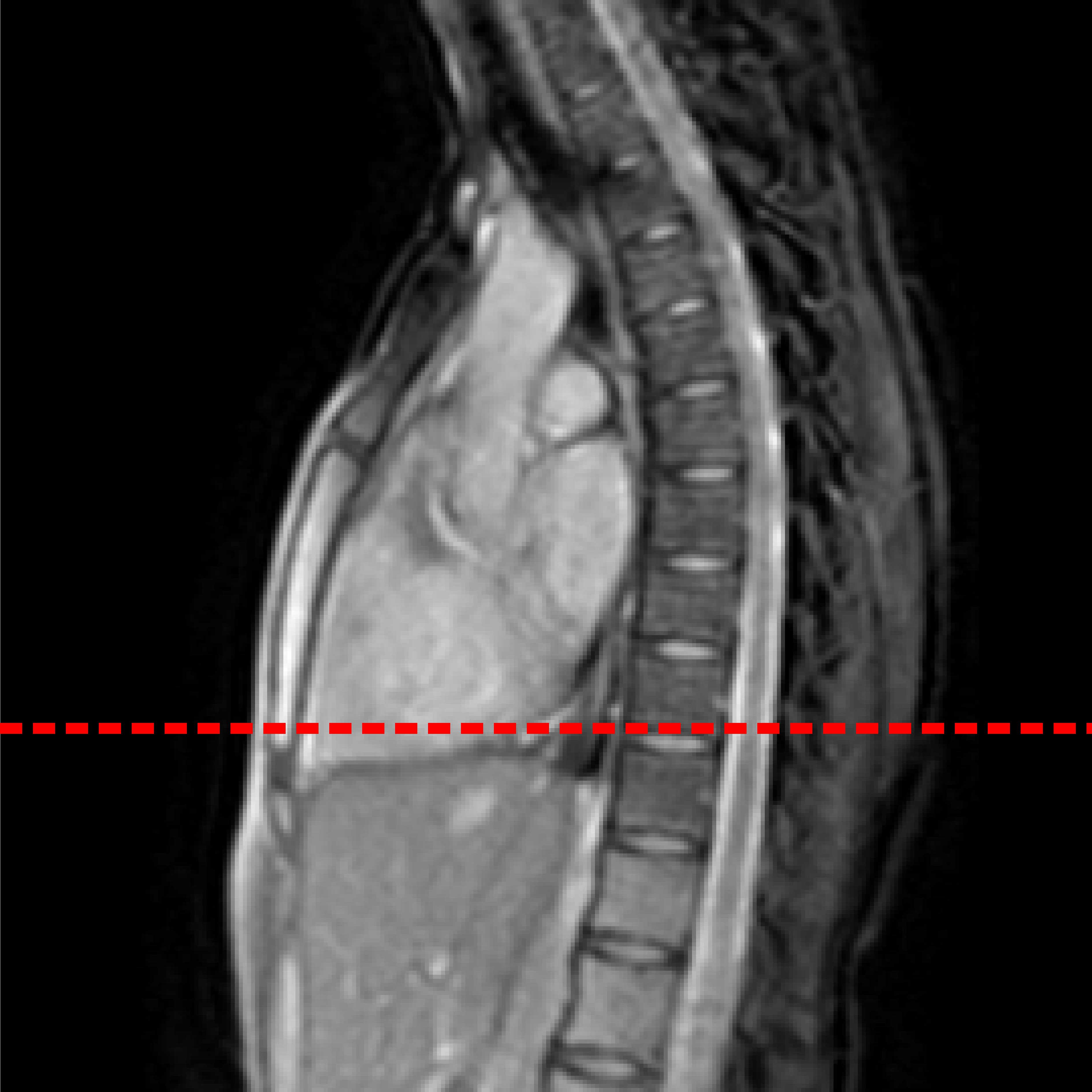} }% 
    \subfloat[Predicted image \\ $t\!=\!58.4\text{s}$, $h\!=\!0.3\text{s}$ \\ \acs{RTRL}]{\includegraphics[width=.125\textwidth]{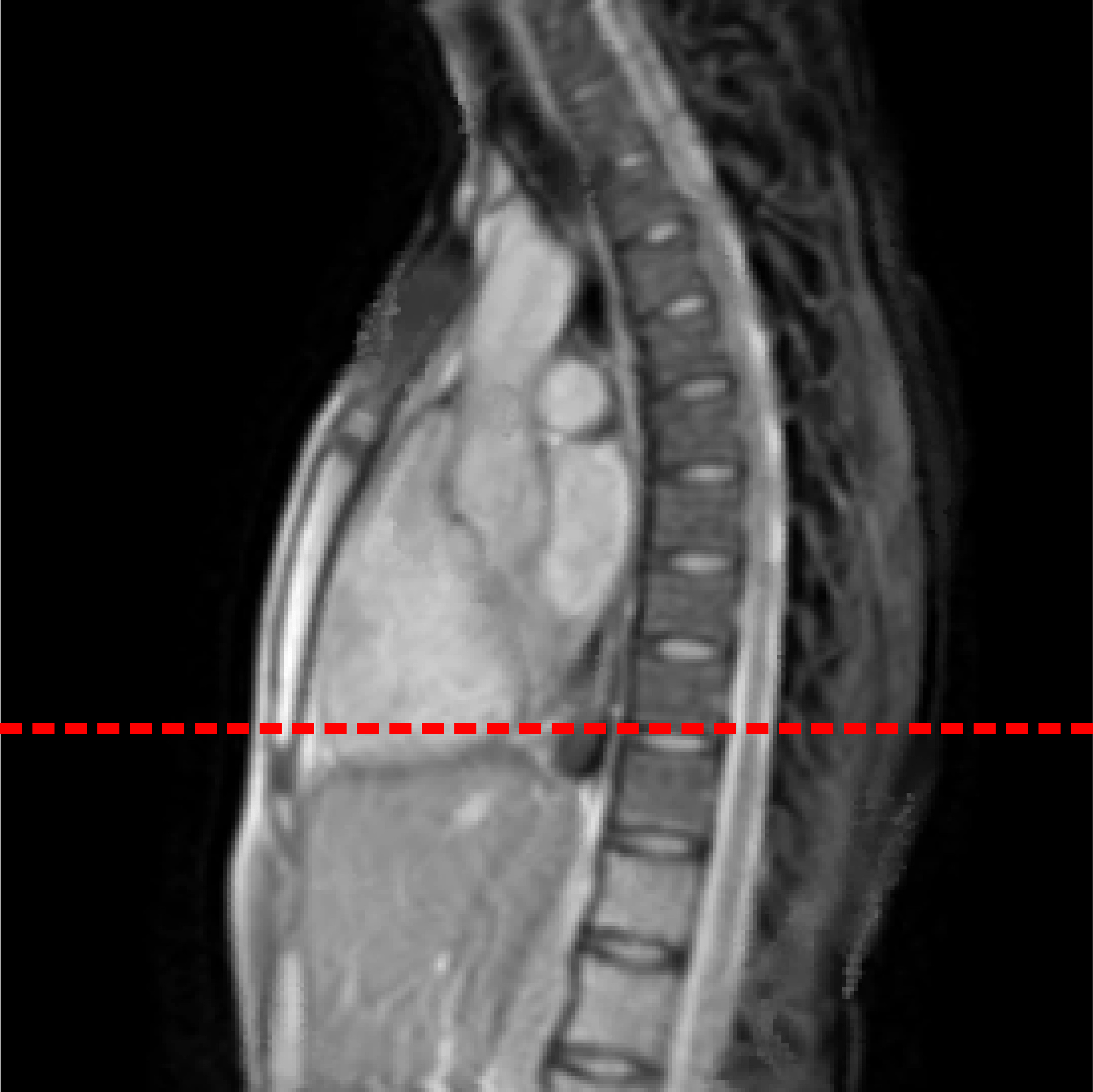} }%    
    \subfloat[Intensity error \\ $t=58.4\text{s}$, $h=0.3\text{s}$ \\ \acs{RTRL}]{\includegraphics[width=.145\textwidth, height=.125\textwidth]{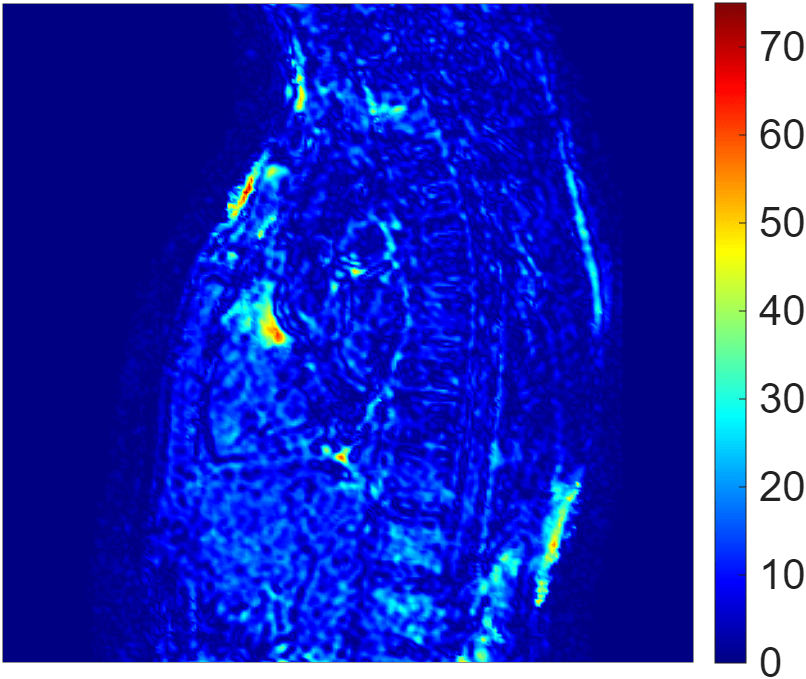} }%    
    \subfloat[\acs{DVF} error (in mm) \\ $t=58.4\text{s}$, $h=0.3\text{s}$ \\ \acs{RTRL}]{\includegraphics[width=.145\textwidth, height=.125\textwidth]{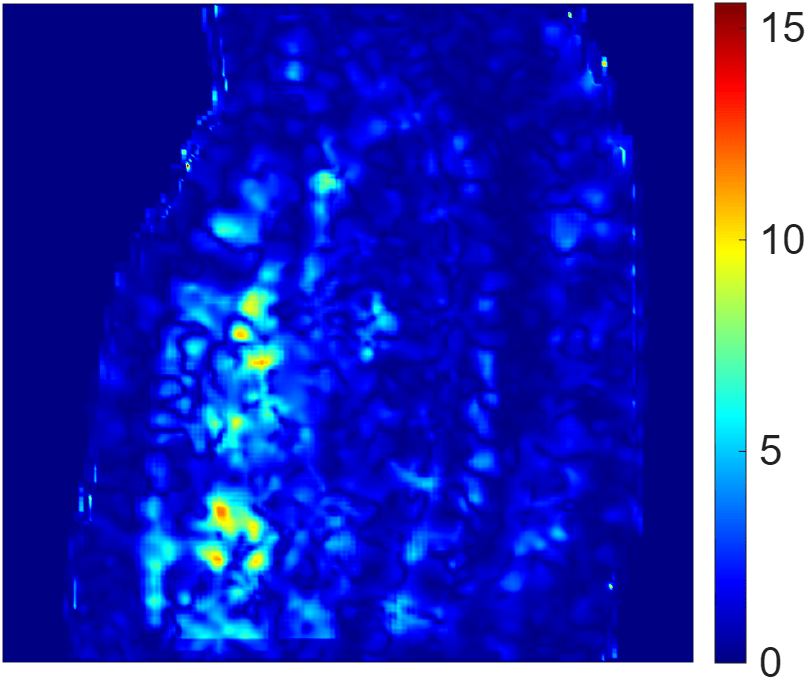} }% 	
    \subfloat[Predicted image \\ $t\!=\!58.4\text{s}$, $h\!=\!0.3\text{s}$ \\ population transformer]{\includegraphics[width=.125\textwidth]{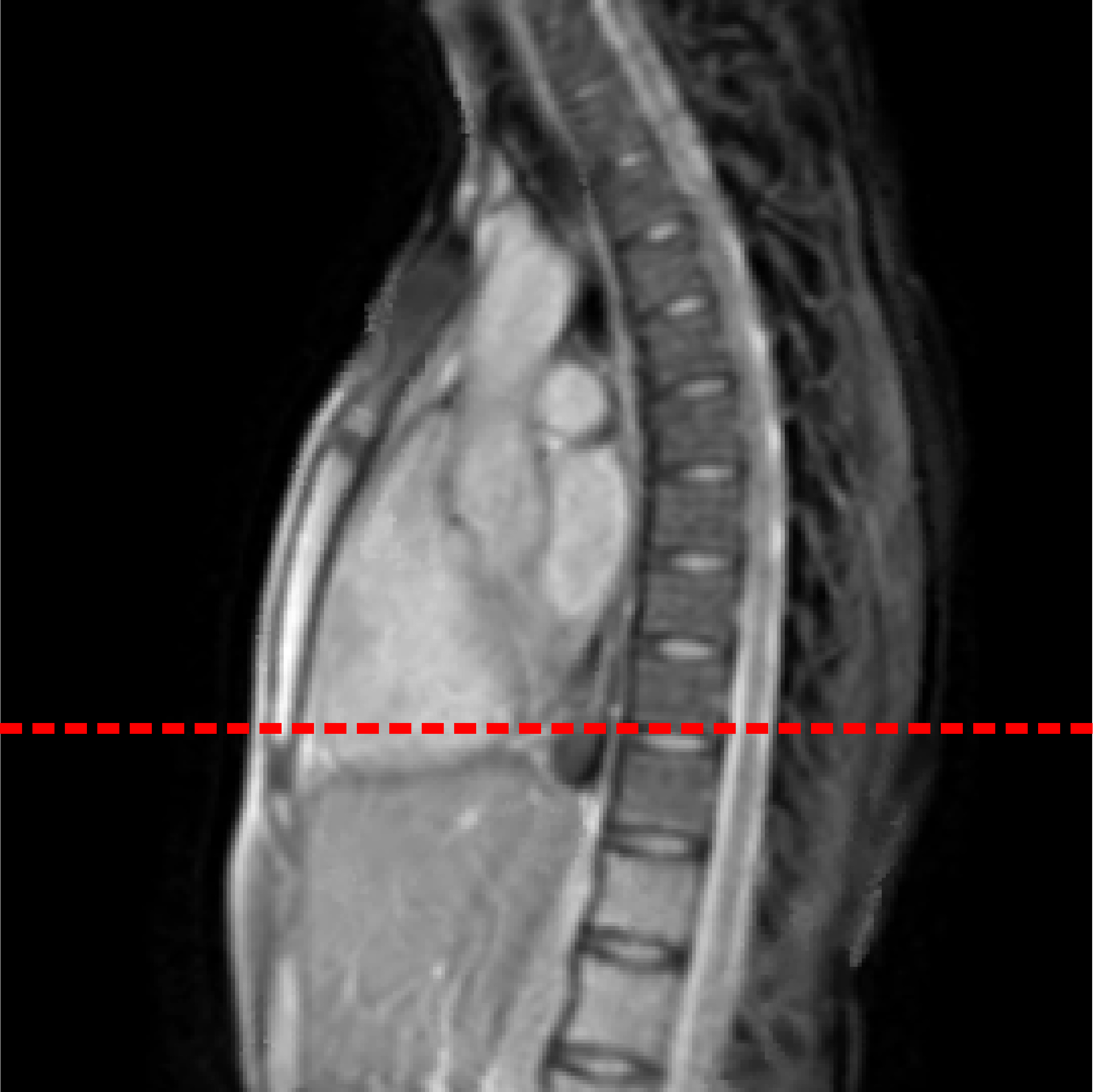} }%    
    \subfloat[Intensity error \\ $t=58.4\text{s}$, $h=0.3\text{s}$ \\ population transformer]{\includegraphics[width=.145\textwidth, height=.125\textwidth]{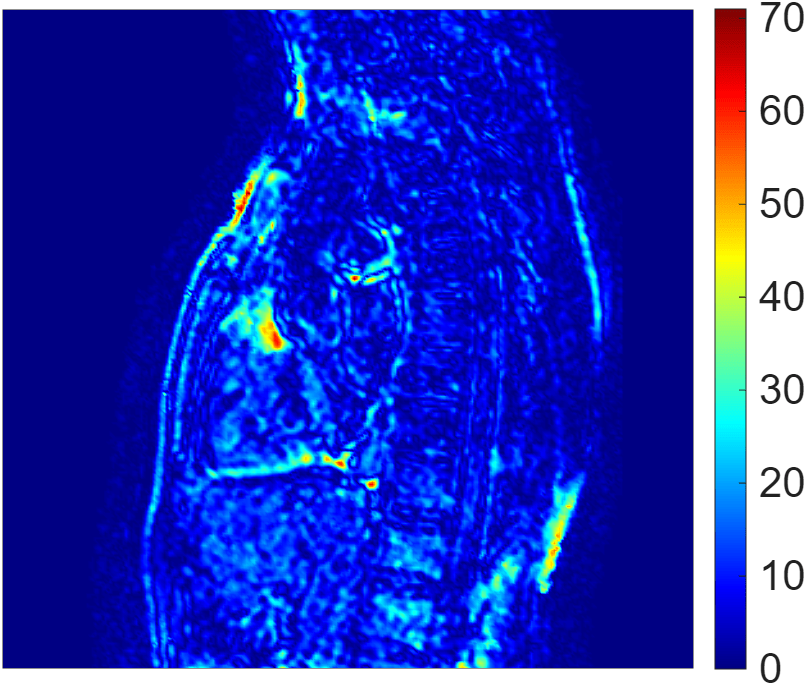} }%        
    \subfloat[\acs{DVF} error (in mm) \\ $t=58.4\text{s}$, $h=0.3\text{s}$ \\ population transformer]{\includegraphics[width=.145\textwidth, height=.125\textwidth]{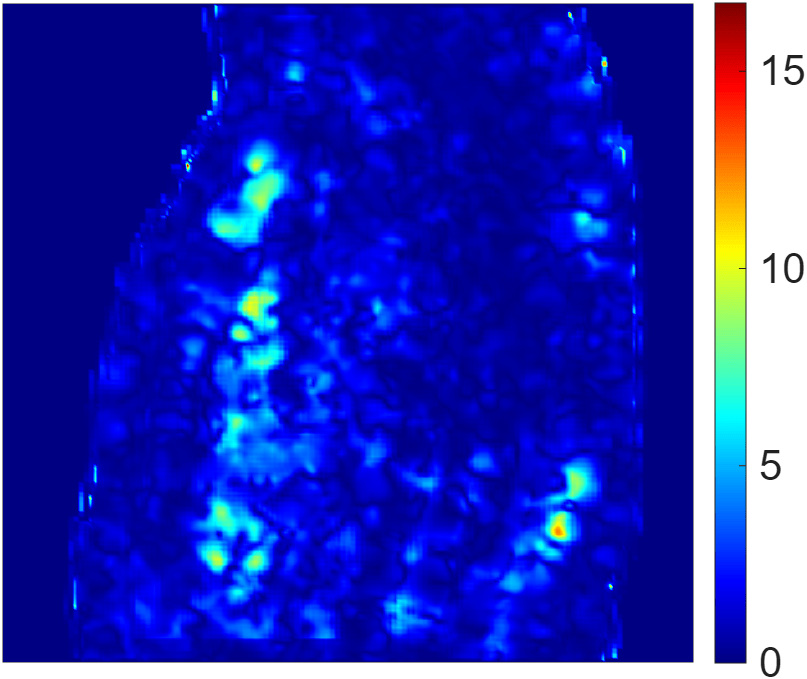} }% 	
	\\   
% Second row: prediction at t = 188 (58.4s) with h=7 (2.2s)
    \hspace{.125\textwidth}
    \subfloat[Predicted image \\ $t\!=\!58.4\text{s}$, $h\!=\!2.2\text{s}$ \\ \acs{RTRL}]{\includegraphics[width=.125\textwidth]{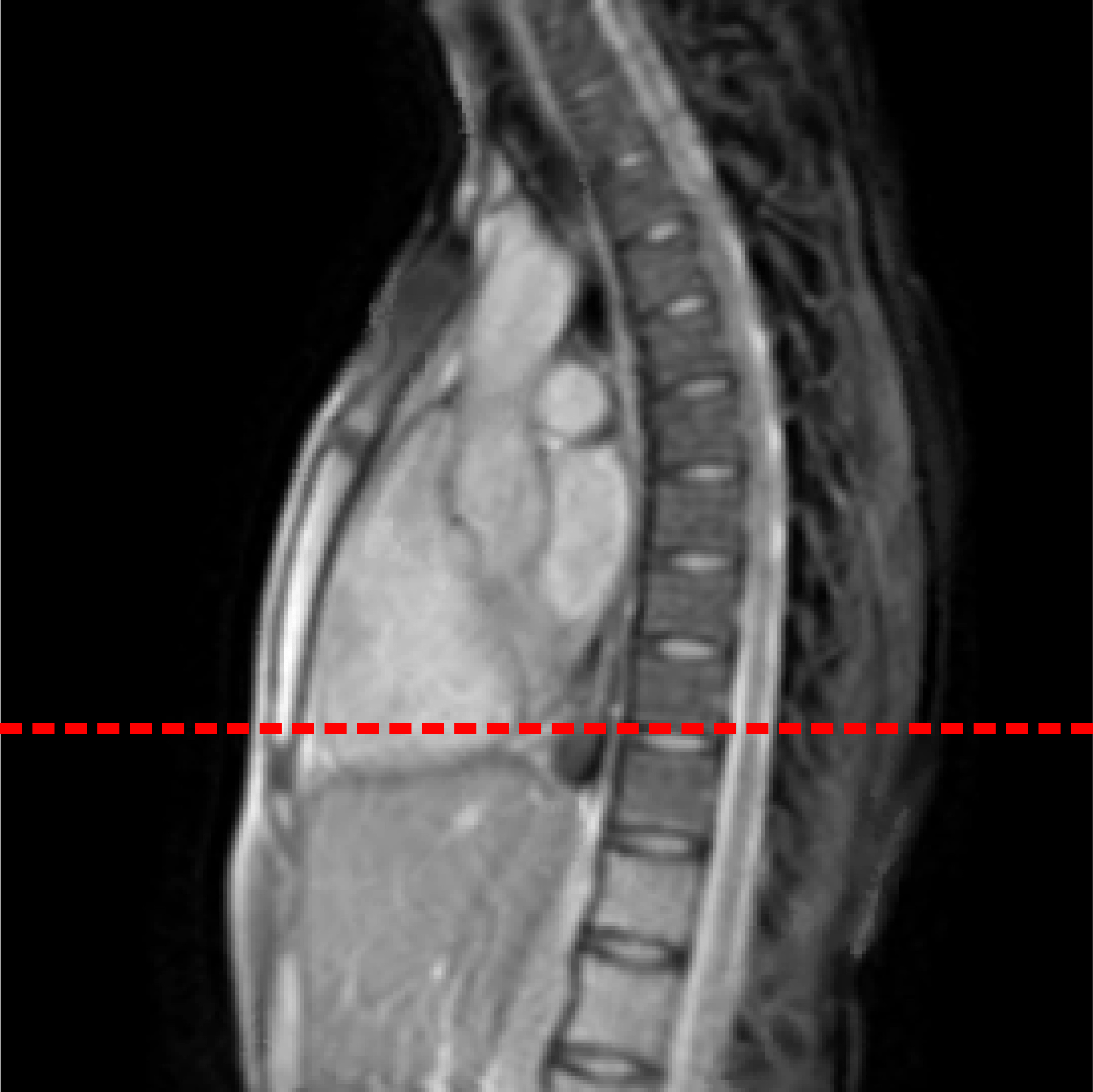} }%    
    \subfloat[Intensity error \\ $t=58.4\text{s}$, $h=2.2\text{s}$ \\ \acs{RTRL}]{\includegraphics[width=.145\textwidth, height=.125\textwidth]{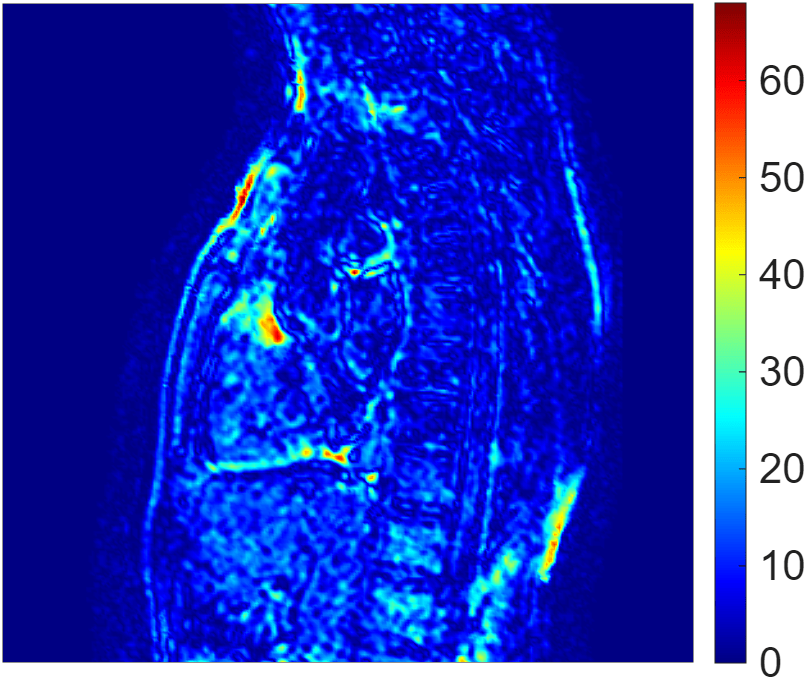} }%    
    \subfloat[\acs{DVF} error (in mm) \\ $t=58.4\text{s}$, $h=2.2\text{s}$ \\ \acs{RTRL}]{\includegraphics[width=.145\textwidth, height=.125\textwidth]{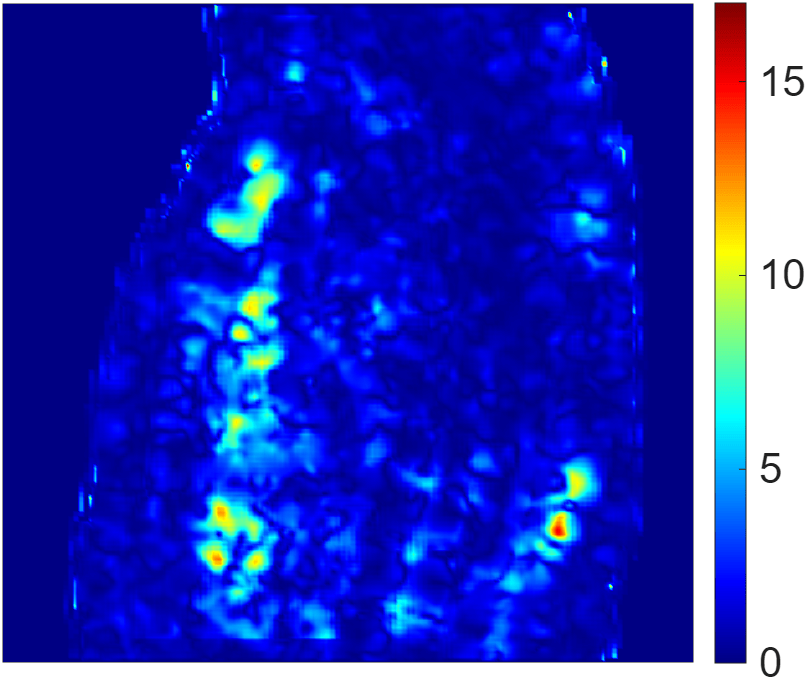} }% 	
    \subfloat[Predicted image \\ $t\!=\!58.4\text{s}$, $h\!=\!2.2\text{s}$ \\ population transformer]{\includegraphics[width=.125\textwidth]{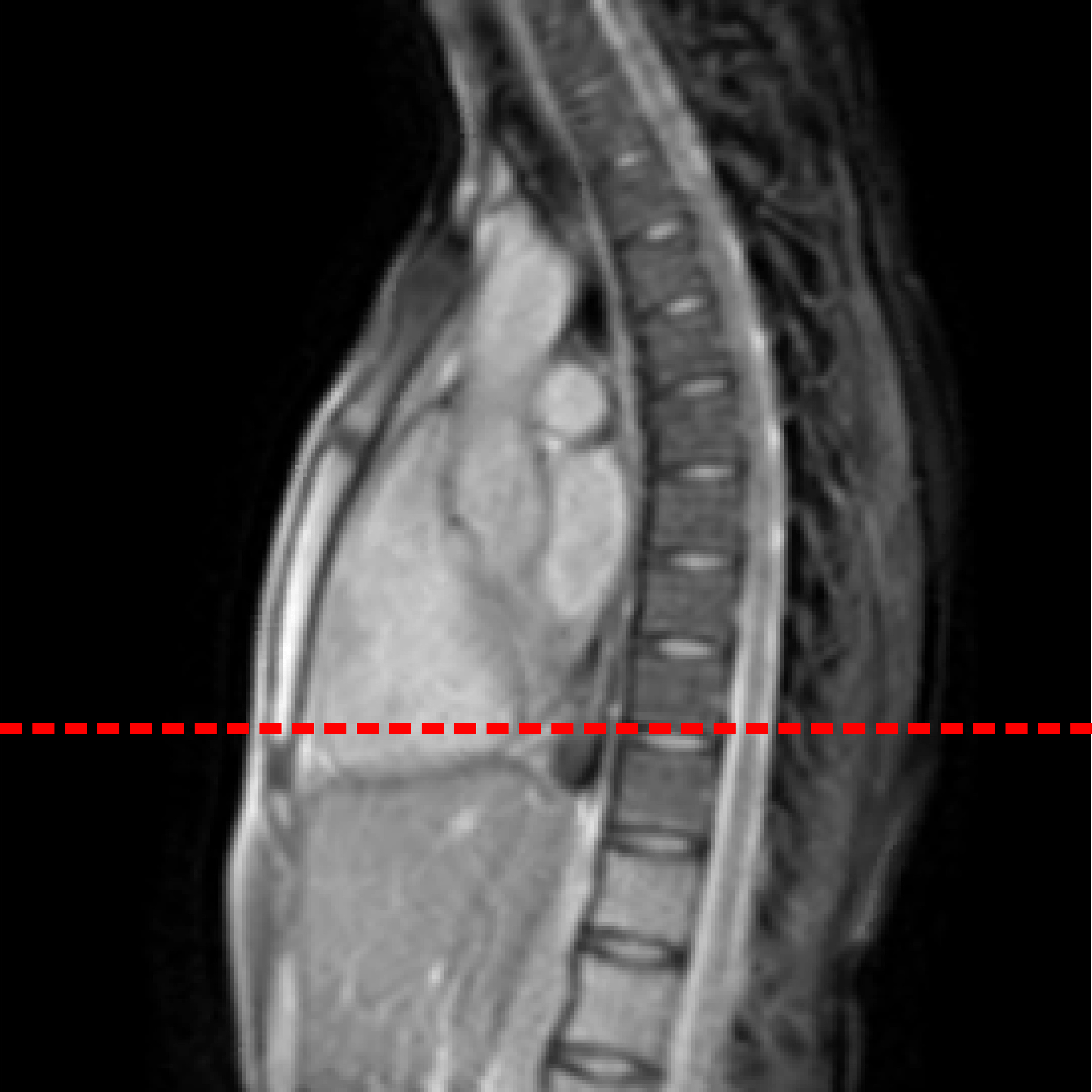} }%    
    \subfloat[Intensity error \\ $t=58.4\text{s}$, $h=2.2\text{s}$ \\ population transformer]{\includegraphics[width=.145\textwidth, height=.125\textwidth]{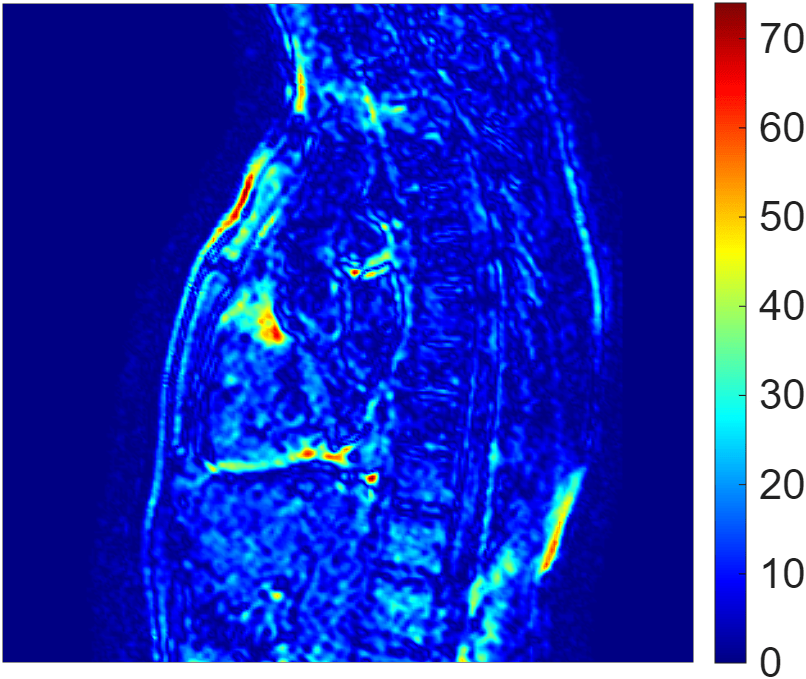} }%        
    \subfloat[\acs{DVF} error (in mm) \\ $t=58.4\text{s}$, $h=2.2\text{s}$ \\ population transformer]{\includegraphics[width=.145\textwidth, height=.125\textwidth]{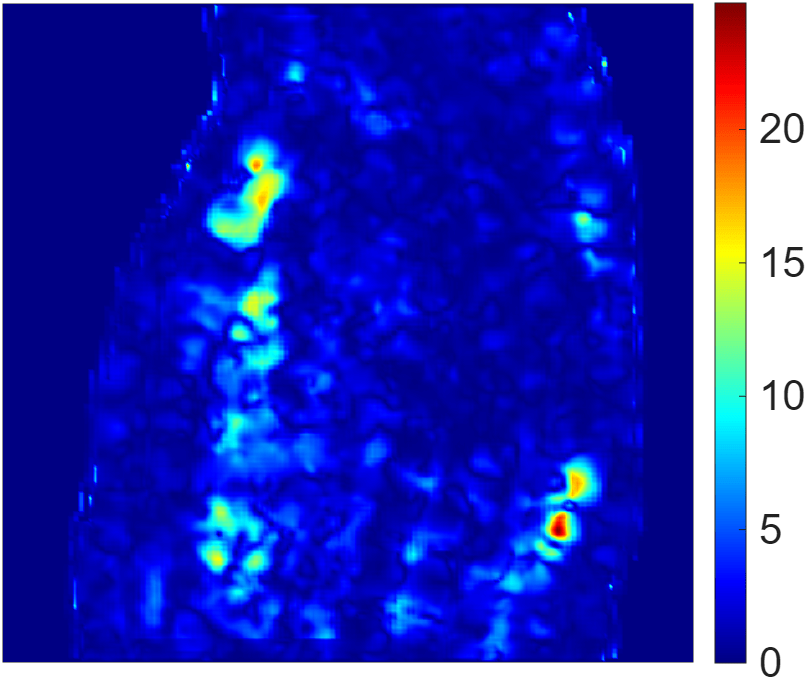} }% 
    \\
% Third row: prediction at t = 196 (60.9s) with h=1	(0.3s)
    \subfloat[Ground-truth \\ image, $t=60.9\text{s}$ \\ (inspiration)]{\includegraphics[width=.125\textwidth]{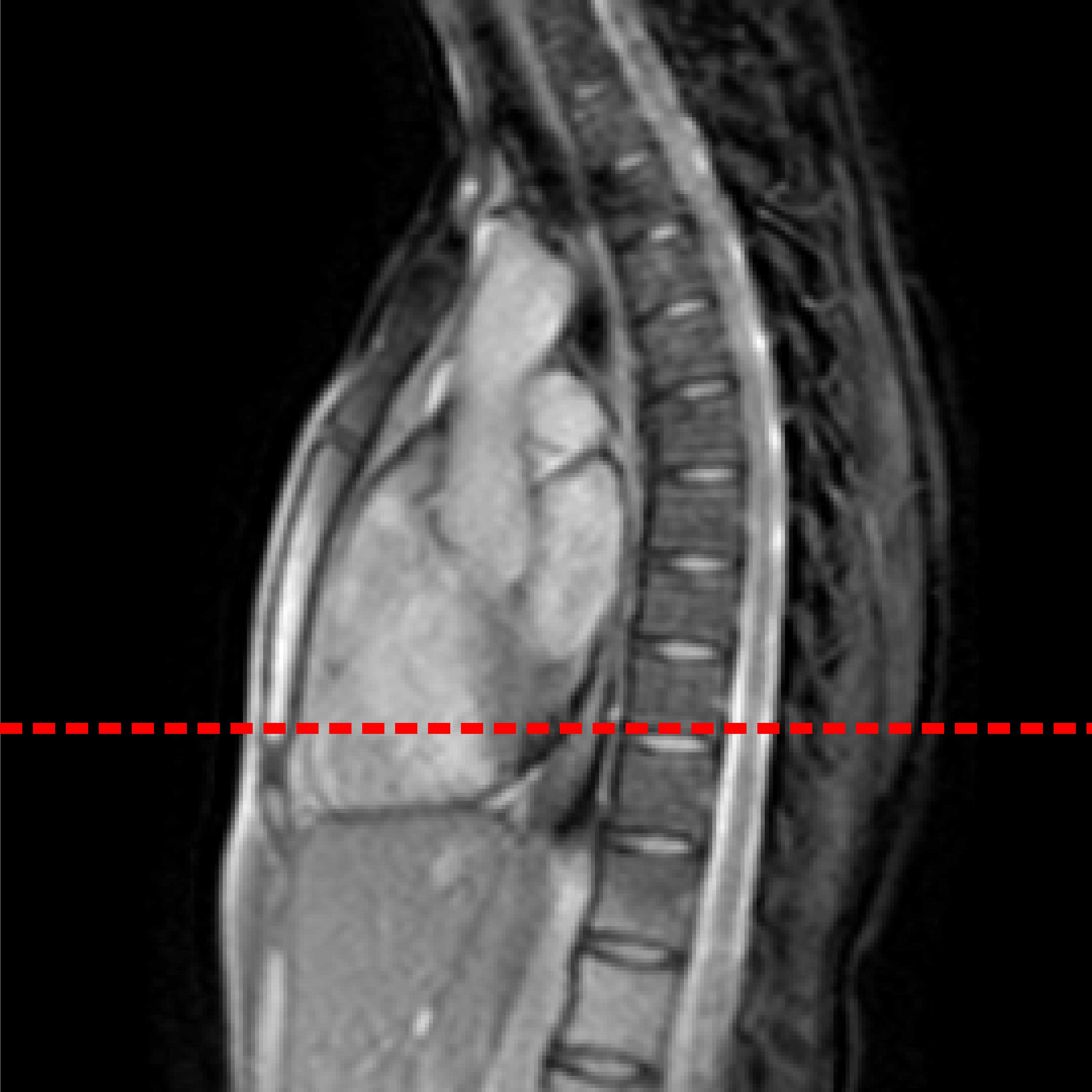} }% 
    \subfloat[Predicted image \\ $t\!=\!60.9\text{s}, h\!=\!0.3\text{s}$ \\ \acs{RTRL}]{\includegraphics[width=.125\textwidth]{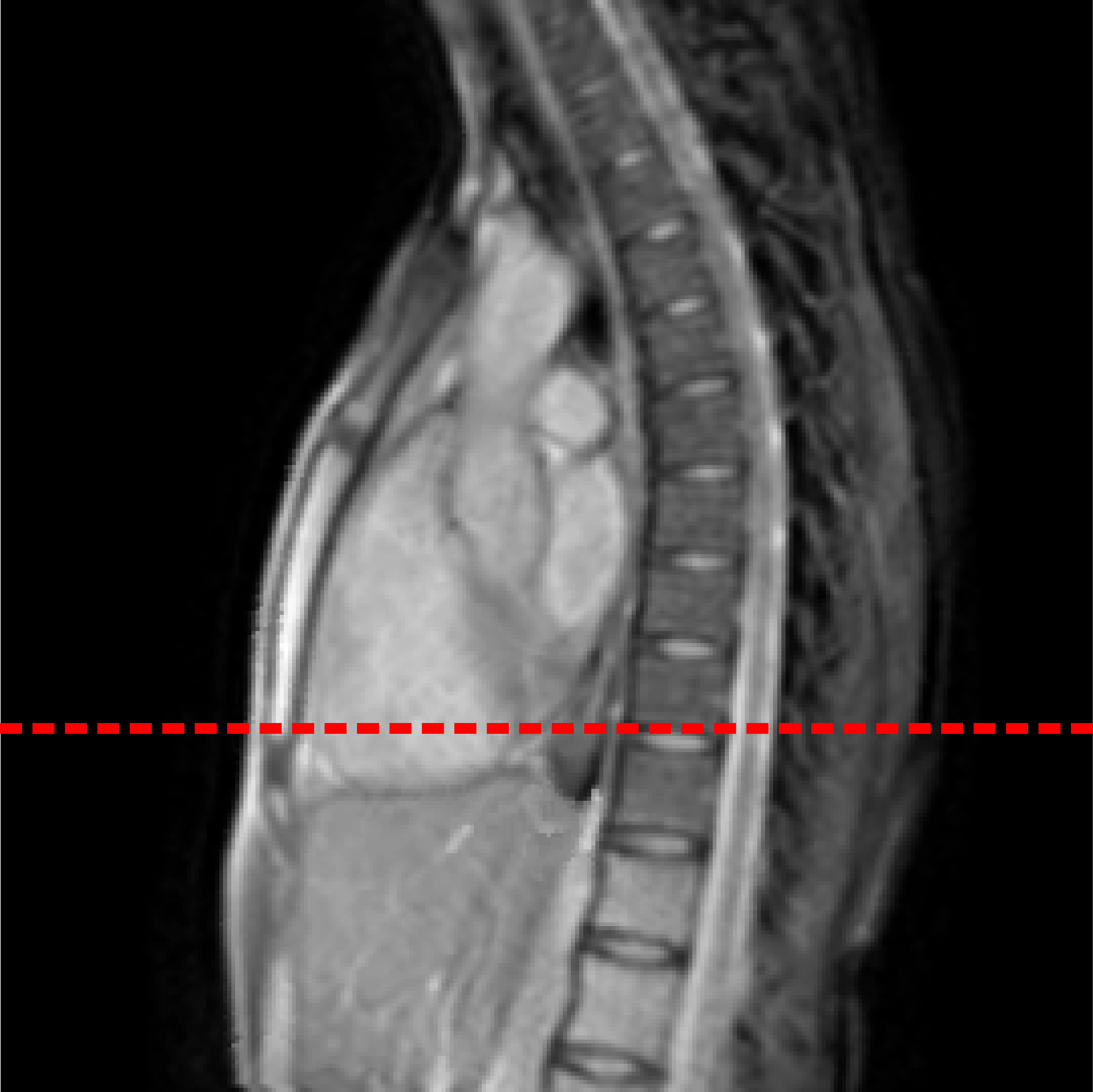} }%    
    \subfloat[Intensity error \\ $t=60.9\text{s}$, $h=0.3\text{s}$ \\ \acs{RTRL}]{\includegraphics[width=.145\textwidth, height=.125\textwidth]{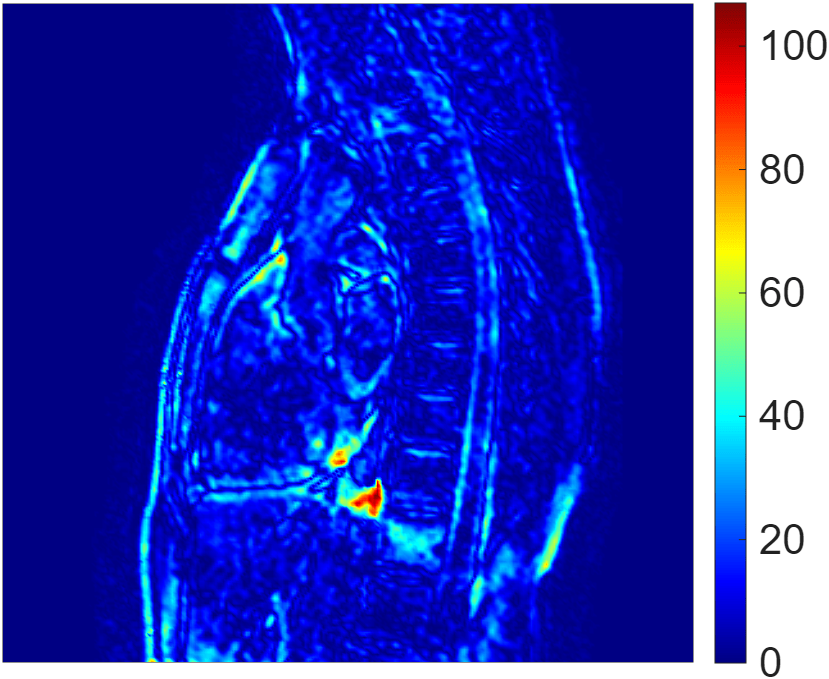} }%      
	\subfloat[\acs{DVF} error (in mm) \\ $t=60.9\text{s}$, $h=0.3\text{s}$ \\ \acs{RTRL}]{\includegraphics[width=.145\textwidth, height=.125\textwidth]{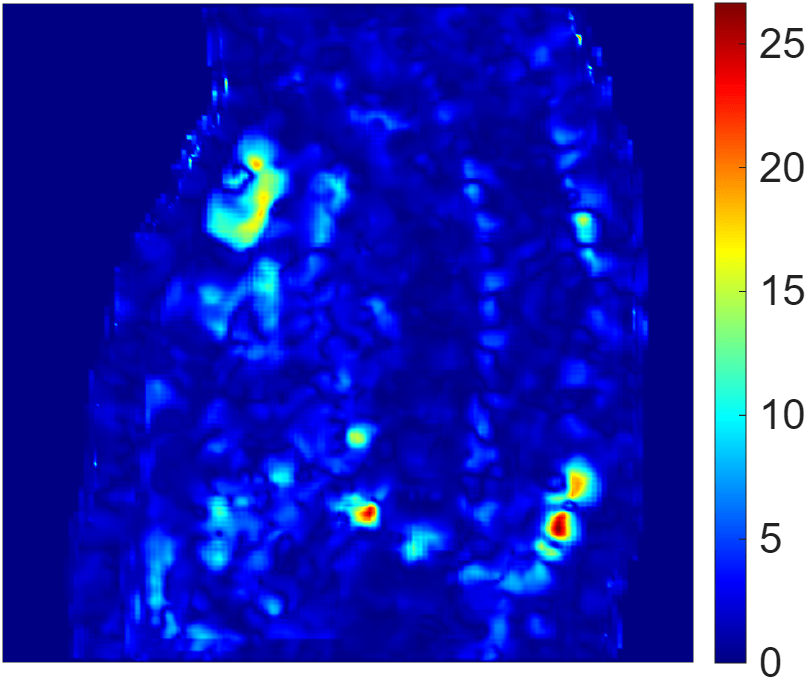} }%      
    \subfloat[Predicted image \\ $t\!=\!60.9\text{s}, h\!=\!0.3\text{s}$ \\ population transformer]{\includegraphics[width=.125\textwidth]{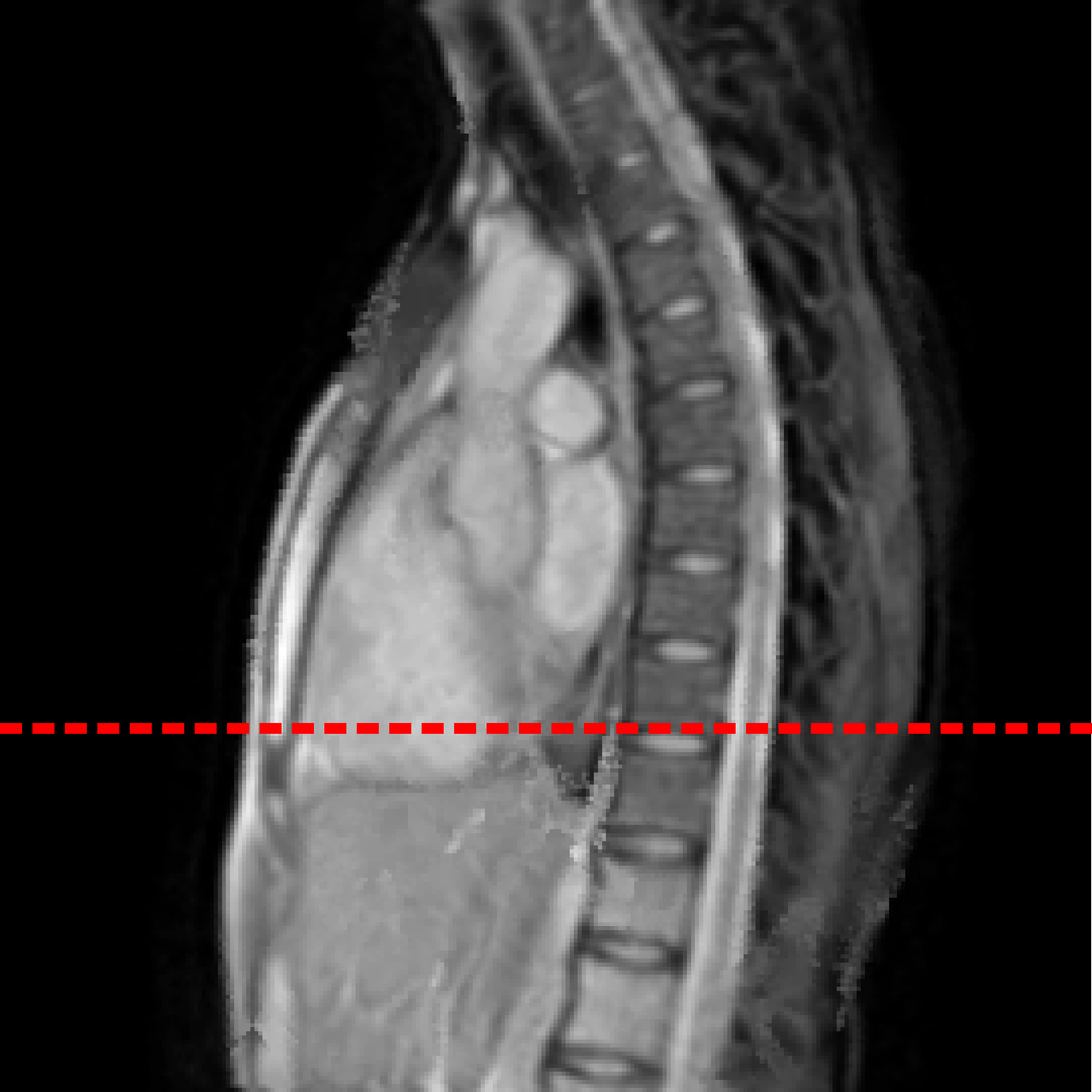} }%    
    \subfloat[Intensity error \\ $t=60.9\text{s}$, $h=0.3\text{s}$ \\ population transformer]{\includegraphics[width=.145\textwidth, height=.125\textwidth]{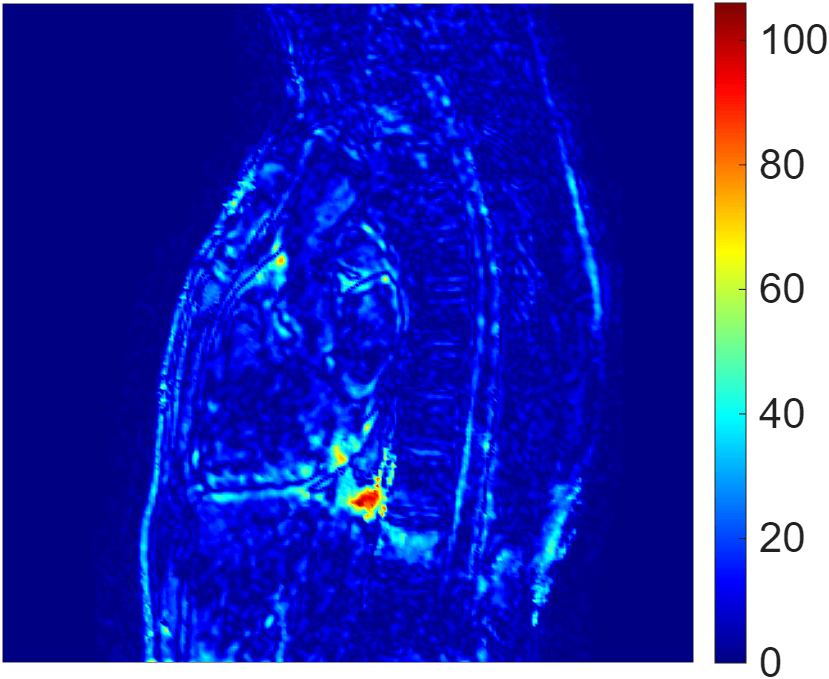} }%     
	\subfloat[\acs{DVF} error (in mm) \\ $t=60.9\text{s}$, $h=0.3\text{s}$ \\ population transformer]{\includegraphics[width=.145\textwidth, height=.125\textwidth]{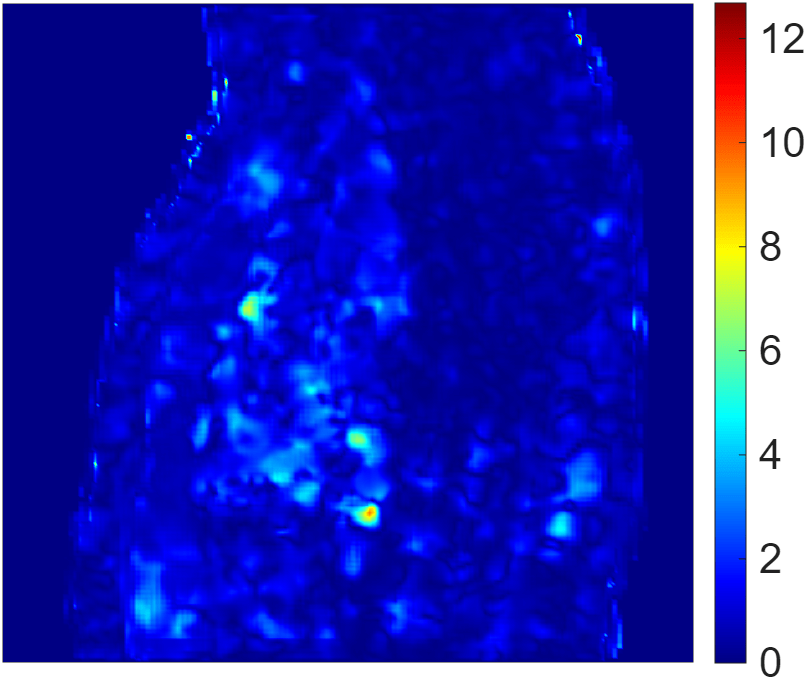} }%     
	\\
% Fourth row: prediction at t = 196 (60.9s) with h=7 (2.2s)
    \hspace{.125\textwidth}
    \subfloat[Predicted image \\ $t\!=\!60.9\text{s}, h\!=\!2.2\text{s}$ \\ \acs{RTRL}]{\includegraphics[width=.125\textwidth]{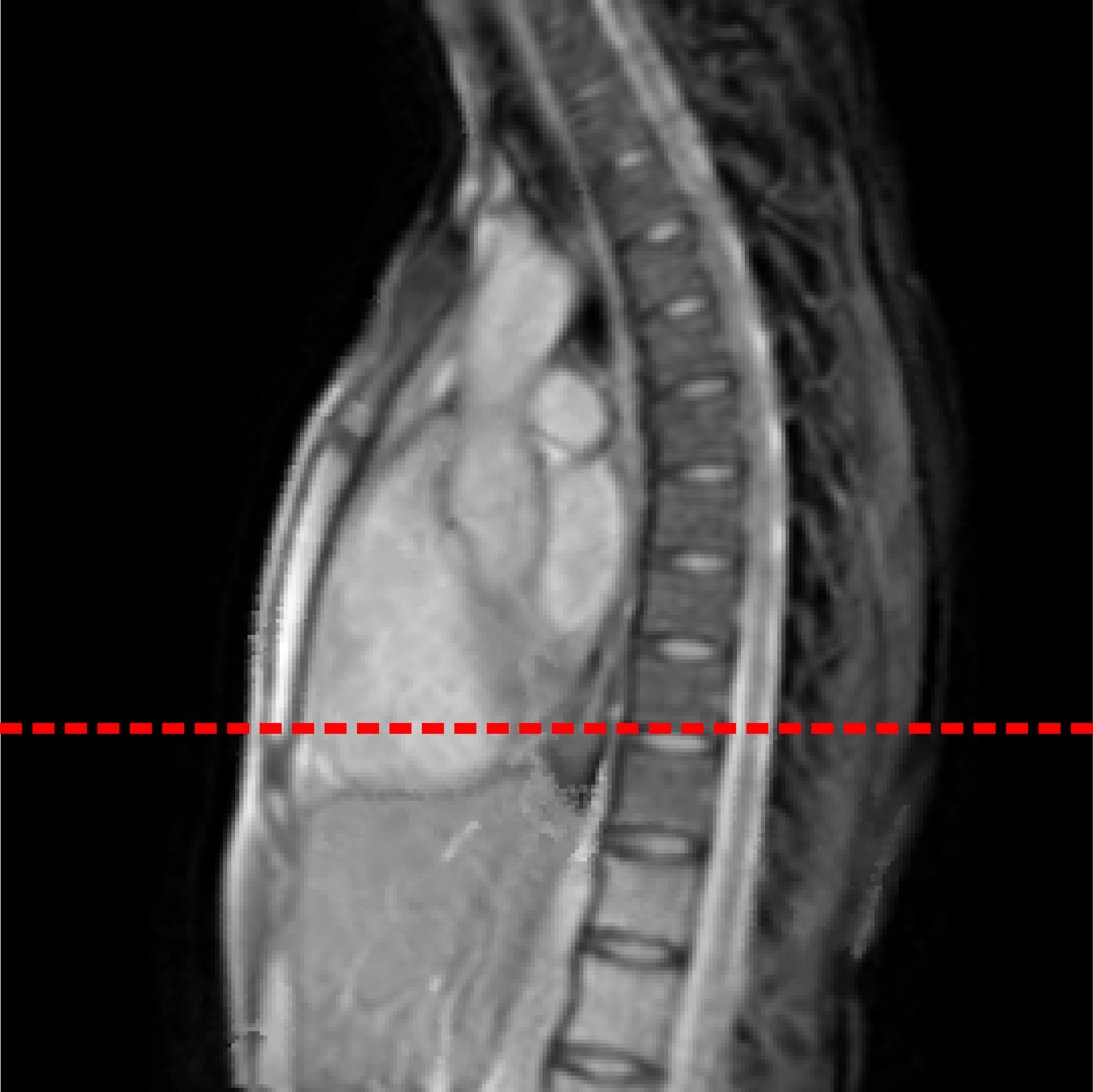} }%    
    \subfloat[Intensity error \\ $t=60.9\text{s}$, $h=2.2\text{s}$ \\ \acs{RTRL}]{\includegraphics[width=.145\textwidth, height=.125\textwidth]{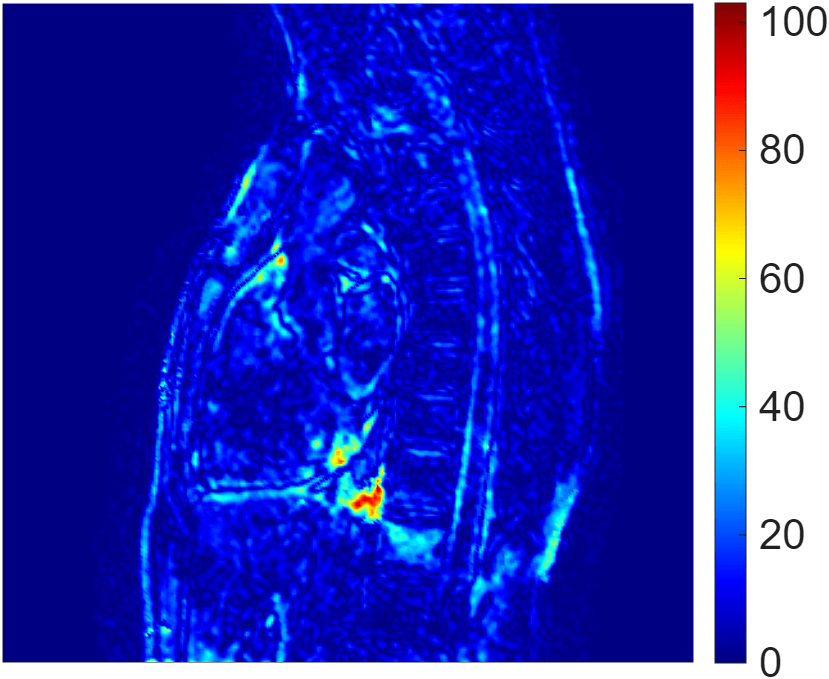} }%      
	\subfloat[\acs{DVF} error (in mm) \\ $t=60.9\text{s}$, $h=2.2\text{s}$ \\ \acs{RTRL}]{\includegraphics[width=.145\textwidth, height=.125\textwidth]{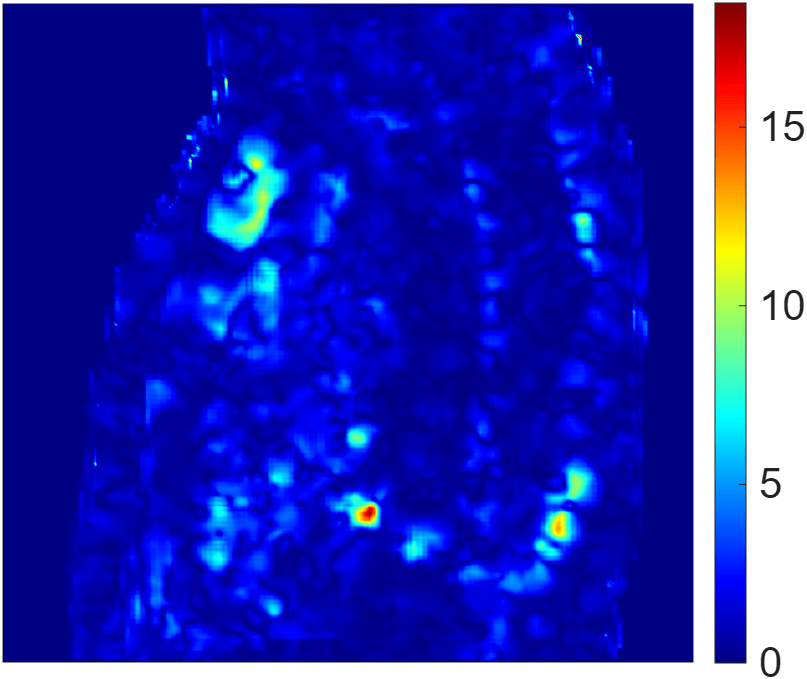} }%      
    \subfloat[Predicted image \\ $t\!=\!60.9\text{s}, h\!=\!2.2\text{s}$ \\ population transformer]{\includegraphics[width=.125\textwidth]{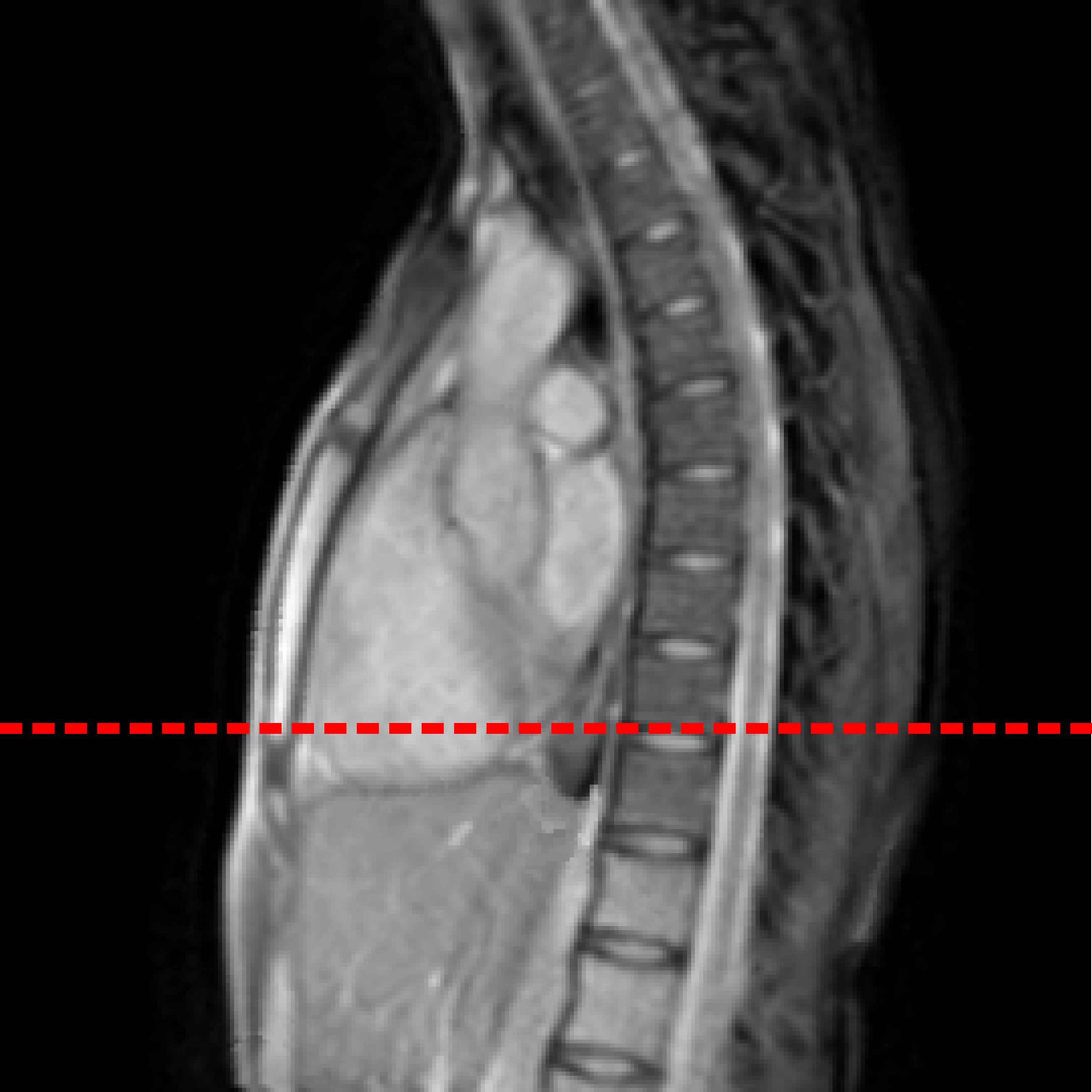} }%    
    \subfloat[Intensity error \\ $t=60.9\text{s}$, $h=2.2\text{s}$ \\ population transformer]{\includegraphics[width=.145\textwidth, height=.125\textwidth]{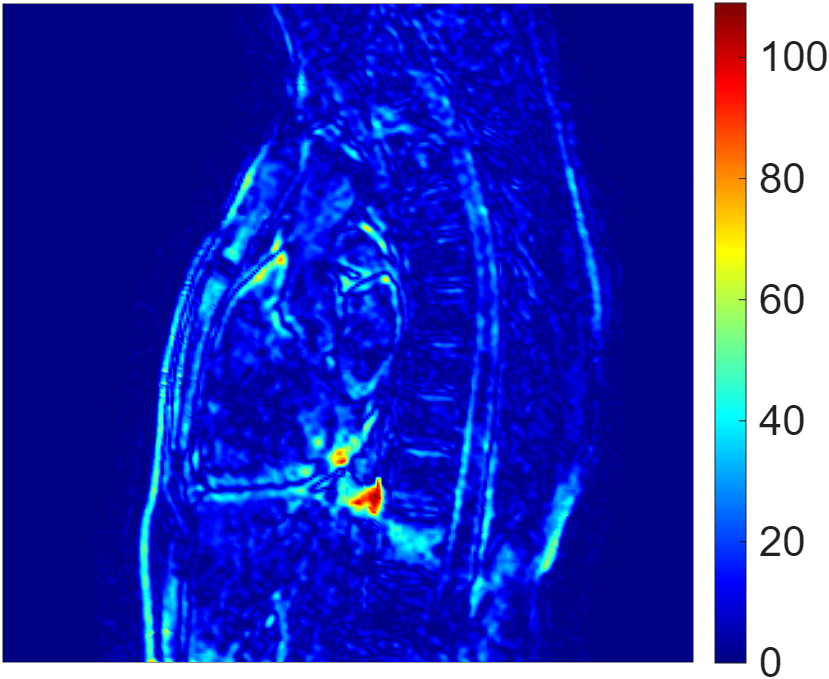} }%     
	\subfloat[\acs{DVF} error (in mm) \\ $t=60.9\text{s}$, $h=2.2\text{s}$ \\ population transformer]{\includegraphics[width=.145\textwidth, height=.125\textwidth]{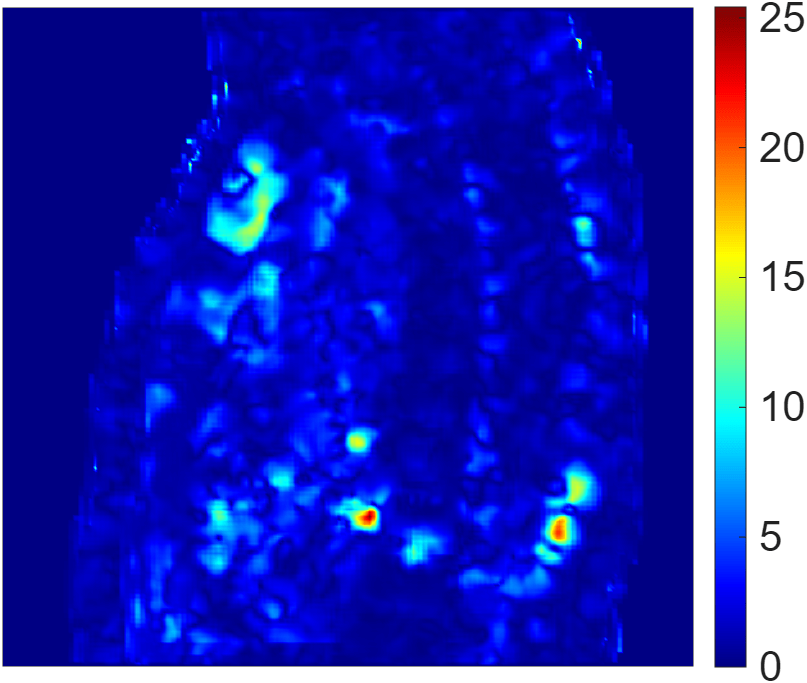} }%   	
    \caption{Original frames in sequence 1 of the ETH Zürich dataset and corresponding predictions using \pgls{RNN} trained with \acs{RTRL} (columns 2--4) and a population transformer trained on the \gls{OvGU} data (columns 5--7) for horizons $h=0.31\text{s}$ and $h=2.20\text{s}$ at two specific time steps in the test set---corresponding to an \gls{EE} phase (first two rows, frame 188) and an \gls{EI} phase (last two rows, frame 196)---along with the associated pixel-wise intensity and Euclidean deformation-error maps. Hyperparameters, including $n_{\text{cp}}$ for \gls{RTRL}, were optimized for each value of $h$ via grid search on the validation set. A dashed horizontal red line is superimposed on the ground-truth and predicted frames at the same height to help assess the accuracy of the predicted diaphragm position. The corresponding \gls{PCA} weights are represented by round markers in Fig. \ref{fig:PCA weights pred RTRL vs pop transformer} (both figures depict the same prediction runs).}
    \label{fig:next frame pred sq 1 RTRL vs pop transformer}
\end{figure*}

% Diaphragm position in sequence 1 and link with the prediction of the PCA weights
In the predicted ETH Zürich frames, the breathing-induced organ deformations and the \gls{SI} motion of the diaphragm and vessels were visually reproduced reasonably well, though not always with perfect accuracy. In sequence 1, the diaphragm appeared more elevated in the predictions from \gls{RTRL} and the population transformer at $t=60.9\text{s}$, corresponding to the \gls{EI} phase, than in the ground-truth image, for both $h=0.31\text{s}$ and $h=2.20\text{s}$ (Fig. \ref{fig:next frame pred sq 1 RTRL vs pop transformer}). Conversely, its predicted boundary at $t=58.4\text{s}$, corresponding to the immediately preceding \gls{EE} phase, was slightly inferior to the ground-truth position for both models at $h=2.20\text{s}$, and also for the transformer at $h=0.31\text{s}$, likely due to predictions of $w_2(t)$ exceeding the reference value (Figs. \ref{fig:2nd PCA weight sq 1 RTRL prediction}--\ref{fig:2nd PCA weight sq 1 pop transformer prediction}). Indeed, given the general downwards direction of the second-order component (Fig. \ref{fig:second principal component sq 1 ETH}), a negative estimated value of $w_2(t)$ that is lower in absolute value produces less upwards displacement (Eq. \ref{eq:PCA weight forecasting}). Likewise, we conjecture that the larger discrepancy at the \gls{EI} phase was primarily driven by overpredicted values of $w_3(t)$ (Figs. \ref{fig:3rd PCA weight sq 1 RTRL prediction}--\ref{fig:3rd PCA weight sq 1 pop transformer prediction}), whose associated principal-component vectors mostly pointed in the superior direction (Fig. \ref{fig:third principal component sq 1 ETH}), although interactions with the second-order component also influenced the final displacement. 

\begin{figure*}[htbp]
    \centering

    \subfloat[Sequence 1 \\ mean ground-truth \\ intensity]{%
        \includegraphics[width=.145\textwidth]{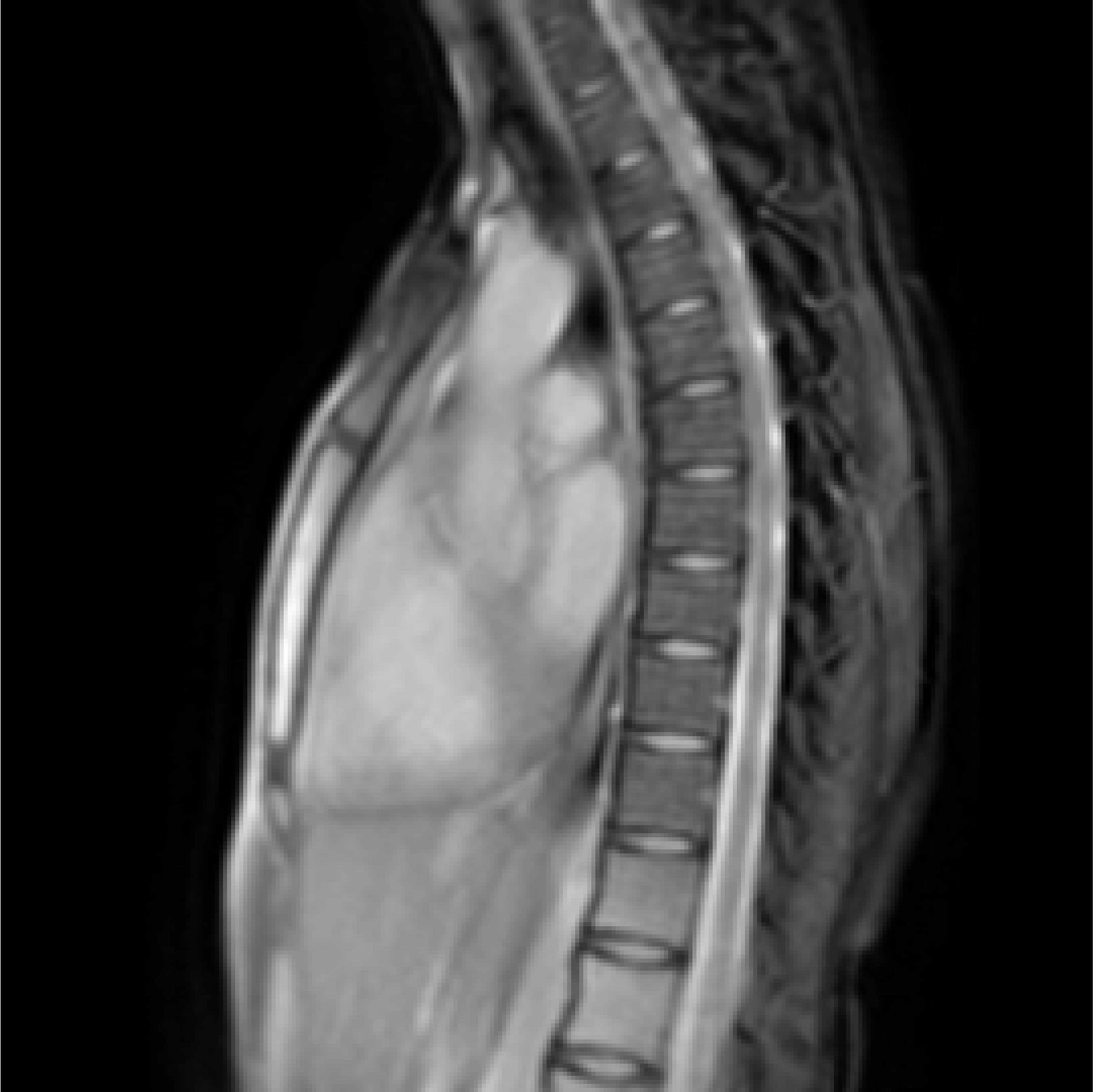}}%
    \hfill
    \subfloat[Sequence 1 \\ mean intensity error]{%
        \includegraphics[width=.165\textwidth,height=.142\textwidth]{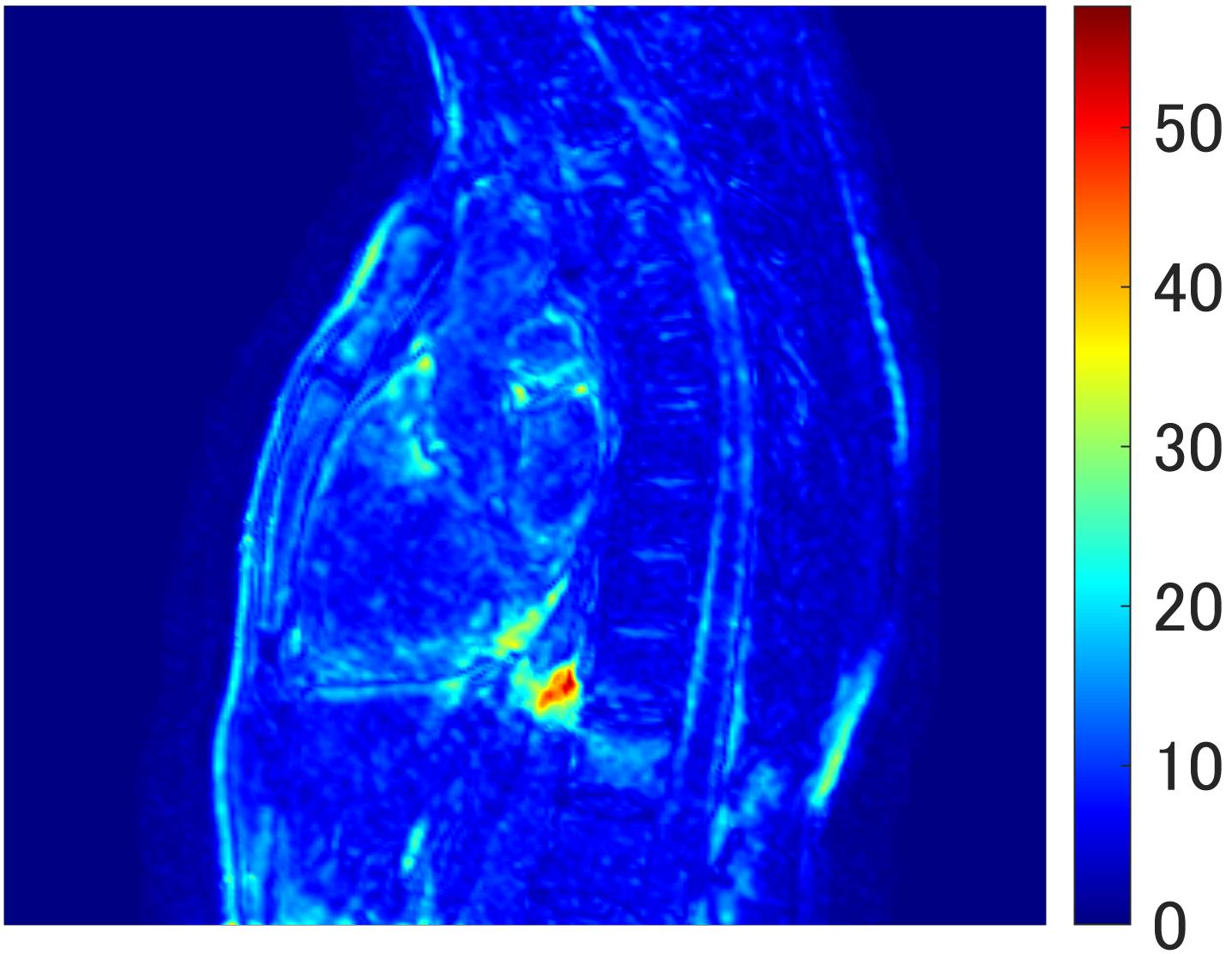}\label{fig:mean test intensity error sq 1}}%
    \hfill
    \subfloat[Sequence 1 - mean deformation error (mm)]{%
        \includegraphics[width=.165\textwidth,height=.142\textwidth]{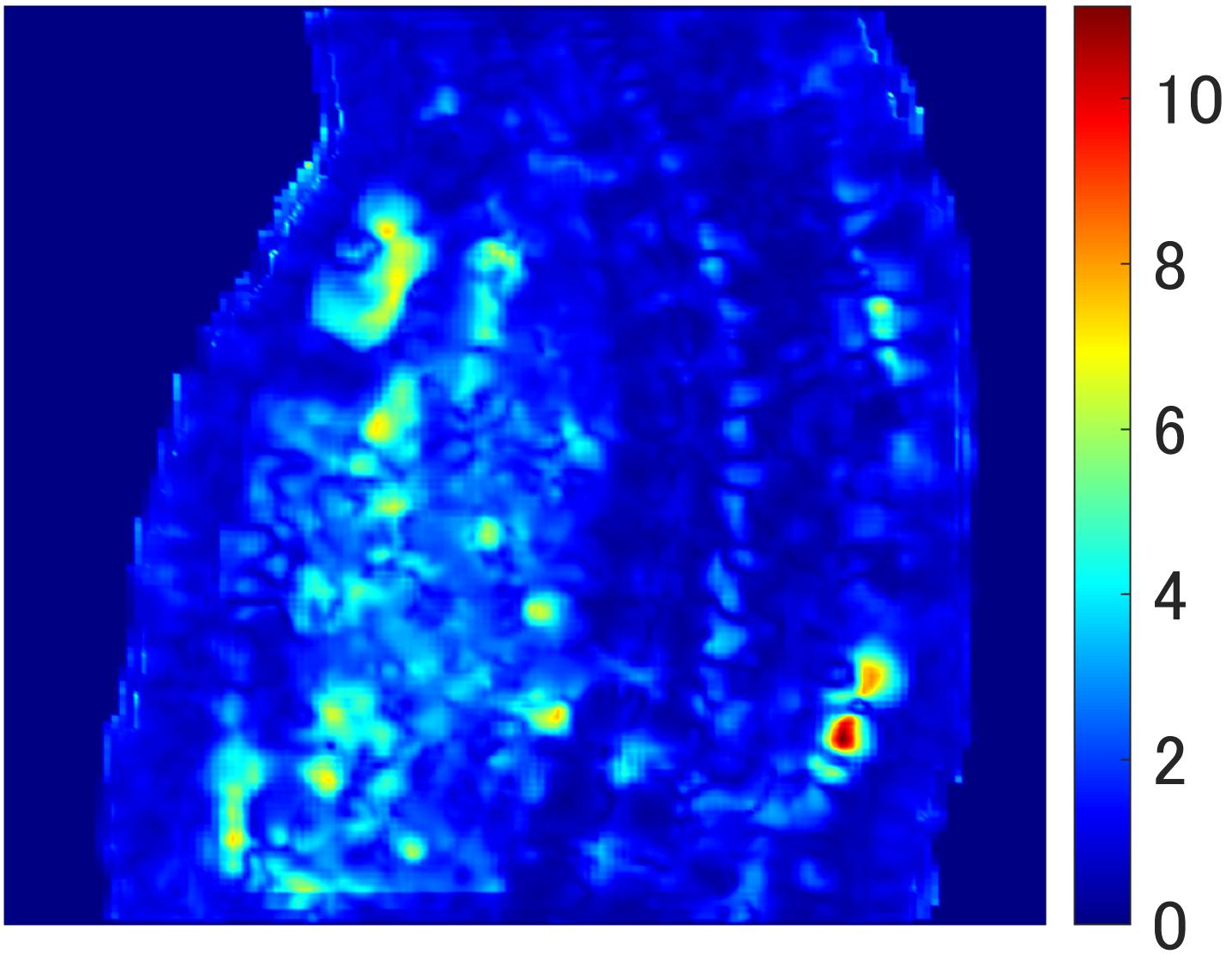}}%
    \hfill
    \subfloat[Sequence 2 \\ mean ground-truth \\ intensity]{%
        \includegraphics[width=.145\textwidth]{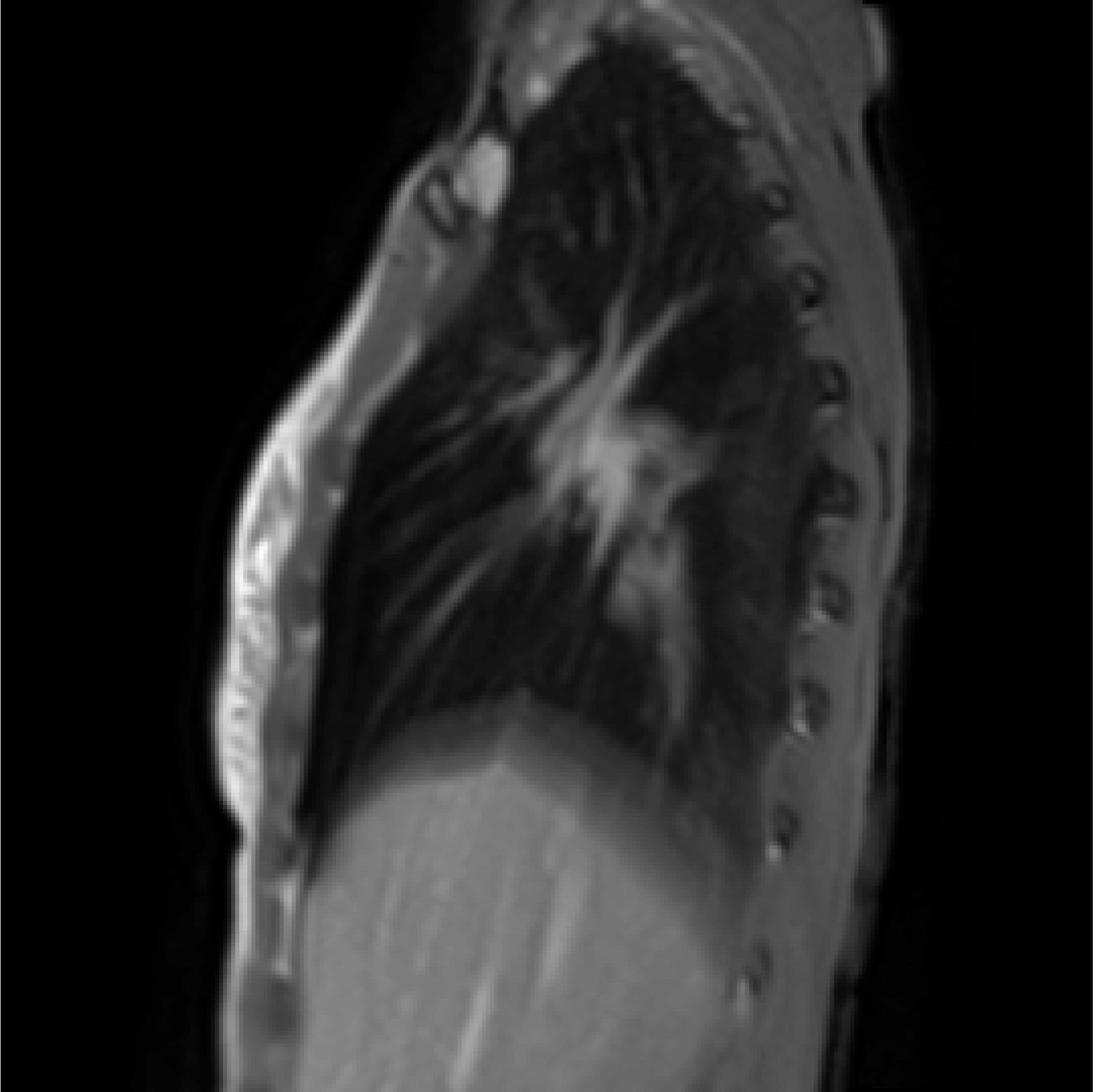}\label{fig:sq 2 mean test intensity}}%
    \hfill
    \subfloat[Sequence 2 \\ mean intensity error]{%
        \includegraphics[width=.165\textwidth,height=.142\textwidth]{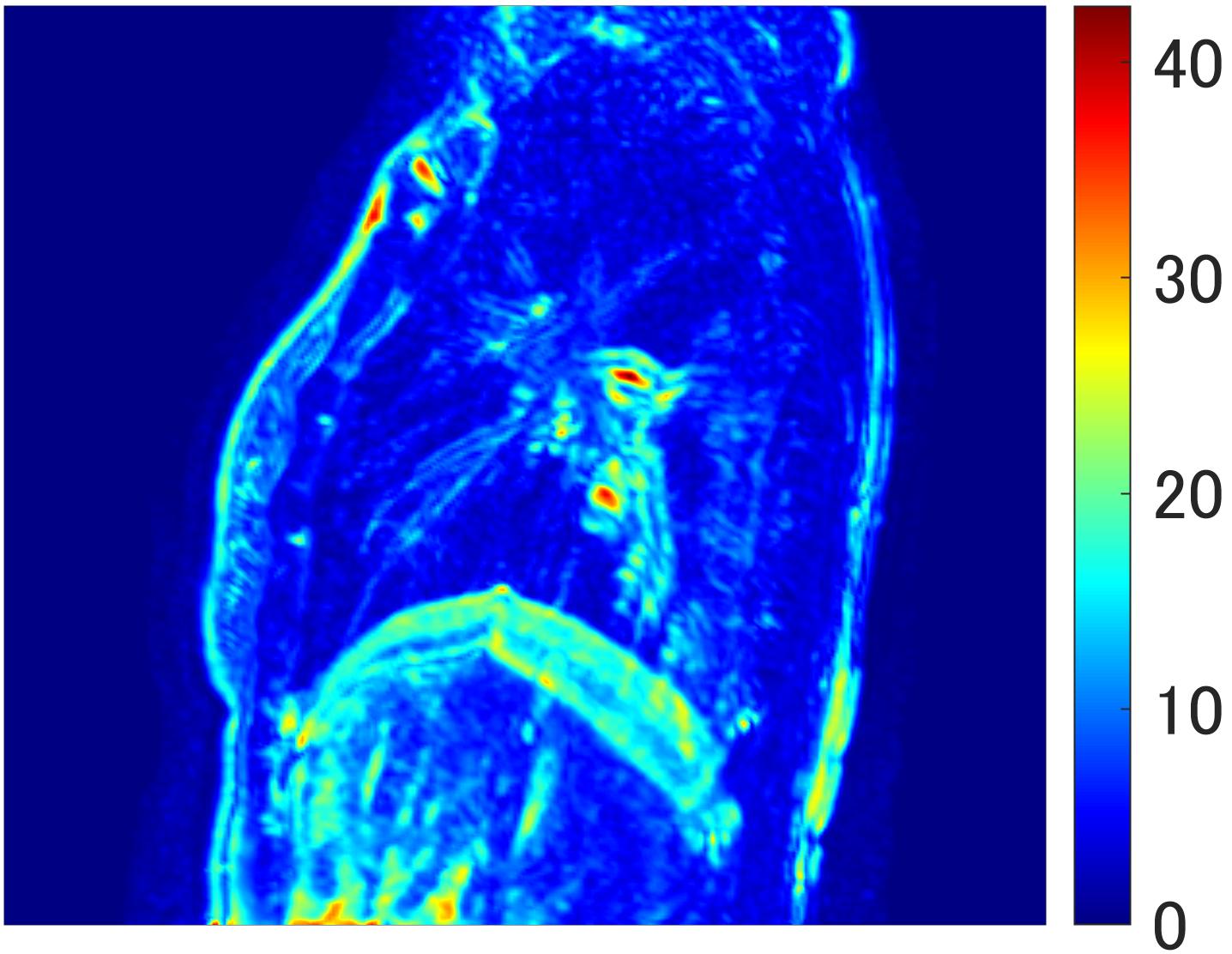}\label{fig:sq 2 mean test intensity error}}%
    \hfill
    \subfloat[Sequence 2 - mean deformation error (mm)]{%
        \includegraphics[width=.165\textwidth,height=.142\textwidth]{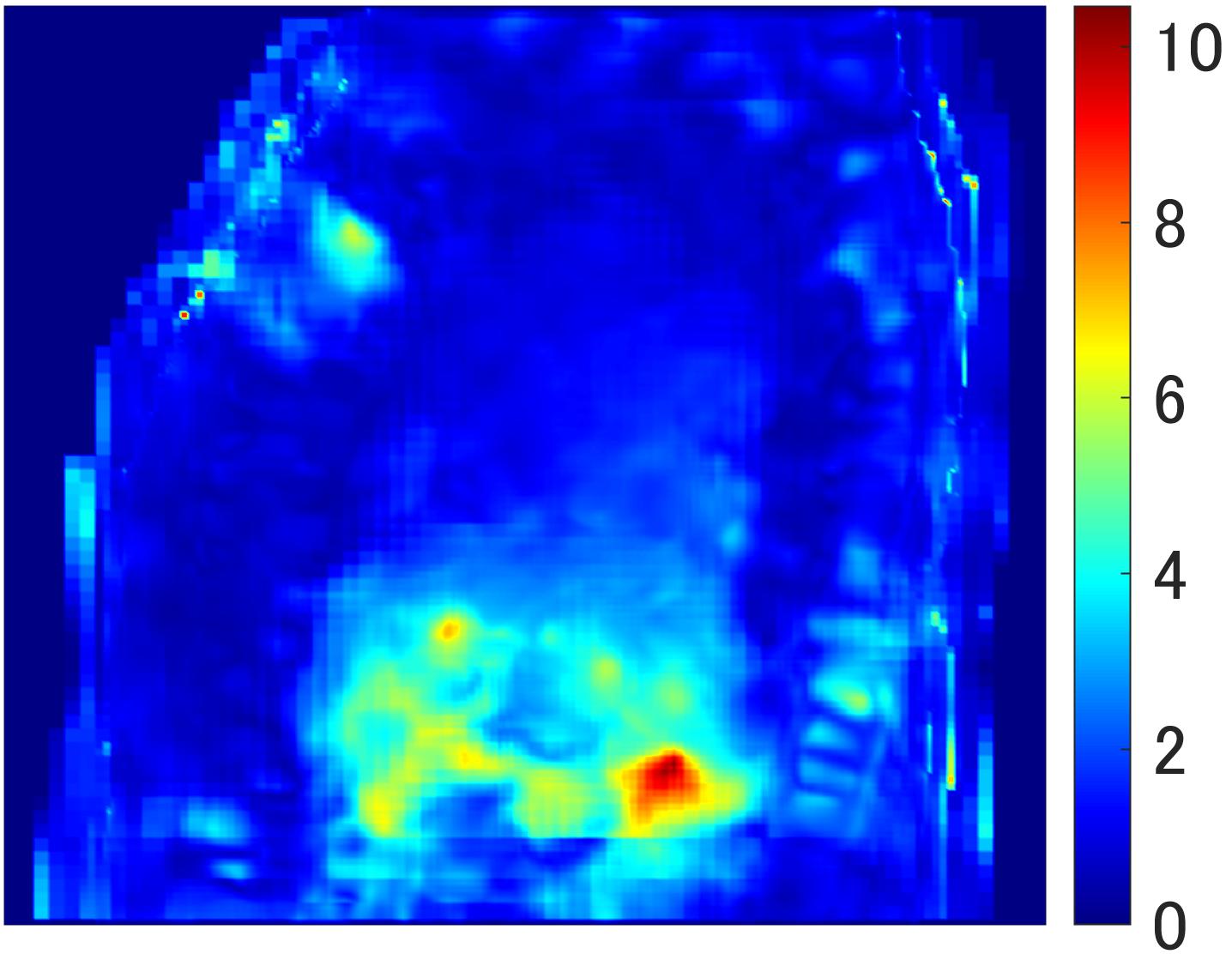}\label{fig:sq 2 mean test deformation error}}%
    \\[1ex]
    \subfloat[Sequence 3 \\ mean ground-truth \\ intensity]{%
        \includegraphics[width=.145\textwidth]{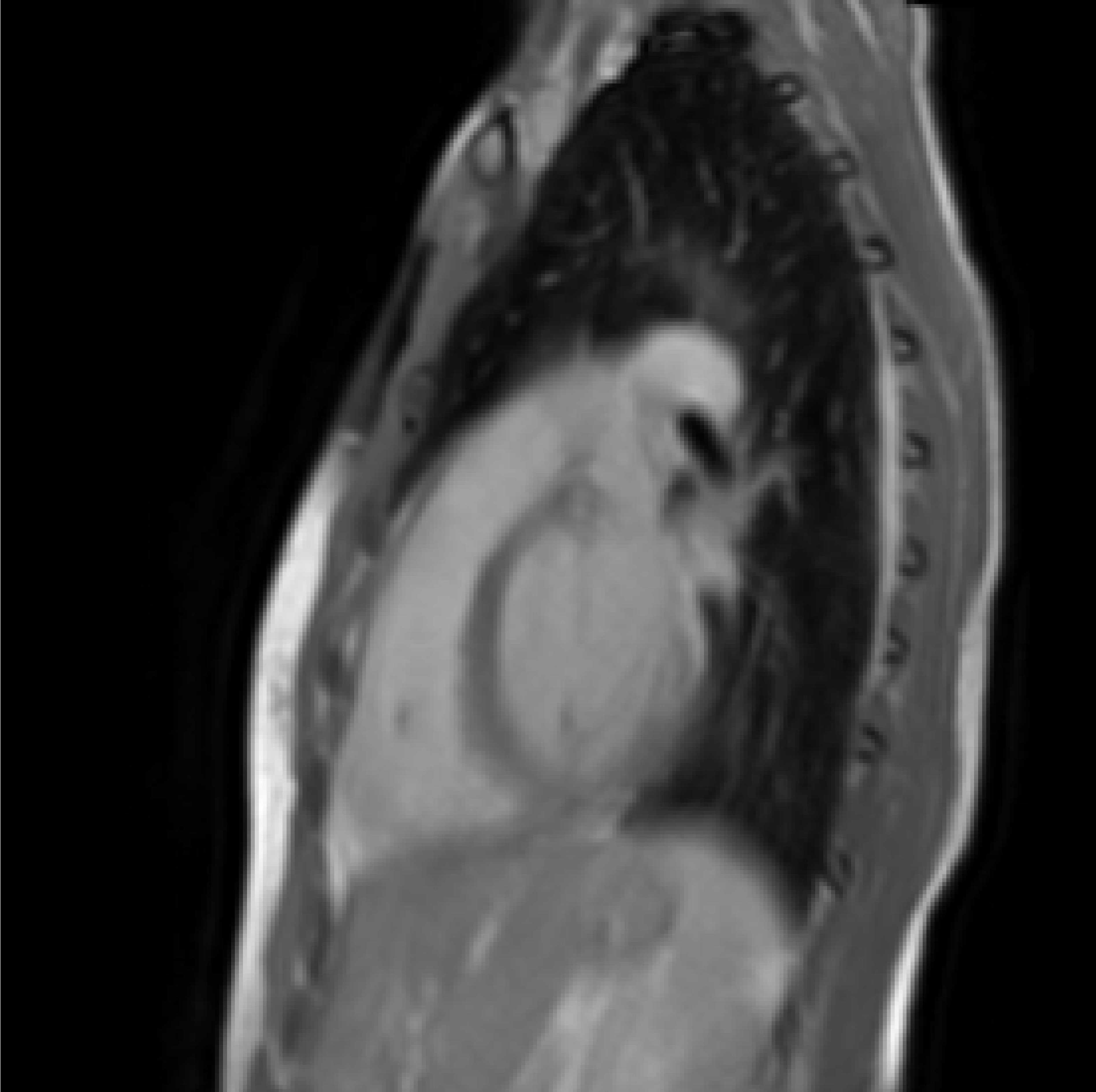}}%
    \hfill
    \subfloat[Sequence 3 \\ mean intensity error]{%
        \includegraphics[width=.165\textwidth,height=.142\textwidth]{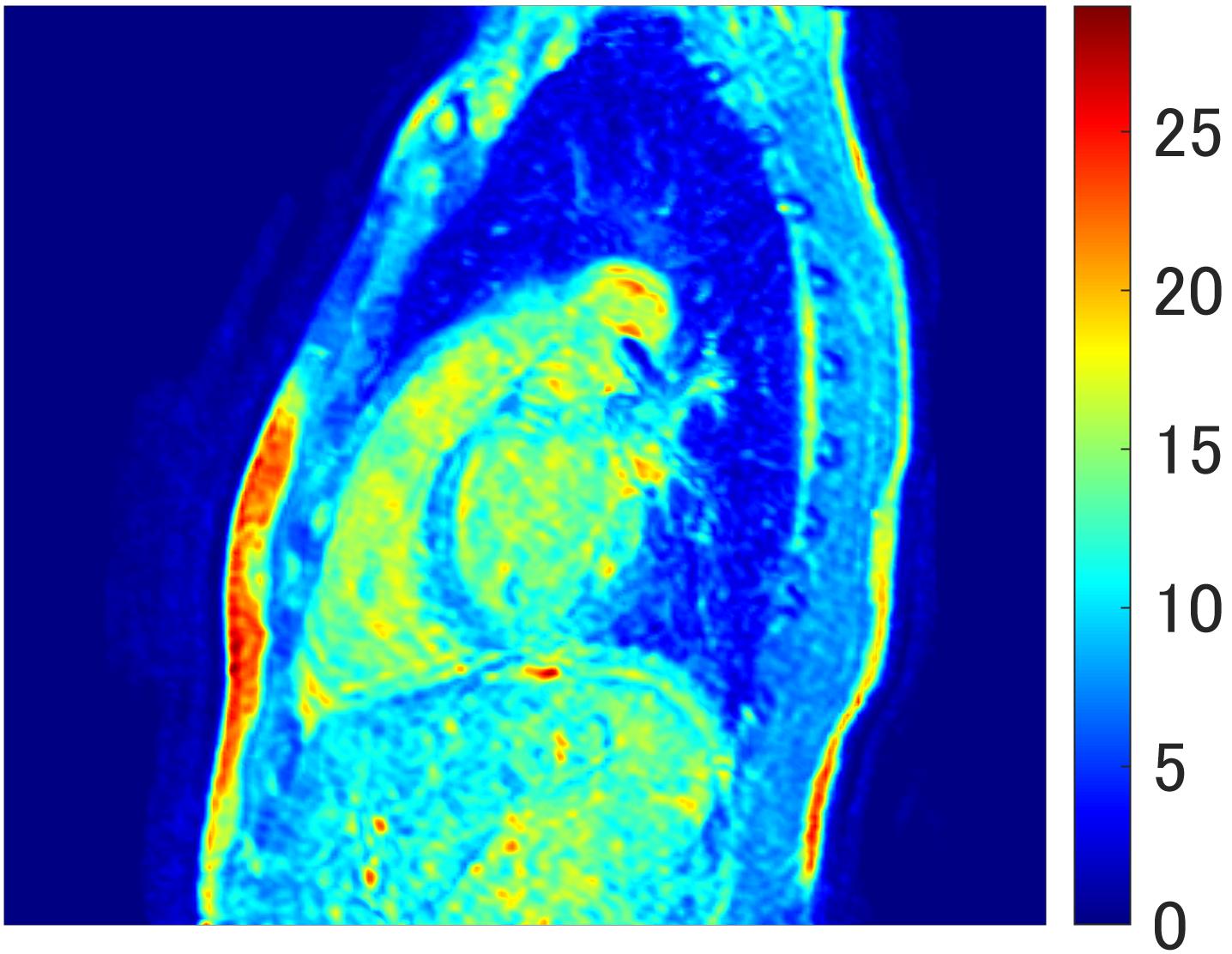}}%
    \hfill
    \subfloat[Sequence 3 - mean deformation error (mm)]{%
        \includegraphics[width=.165\textwidth,height=.142\textwidth]{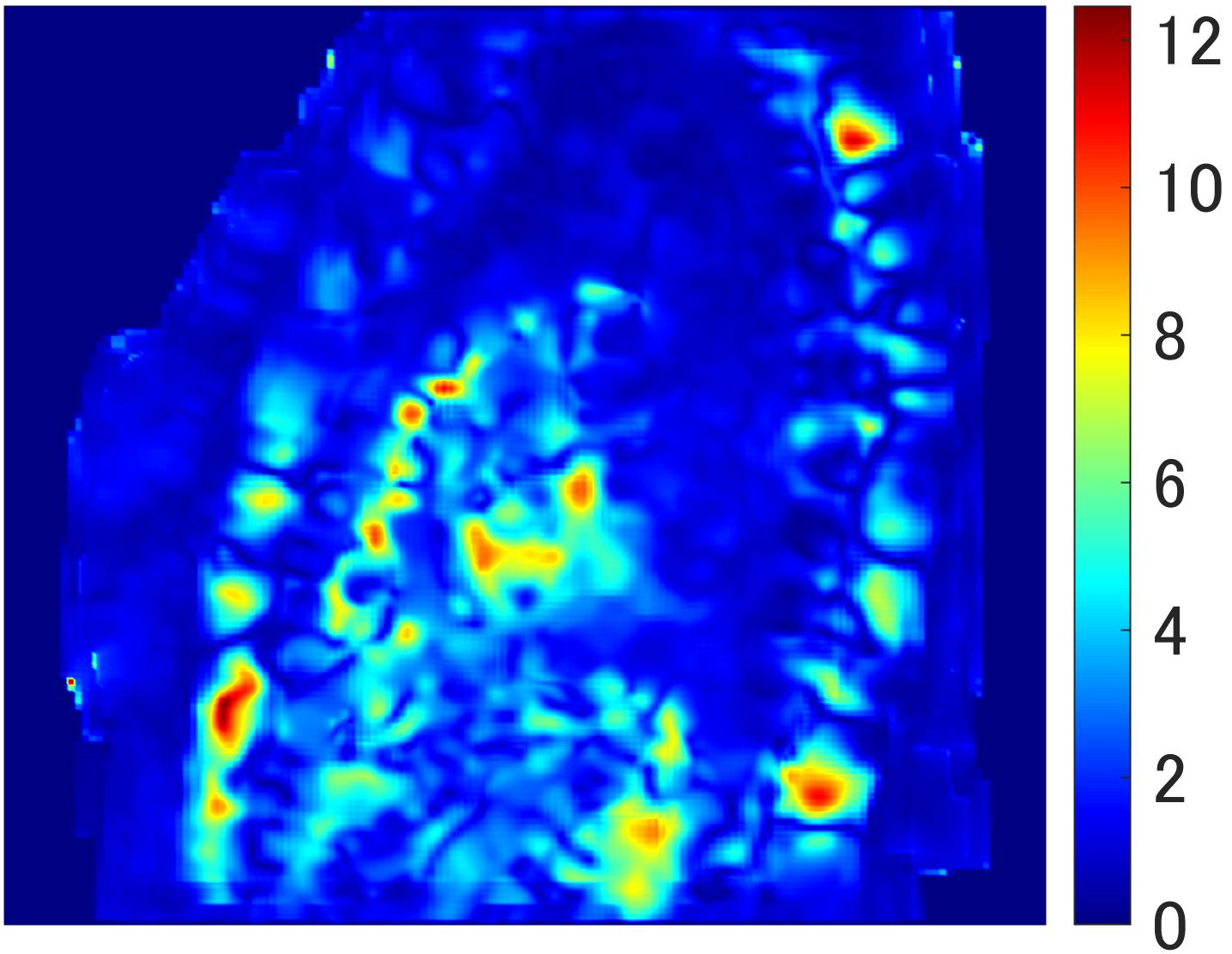}}%
    \hfill
    \subfloat[Sequence 4 \\ mean ground-truth \\ intensity]{%
        \includegraphics[width=.145\textwidth]{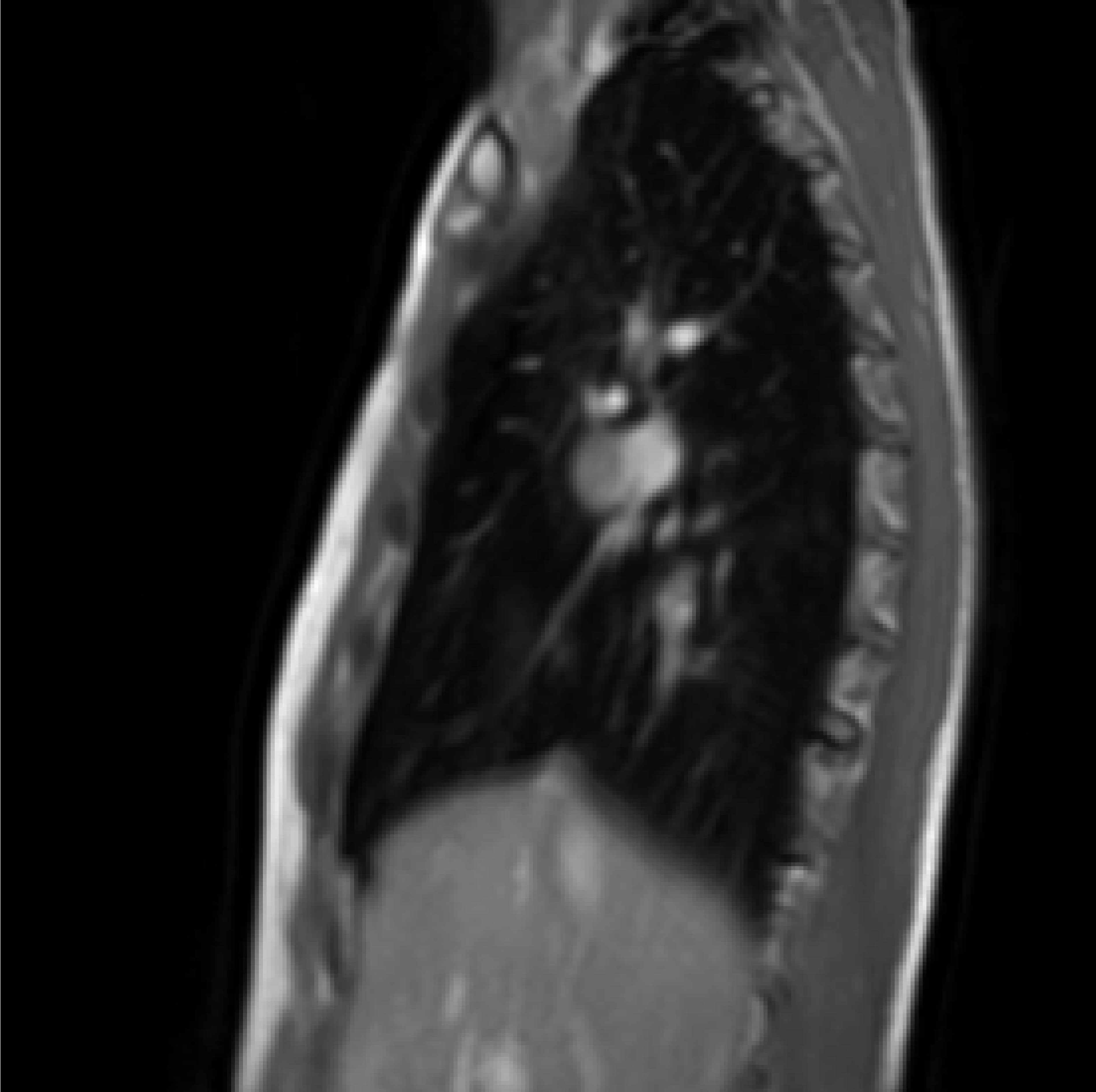}}%
    \hfill
    \subfloat[Sequence 4 \\ mean intensity error]{%
        \includegraphics[width=.165\textwidth,height=.142\textwidth]{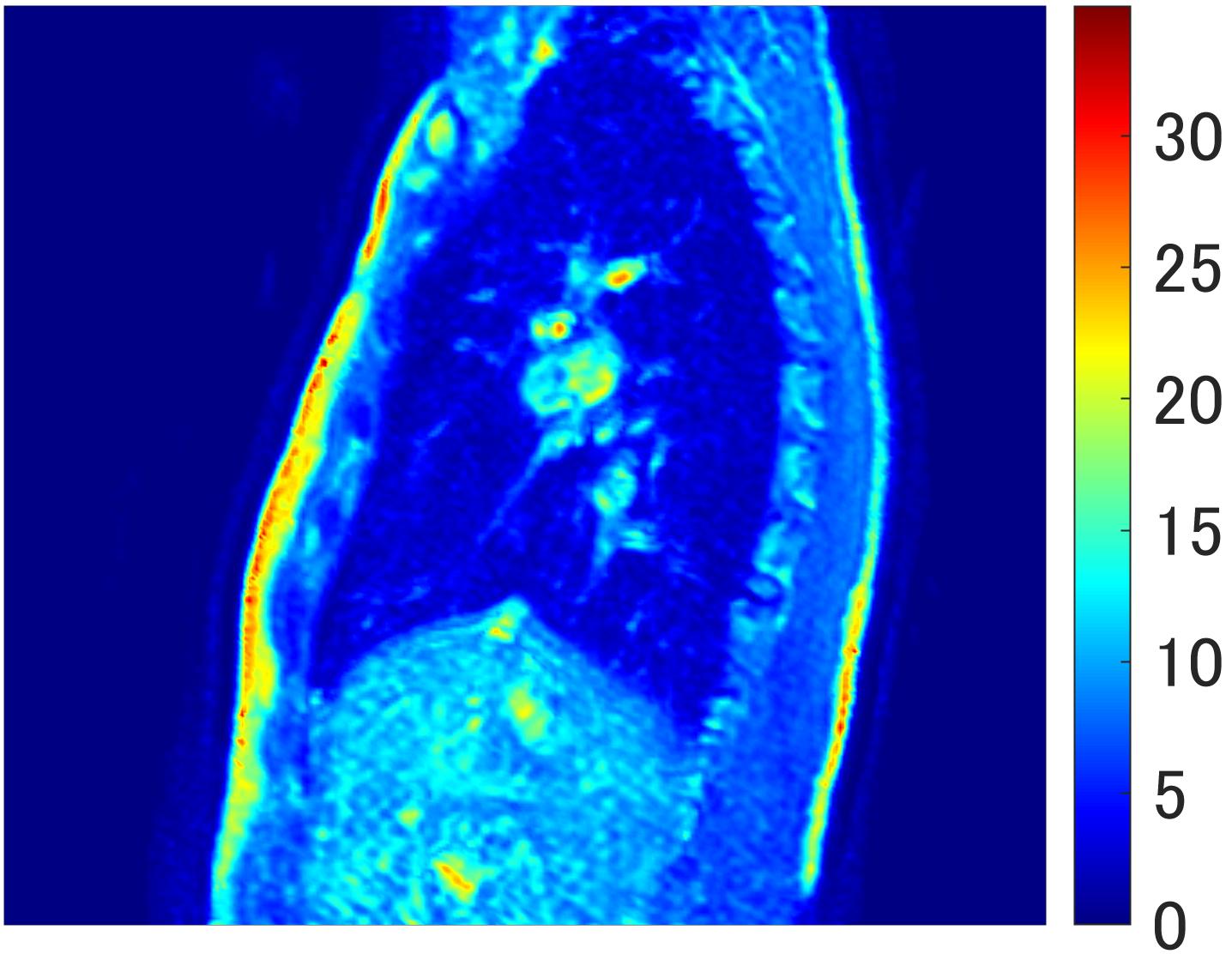}}%
    \hfill
    \subfloat[Sequence 4 - mean deformation error (mm)]{%
        \includegraphics[width=.165\textwidth,height=.142\textwidth]{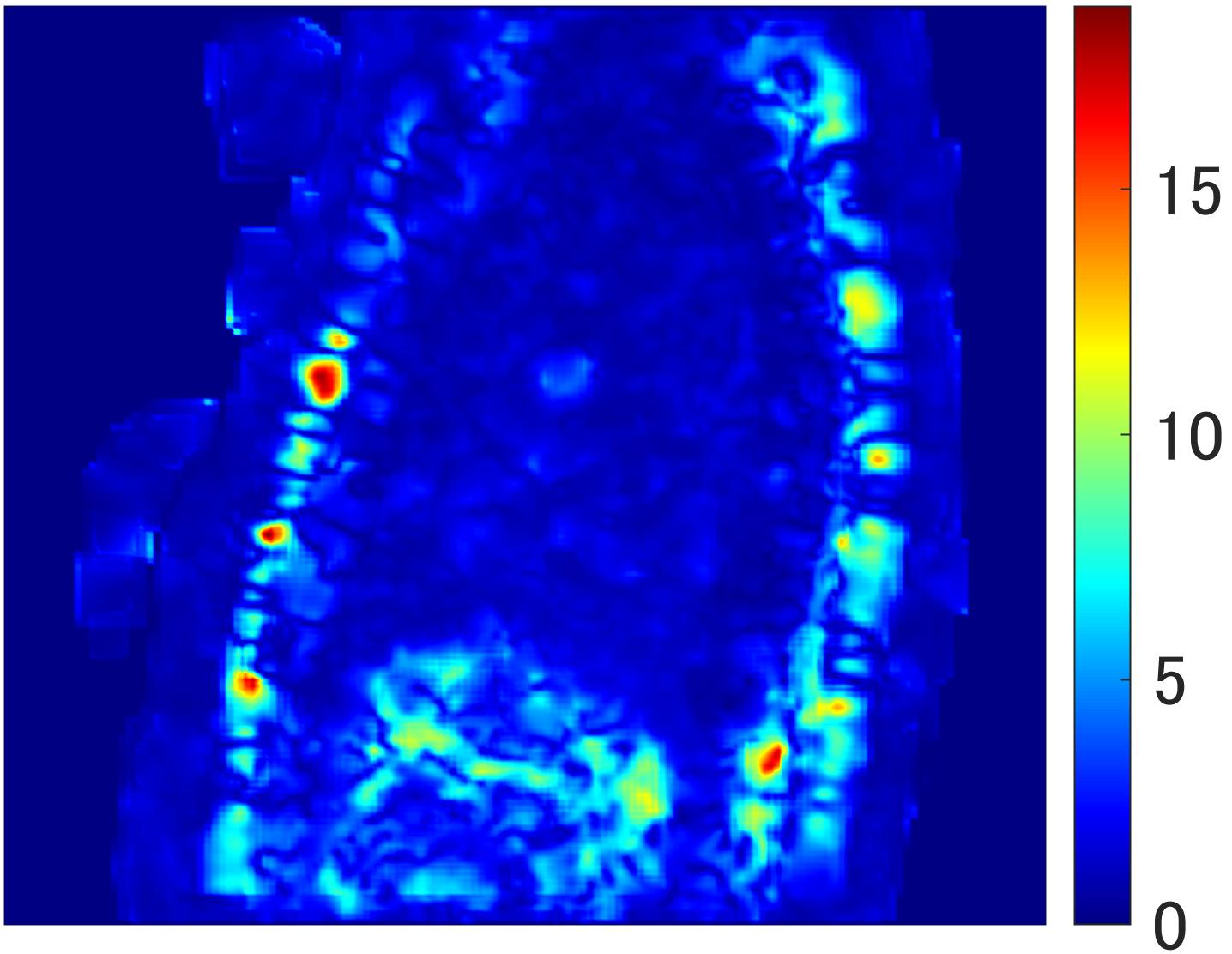}}%

    \caption{Prediction error maps for \pgls{RNN} trained with \acs{SnAp-1} at $h=1.88\text{s}$, averaged over the test-set time points and 5 runs to account for stochasticity, along with the mean ground-truth intensity across the test set, for the four ETH Zürich sequences. Hyperparameters, including $n_{\text{cp}}$, were optimized for each sequence individually via grid search on the validation set.}
    \label{fig:next frame pred all sq SnAp-1 h=6}
\end{figure*}

% Instantaneous errors for the prediction of sequence 1
In sequence 1, at $t=58.4\text{s}$ (\gls{EE}), large instantaneous intensity errors were concentrated near the superior wall of the right ventricle, whereas the corresponding deformation errors were more homogeneous across that ventricle and the pancreas. At $t=60.9\text{s}$ (\gls{EI}), both error types were more pronounced, with higher frame-wise maxima, particularly in the small lung region posterior to the heart, due to out-of-plane motion. Mean intensity errors across the test set in sequence 1 were most prominent around that posterior air pocket. They were moderate along sharp anatomical boundaries, including the diaphragm and the superior wall of the right ventricle (cf. \gls{SnAp-1} errors in Fig. \ref{fig:mean test intensity error sq 1}). Instantaneous \gls{DVF} errors broadly matched intensity-error patterns in several regions for \gls{RTRL} and the population transformer, for instance, in lumbar and sternal areas. However, they were more spatially diffuse, reflecting the locally uniform nature of the displacement fields (Figs. \ref{fig:Geometrical viewpoint}, \ref{fig:DVF sq 1 ETH inspiration}--\ref{fig:DVF sq 1 ETH expiration}). Notably, inaccuracies in the predicted deformations within low-contrast areas did not necessarily lead to large intensity errors. Conversely, local intensity mismatches may be attributed to a sharp contrast between adjacent organs, such as at the diaphragm boundary, rather than acute and localized motion error.

% Spatial errors averaged over the test set
The spatial diffuseness of \gls{DVF} errors and their only moderate correspondence with intensity errors could also be observed in mean test-set error maps for \gls{SnAp-1} across all four sequences (Fig. \ref{fig:next frame pred all sq SnAp-1 h=6}). In sequences 2 and 4, moderate-to-high intensity errors near the central pulmonary vessels did not correlate with substantial geometrical errors. Notably, mean intensity errors were relatively strong near the diaphragm and the intrahepatic vessels in sequence 2 (Fig. \ref{fig:sq 2 mean test intensity error}). This likely results from fast \gls{SI} liver motion relative to the sampling frequency, as indicated by blurring near the diaphragm in the test-set mean-intensity map. Specifically, in this acquisition, deformation errors were concentrated near the posterior part of the liver (Fig. \ref{fig:sq 2 mean test deformation error}), where blurring in the mean-intensity map was most pronounced and motion amplitude was high. Across sequences, noise and temporal variations in local tissue texture also produced moderate, spatially diffuse intensity errors within tissue and moving organs, which were particularly visible in sequences 3 and 4. Lastly, all sequences exhibited moderate-to-large errors near the skin surface, subcutaneous fat, and osseous structures (the rib cage and spine), due to strong tissue--bone--air contrast and out-of-plane motion.

% Rk: I could have added local SSIM maps 

\subsubsubsection{Visual assessment on the OvGU dataset}

% Sequence by sequence assessment of organ position together with interpretation in the PCA space
In the predicted \gls{OvGU} frames, the overall positions and texture of anatomical structures, including the vessels, appeared generally correct despite substantial noise in the original acquisitions and the relatively long horizon for qualitative assessment ($h = 1.66\text{s}$; Fig. \ref{fig:visual comparison of ROI prediction with SnAp-1 and transformer at h=10 on OvGU dataset}). In sequence 1, the liver location in the predicted image at $t= 81.7\text{s}$ (\gls{EI}) seemed relatively accurate for \gls{SnAp-1} but appeared gradually more elevated in the predictions from \gls{UORO}, \gls{DNI}, and the population transformer, for which the mismatch with the ground truth was the strongest. This trend is consistent with the decreasing positive value of the predicted first-order \gls{PCA} weight for these algorithms, in the same order (corresponding to an increasing error; Fig. \ref{fig:1st PCA component prediction in Sq 1 of OgVU dataset}), at a time point when breathing irregularities in amplitude and phase were relatively strong, combined with the predominantly inferior direction of the first-order principal deformations (Fig. \ref{fig:1st PCA component Sq 1 OvGU dataset}). At the immediately preceding \gls{EE} phase ($t = 79.2\text{s}$), the transformer prediction placed the liver slightly higher than in the ground-truth frame, again in line with an underestimation of the negative first-order \gls{PCA} weight. Likewise, in sequence 6, inaccurate \gls{PCA}-coefficient prediction caused the liver position, as estimated by the transformer, to be slightly inferior to the ground truth at $t = 82.7\text{s}$ (\gls{EE}; Fig. \ref{fig:1st PCA component prediction in Sq 6 of OgVU dataset}).
% A similar interpretation could also explain the slightly higher position in the liver in the images predicted by the transformer $t = 82.7\text{s}$ in sequence 6.
%Likewise, the slightly higher position in the liver in the images predicted by the transformer...
In sequence 4, the liver position was also challenging to forecast, especially near the deep inspiration at $t = 51.3\text{s}$. At that time point, the three \gls{RNN}-based predictors estimated the first-order \gls{PCA} weight reasonably well (Fig. \ref{fig:1st PCA component prediction in Sq 4 of OgVU dataset}), and the general shape and relatively inferior location of the organs were broadly faithful to the ground truth. However, organ contours and texture were noticeably blurred; \gls{PCA}-based spatial modeling may lack robustness when estimating the state of anatomical structures that lie unusually far from their average position within the training-set frames. Conversely, the image predicted via the transformer appeared sharper, with better-preserved texture. Nonetheless, the liver in that image was positioned too high because of inaccurate \gls{PCA}-weight forecasts. Notably, in this sequence, the underestimated first-order coefficient for \gls{SnAp-1} at the immediately succeeding \gls{EI} phase ($t=55.0\text{s}$), together with the moderate horizontal alignment of the first principal \gls{DVF} around the top of the liver (Fig. \ref{fig:1st PCA component Sq 4 OvGU dataset}), yielded a diaphragm shape that was more curved than in the ground-truth frame. %, where that anatomical boundary appeared flatter. % Likewise, in sequence 6, the liver in the frame predicted by the transformer at $t = 82.7\text{s}$ (\gls{EE}) was slightly lower than in the original image, consistent with an overestimated negative first-order \gls{PCA} score at that time point (Fig. \ref{fig:1st PCA component prediction in Sq 6 of OgVU dataset}) and with the predominantly inferior direction of the first principal \gls{DVF} in that region (Fig. \ref{fig:1st PCA component Sq 6 OvGU dataset}).

% General comments on vessel flickering and general qualitative motion assessment
Out-of-plane motion, particularly noticeable in sequence 6, and vessel flickering due to blood flow were not always predicted accurately, occasionally producing relatively unnatural deformations. Specifically, in sequences 1 and 2, tissue surrounding large vessels in the generated images seemed to contract and expand, intermittently masking bright vascular structures visible in the reference frame at $t_1$, thereby mimicking flickering. This effect arose because all relevant content was assumed to be present in the reference image and to remain unchanged over time, and motion was modeled as strictly planar. By contrast, motion generally appeared more natural in the predicted ETH Zürich sequences due to more favorable acquisition characteristics (lower noise and higher, more stable contrast).
% Maybe more natural closure? "Overall, these artifacts were less apparent in the ETH Zürich dataset, where..."

% New figure showing predicted vs original for two different states and four images
% The space between } and }% is very important to preserve the same amount of white spacing
\newlength{\myfigwidth}
\setlength{\myfigwidth}{.084\textwidth}
\begin{figure*}[pos=htbp, align=\centering]
    \captionsetup[subfigure]{labelformat=empty, font=scriptsize}
    \captionsetup[subfloat]{farskip=2pt, captionskip=1pt} % https://tex.stackexchange.com/questions/162824/vertical-spacing-between-subfloat
    \centering
    \subfloat[Sequence 1\\* \text{\small $t=t_{1}$} \\* reference]{\includegraphics[width=\myfigwidth, height=2.012\myfigwidth]{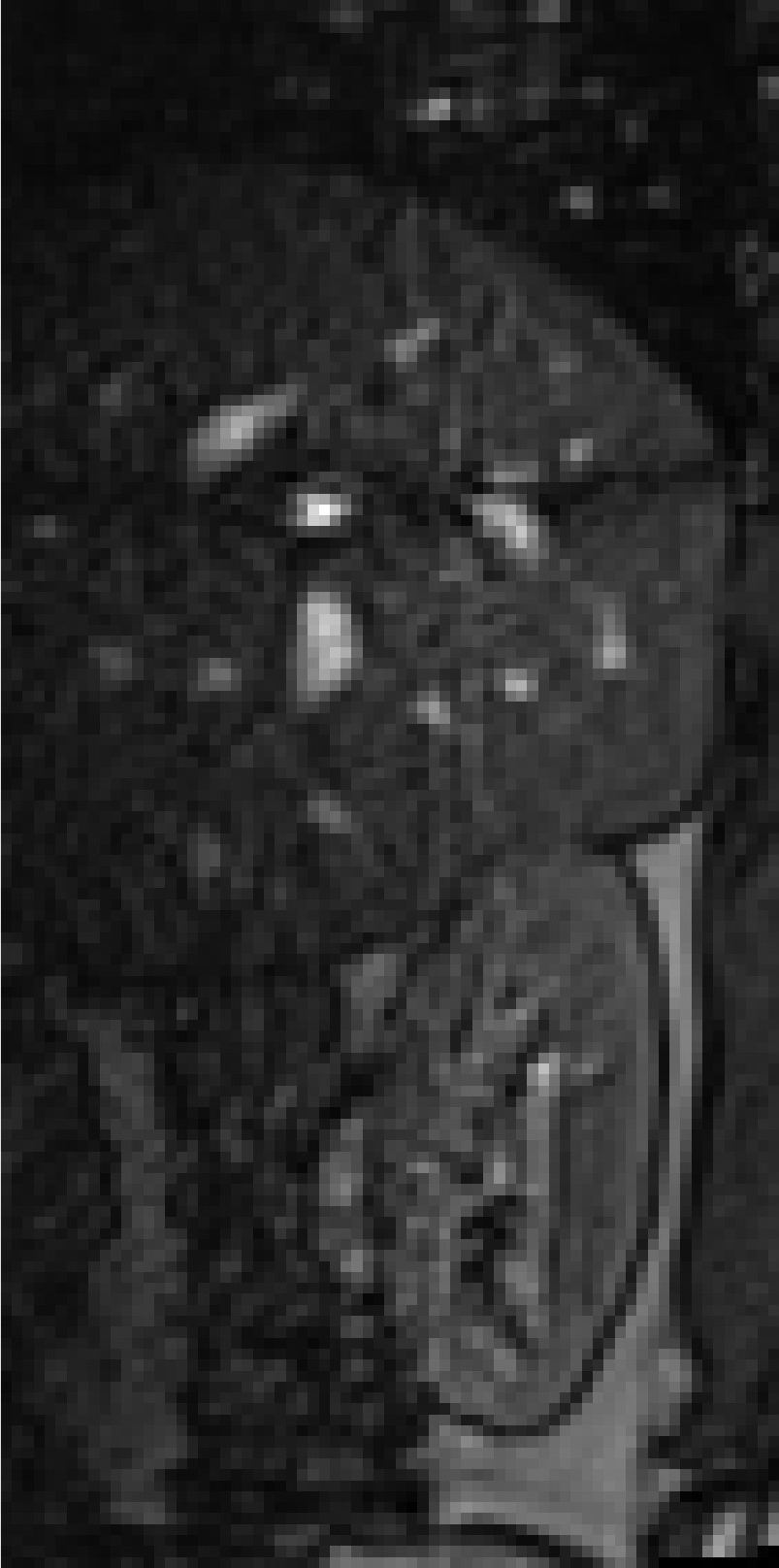} }%
    \subfloat[Sequence 1\\* \text{\small $t_{491}$} (81.7s)\\* ground truth]{\includegraphics[width=\myfigwidth, height=2.012\myfigwidth]{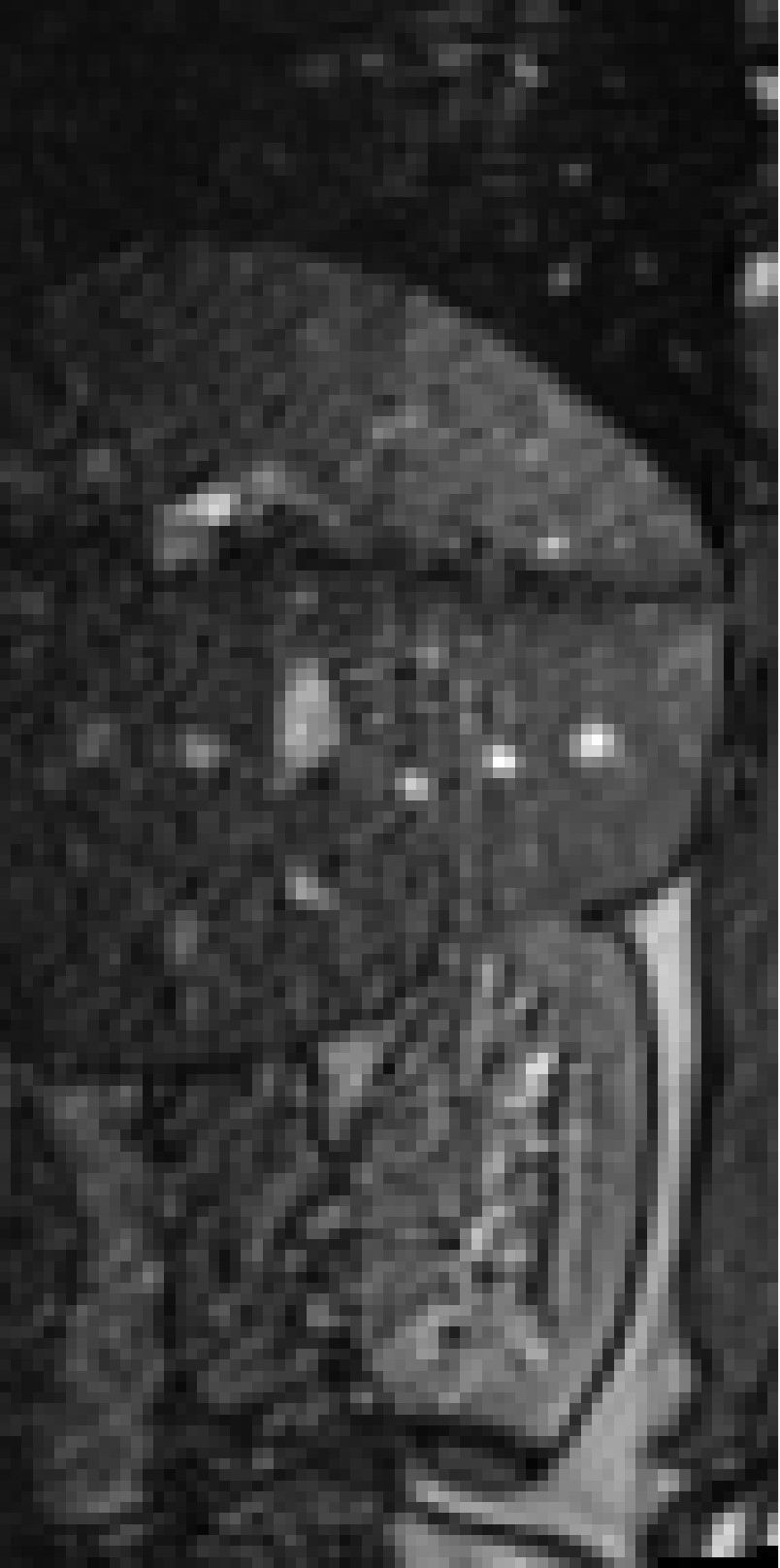} }%  
    \subfloat[Sequence 1\\* \text{\small $t_{491}$} (81.7s)\\* UORO]{\includegraphics[width=\myfigwidth, height=2.012\myfigwidth]{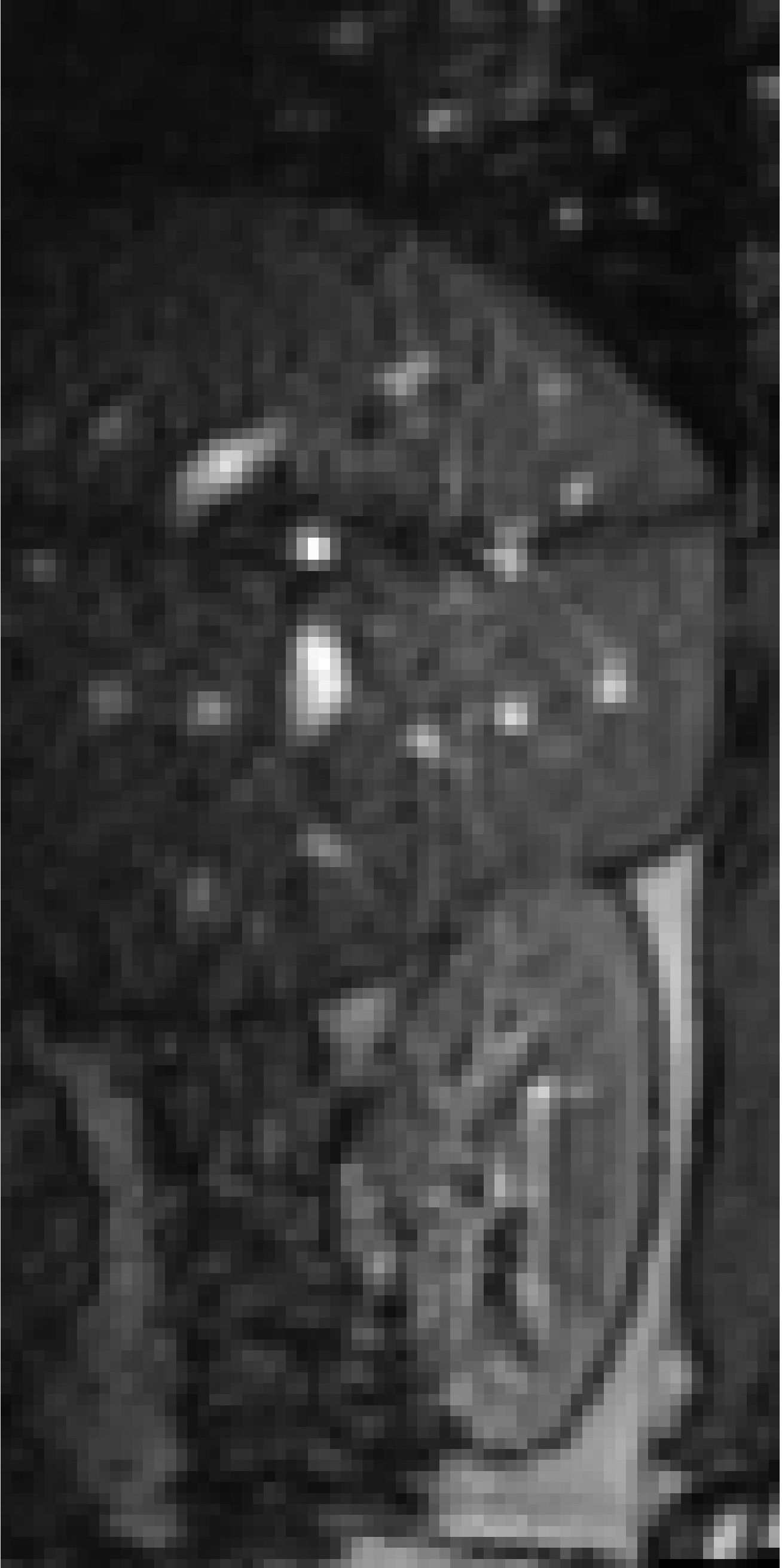} }% 
    \subfloat[Sequence 1\\* \text{\small $t_{491}$} (81.7s)\\* SnAp-1]{\includegraphics[width=\myfigwidth, height=2.012\myfigwidth]{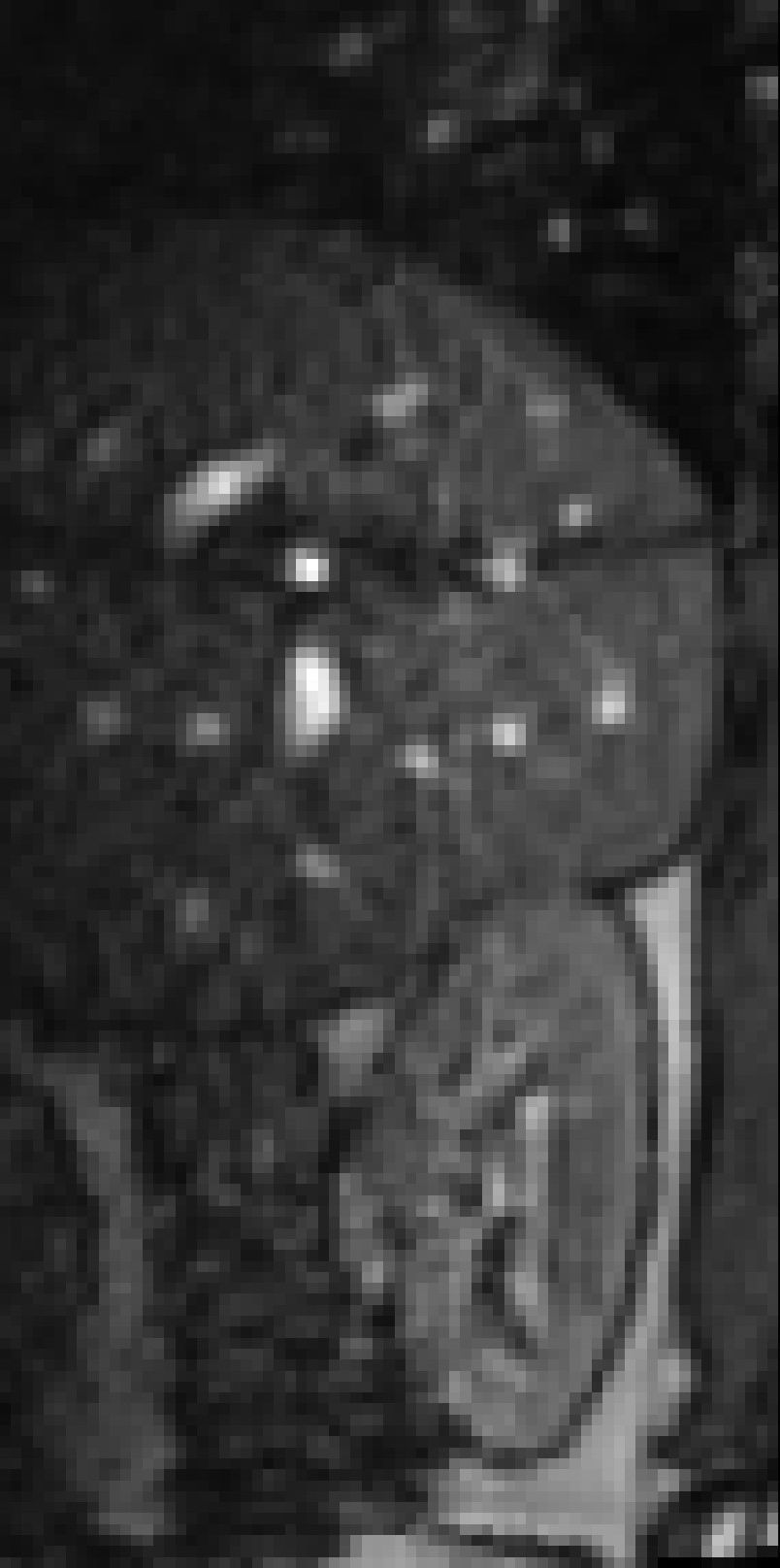} }%        
    \subfloat[Sequence 1\\* \text{\small $t_{491}$} (81.7s)\\* DNI]{\includegraphics[width=\myfigwidth, height=2.012\myfigwidth]{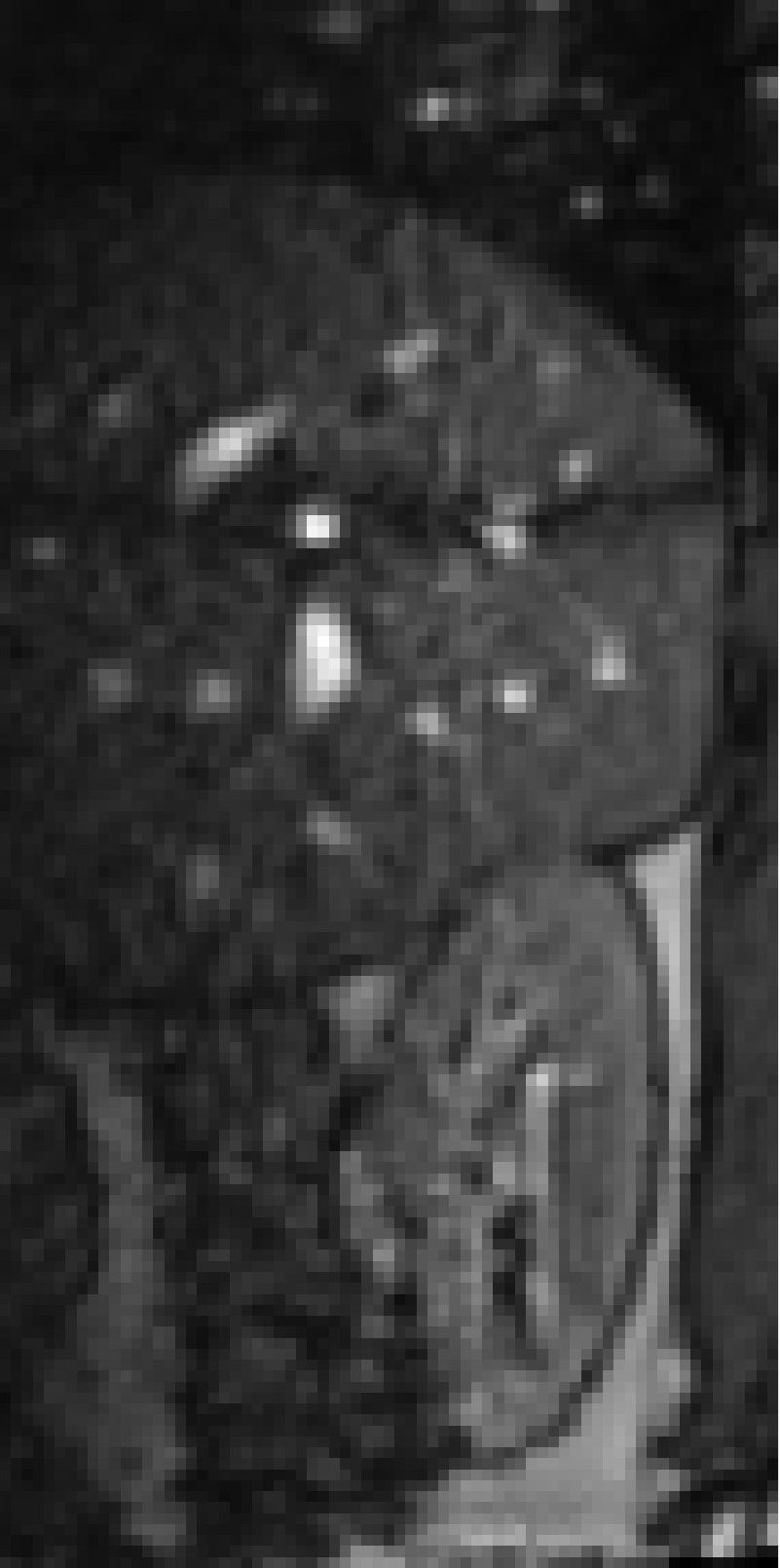} }%  
    \subfloat[Sequence 1\\* \text{\small $t_{491}$} (81.7s)\\* subj-specific \\* transformer]{\includegraphics[width=\myfigwidth, height=2.012\myfigwidth]{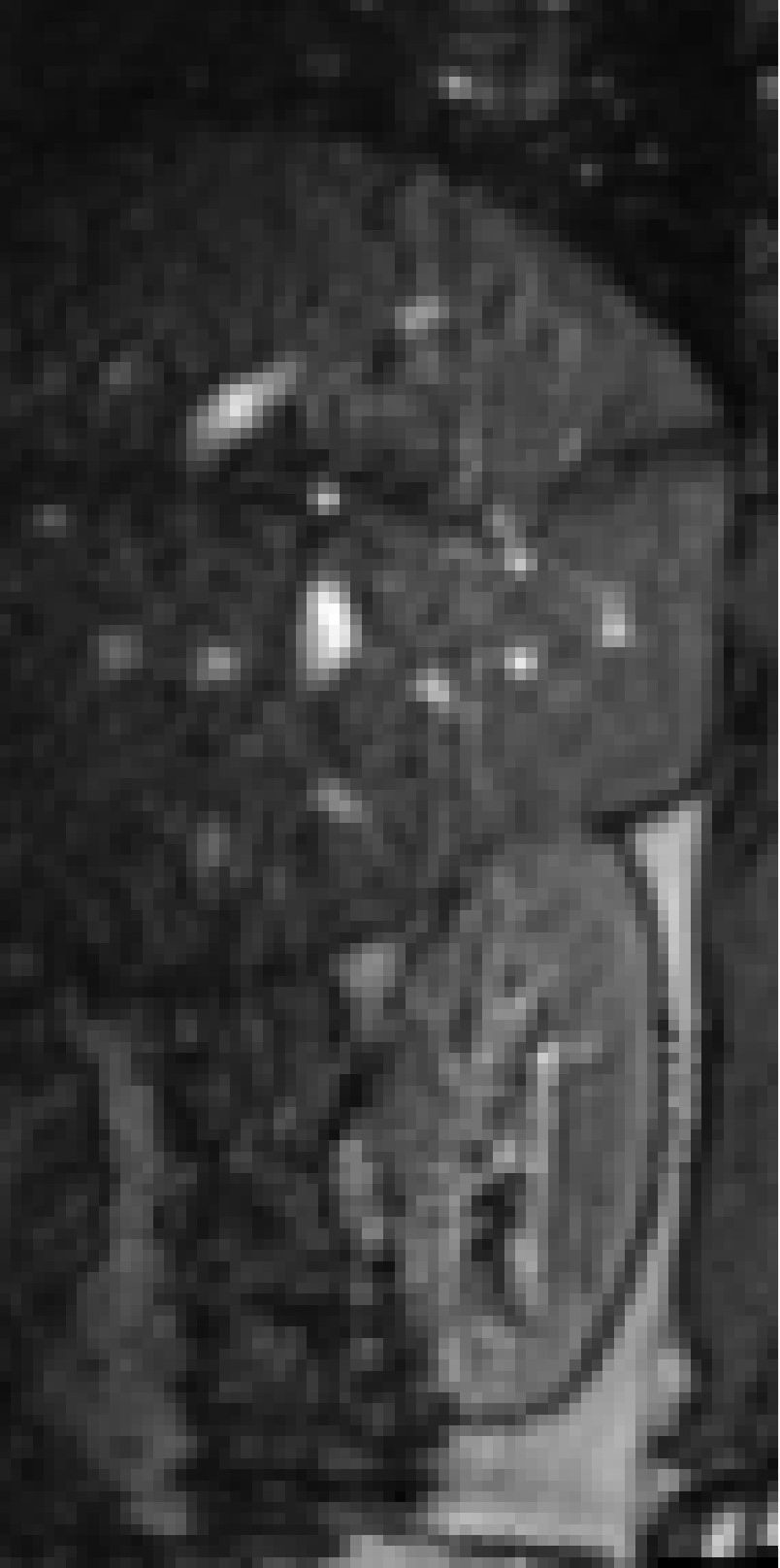} }%     
    \subfloat[Sequence 1\\* \text{\small $t_{476}$} (79.2s)\\* ground truth]{\includegraphics[width=\myfigwidth, height=2.012\myfigwidth]{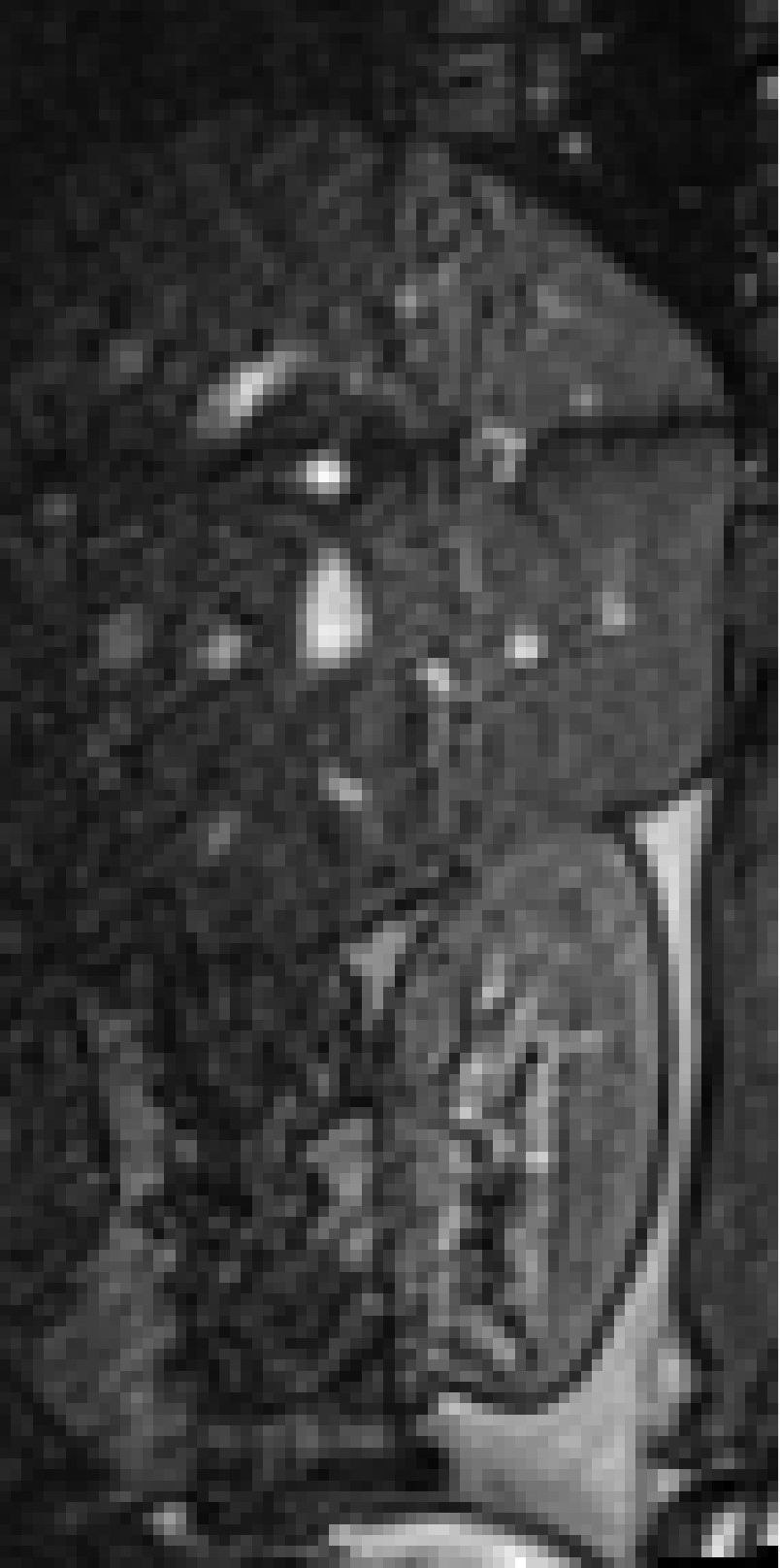} }% 
    \subfloat[Sequence 1\\* \text{\small $t_{476}$} (79.2s)\\* UORO]{\includegraphics[width=\myfigwidth, height=2.012\myfigwidth]{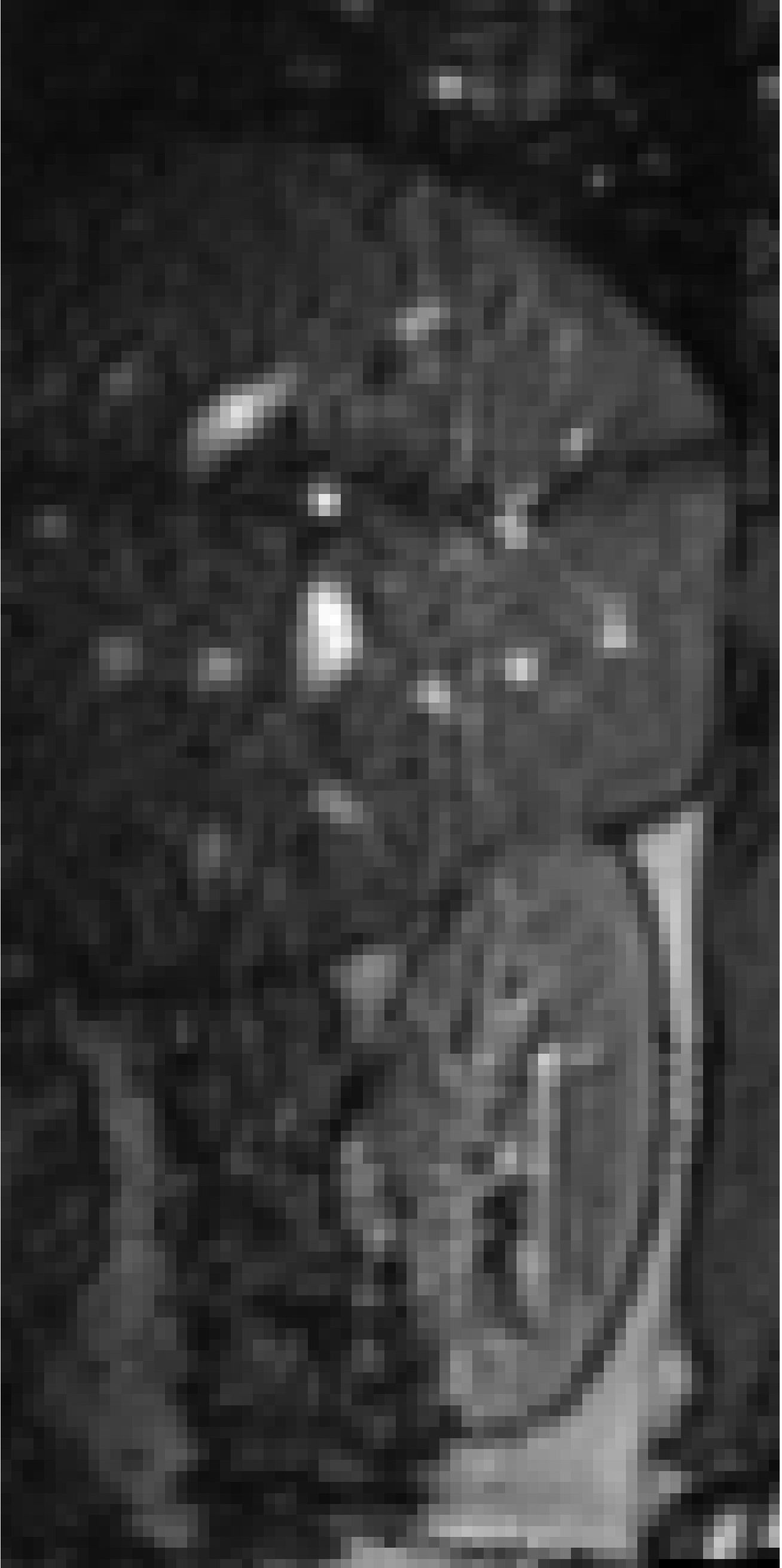} }%  
    \subfloat[Sequence 1\\* \text{\small $t_{476}$} (79.2s)\\* SnAp-1]{\includegraphics[width=\myfigwidth, height=2.012\myfigwidth]{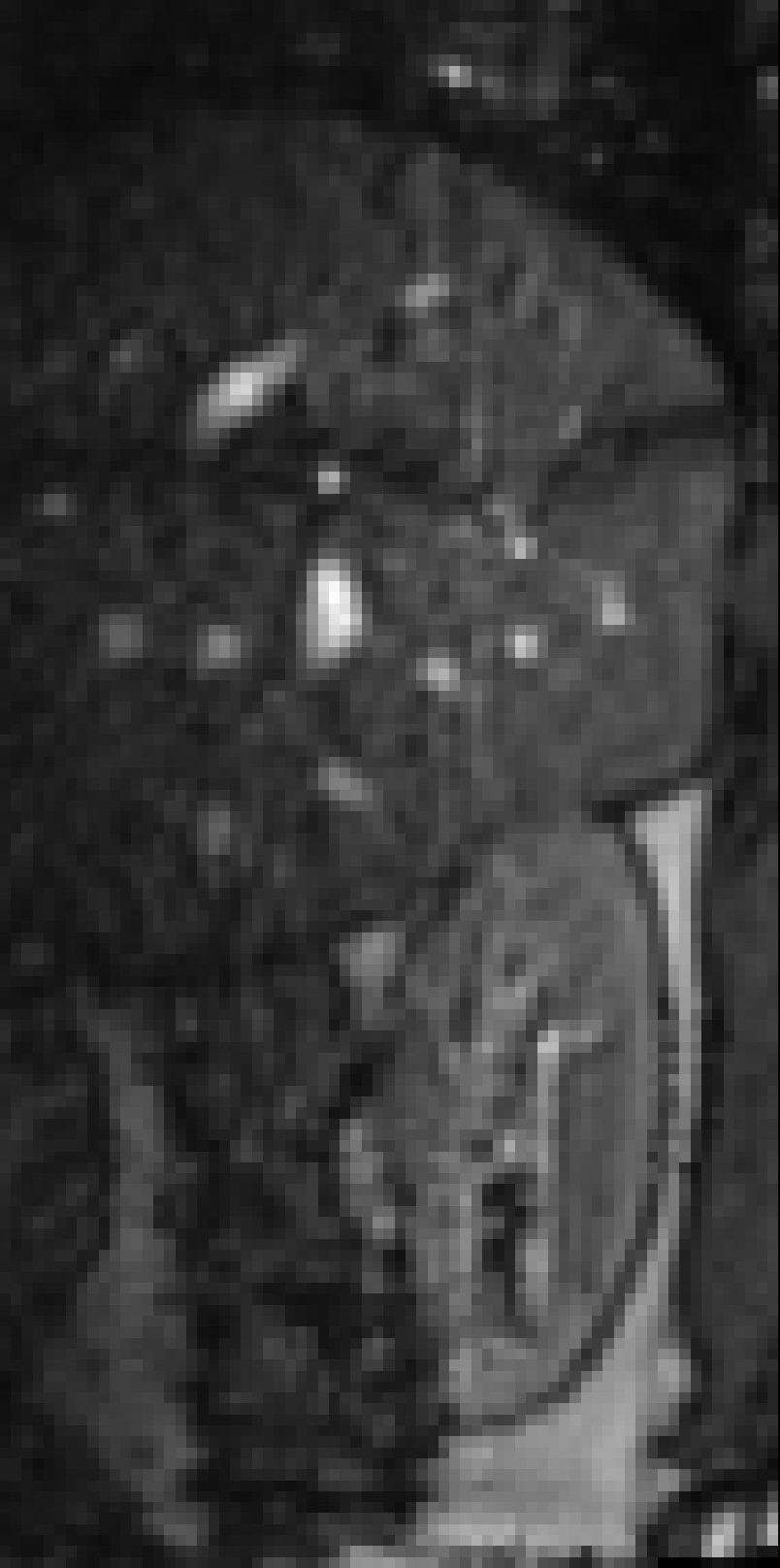} }%   
    \subfloat[Sequence 1\\* \text{\small $t_{476}$} (79.2s)\\* DNI]{\includegraphics[width=\myfigwidth, height=2.012\myfigwidth]{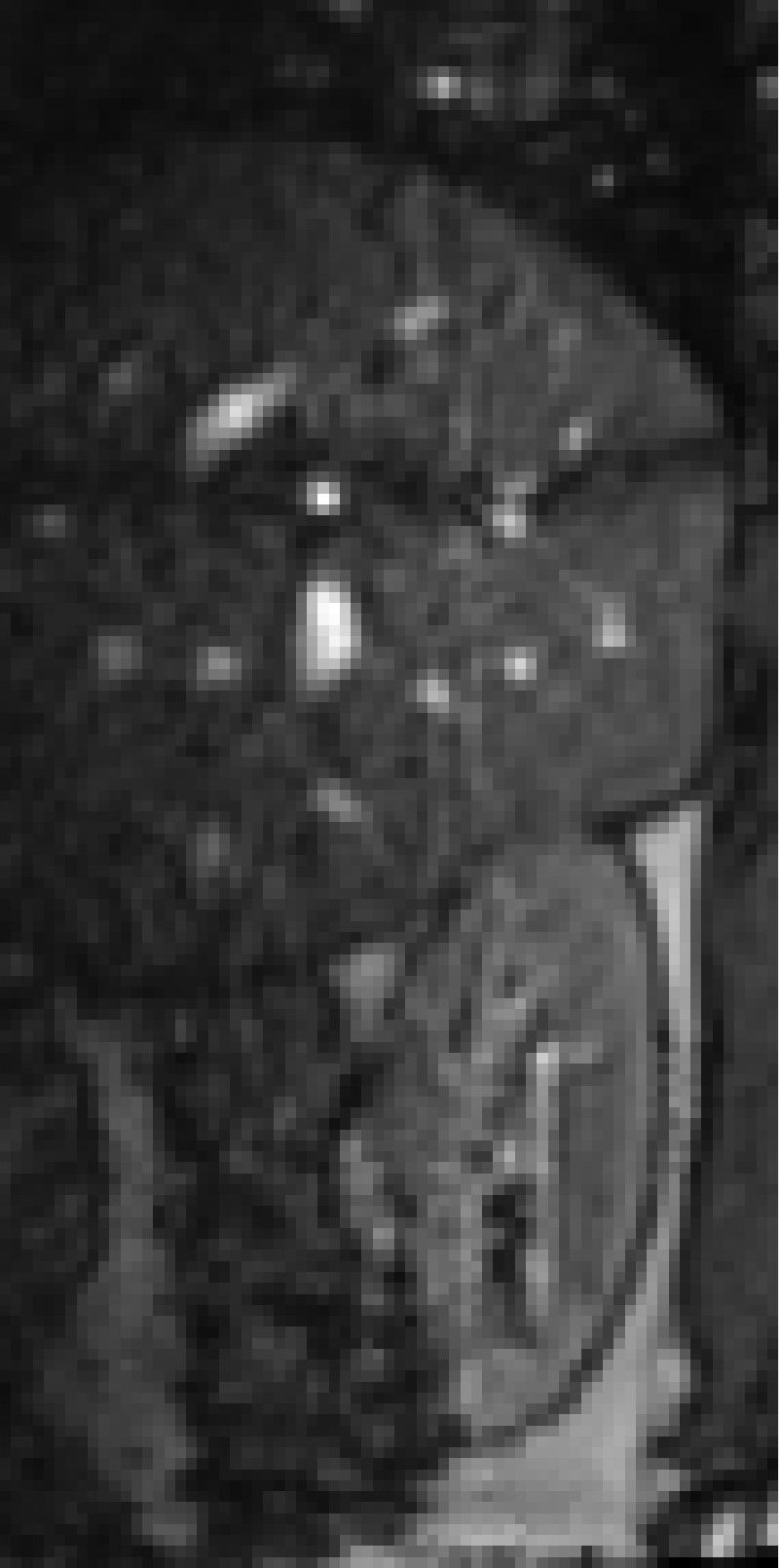} }%      
    \subfloat[Sequence 1\\* \text{\small $t_{476}$} (79.2s)\\* subj-specific \\* transformer]{\includegraphics[width=\myfigwidth, height=2.012\myfigwidth]{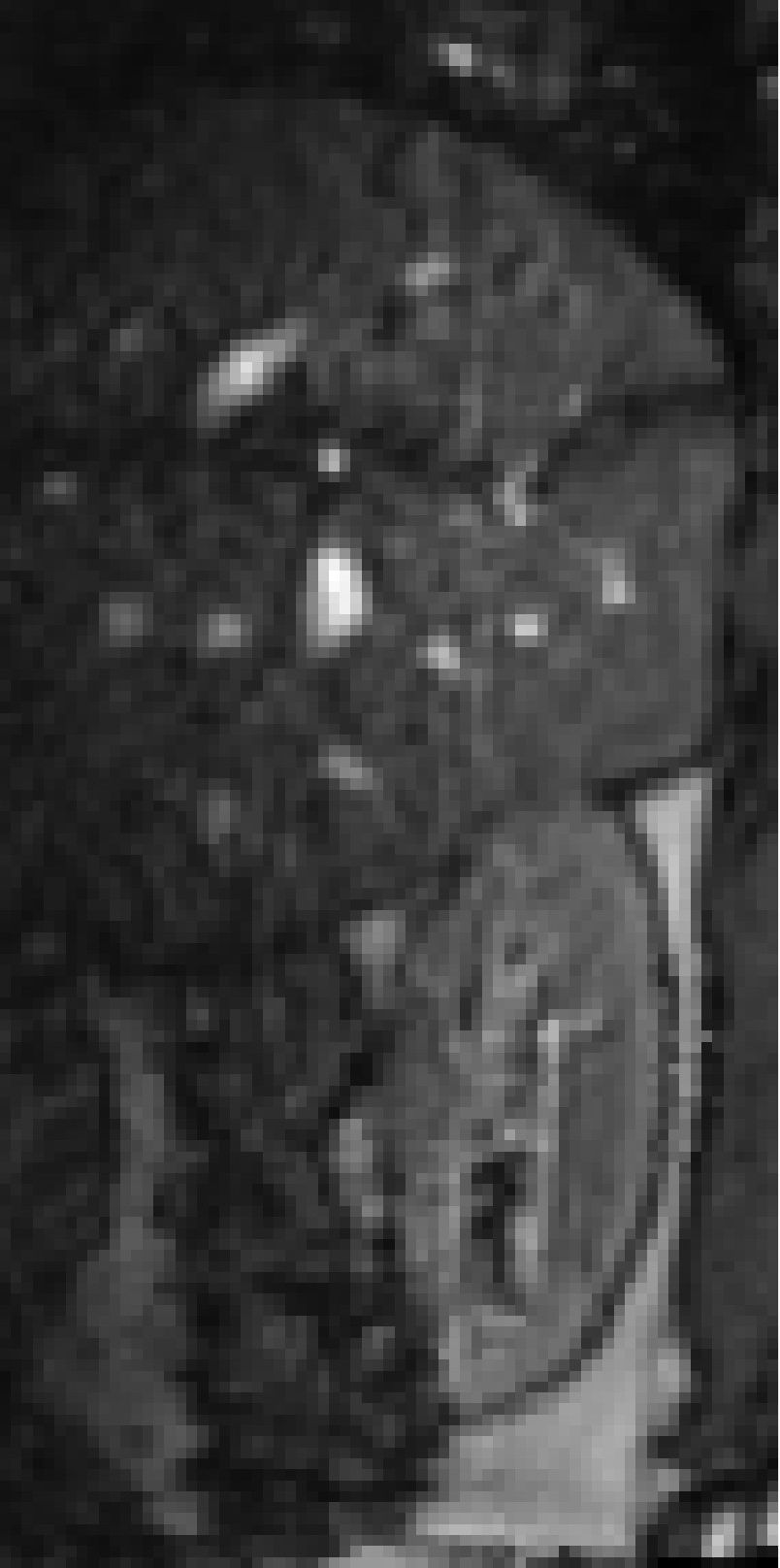} }%   
    \hfill
    \subfloat[Sequence 2 \\* \text{\small $t=t_{1}$} \\* reference]{\includegraphics[width=\myfigwidth, height=2.061\myfigwidth]{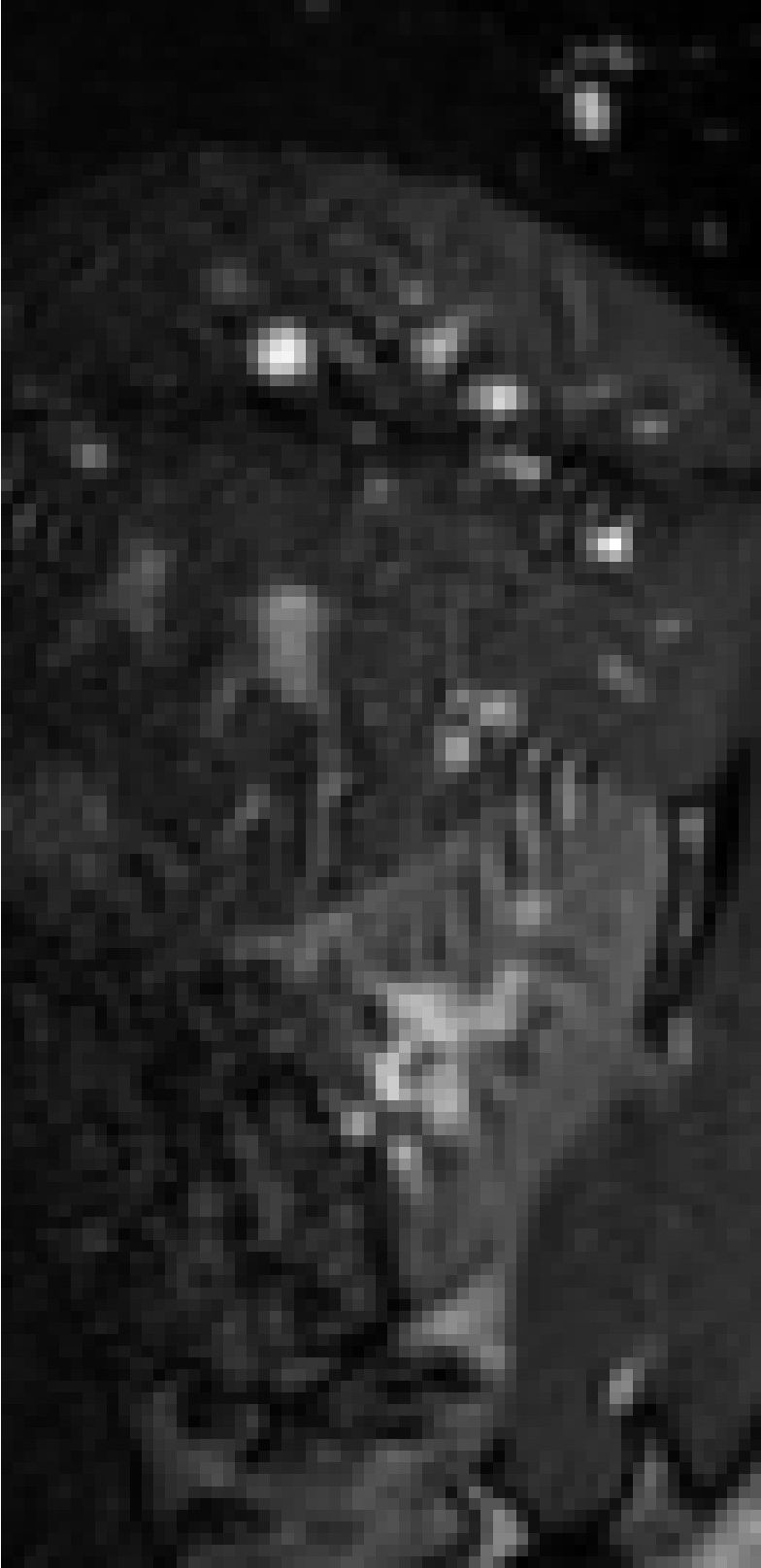} }%     
    \subfloat[Sequence 2\\* \text{\small $t_{437}$} (72.7s)\\* ground truth]{\includegraphics[width=\myfigwidth, height=2.061\myfigwidth]{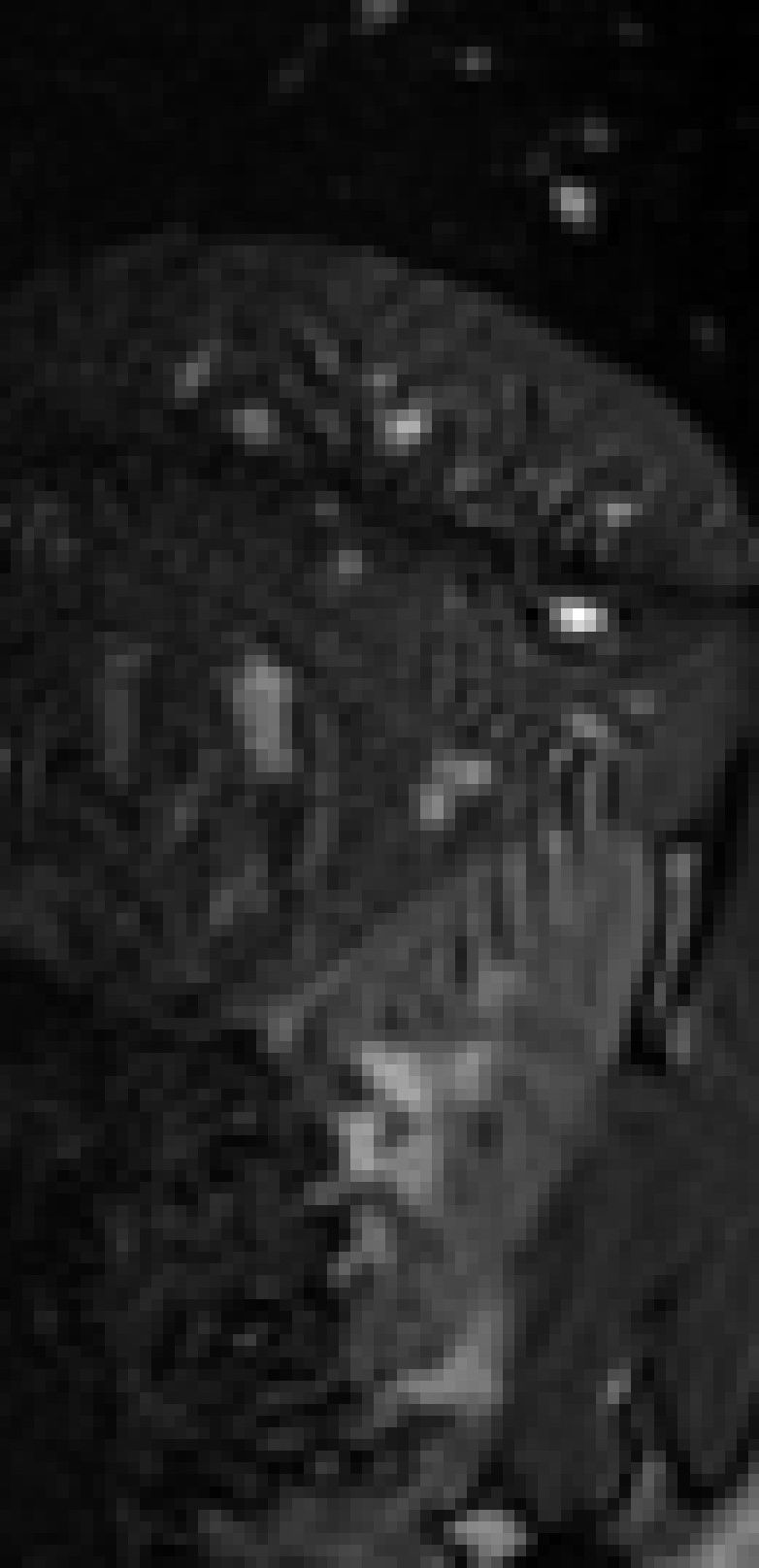} }% 
    \subfloat[Sequence 2\\* \text{\small $t_{437}$} (72.7s)\\* UORO]{\includegraphics[width=\myfigwidth, height=2.061\myfigwidth]{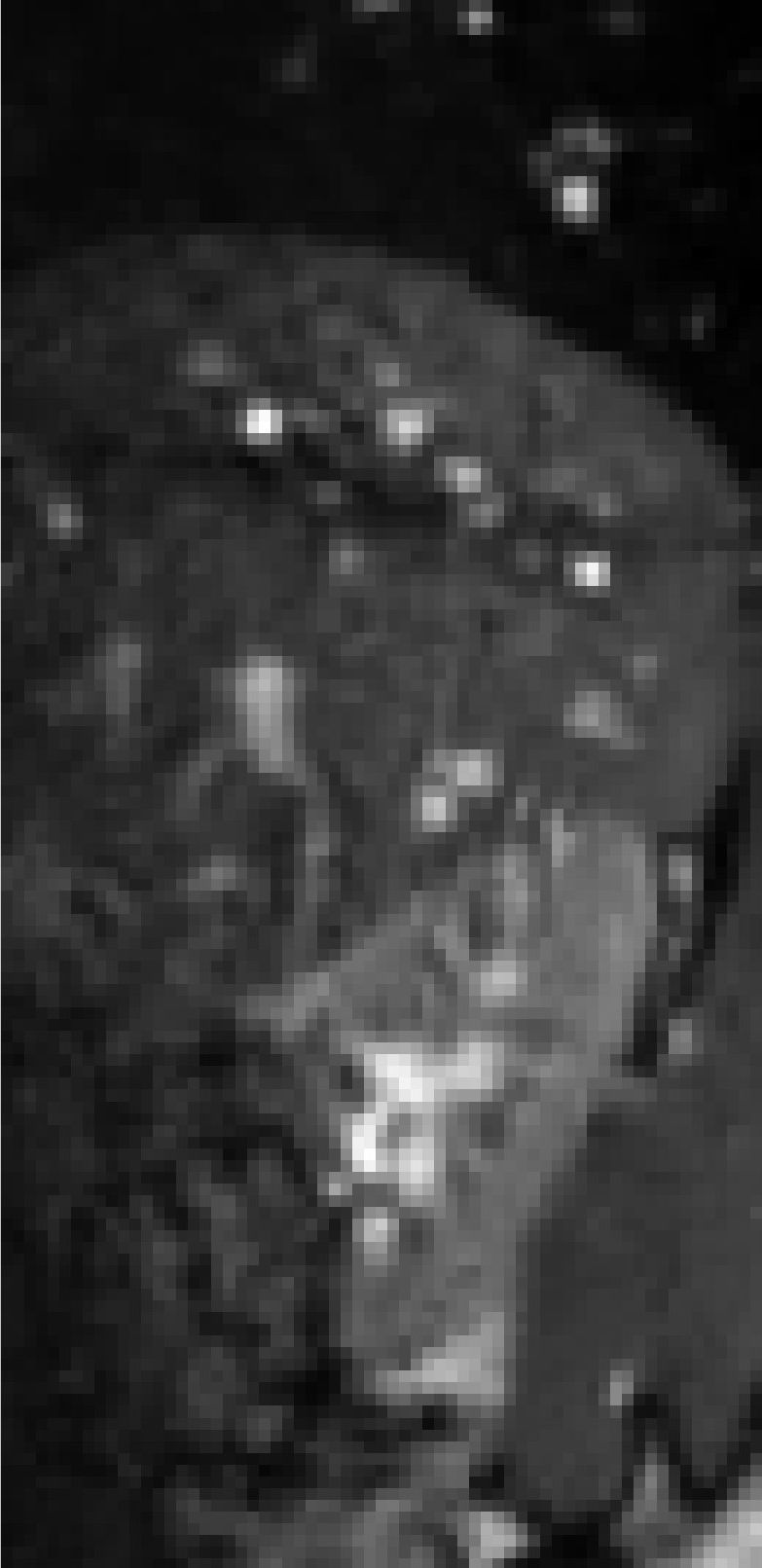} }%   
    \subfloat[Sequence 2\\* \text{\small $t_{437}$} (72.7s)\\* SnAp-1]{\includegraphics[width=\myfigwidth, height=2.061\myfigwidth]{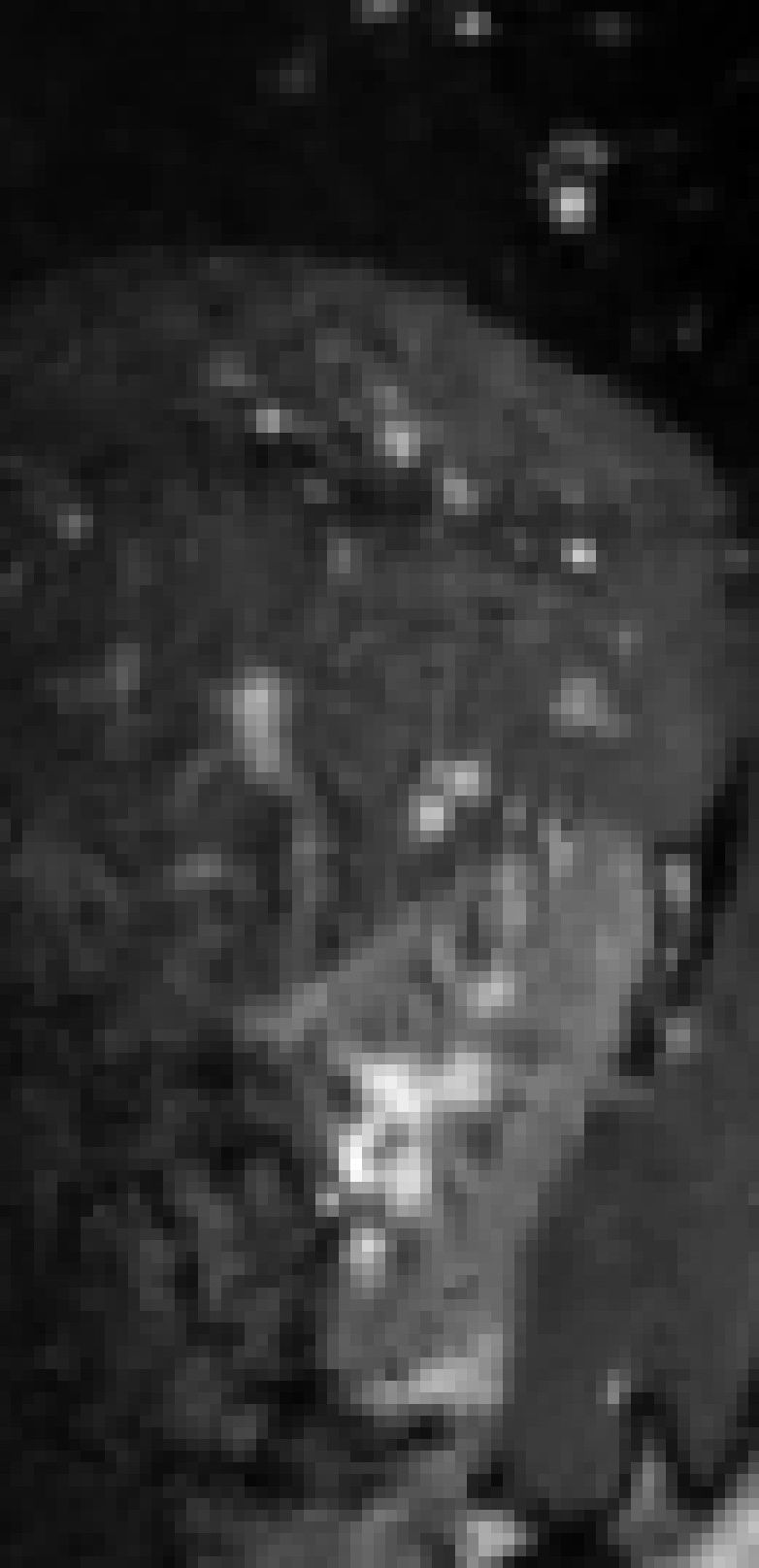} }%   
    \subfloat[Sequence 2\\* \text{\small $t_{437}$} (72.7s)\\* DNI]{\includegraphics[width=\myfigwidth, height=2.061\myfigwidth]{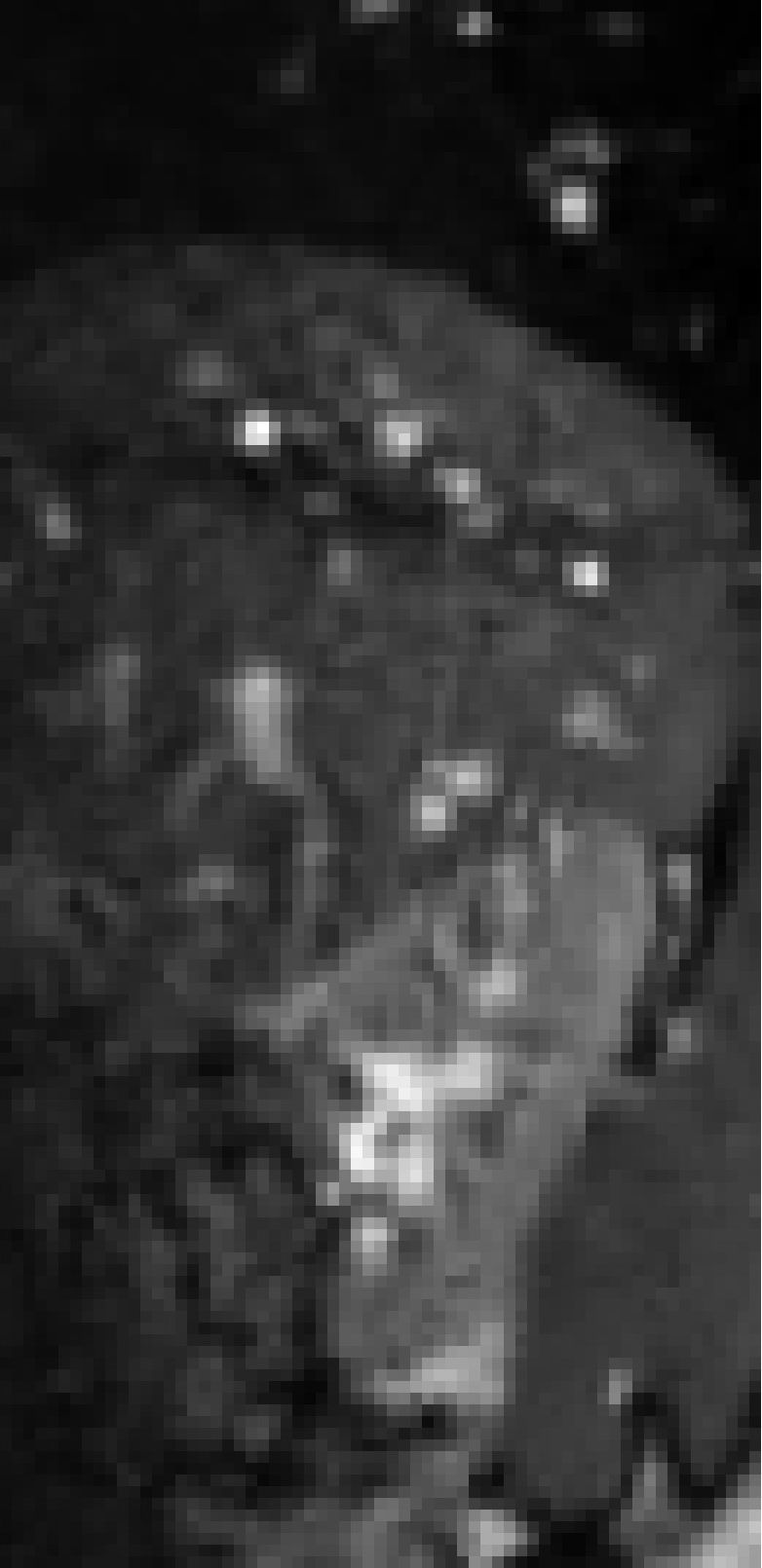} }%     
    \subfloat[Sequence 2\\* \text{\small $t_{437}$} (72.7s)\\* subj-specific \\* transformer]{\includegraphics[width=\myfigwidth, height=2.061\myfigwidth]{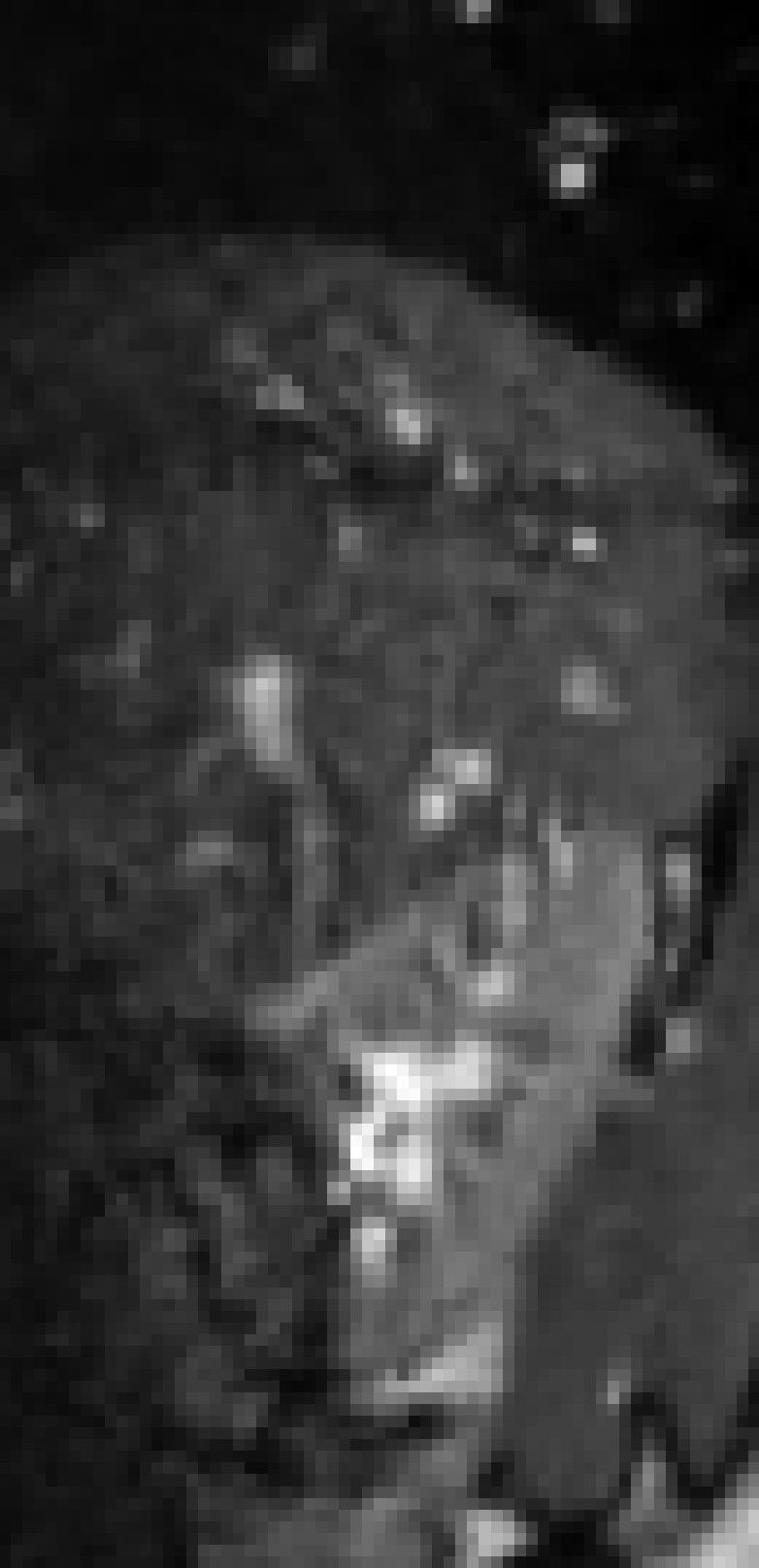} }%     
    \subfloat[Sequence 2\\* \text{\small $t_{496}$} (82.5s)\\* ground truth]{\includegraphics[width=\myfigwidth, height=2.061\myfigwidth]{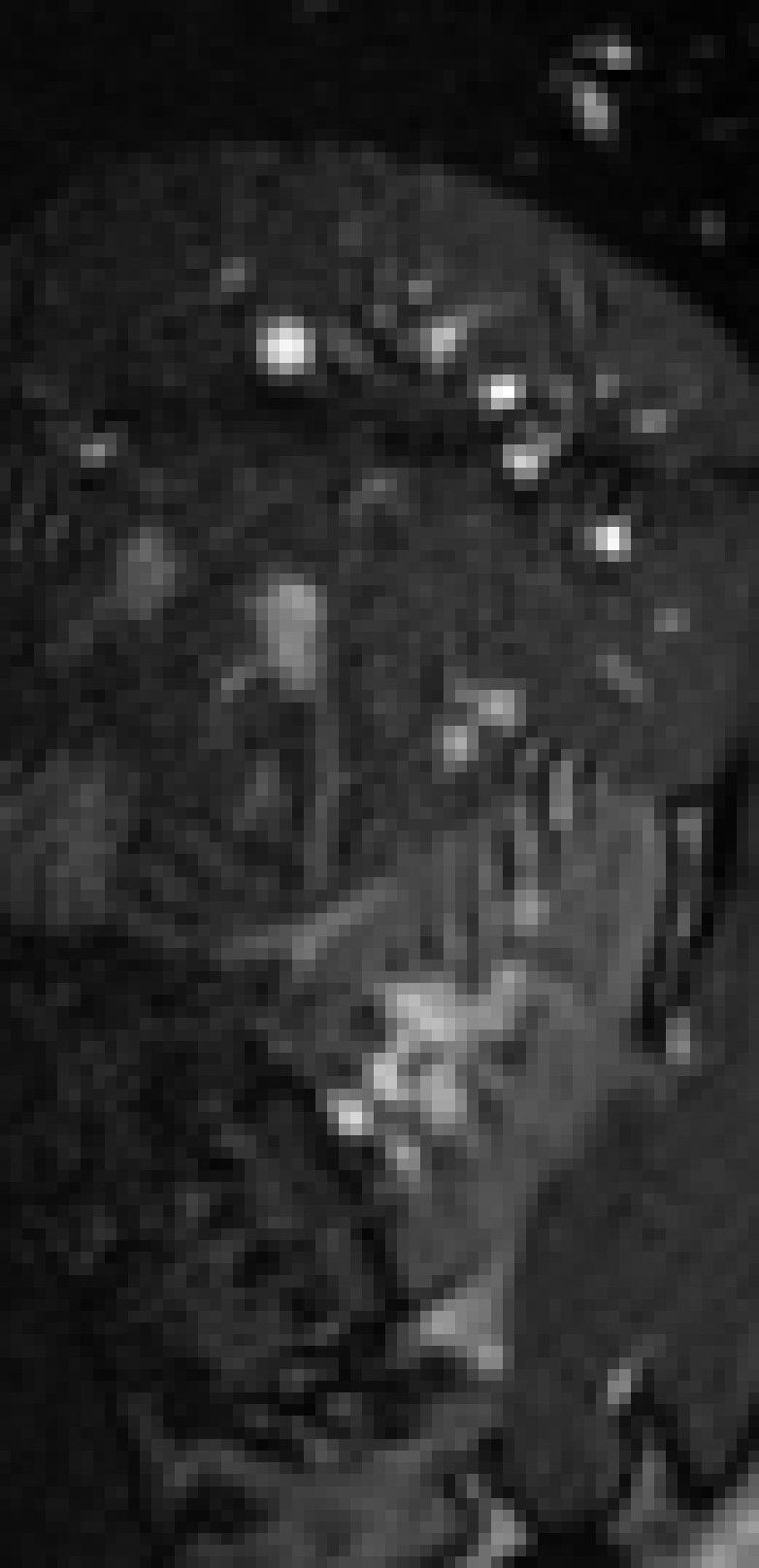} }% 
    \subfloat[Sequence 2\\* \text{\small $t_{496}$} (82.5s)\\* UORO]{\includegraphics[width=\myfigwidth, height=2.061\myfigwidth]{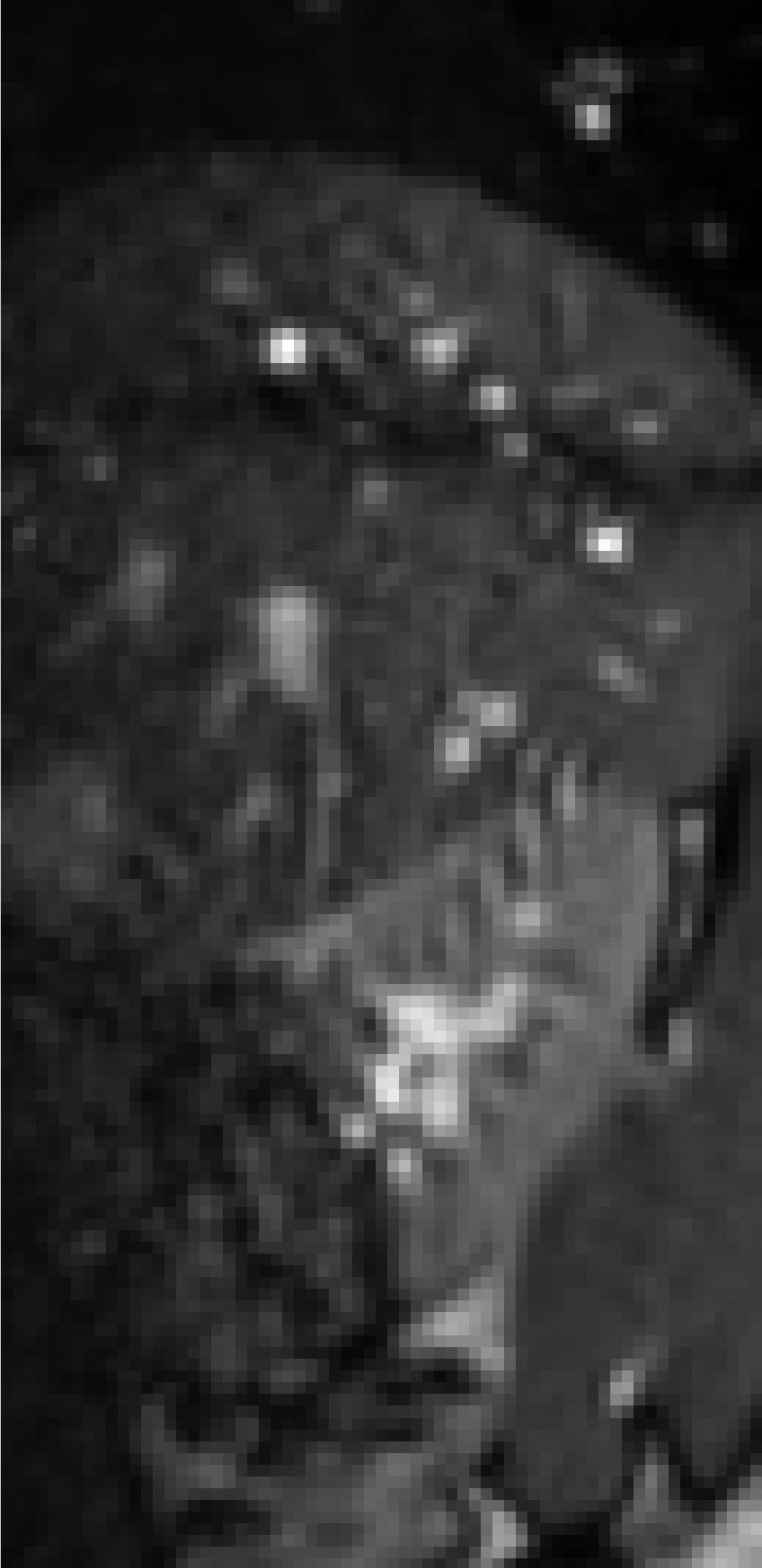} }%  
    \subfloat[Sequence 2\\* \text{\small $t_{496}$} (82.5s)\\* SnAp-1]{\includegraphics[width=\myfigwidth, height=2.061\myfigwidth]{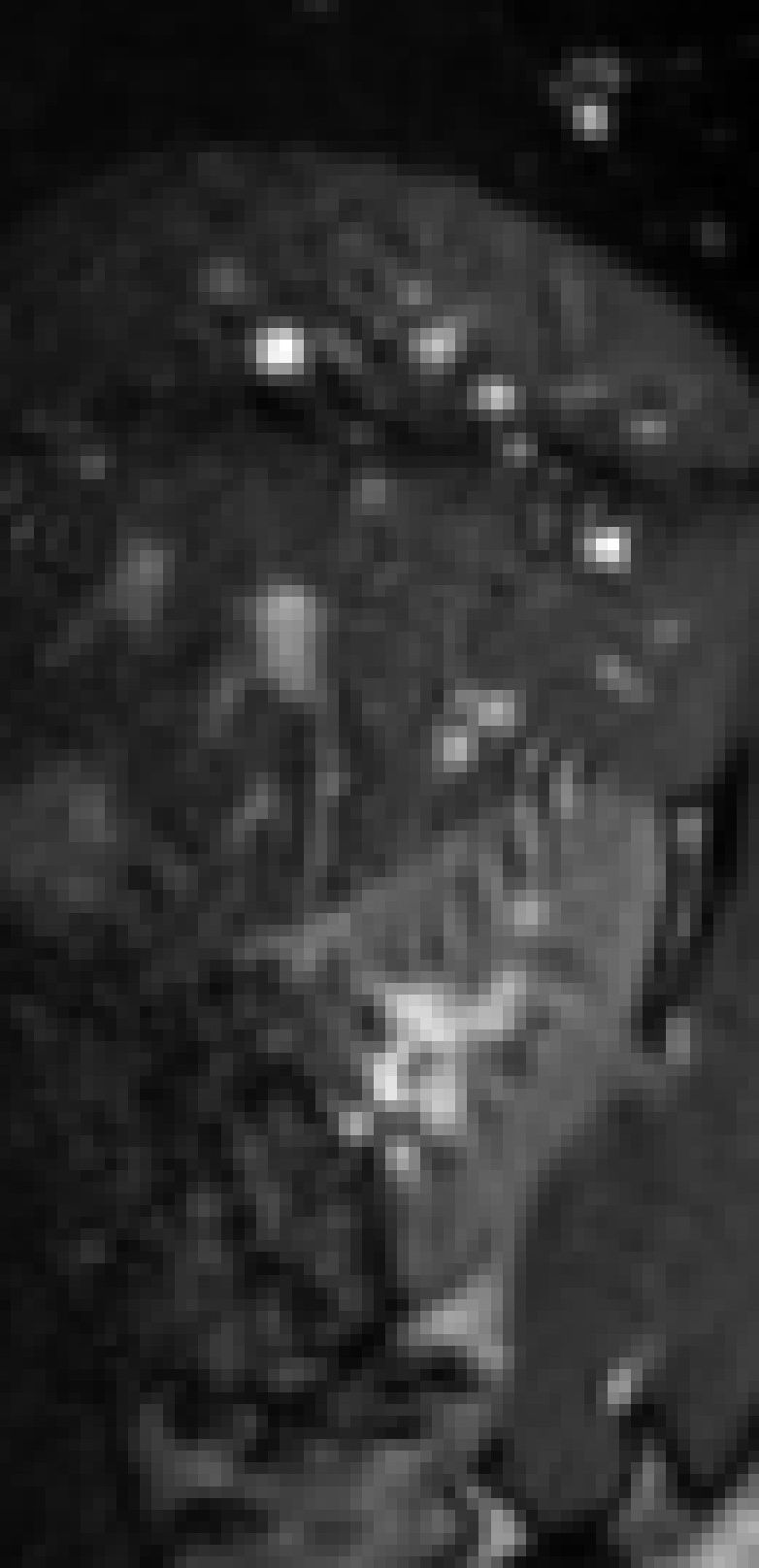} }%   
    \subfloat[Sequence 2\\* \text{\small $t_{496}$} (82.5s)\\* DNI]{\includegraphics[width=\myfigwidth, height=2.061\myfigwidth]{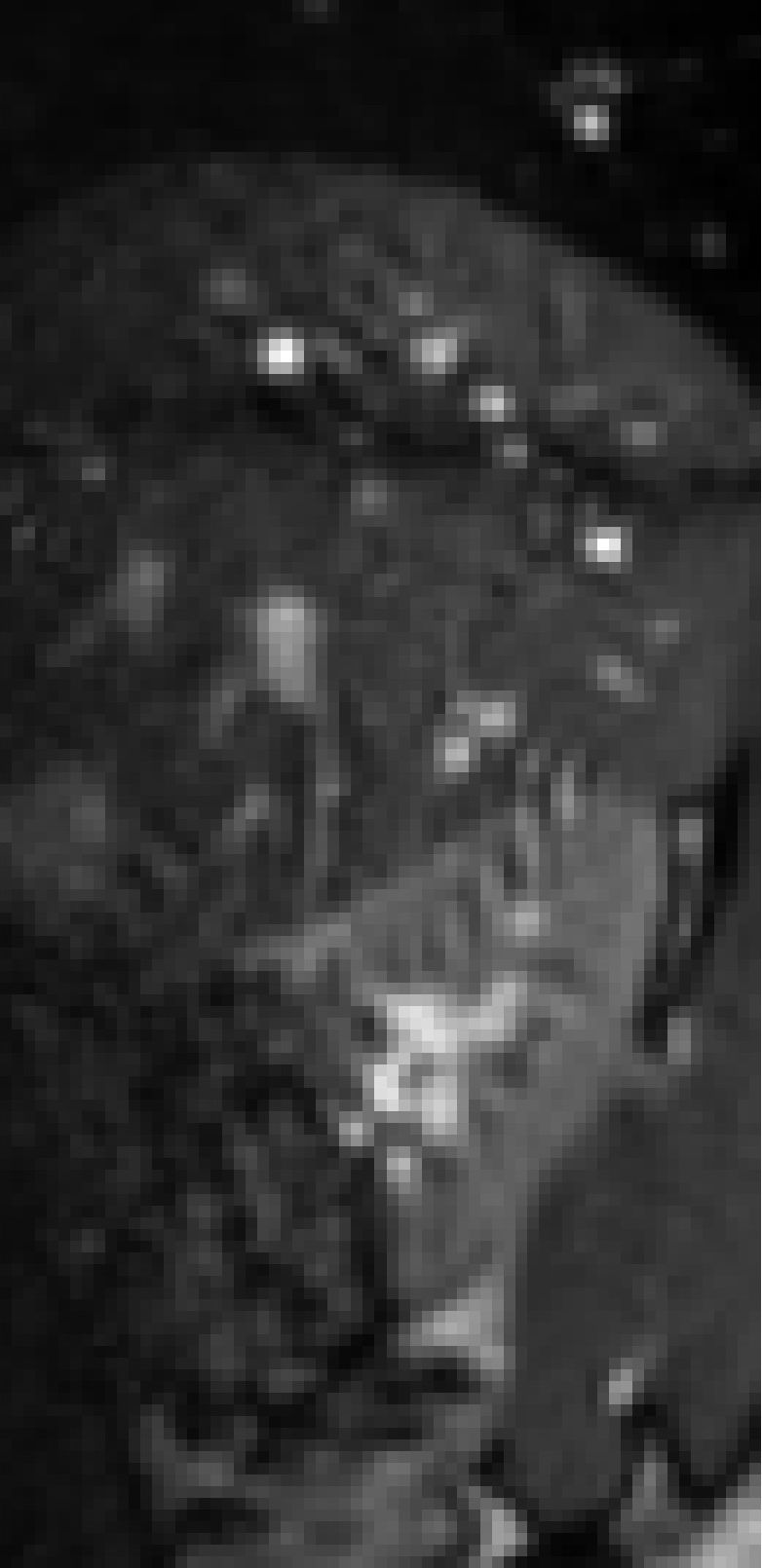} }%     
    \subfloat[Sequence 2\\* \text{\small $t_{496}$} (82.5s)\\* subj-specific \\* transformer]{\includegraphics[width=\myfigwidth, height=2.061\myfigwidth]{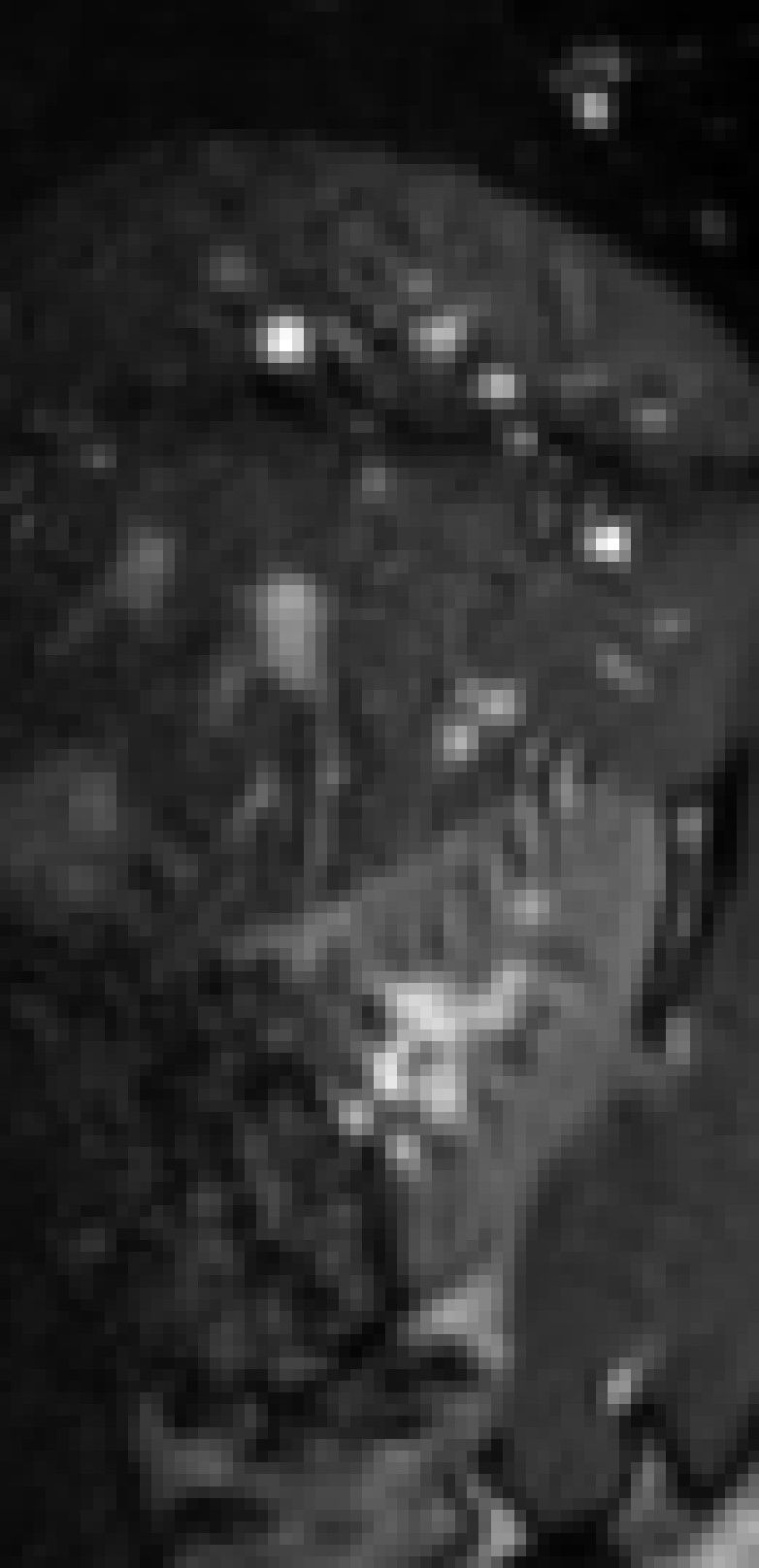} }%          
    \hfill
    \subfloat[Sequence 4 \\* \text{\small $t=t_{1}$} \\* reference]{\includegraphics[width=.084\textwidth]{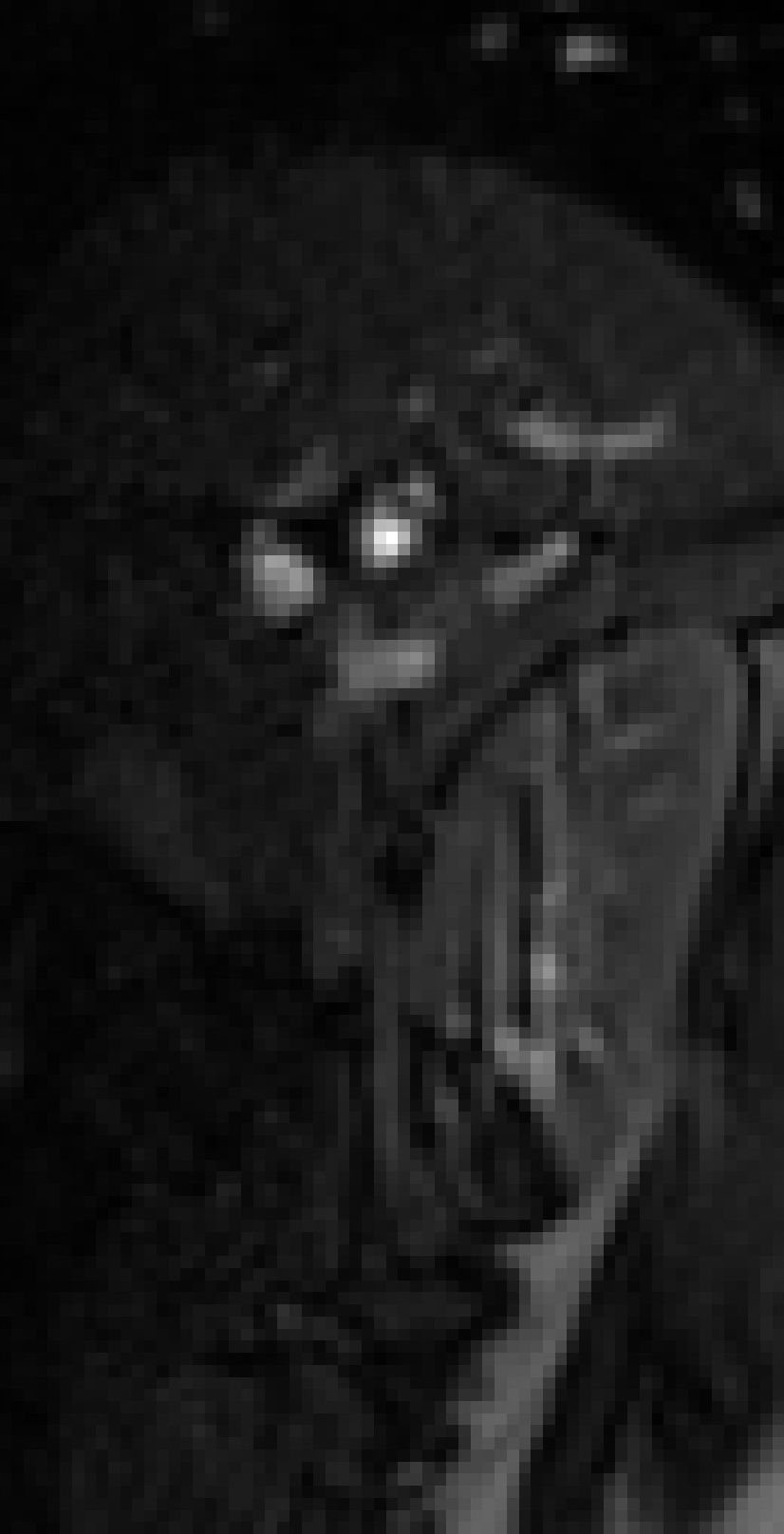} }%     
    \subfloat[Sequence 4\\* \text{\small $t_{309}$} (51.3s)\\* ground truth]{\includegraphics[width=\myfigwidth, height=1.956\myfigwidth]{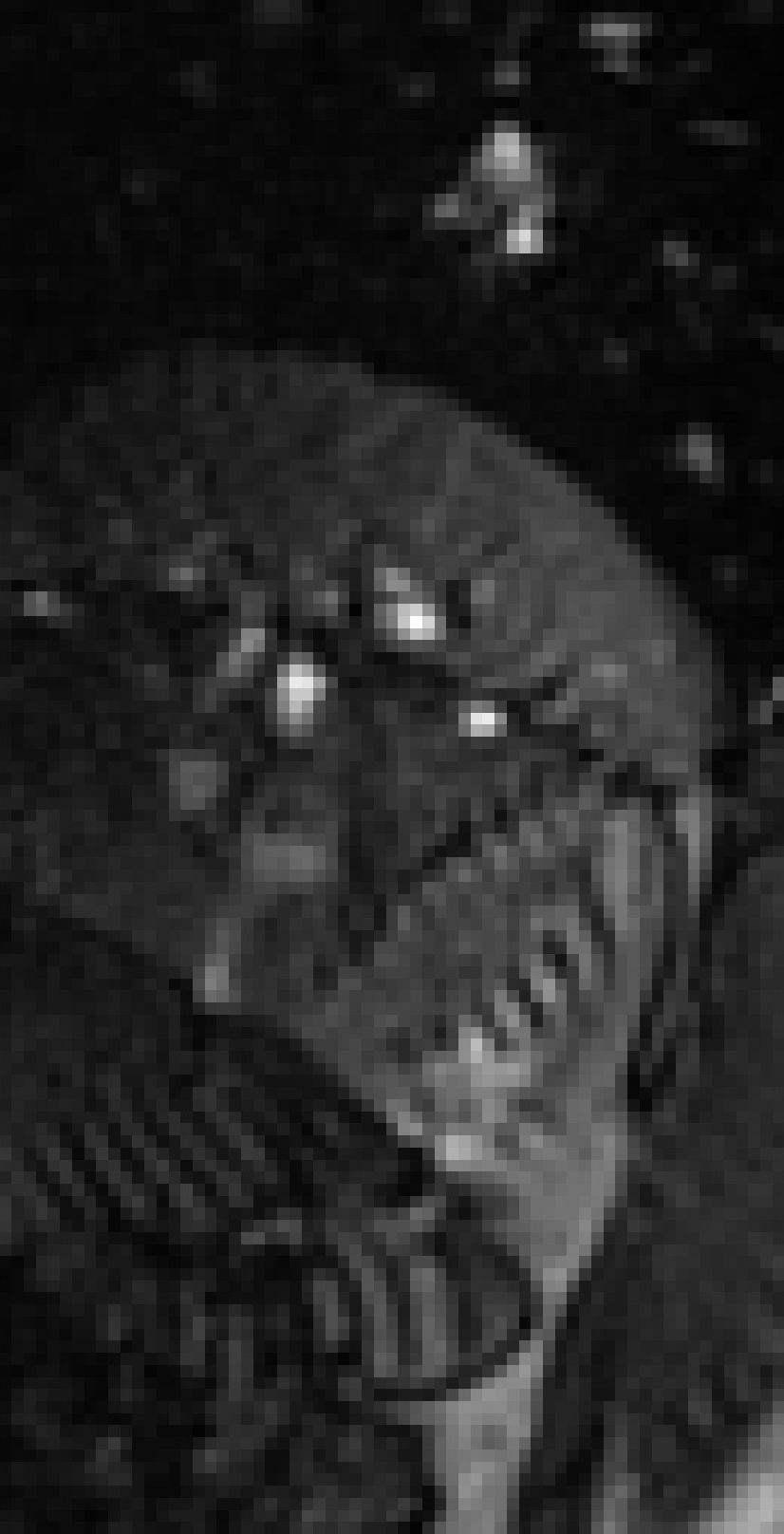} }% 
    \subfloat[Sequence 4\\* \text{\small $t_{309}$} (51.3s)\\* UORO]{\includegraphics[width=\myfigwidth, height=1.956\myfigwidth]{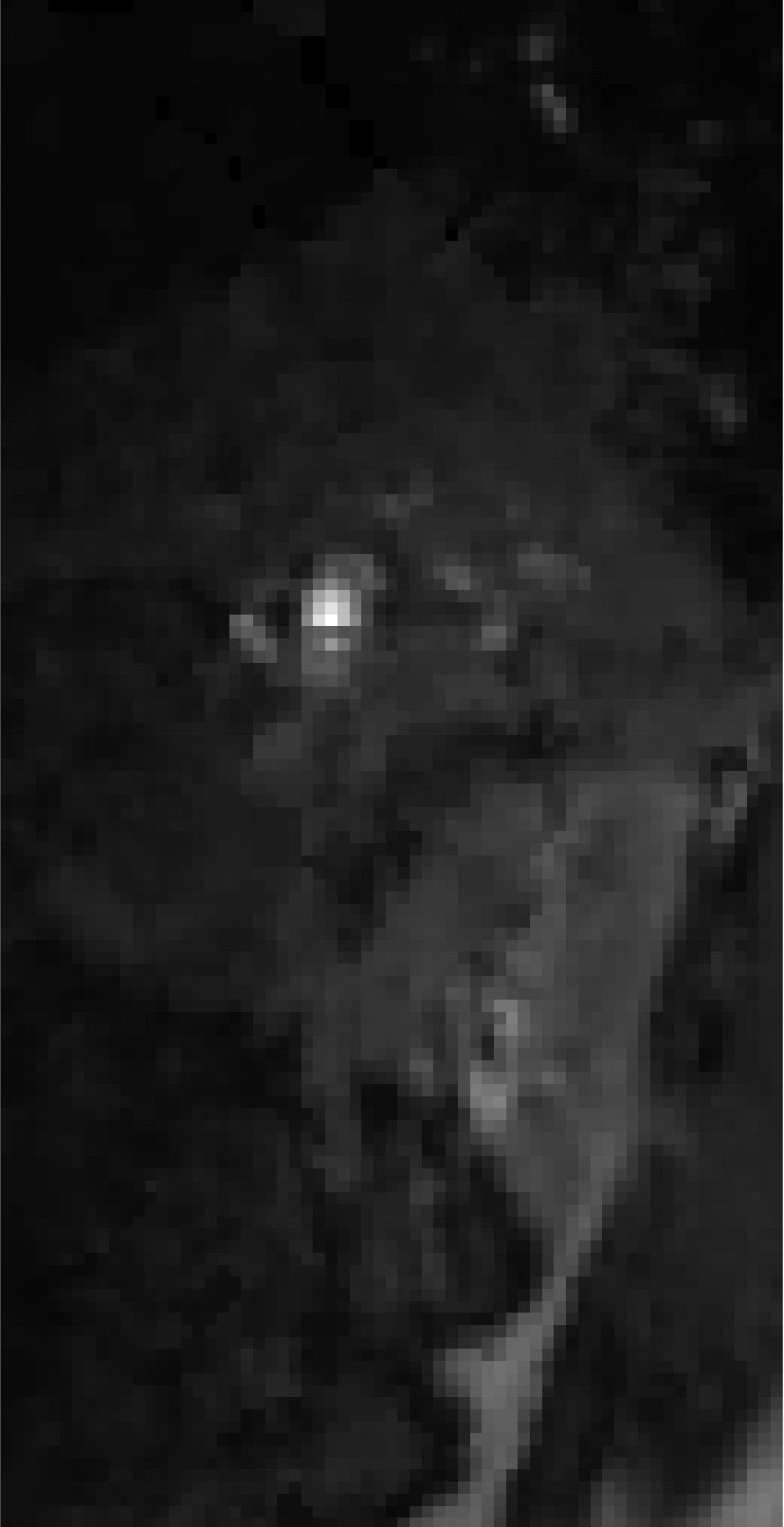} }%  
    \subfloat[Sequence 4\\* \text{\small $t_{309}$} (51.3s)\\* SnAp-1]{\includegraphics[width=\myfigwidth, height=1.956\myfigwidth]{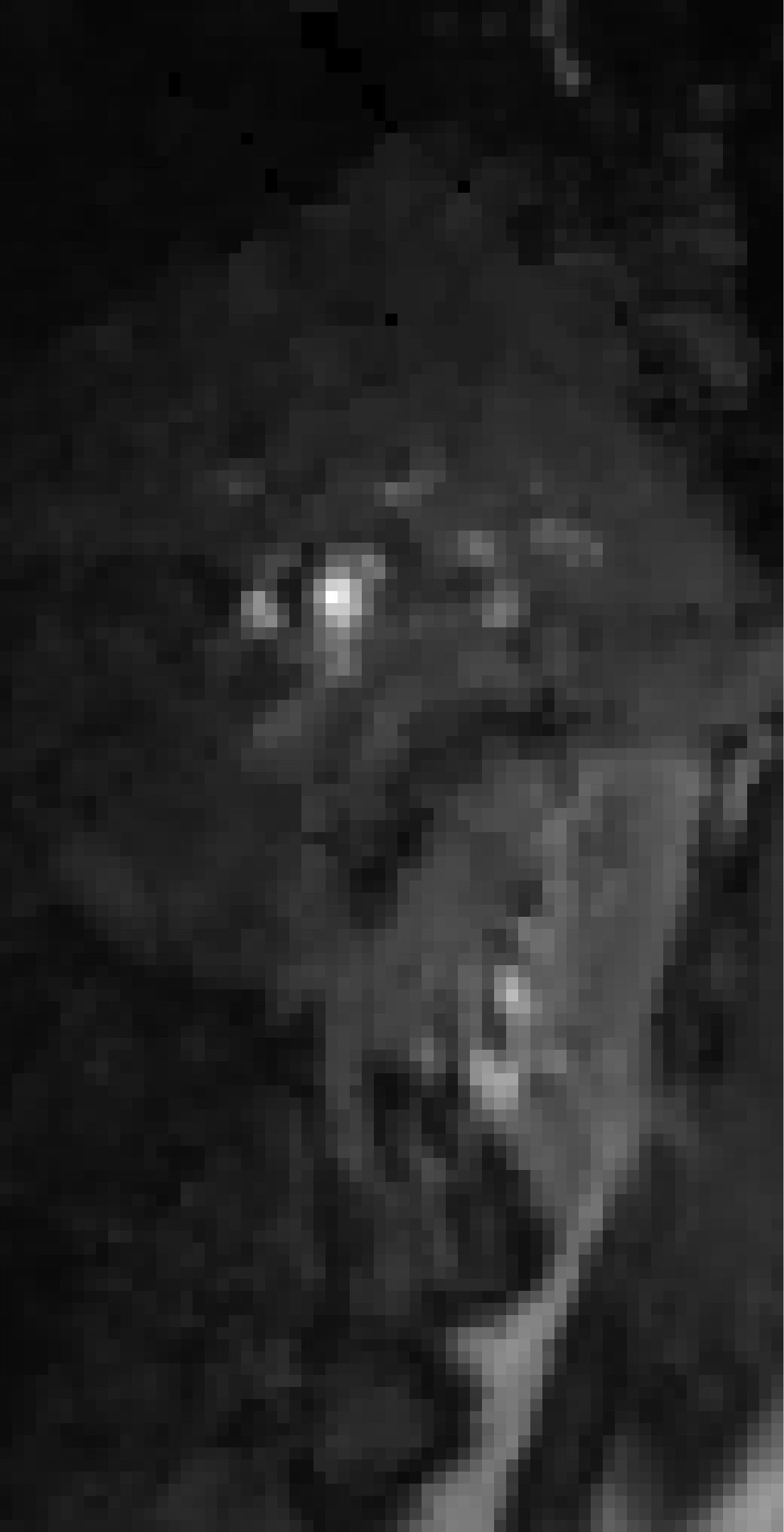} }%   
    \subfloat[Sequence 4\\* \text{\small $t_{309}$} (51.3s)\\* DNI]{\includegraphics[width=\myfigwidth, height=1.956\myfigwidth]{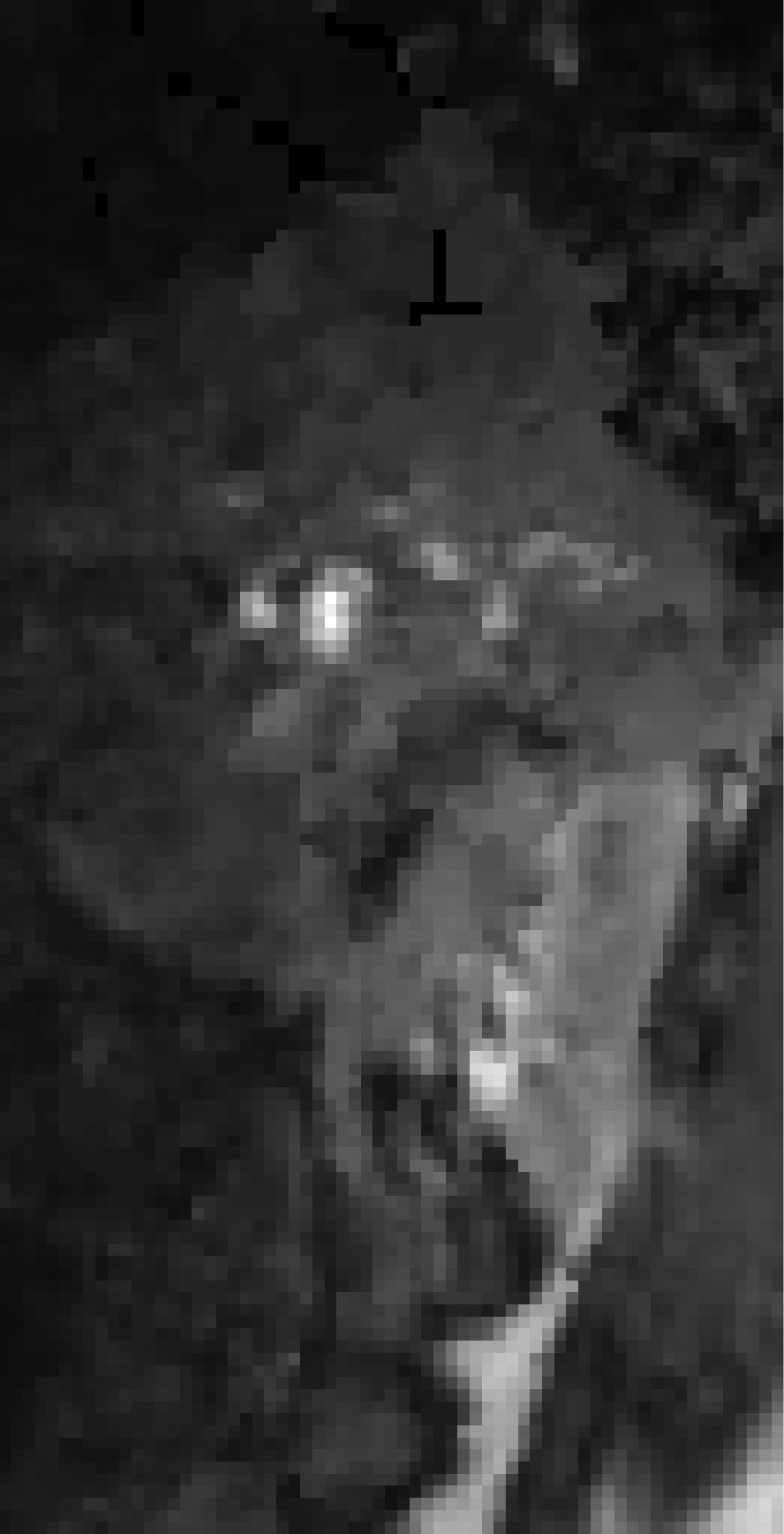} }%       
    \subfloat[Sequence 4\\* \text{\small $t_{309}$} (51.3s)\\* subj-specific \\* transformer]{\includegraphics[width=\myfigwidth, height=1.956\myfigwidth]{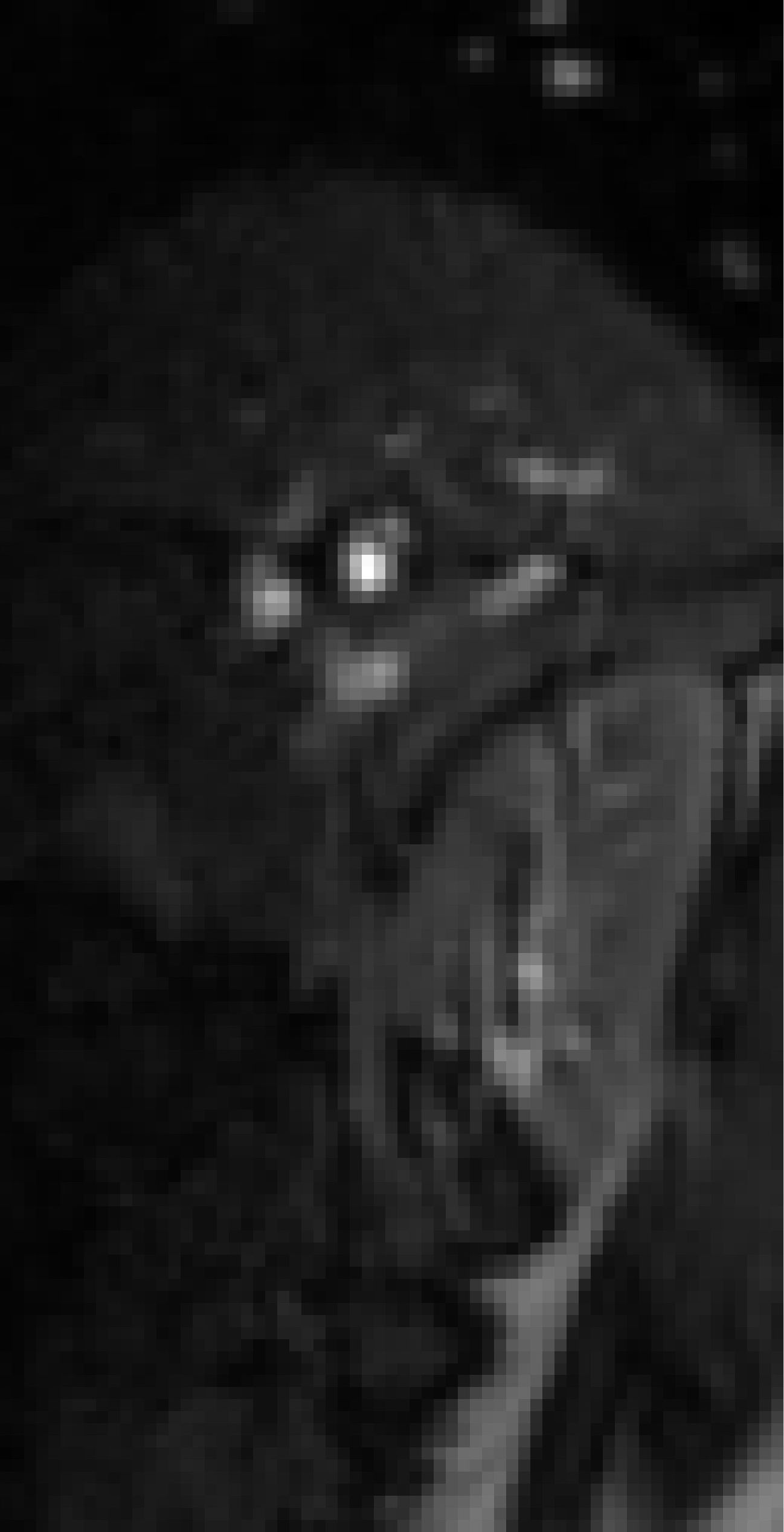} }%     
    \subfloat[Sequence 4\\* \text{\small $t_{331}$} (55.0s)\\* ground truth]{\includegraphics[width=\myfigwidth, height=1.956\myfigwidth]{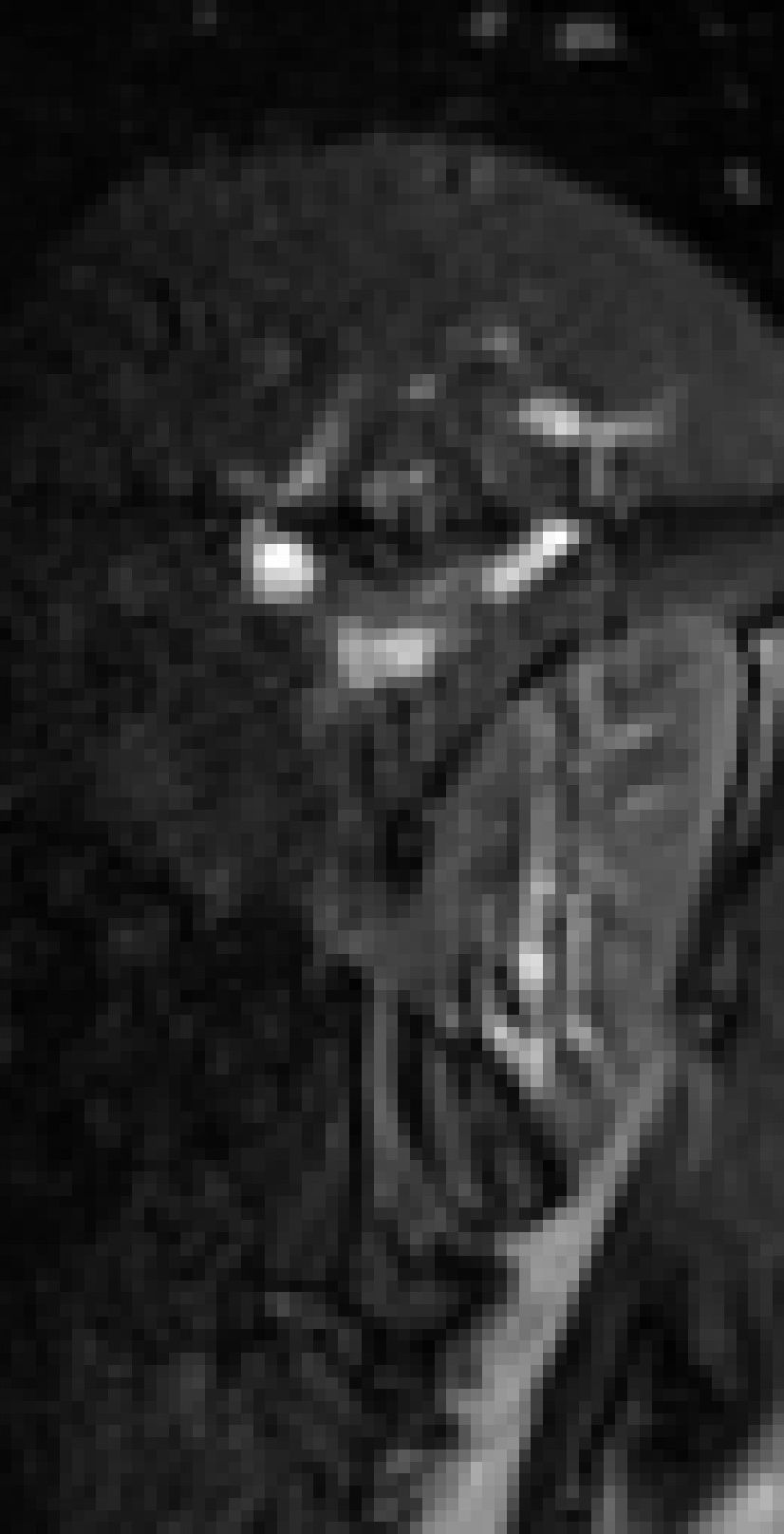} }% 
    \subfloat[Sequence 4\\* \text{\small $t_{331}$} (55.0s)\\* UORO]{\includegraphics[width=\myfigwidth, height=1.956\myfigwidth]{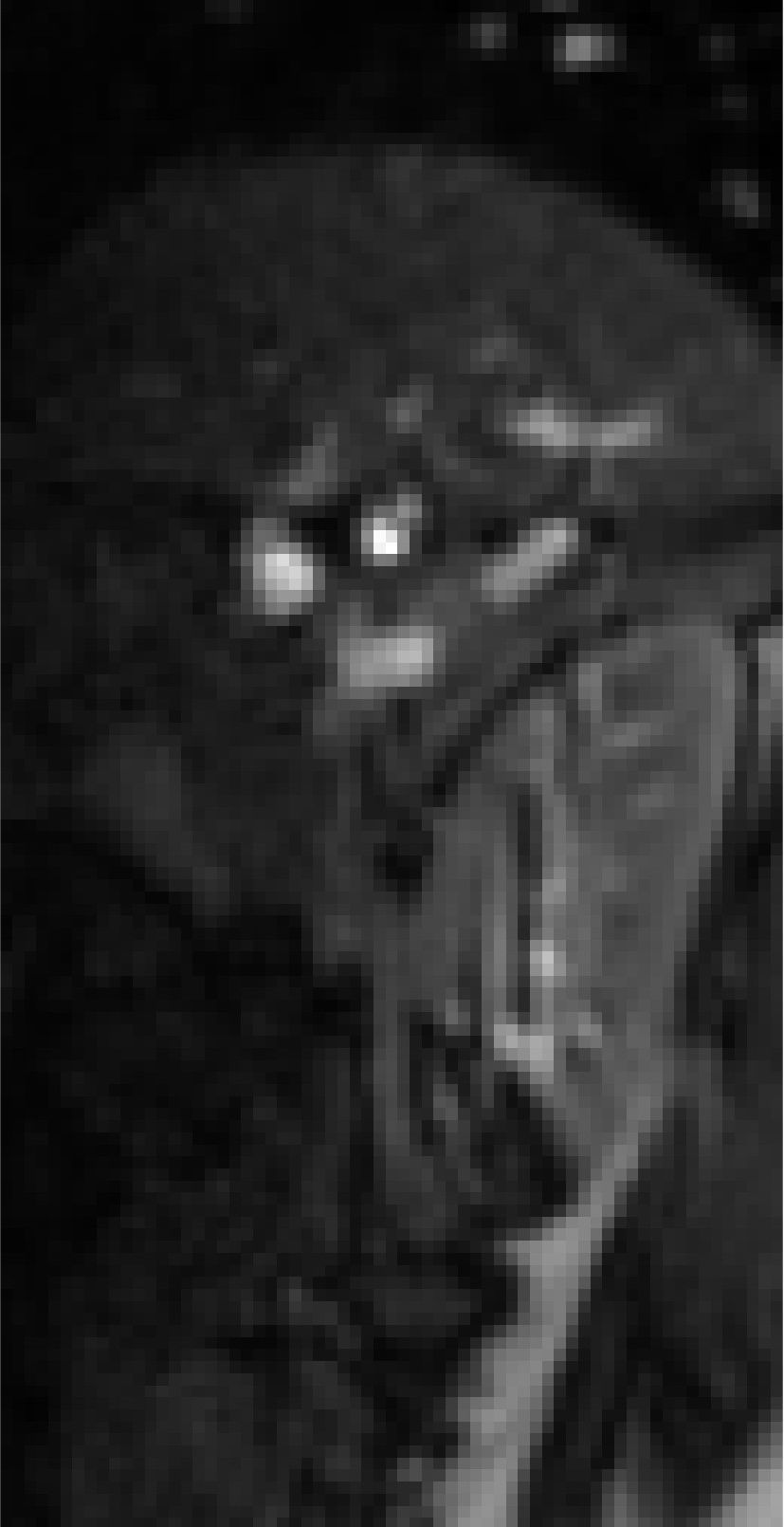} }%   
    \subfloat[Sequence 4\\* \text{\small $t_{331}$} (55.0s)\\* SnAp-1]{\includegraphics[width=\myfigwidth, height=1.956\myfigwidth]{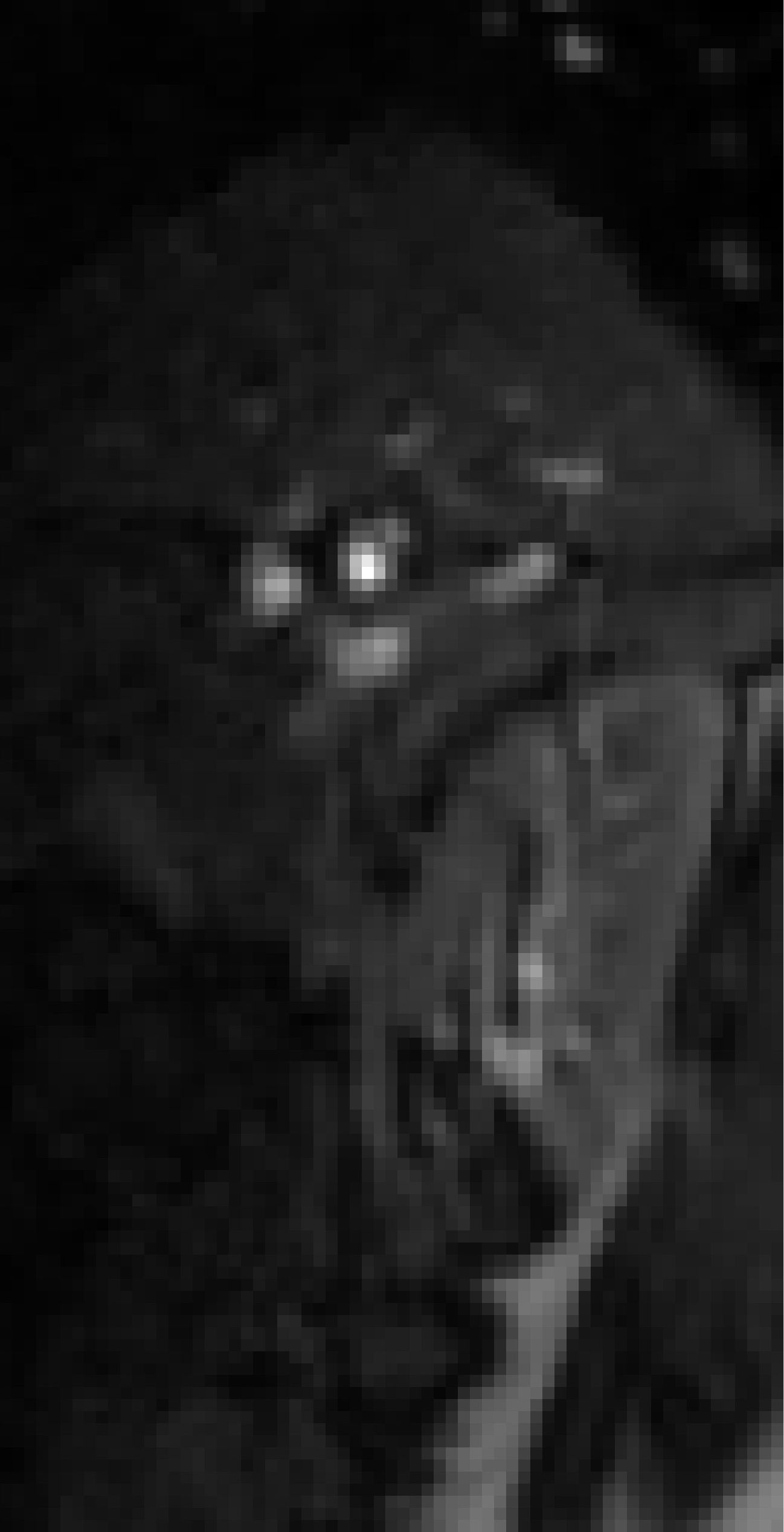} }%   
    \subfloat[Sequence 4\\* \text{\small $t_{331}$} (55.0s)\\* DNI]{\includegraphics[width=\myfigwidth, height=1.956\myfigwidth]{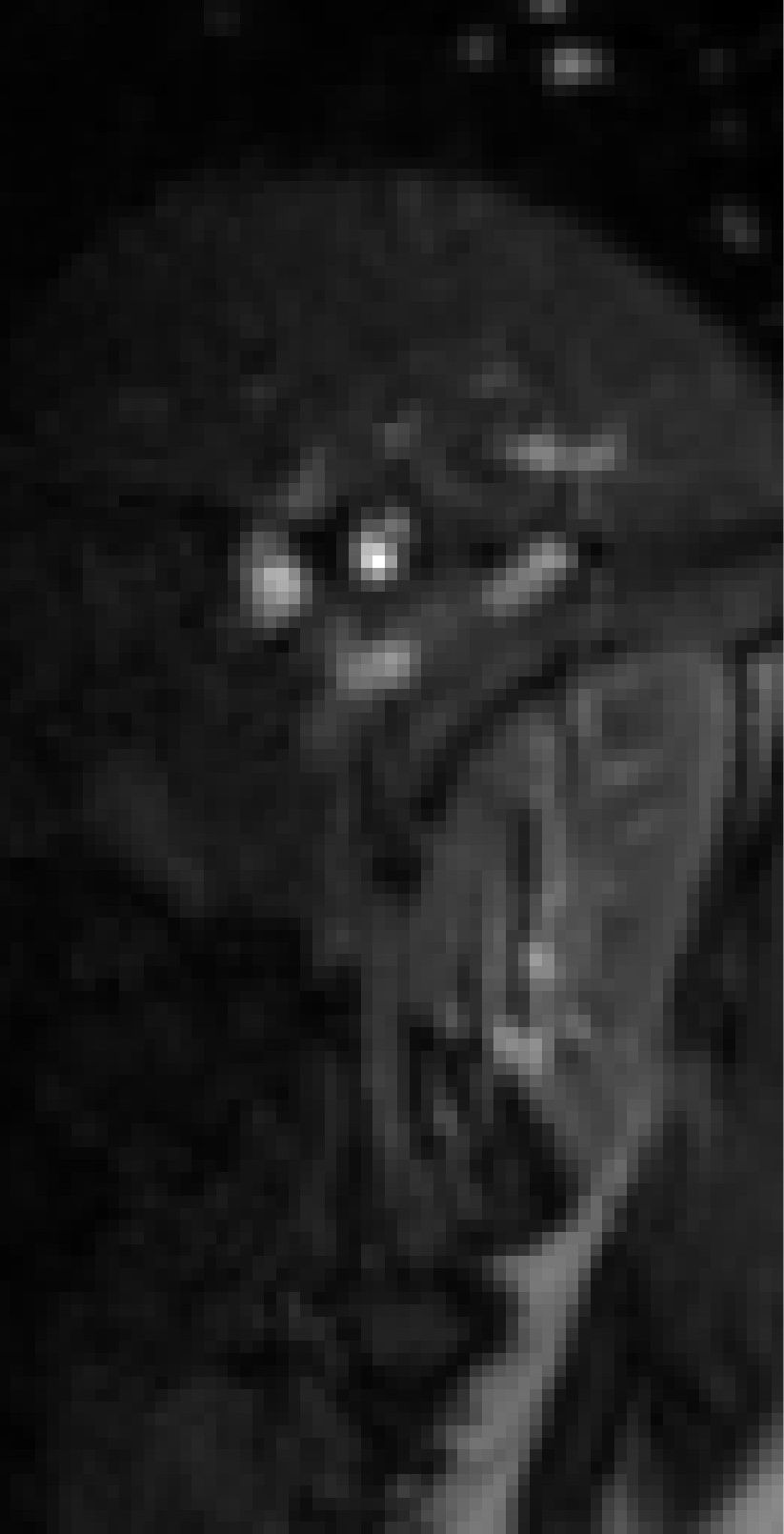} }%      
    \subfloat[Sequence 4\\* \text{\small $t_{331}$} (55.0s)\\* subj-specific \\* transformer]{\includegraphics[width=\myfigwidth, height=1.956\myfigwidth]{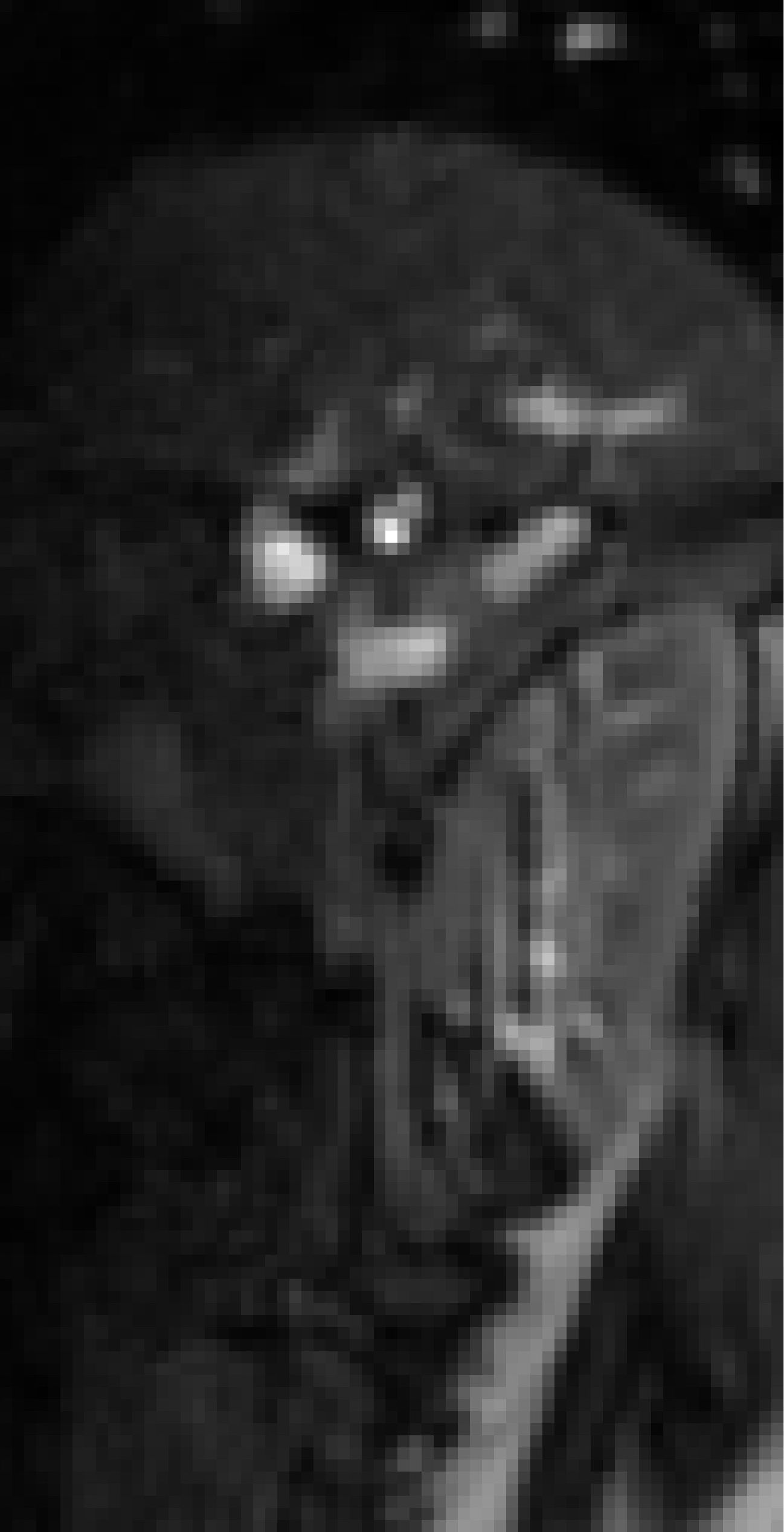} }%    
    \hfill
    \subfloat[Sequence 5 \\* \text{\small $t=t_{1}$} \\* reference]{\includegraphics[width=\myfigwidth, height=1.654\myfigwidth]{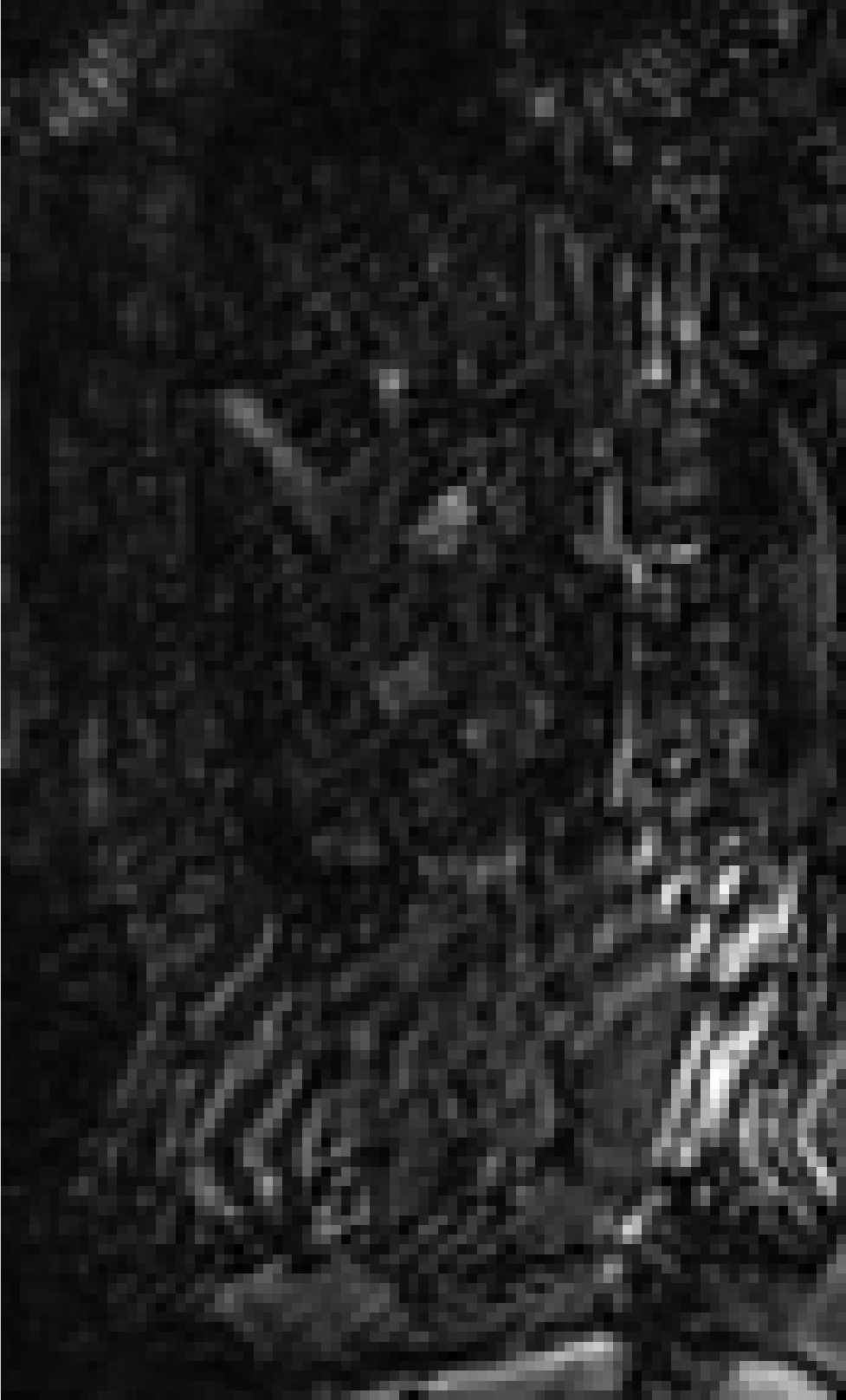} }%     
    \subfloat[Sequence 5\\* \text{\small $t_{423}$} (70.3s)\\* ground truth]{\includegraphics[width=\myfigwidth, height=1.654\myfigwidth]{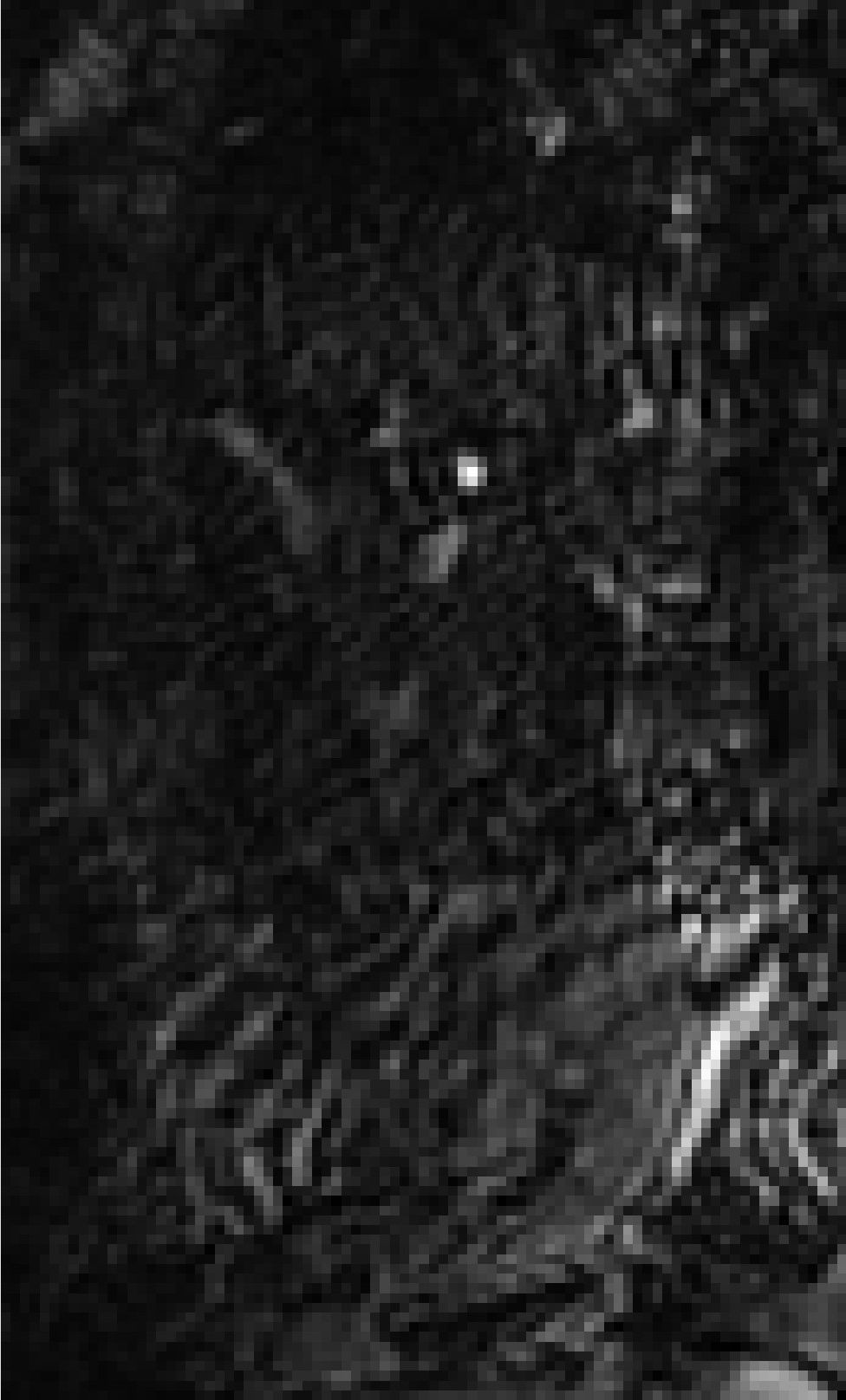} }% 
    \subfloat[Sequence 5\\* \text{\small $t_{423}$} (70.3s)\\* UORO]{\includegraphics[width=\myfigwidth, height=1.654\myfigwidth]{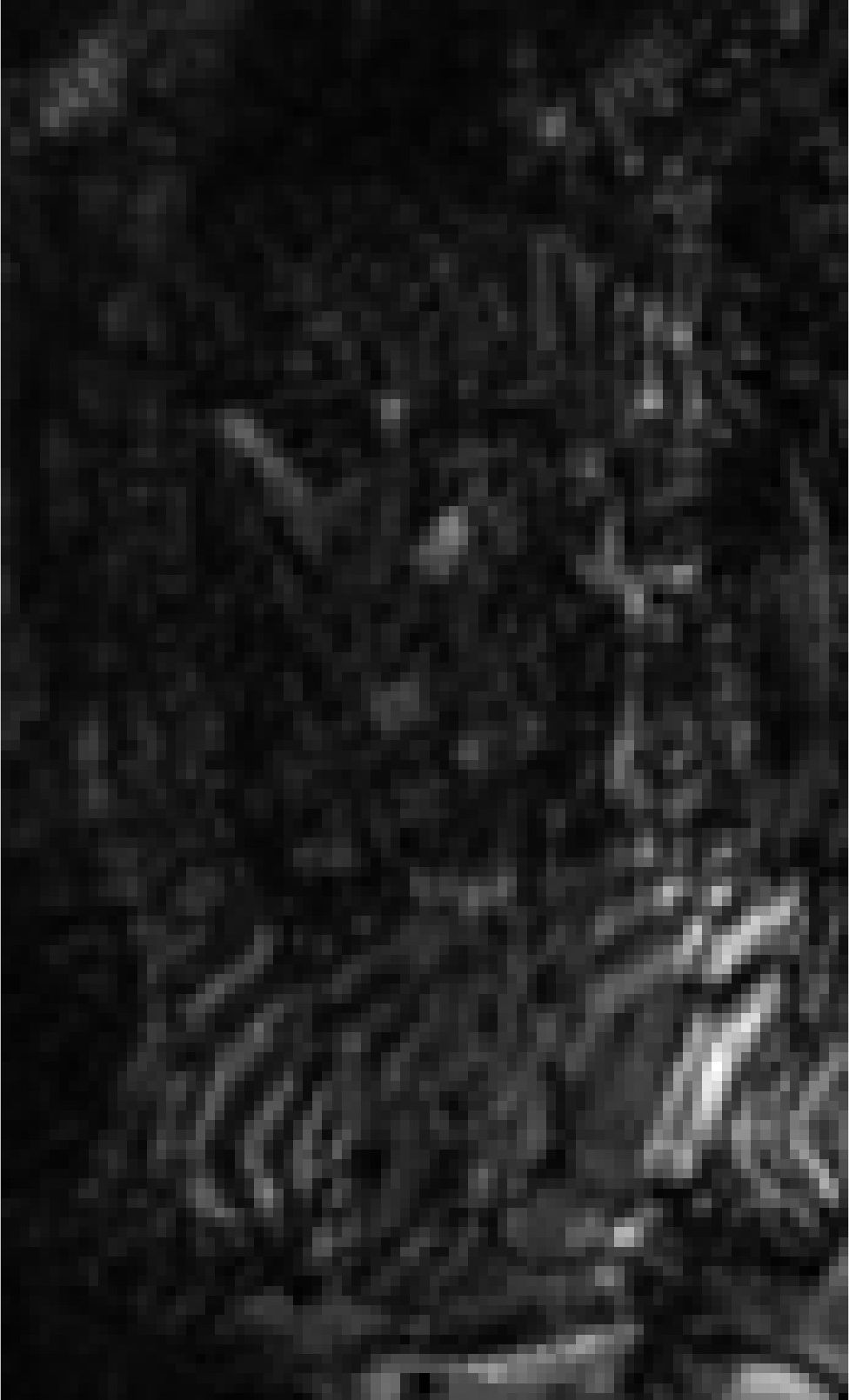} }% 
    \subfloat[Sequence 5\\* \text{\small $t_{423}$} (70.3s)\\* SnAp-1]{\includegraphics[width=\myfigwidth, height=1.654\myfigwidth]{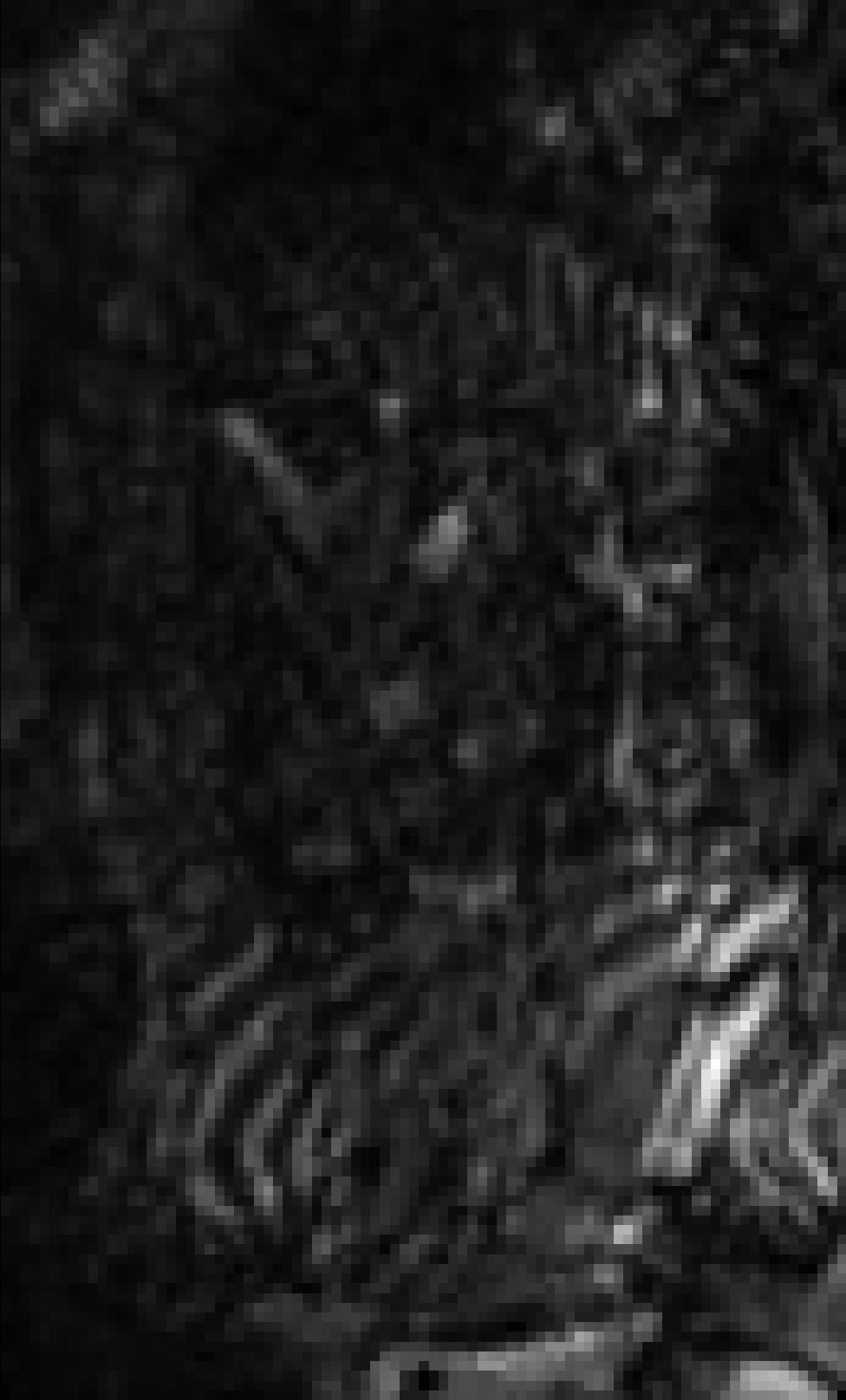} }%   
    \subfloat[Sequence 5\\* \text{\small $t_{423}$} (70.3s)\\* DNI]{\includegraphics[width=\myfigwidth, height=1.654\myfigwidth]{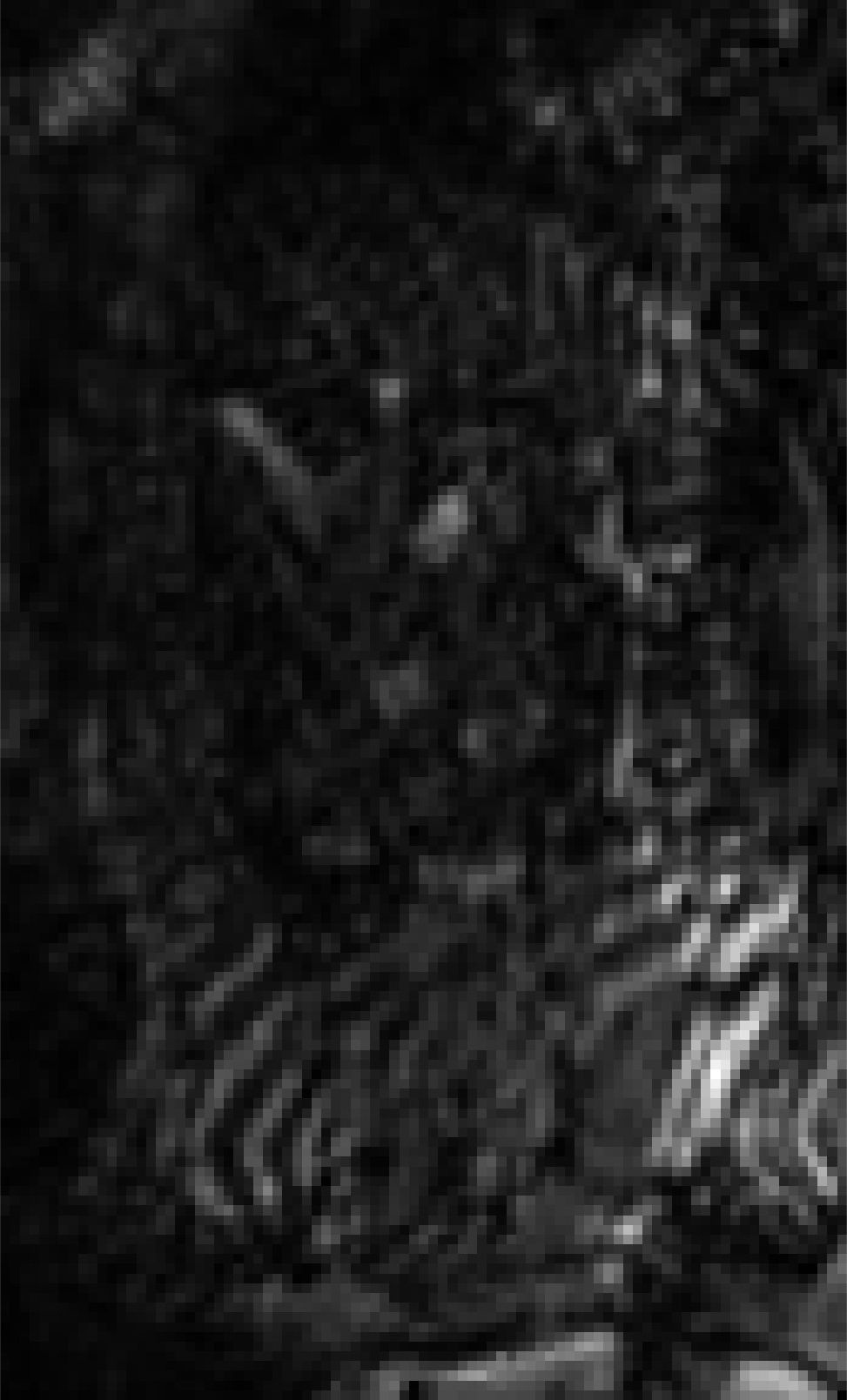} }%      
    \subfloat[Sequence 5\\* \text{\small $t_{423}$} (70.3s)\\* subj-specific \\* transformer]{\includegraphics[width=\myfigwidth, height=1.654\myfigwidth]{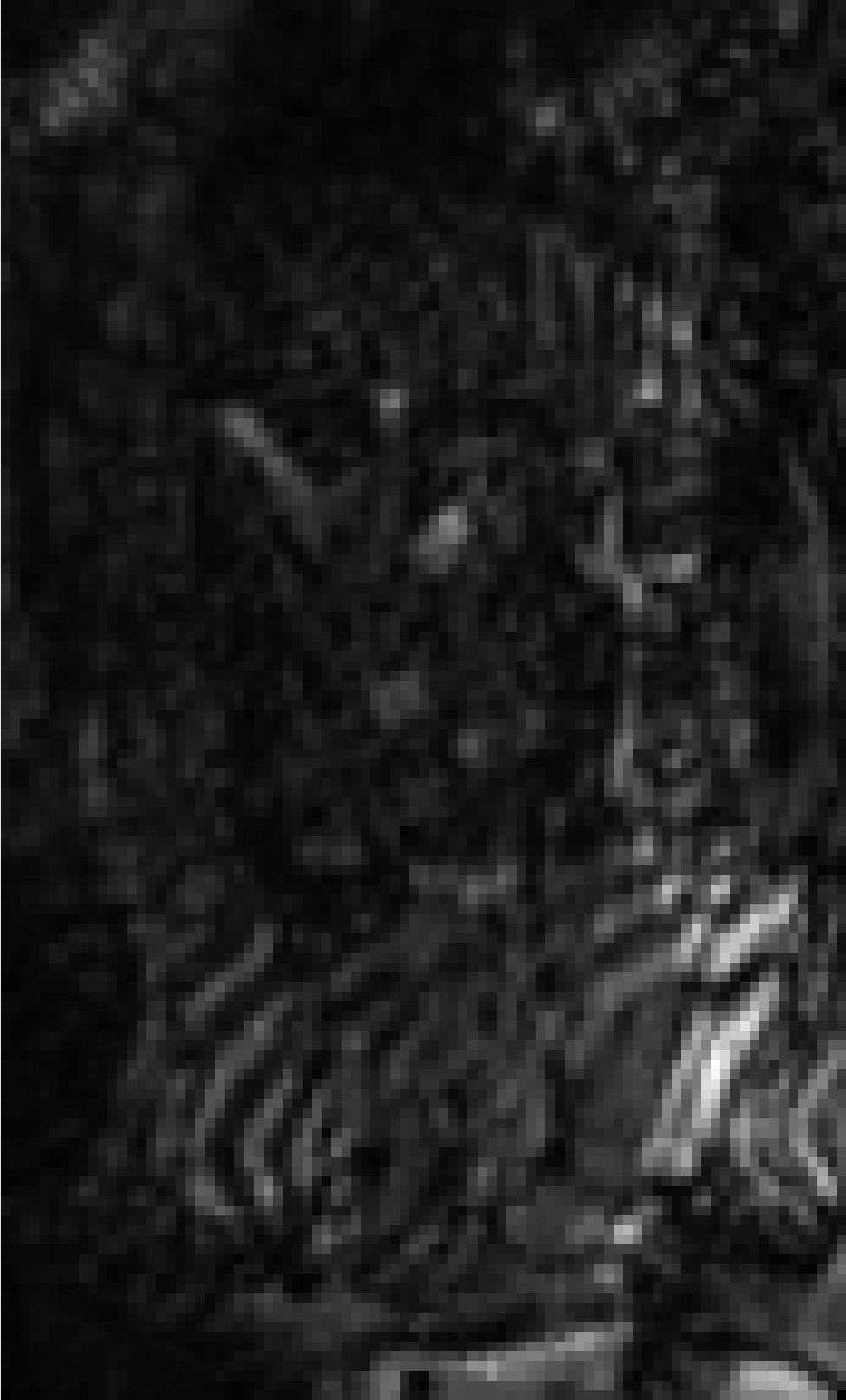} }%     
    \subfloat[Sequence 5\\* \text{\small $t_{415}$} (69.0s)\\* ground truth]{\includegraphics[width=\myfigwidth, height=1.654\myfigwidth]{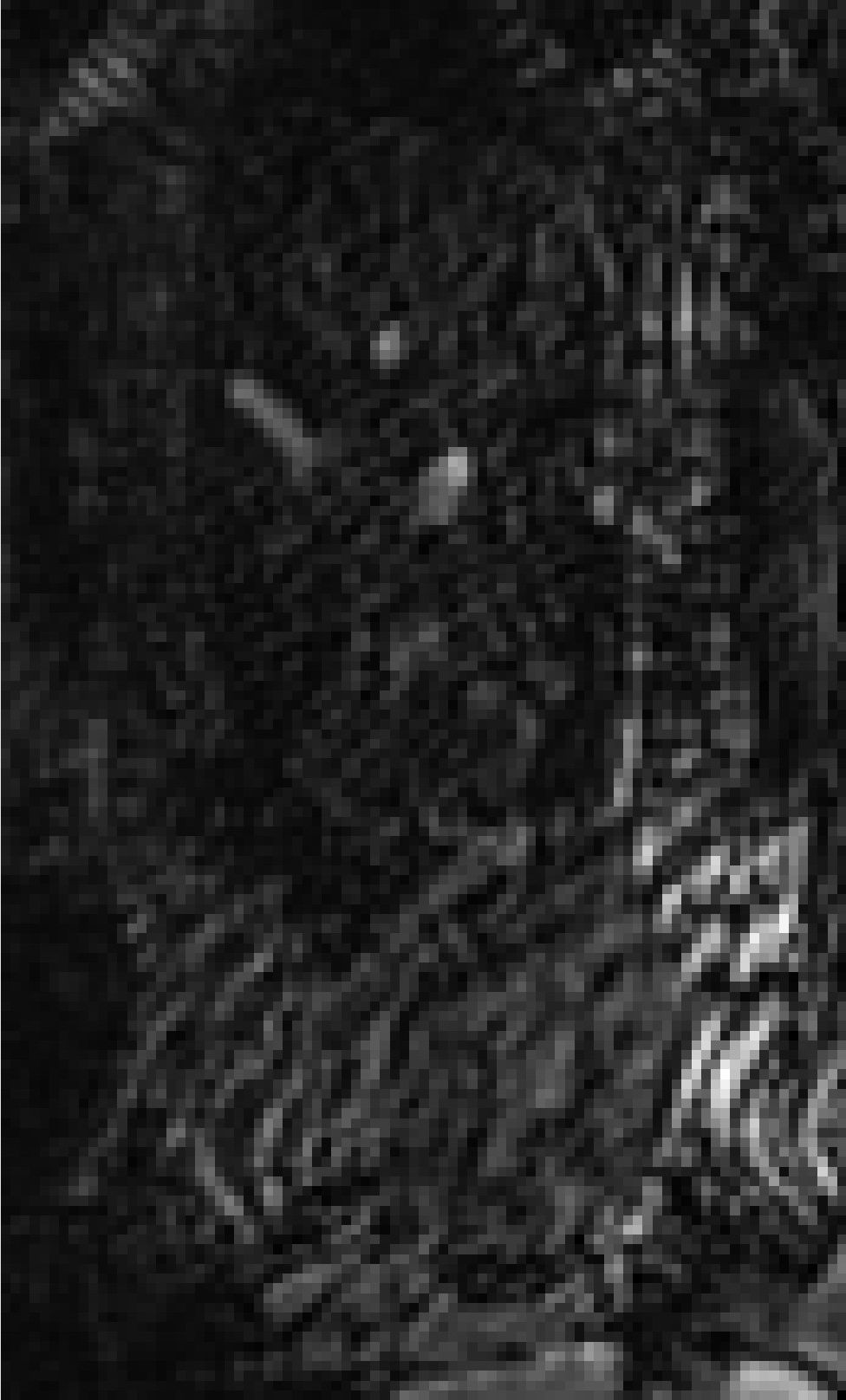} }% 
    \subfloat[Sequence 5\\* \text{\small $t_{415}$} (69.0s)\\* UORO]{\includegraphics[width=\myfigwidth, height=1.654\myfigwidth]{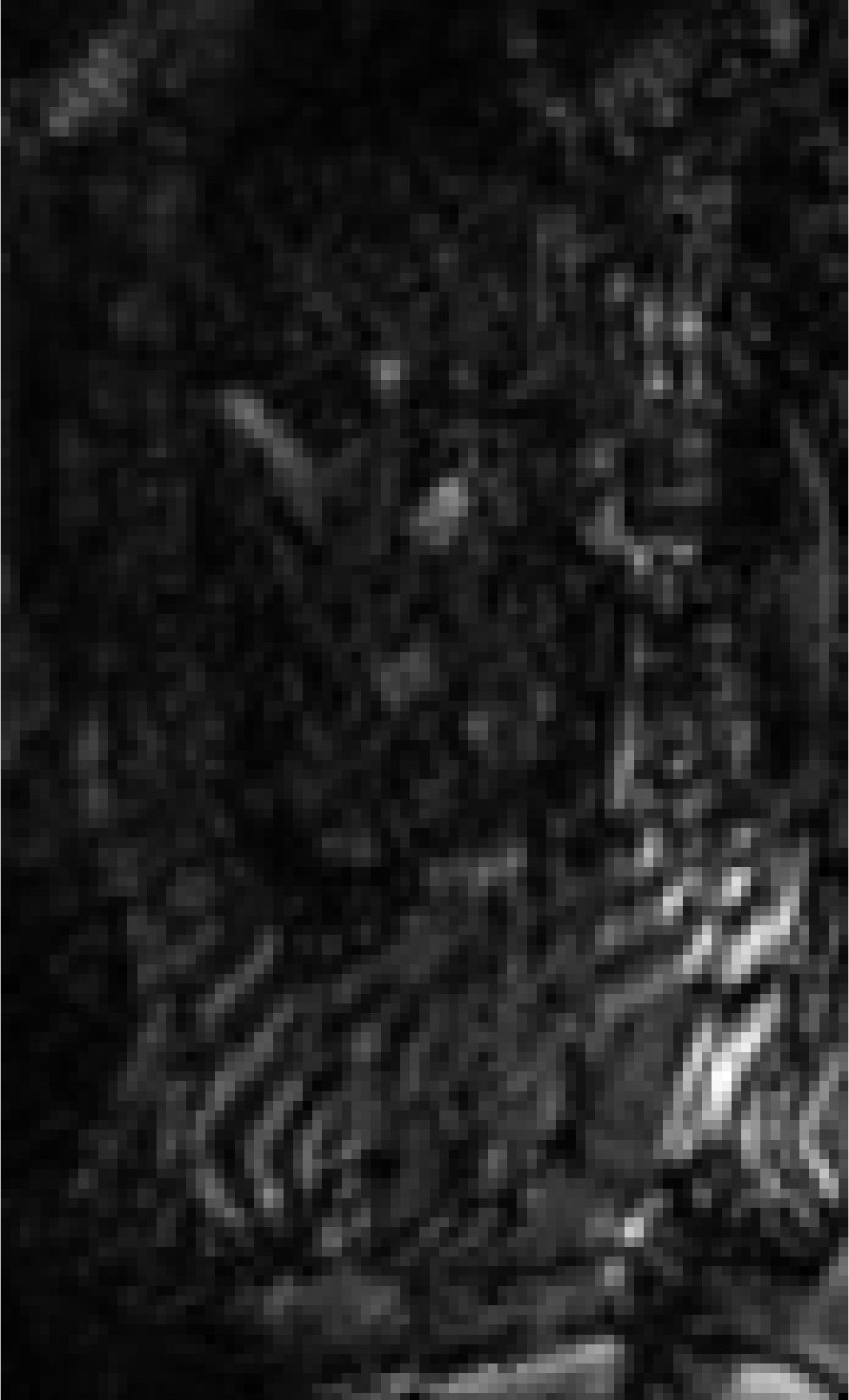} }%  
    \subfloat[Sequence 5\\* \text{\small $t_{415}$} (69.0s)\\* SnAp-1]{\includegraphics[width=\myfigwidth, height=1.654\myfigwidth]{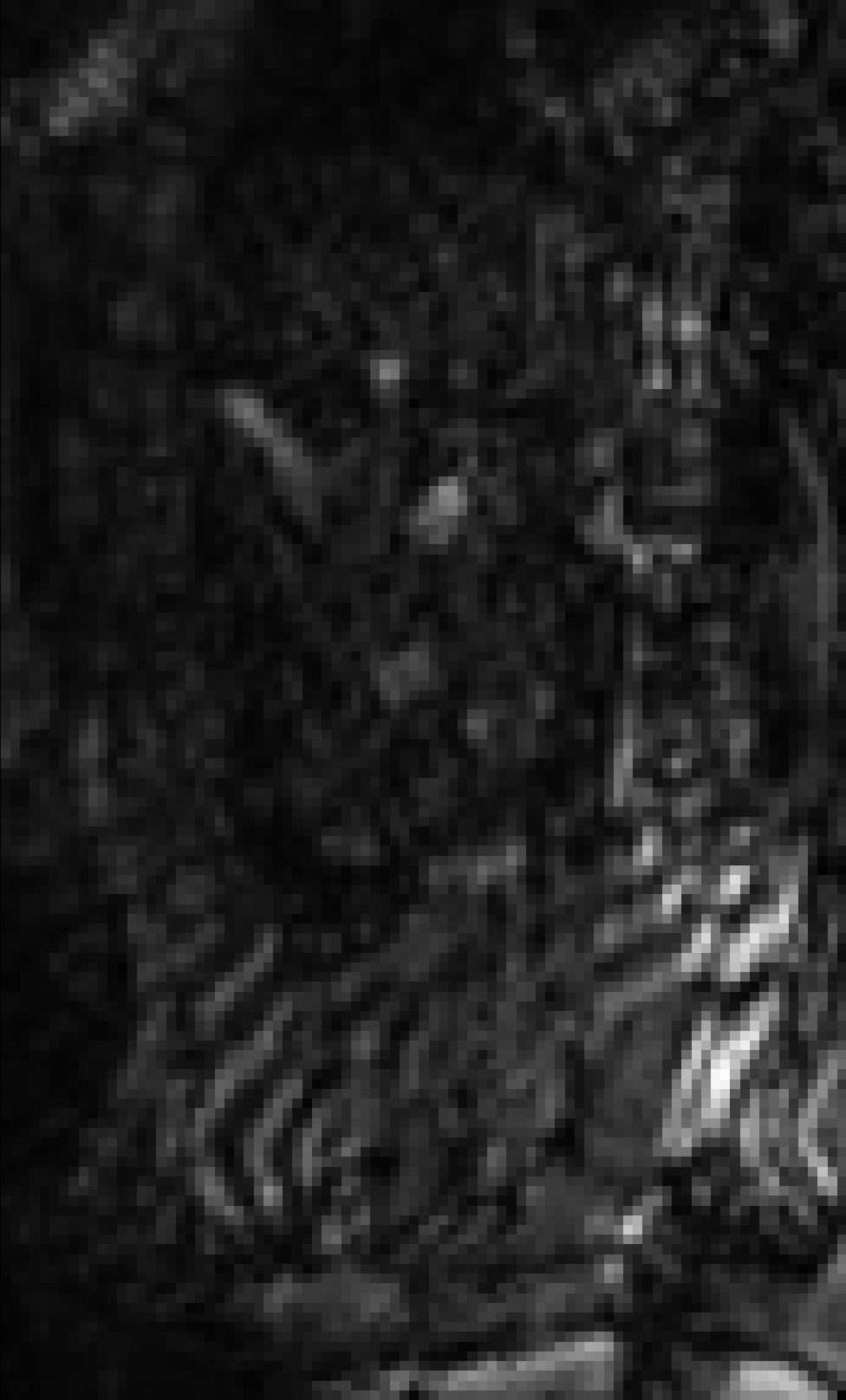} }%   
    \subfloat[Sequence 5\\* \text{\small $t_{415}$} (69.0s)\\* DNI]{\includegraphics[width=\myfigwidth, height=1.654\myfigwidth]{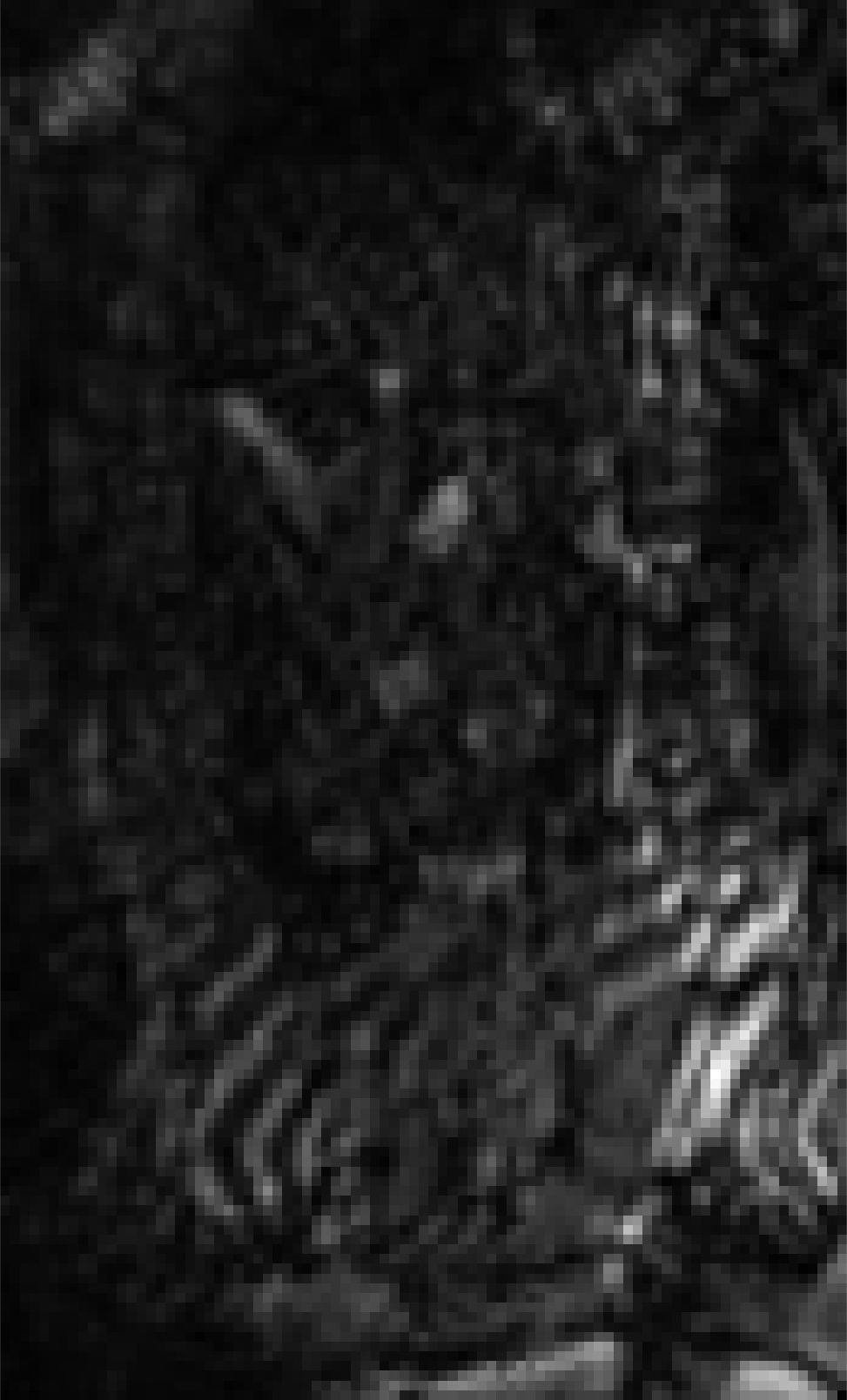} }%       
    \subfloat[Sequence 5\\* \text{\small $t_{415}$} (69.0s)\\* subj-specific \\* transformer]{\includegraphics[width=\myfigwidth, height=1.654\myfigwidth]{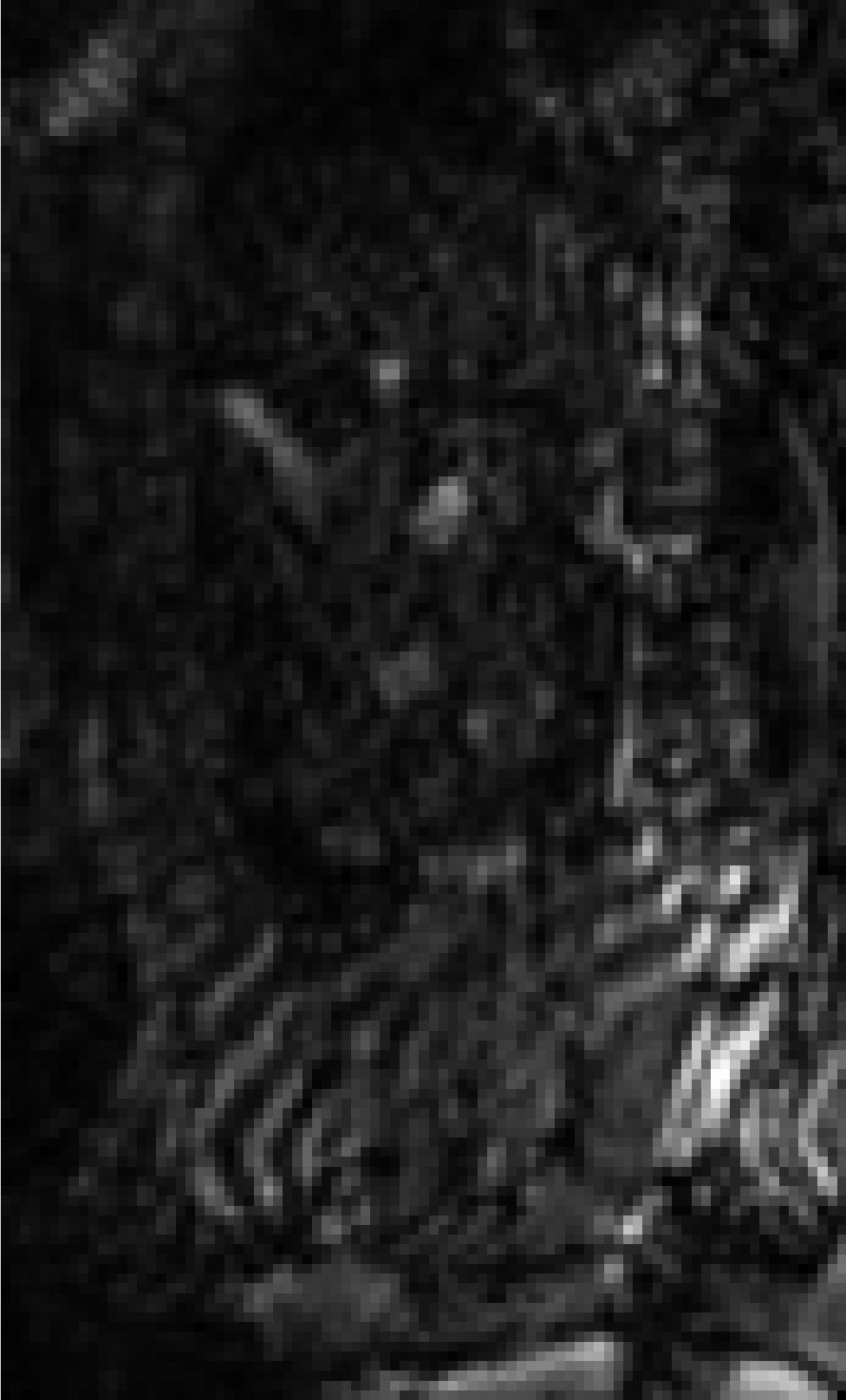} }%                  
    \hfill
    \subfloat[Sequence 6 \\* \text{\small $t=t_{1}$} \\* reference]{\includegraphics[width=\myfigwidth, height=1.942\myfigwidth]{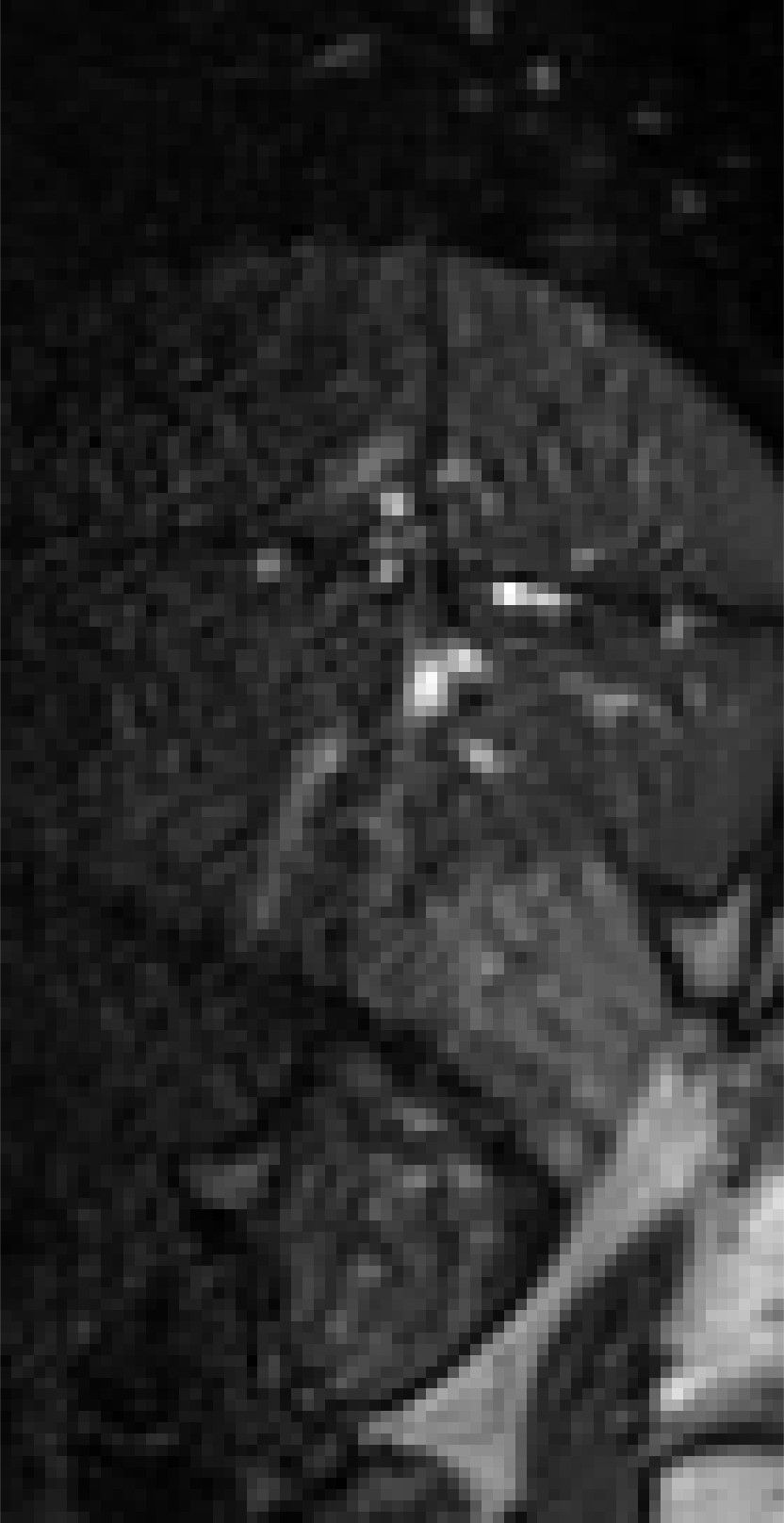} }%     
    \subfloat[Sequence 6\\* \text{\small $t_{392}$} (65.2s)\\* ground truth]{\includegraphics[width=\myfigwidth, height=1.942\myfigwidth]{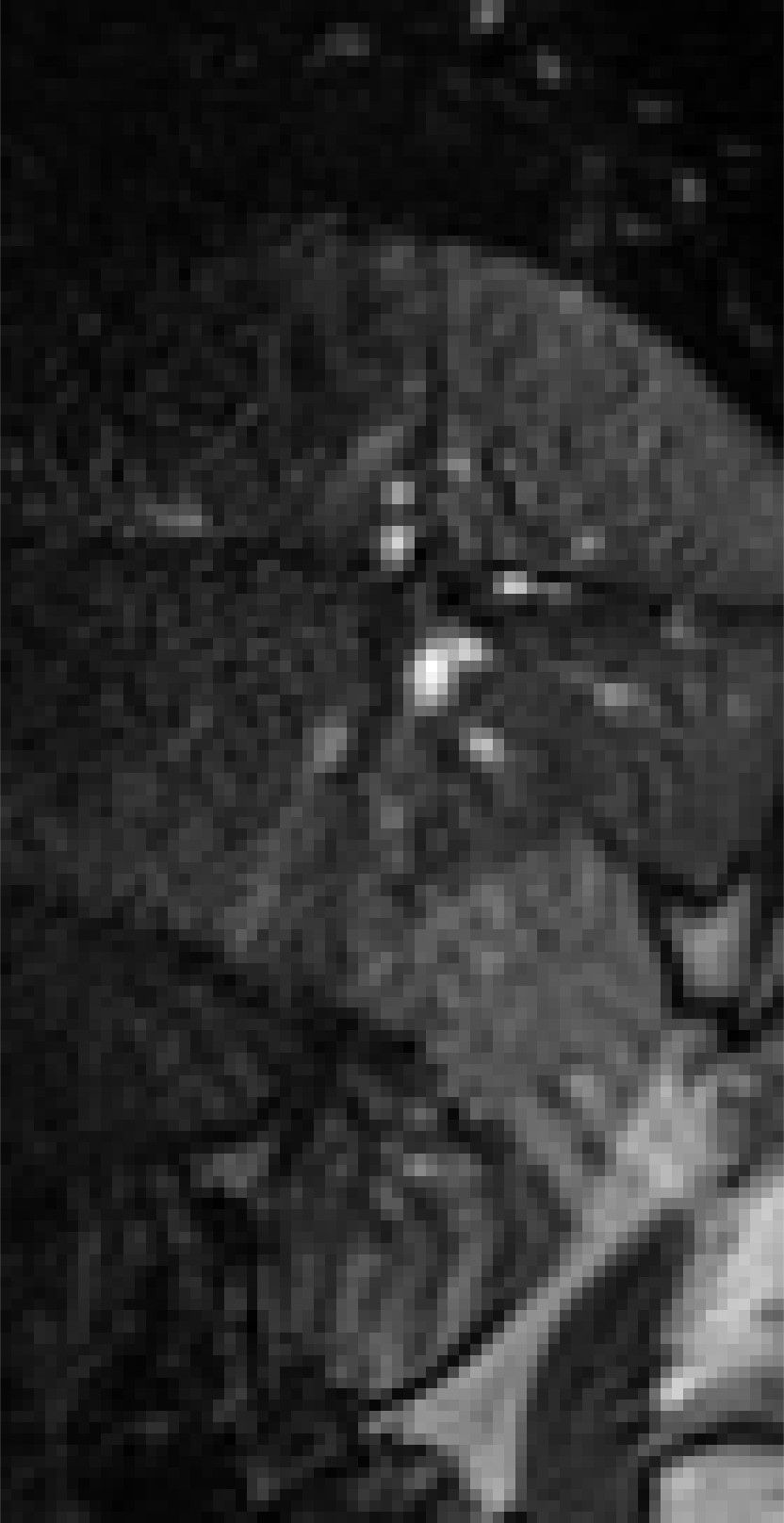} }% 
    \subfloat[Sequence 6\\* \text{\small $t_{392}$} (65.2s)\\* UORO]{\includegraphics[width=\myfigwidth, height=1.942\myfigwidth]{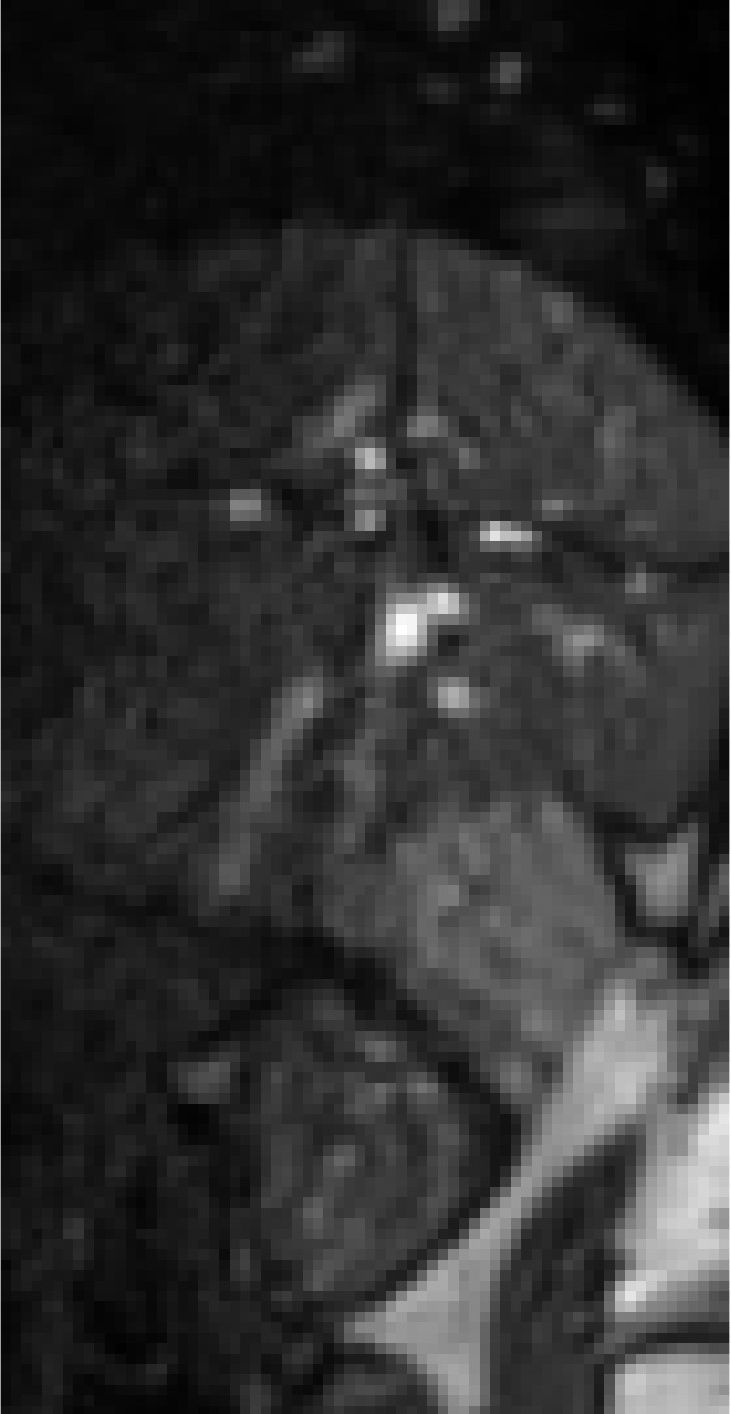} }%   
    \subfloat[Sequence 6\\* \text{\small $t_{392}$} (65.2s)\\* SnAp-1]{\includegraphics[width=\myfigwidth, height=1.942\myfigwidth]{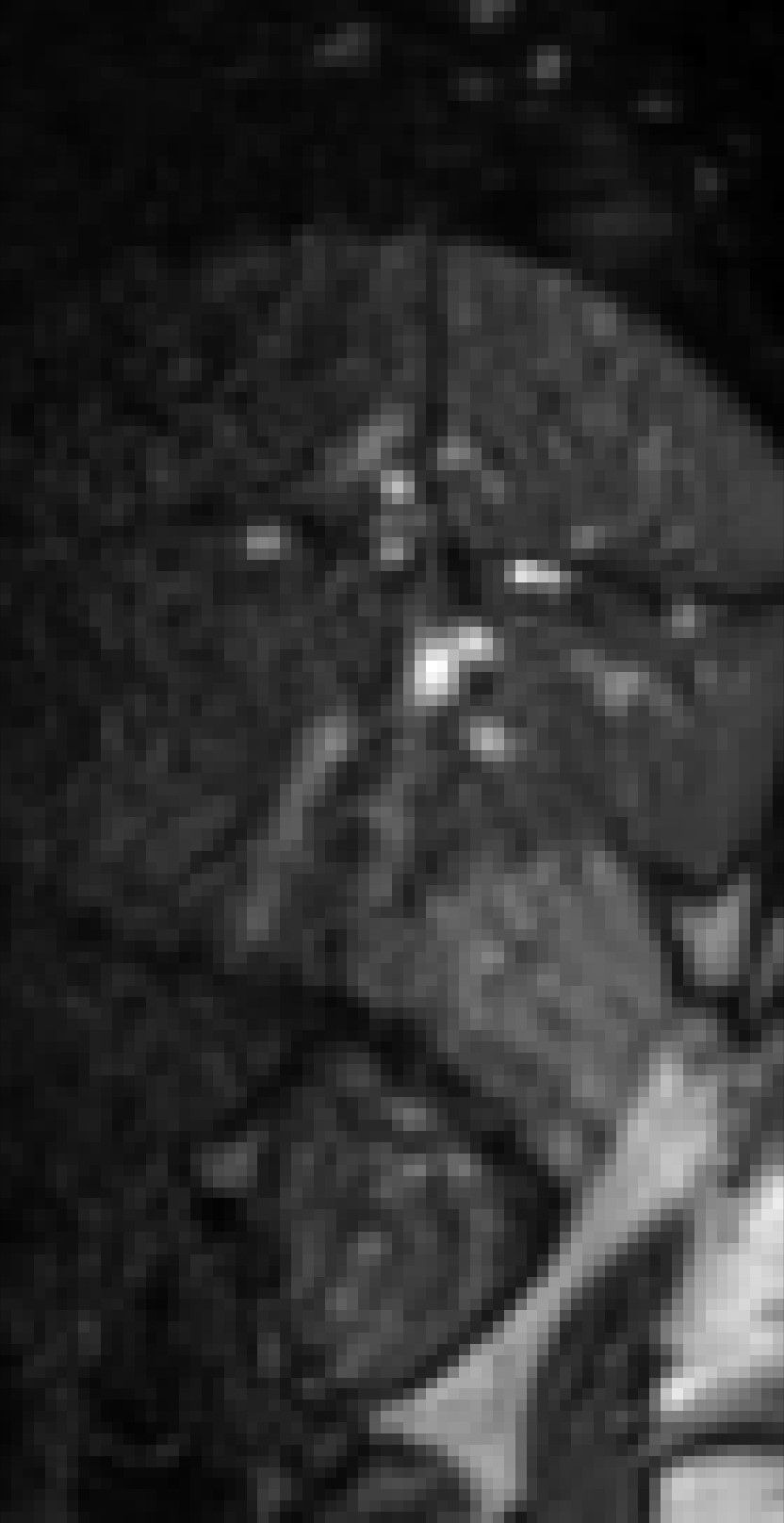} }%   
    \subfloat[Sequence 6\\* \text{\small $t_{392}$} (65.2s)\\* DNI]{\includegraphics[width=\myfigwidth, height=1.942\myfigwidth]{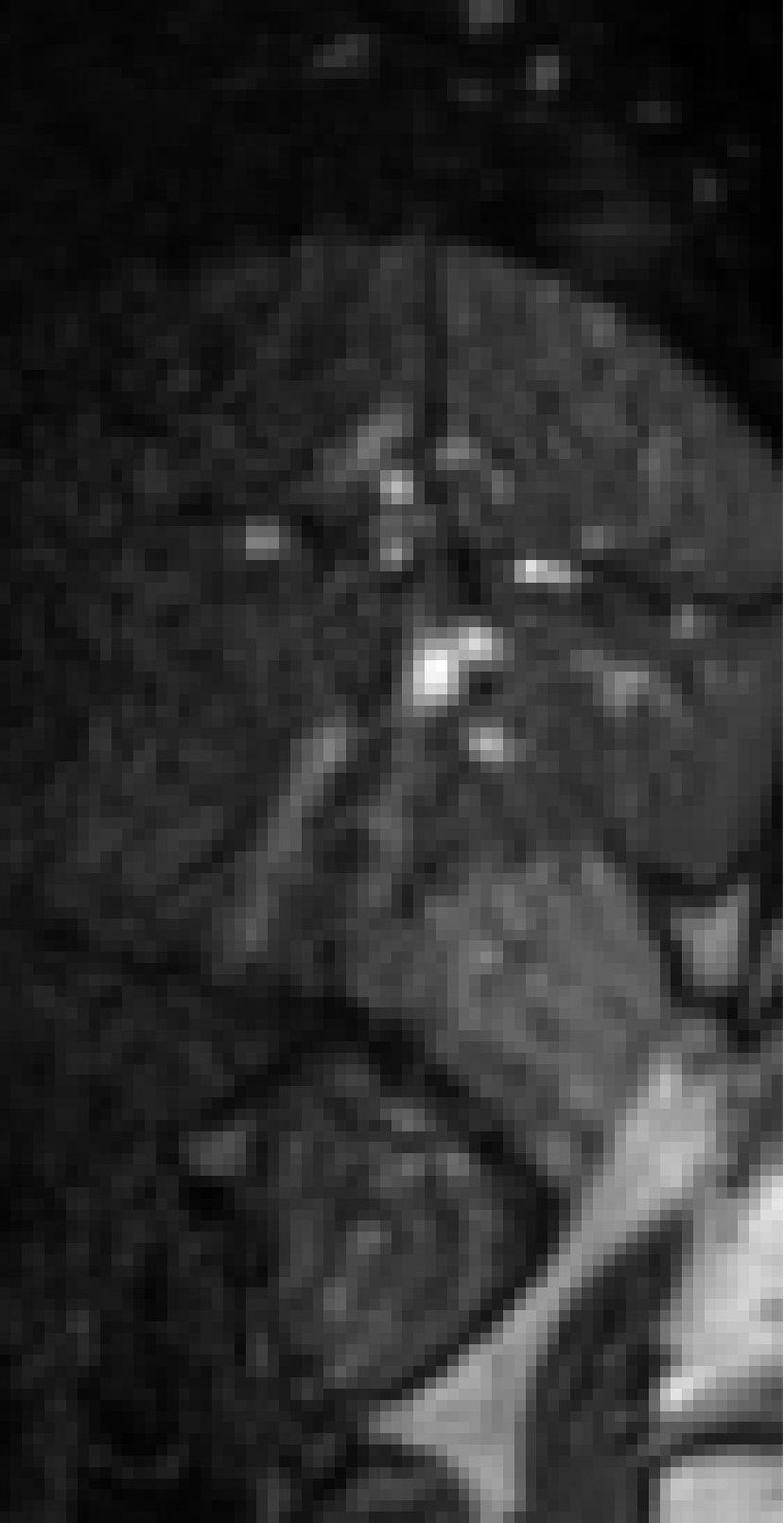} }%     
    \subfloat[Sequence 6\\* \text{\small $t_{392}$} (65.2s)\\* subj-specific \\* transformer]{\includegraphics[width=\myfigwidth, height=1.942\myfigwidth]{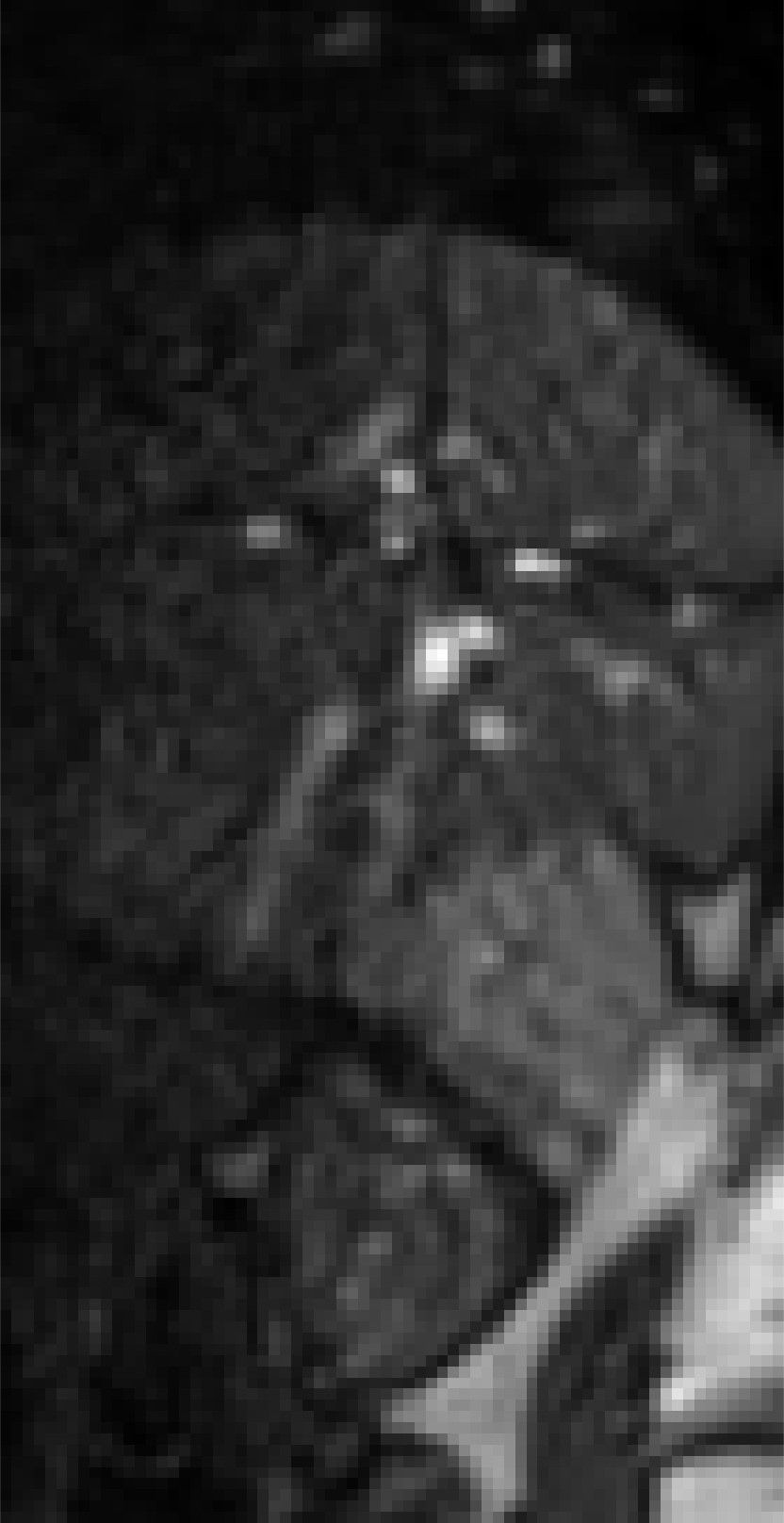} }%     
    \subfloat[Sequence 6\\* \text{\small $t_{497}$} (82.7s)\\* ground truth]{\includegraphics[width=\myfigwidth, height=1.942\myfigwidth]{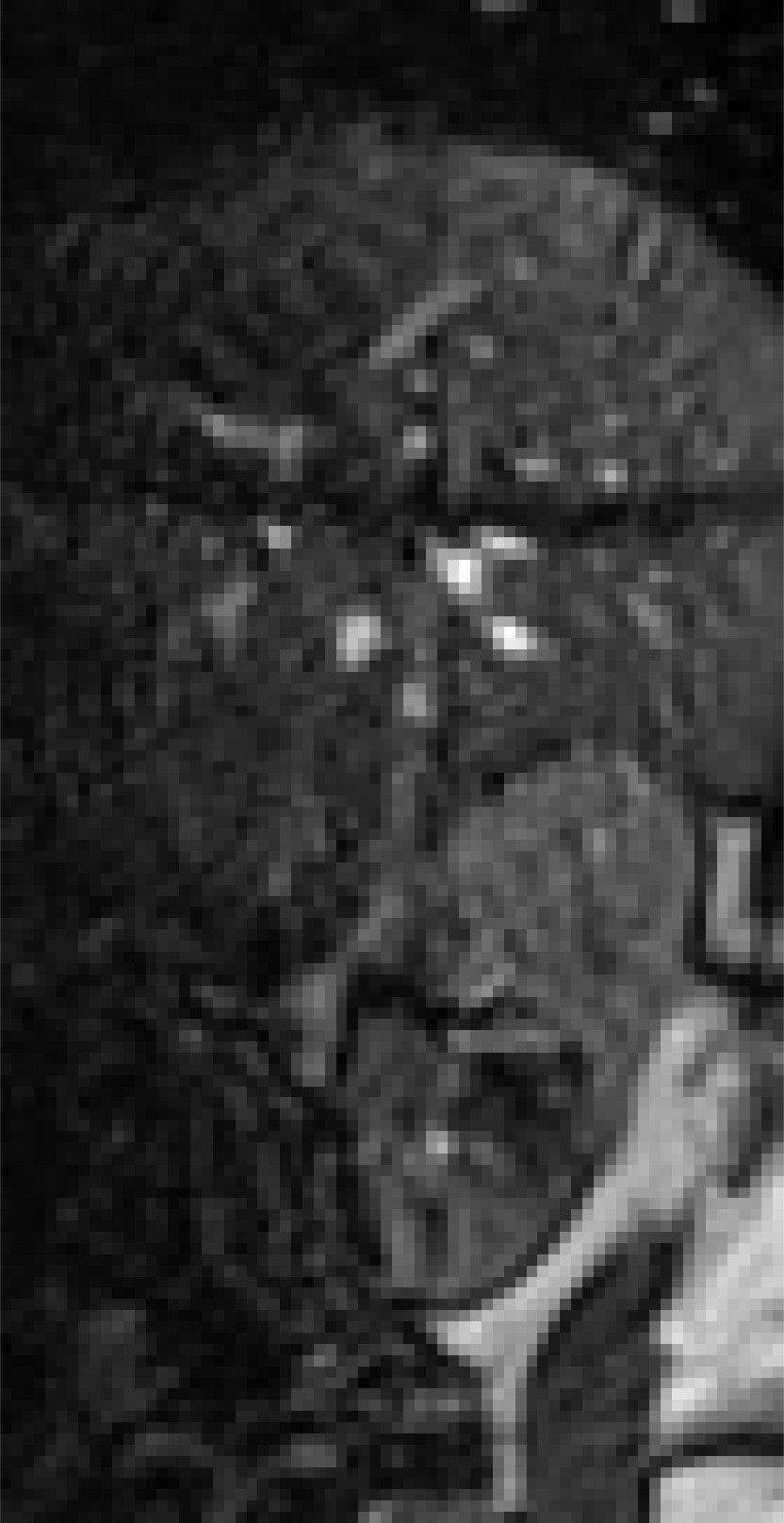} }% 
    \subfloat[Sequence 6\\* \text{\small $t_{497}$} (82.7s)\\* UORO]{\includegraphics[width=\myfigwidth, height=1.942\myfigwidth]{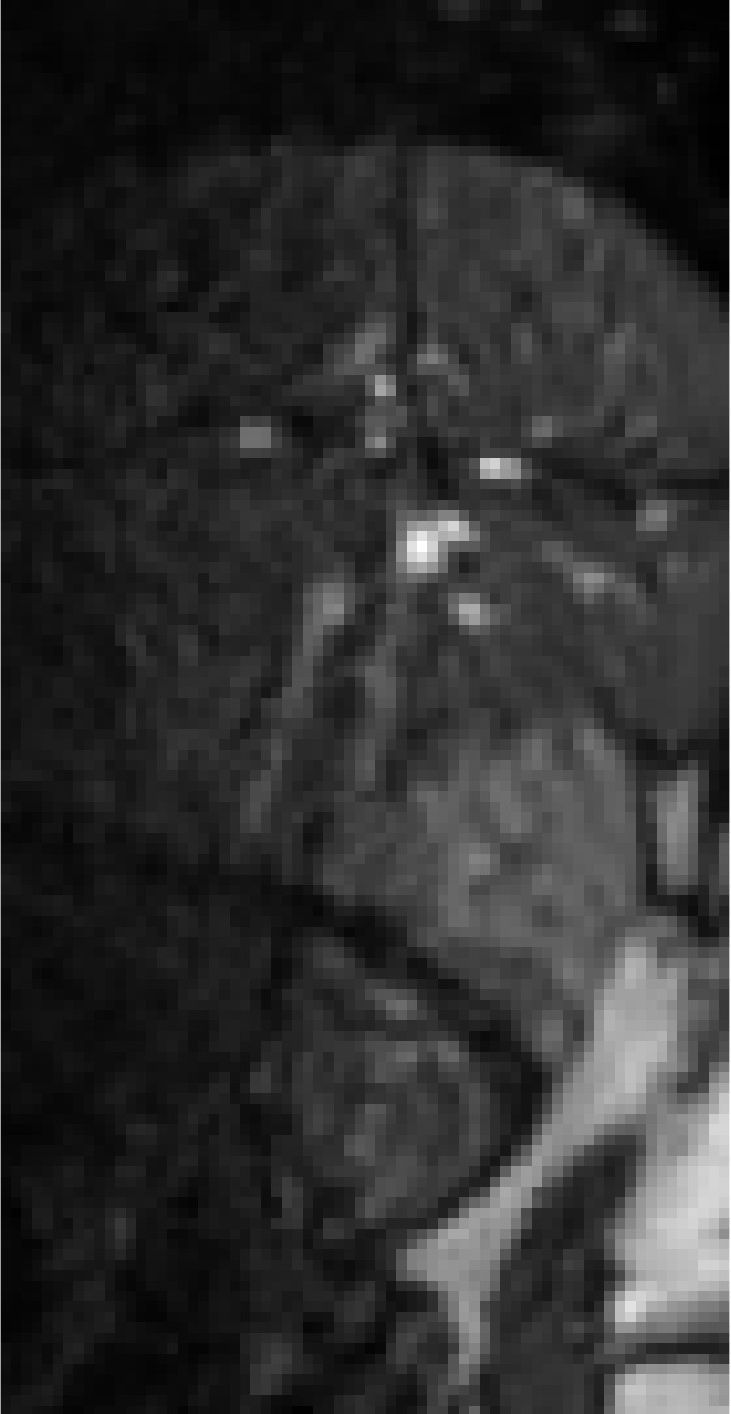} }%   
    \subfloat[Sequence 6\\* \text{\small $t_{497}$} (82.7s)\\* SnAp-1]{\includegraphics[width=\myfigwidth, height=1.942\myfigwidth]{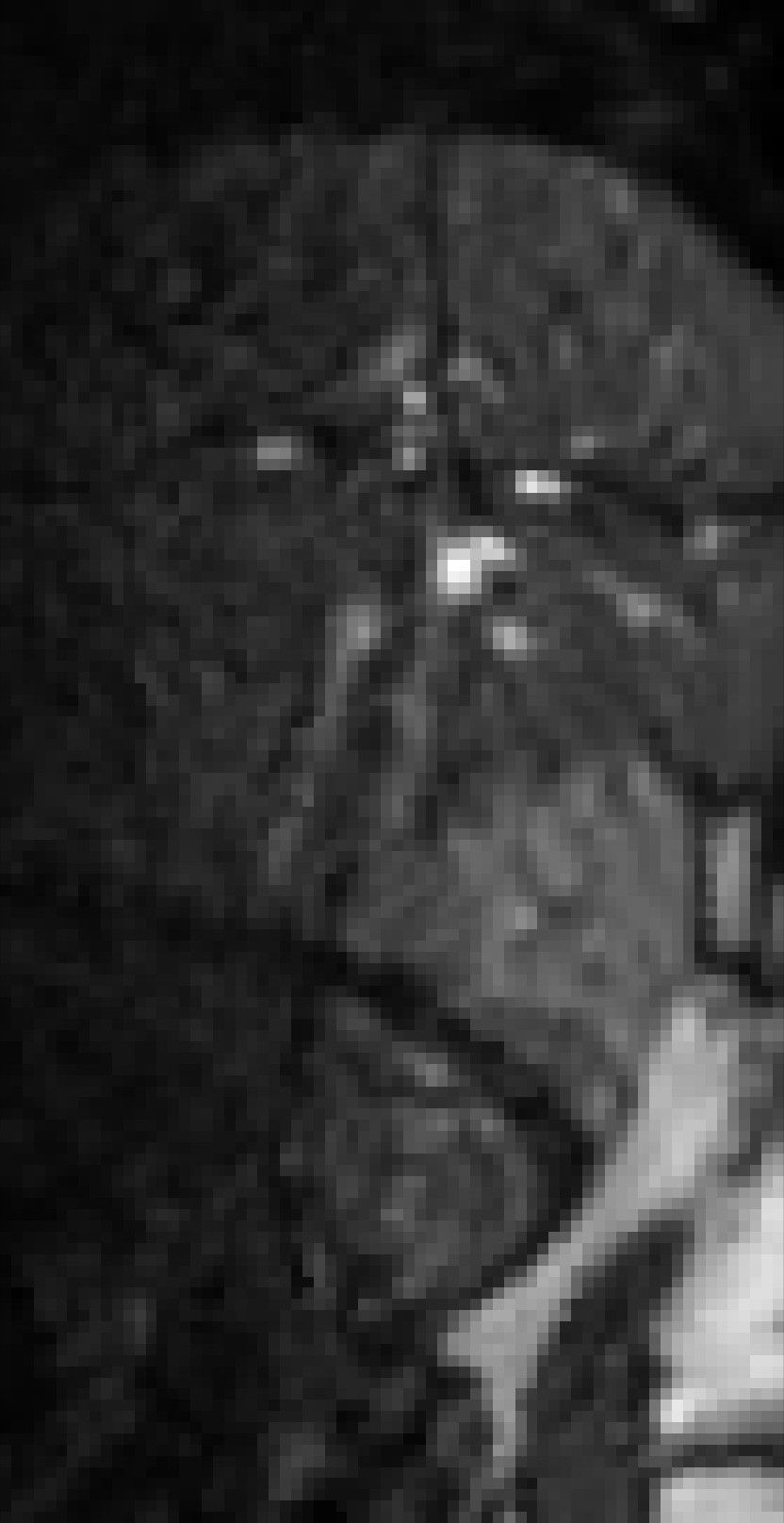} }%   
    \subfloat[Sequence 6\\* \text{\small $t_{497}$} (82.7s)\\* DNI]{\includegraphics[width=\myfigwidth, height=1.942\myfigwidth]{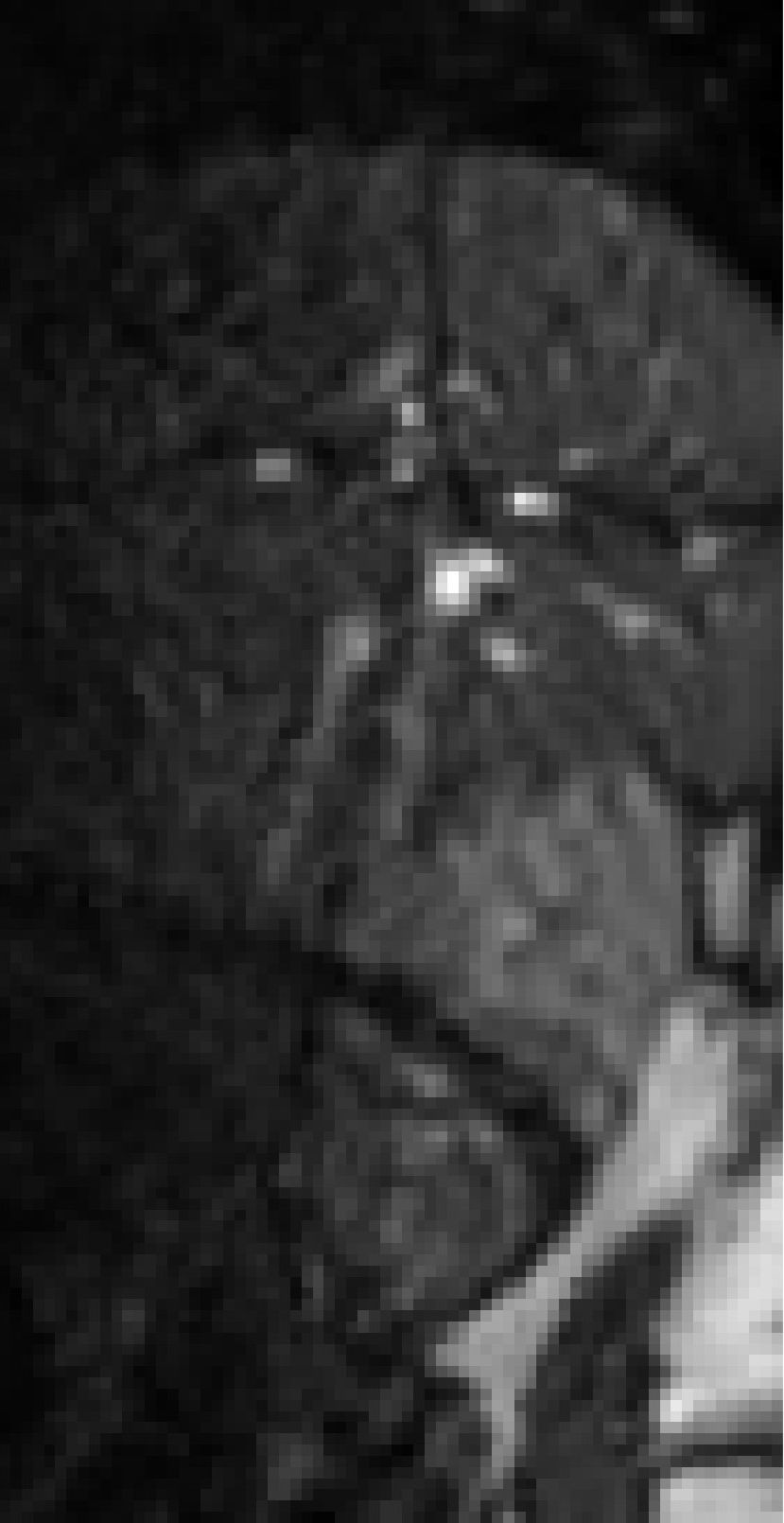} }%       
    \subfloat[Sequence 6\\* \text{\small $t_{497}$} (82.7s)\\* subj-specific \\* transformer]{\includegraphics[width=\myfigwidth, height=1.942\myfigwidth]{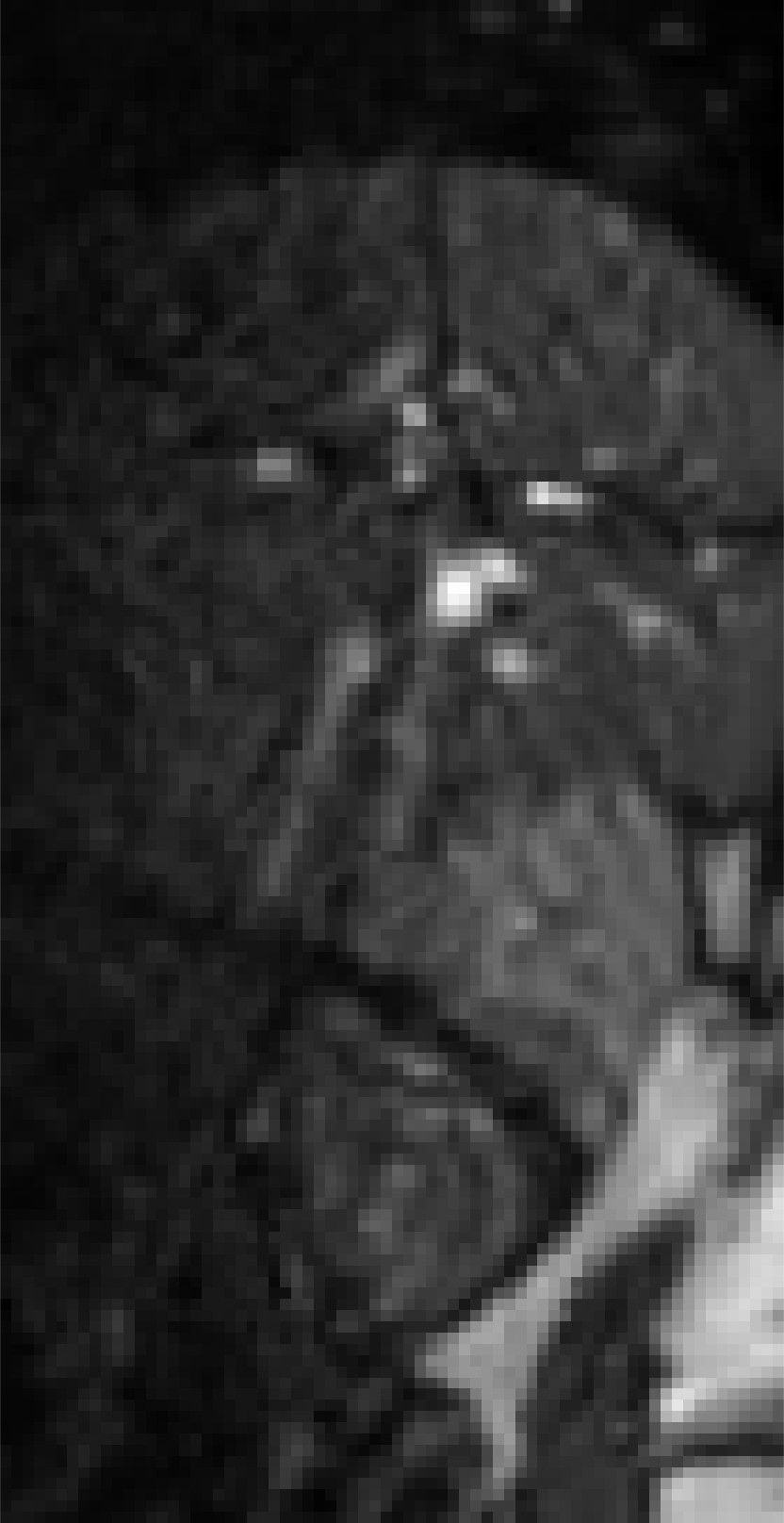} }%                      
    \caption{Ground-truth and predicted \acs{ROI} images obtained using \pgls{RNN} trained with \gls{UORO}, \gls{SnAp-1}, and \gls{DNI}, and a subject-specific transformer, at an \gls{EI} phase (columns 2--6) and an \gls{EE} phase (columns 7--11), along with the reference frame used for warping (column 1), for five representative sequences in the \acs{OvGU} dataset. The horizon is set to $h=1.66\text{s}$, and the hyperparameters, including $n_{\text{cp}}$, are those optimal for this value of $h$ and each sequence individually. The displayed cross-sections from sequence 4 do not belong to the test set (which corresponds to $t \geq t_{341}$) but illustrate forecasting under high motion variability around a deep-inspiration time point. The prediction runs for sequences 1, 4, and 6 are the same as those represented in Fig. \ref{fig:1st cpt prediction (SnAp-1 vs vs UORO vs transformer) Magdeburg}.}
    \label{fig:visual comparison of ROI prediction with SnAp-1 and transformer at h=10 on OvGU dataset}
\end{figure*}
% add a vertical line?

%\subsubsection{Time performance}
% There could be a discussion here if needed

\section{Discussion}

% Important: I may want to add somewhere: "Although mean DVF errors are reported to 2 decimals for consistency and comparability, differences below ~0.1 mm should not be over-interpreted, given the underlying image resolution and registration uncertainty.”" 

\subsection{Comparison with prior results on spatiotemporal respiratory motion forecasting}\label{section: comparison with previous works MRI}
%: do I need "on spatiotemporal respiratory motion forecasting" in the title? That seems very long...

\begin{table*}[pos=htbp, align=\centering]
\scriptsize% \footnotesize
\setlength{\tabcolsep}{0.9pt} % minimum that does not cross the margin
\begin{center}
\begin{tabular}{lllllllll}
\hline
Work                           & Prediction                          & Video prediction & Training     & Imaging                                         & Sampling  & Amount             & Response  & Prediction                   \\
\vspace{0.1cm}                 & method              & model type \cite{oprea2020review}& scheme       & data                                            & rate      & of data                 & time $h$  & performance                  \\
\hline
\cite{chhatkuli2015dynamic, ritu2016development}&\acs{PCA} applied to& Direct pixel     & Subject      & 1) \acs{kV} fluoroscopy (physical                         & 1) 10Hz   & 1) 1 seq. (repeated  & 1) 0.10s  & 1) correlation: $r=0.998$, \rule{0pt}{2.6ex}\\
                               & pixel intensities;                  & synthesis        & -specific    & phantom); 2) coronal                                       &           & 24-frame cycle)      &           & \acs{SSIM}: 0.971            \\
                               & prediction with                     &                  &              & \acs{4DCT} cross-sections                       & 2) 2Hz    & 2) 1 seq. (repeated  & 2) 0.50s  & 2) $r=0.999$, \acs{SSIM}: 0.995\\
\vspace{0.1cm}                 & \acs{MSSA}                          &                  &              & from a lung cancer patient                      &           & 28-frame cycle)      &           &              \\
\cite{pham2019predicting}      & \acs{PCA} applied to                & Prediction space & Subject      & 1) \acs{4D} \acs{XCAT} phantom data                 & 1) 8Hz    & 1) 5 sequences       & 1) 0.13s, & 1) \acs{3D} (simulated) tumor       \\
                               & \glspl{DVF}; prediction             & factorization    & -specific    & 2) \acs{4D} cine \acs{MRI} (plus on-              &           & (2 min each)         & 0.63s     & \acs{TE}: 1.55mm, 2.60mm            \\
                               & with adaptive-                      & with explicit    &              & board sagittal \acs{MRI}) of                        & 2) 3Hz    & 2) 1 sequence        & 2) 0.33s  & 2) Tumor \acs{TE}: 1.58mm (\acs{SI}),\\
\vspace{0.1cm}                 & boosted \acsp{MLP}                  & transformations  &              & a liver cancer patient              &           & of 1 min             &           & 1.90mm (\acs{AP})                \\
\cite{nabavi2020respiratory}   & \acs{PredNet} \citep{lotter2017deep}& Direct pixel     & Population   & x/y/z cross-sections                            & Not     & One 10-phase \acs{4DCT}& Not       & \acs{SSIM}: 0.935--0.951   \\
                               &                                     & synthesis        &              & from \acs{4DCT} data of lung                         & reported  & sequence for each    & reported  &                  \\
\vspace{0.1cm}                 &                                     &                  &              & cancer patients \cite{vandemeulebroucke2007popi}&           & of the 6 patients    &           &                              \\
\cite{romaguera2020prediction} & Voxelmorph-like                     & Prediction space & Population   & 1) sagittal liver cine \acs{MRI}                & 1) 3.13Hz & 1) 12 sequences      & 1) 0.32s  & 1) local \acs{NCC}: 0.95--0.97,      \\
                               & recurrent \acs{AE}                  & factorization    &              & from healthy subjects                           &           & (50 frames each)        & to 1.60s  & \acs{TE}: 0.45--0.77mm     \\
                               & and spatial                         & with explicit    &              & 2) sagittal cross-sections                      & 2) 2.5Hz  & 2) 10 sequences      & 2) 0.40s  & 2) \acs{TE}: 0.28--0.42mm    \\
\vspace{0.1cm}                 & transformer                         & transformations  &              & from chest \acs{4DCT} data \cite{castillo2009four}&           & (10 frames each)         & to 2.00s  &                    \\
\cite{romaguera2023conditional}& Conditional-based                   & Prediction space & Population   & Right-hemidiaphragm                        & 2.22 Hz   & 80 reconstructed     & 0.45s     & \acs{2D} frame-forecasting \acs{SSIM}:\\
                               & transformer                         & factorization    &              & \acs{4D}-\acs{MRI} (sagittal slices                              &           & \acs{4D}-\acs{MRI} sequences &   & 0.78 $\pm$ 0.11,              \\
                               & with learnable                      & with explicit    &              & covering the liver and                           &           & of 31 frames for     &           & \acs{TE} in generated \acs{3D} frames:\\
\vspace{0.1cm}                 & queries                             & transformations  &              & part of the lungs)                           &           & each of 25 subjects  &           &  1.56 $\pm$ 1.13mm  \\
\hline
This                            & \acs{PCA} applied to                & Prediction space & Subject      & 1) ETH Zürich data:                             & 1) 3.18Hz & 1) 4 sequences       & 1) 0.31s  & \textit{\acs{SnAp-1} (\acs{OvGU}, full-frame)} \rule{0pt}{2.6ex}\\
work                           & Lucas--Kanade                       & factorization    & -specific    & sagittal chest slices                           &           & (2 per subject,         & to 2.20s  & Correlation $r$:       \\
                               & optical-flow field;                 & with explicit    & (except for  & from \acs{4D} cine \acs{MRI}                    &           & 200 frames each) &           & 1) 0.987 2) 0.942 \\
                               & prediction with                     & transformations  & population   & 2) \acs{OvGU} data: & 2) 6.0Hz  & 2) 8 sequences, each & 2) 0.17s  & \acs{SSIM}: 1) 0.899 2) 0.660     \\
                               & \glspl{RNN} trained                 &                  & transformers)& sagittal liver cine-\acs{MR}                          &           & from 1 subject and & to 2.17s  & Mean \acs{DVF} error:         \\                               
                               & online/transformers                  &                  &              & navigator acquisitions                          &           & containing 498 frames &           & 1) 1.41mm 2) 2.73mm    \\[0.05cm]                                                              
\hline  
\end{tabular}
\end{center}
\caption{Comparison with previous studies on frame forecasting in chest and liver imaging. The results for this work are reported in Table \ref{table:frame pred perf}. We selected \acs{SnAp-1} as a representative online training algorithm for \glspl{RNN} and the full-frame evaluation setting, matching how performance metrics are reported for the other studies.}
\label{table:comparison_with litterature frame pred}
\end{table*} 

%Table \ref{table:comparison_with litterature frame pred} compares our \acs{PCA}-based model with the literature regarding next-frame prediction in \gls{2D} dynamic chest imaging. 
Comparison with prior work is challenging, as each study relies on a distinct dataset (Table \ref{table:comparison_with litterature frame pred}). Specifically, the imaging modality, anatomical regions imaged, spatial resolution, brightness, contrast, sampling frequency, and noise characteristics, as well as respiratory motion amplitude, frequency, and regularity, differ across studies. Furthermore, experimental choices---such as the response time and data partitioning into training, validation, and test sets---also vary, further complicating direct comparison.

% CT studies: Chhatkuli (PCA + MSSA), Nabavi (PredNet)
\citeauthor{chhatkuli2015dynamic} reported higher $r$ values and \glspl{SSIM} than those achieved in our study \cite{chhatkuli2015dynamic, ritu2016development}. However, they simulated artificially regular breathing by looping identical respiratory cycles. Moreover, they applied \gls{PCA} directly to raw intensities rather than to motion fields (e.g., optical-flow vectors). This strategy may be less effective at capturing motion-variability components and local deformation patterns. Indeed, up to twenty principal components were required for accurate prediction of \gls{kV} fluoroscopic images in \cite{chhatkuli2015dynamic}. We conjecture that such direct pixel-synthesis architectures (including \gls{PredNet} \cite{nabavi2020respiratory}) are more prone to artifacts and inconsistencies when faced with large deformations not encountered during training or when operating at long horizons. In particular, \citeauthor{oprea2020review} argued that methods directly modeling raw pixel intensities "often lead to the regression-to-the-mean problem, visually represented as blurriness" \cite{oprea2020review}. Similarly, higher \glspl{SSIM} and lower target \glspl{TE} were reported in \cite{nabavi2020respiratory, romaguera2020prediction}, which also relied on \gls{4DCT} data. However, \gls{4DCT} captures motion averaged over several respiratory cycles, reducing apparent variability and precluding its use for real-time intrafractional motion compensation.

% geometrical/endpoint errors
Landmark \glspl{TE} in the literature span from sub-millimeter values to about 3mm, partially overlapping with the magnitude of our geometrical errors. Both metrics reflect different aspects of performance: the former assesses motion prediction locally, as it depends on a tracked anatomical point, whereas the latter quantifies accuracy across the entire image or a specific region. Specifically, \glspl{TE} can vary substantially even within a single patient and time point, as motion amplitude is typically greater near the diaphragm than the lung apex. Those reported in \cite{romaguera2020prediction} were relatively small (0.45--0.77mm), partly because only right-hemidiaphragm images were selected to avoid cardiac motion. This setting contrasts with our evaluation on the ETH Zürich sagittal slices, some of which include the heart and adjacent thoraco-abdominal structures. The \gls{3D} geometrical errors achieved by the transformer-based model in \cite{romaguera2023conditional} (1.25$\pm$0.74mm at $h=0.45\text{s}$) lie in a similar range to the \gls{2D} geometrical errors of our sequence-specific transformer on the ETH Zürich data (1.37$\pm$0.13mm at $h=0.63\text{s}$; Fig. \ref{fig:next frame pred perf vs hrz on ETH}). However, these values are not directly comparable, given the different problem settings---simultaneous \gls{4D} volume reconstruction from partial observations and forecasting versus \gls{2D} frame prediction---and the influence of large, static background regions in \gls{3D} images, which may lower average errors.
% Concerning last study (double check): Furthermore, the sequences used in that study seemed more diverse than ours regarding patient anatomies and breathing characteristics (e.g., the authors mentioned irregularities such as small apneas).

% comparison with other studies in general: performance decrease with horizon
Prior work on respiratory motion forecasting shows that accuracy tends to decrease as $h$ increases, confirming the trend observed in our results. In \cite{pham2019predicting}, the tumor center-of-mass \gls{TE} in \acs{XCAT} phantom images increased from 1.55mm to 2.60mm, as $h$ increased from 0.13s to 0.63s. The same study also reported an increase in the \gls{nRMSE} of \gls{PCA}-weight forecasts with $h$, similar to the trend shown in Fig. \ref{fig:signal pred error 3 PCA cpts vs horizon}. Likewise, prediction of the tumor \gls{SI} coordinate in cine \gls{MRI} with dynamically retrained \glspl{LSTM} in \cite{lombardo2022offline} produced \glspl{RMSE} that increased from 0.48mm at $h=0.25\text{s}$ to 2.20mm at $h=0.75\text{s}$. More broadly, works on natural video forecasting also report a clear decrease in frame-prediction performance as $h$ increases \cite{finn2016unsupervised, villegas2017decomposing, babaeizadeh2017stochastic}.
% % Rk: in Pham the nRMSE is scaled differently, i.e. normalization with the maximum - minimum

% Explanation of relatively low metrics in light of registration
On the \gls{OvGU} dataset, the best full-frame \gls{SSIM} (0.66$\pm$0.03, attained by \gls{SnAp-1}; Table \ref{table:frame pred perf}) was about 15\% lower than the mean \gls{SSIM} reported in \cite{romaguera2023conditional} (0.78 $\pm$ 0.11), although \glspl{CI} overlapped. Our relatively low \glspl{SSIM} mainly reflect the high noise, low contrast, and irregular motion in the \gls{OvGU} frames, as already noted in Section \ref{section:image prediction accuracy}. Crucially, warping the initial frame using the optical-flow field---our oracle baseline---yielded \pgls{SSIM} of only 0.63 in the \glspl{ROI}. Thus, the \gls{ROI}-based \gls{SSIM} of \gls{SnAp-1} (0.48%; Table \ref{table:frame pred perf}
) was approximately 75\% of this upper bound corresponding to perfect forecasting of the reference \gls{DVF}. In other words, the registration method, rather than the forecasting algorithm, largely determined the performance ceiling. This suggests that our comparative conclusions regarding the effectiveness of online-trained \glspl{RNN} under limited-data conditions remain valid despite the relatively moderate quality of the \gls{OvGU} frames.

\subsection{Methodological considerations}

% Error quantification
Quantifying performance with multiple metrics allows for fine-grained analysis. For instance, $r$ is sensitive to coherent intensity variations across the image, whereas the \gls{SSIM} emphasizes local similarity, extending the local \gls{NCC} by incorporating luminance and contrast factors. However, as noted in \cite{oprea2020review}, these intensity-based metrics "prefer blurry predictions nearly accommodating the exact ground truth than sharper and plausible but imperfect generations." This has motivated the use of perceptual, feature-based metrics, such as the \glsentrylong{LPIPS} \cite{zhang2018unreasonable} and \glsentrylong{VGG}--based cosine similarity, which compare deep \gls{CNN} features to better align with human visual perception. Nonetheless, in \gls{MR} imaging, where intensities are not physically calibrated (unlike \gls{CT}), geometrical metrics derived from the predicted \glspl{DVF} remain particularly interpretable. They also relate more directly to an intuitive understanding of tumor tracking in radiotherapy.

% Selection of the number of principal components
Our study is the first to address optimal selection of the number of components in the \gls{PCA} respiratory motion model using a validation-based procedure in chest and liver cine-\gls{MRI} forecasting. Prior work on \gls{PCA}-based motion modeling has commonly used two or three principal components (Section \ref{section: intro respiratory motion management in MRgRT}); we also observed on the ETH Zürich dataset that this choice tended to yield high accuracy (Section \ref{section: optimization of nb of PCA cpts}). Indeed, increasing $n_{\text{cp}}$ from 1 to 2 substantially reduced $E_{\text{pred}}(n_{\text{cp}})$ (Fig. \ref{fig:nb PCA cpts optim}); a similar drop in the validation-set geometrical error was also reported in \cite{li2011pca}. Nonetheless, the latter reflected only motion-modeling accuracy, since forecasting was not performed in that study. \citeauthor{pham2019predicting} noted that the third-order coefficient appeared "less predictable" than the first two, which is consistent with its relatively higher noise and lower amplitude in Figs. \ref{fig:DVF principal components and weights sequence 4}--\ref{fig:PCA weights pred RTRL vs pop transformer}, and that "excluding it [had] minimal effect on the final [predicted] volumetric cine \gls{MRI}" \cite{pham2019predicting}. Overall, the \gls{PCA} coefficients in our study resembled those shown in \cite{chhatkuli2015dynamic, pham2019predicting, liu2016prediction}, with the coefficient predominantly associated with respiratory motion exhibiting a relatively smooth, near-sinusoidal pattern. We observed that the optimal value of $n_{\text{cp}}$ depended on both the prediction method and horizon, with high-capacity models and short horizons generally favoring larger values of $n_{\text{cp}}$.

% Spatial modeling with end-to-end deep learning vs temporal modular modeling: comparison of approaches with the literature
Deep learning--based population models for video forecasting, such as those in \cite{nabavi2020respiratory, romaguera2020prediction, romaguera2023conditional}, enable inference on unseen patient data without an explicit registration step. Yet, \citeauthor{romaguera2020prediction} noted that "classical registration approaches still outperform deep learning techniques in several medical imaging applications," underscoring the continued relevance of \gls{DIR}-based pipelines \cite{romaguera2020prediction}. 
% why not citing other studies 
The end-to-end strategy adopted in those studies contrasts with our modular pipeline, which follows the general methodology in \cite{chhatkuli2015dynamic, pham2019predicting}. Such modularity improves interpretability, as each \gls{PCA} component and weight can be inspected individually (Sections \ref{section: PCA breathing motion modeling results} and \ref{section: PCA weight forecasting qualitative eval}). Furthermore, prediction errors in the image space can be traced back to weight dynamics (Section \ref{section: frame forecasting qualitative eval}). Notably, while several studies focused mainly on deep-learning--based spatial modeling, we adopted a complementary approach, emphasizing developments in the time-series forecasting module. Indeed, we investigate online learning algorithms for \glspl{RNN} in thoraco-abdominal \gls{MR} image prediction for the first time, motivated by the need to adapt to non-stationary respiratory signals in real time with scarce data.
% I removed "hybrid framework" because that applies best only to subject-specific + population transformer
% I could explicitly mention "both paradigms" (subject-specific vs. population) even if I already did that in the intro

% Cross-subject PCA-domain predictor integration: overfitting and fairness aspects
Notably, our framework can accommodate cross-subject components, such as the population transformer in this study, enabling robust learning of respiratory patterns from large databases. Population models that forecast motion projections onto the \gls{PCA} subspace may better handle domain shift than end-to-end image-based architectures in low-data settings. Indeed, this reduced low-dimensional representation can prevent overfitting to dataset-specific frame appearance. Moreover, using a common learning space enables a relatively fair comparison between transformers and \glspl{RNN} by requiring both to forecast the same compact signal. This makes differences in performance more directly attributable to their temporal modeling capabilities rather than disparities in input complexity. However, transformers did not perform well in our experiments. The following section discusses several factors that may explain this outcome.

\subsection{Remarks on the performance of RNNs and transformers}

%\subsubsection{Factors explaining the experimental underperformance of transformers in our study}
\subsubsection{Factors influencing transformer performance}\label{section: factors influencing transformer performance}
% Maybe I can reuse my previous sentence, emphasizing online learning benefits: "However, the subject-specific transformer accuracy degraded rapidly as $h$ increased, due to its limited training set (the first 50.5s of each sequence) and lack of adaptation capabilities, contrasting with the \glspl{RNN} algorithms explored, that can leverage the test set as additional training points." 

% Data scarcity (low-data regime, weak inductive biases, resampling limitations)
Data scarcity was likely a major factor limiting the accuracy of lightweight encoder-only transformers in this research. For example, \citeauthor{romaguera2023conditional} reported conditional-based transformer gains over \gls{ConvLSTM} and \gls{ConvGRU} models for sagittal cine-\gls{MRI} prediction (with geometrical errors of 1.25mm, 1.34mm, and 1.60mm for the three models, respectively, at $h=0.45\text{s}$), but their study relied on a large dataset from 25 subjects, with approximately 20 min of data per subject \cite{romaguera2023conditional}. Similarly, \citeauthor{jeong2022clinical} observed that full encoder--decoder transformers outperformed \gls{LSTM}-based baselines in respiratory trace forecasting for $h \geq 0.50\text{s}$, based on a large cohort (442 patients, 540 traces, mean record length of 145s) \cite{jeong2022clinical}. By contrast, our datasets comprise 12 sequences in total, with each acquisition lasting at most 83s. The relative underperformance of transformers in our work is consistent with their weaker inductive biases compared with \glspl{RNN}, which can better generalize in low-data regimes (Section \ref{section: breathing motion pred. with RNNs and attention-based ANNs}). Notably, the accuracy of sequence-specific transformers deteriorated rapidly as $h$ increased (Sections \ref{section: prediction of weights on ETH Zurich data} and \ref{section:image prediction accuracy}), due to their limited training set (the first 50.4s of each sequence), whereas the performance of online-trained \glspl{RNN} was more stable across horizons, since they could use test-set samples as additional training data. When training the population transformer on the \gls{OvGU} data, we downsampled the corresponding \gls{PCA} weights to match the frequency of the test sequences from ETH Zürich, which reduced temporal granularity. Future work could explore multi-start (phase-offset) downsampling, i.e., generating additional training data by resampling the original weights with multiple start indices, to avoid discarding information.

% Performance of population transformers (domain shift)
The domain shift between the ETH Zürich and \gls{OvGU} datasets further limited the performance of the population transformer in our experiments. Indeed, the body parts imaged and the acquisition settings differ between the two datasets, changing the semantic meaning, ordering, and characteristics of the \gls{PCA} modes. For instance, local deformations of the heart or liver dominated the first-order mode in two of the ETH Zürich sequences, whereas overall respiratory motion contributed most to the first component in all \gls{OvGU} acquisitions (Section \ref{section: PCA breathing motion modeling results}). The variety of breathing patterns across subjects exacerbated the lack of shared structure and semantic mismatch between components. Although we augmented \gls{PCA}-weight data during training to increase robustness to ordering, scale, offset, and drift (Section \ref{section: PCA weight cross-validation}), this strategy was not sufficient to fully resolve component misalignment across subjects in our low-data regime. Still, to our knowledge, our study is the first to investigate training and testing with multi-center data for \gls{2D} cine-\gls{MRI} forecasting. Compared with works that used more homogeneous cine-\gls{MRI} acquisition protocols for learning and test-set evaluation, our cross-dataset setting is more challenging for population models.
% By contrast, \citeauthor{romaguera2023conditional} performed training and testing on cine-\acs{MRI} acquired with consistent protocols, easing learning for their transformer model \cite{romaguera2023conditional}.
% probably point to the table here... no because it's the geometrical errors that are maybe most meaningful, and I provided them above.
Future research could explore additional data augmentations, such as \gls{PCA}-coefficient resampling at slightly different frequencies or piecewise time warping between control knots on training examples to simulate global and local variability in breathing rate, respectively. Furthermore, one could investigate alternative optimization algorithms (e.g., AdamW \cite{loshchilov2017decoupled}) or validation objectives to improve generalization. Indeed, the \gls{RMSE} objective in our study upweighted sequences whose \gls{PCA} coefficients exhibited higher variance, whereas averaging the per-sequence validation \gls{nRMSE} in future work would help better balance contributions across subjects.

% Performance of transformers (low SHL, limited grid, use of RMSE)
Capping the transformer history length at $L = 9.5\text{s}$ (broadly, two breathing cycles; Table \ref{table:models comparison}) likely constrained performance, although several studies on motion forecasting in cine-\gls{MRI} used lower \glspl{SHL} (8.0s and 2.25s in \cite{lombardo2022offline} and \cite{romaguera2023conditional}, respectively). \Glspl{RNN} may struggle to capture long-term dependencies due to vanishing gradients and the compression of past information into a finite-dimensional state. Transformers can, in principle, exploit longer context by letting each time step directly attend to any other time step without forced reduction of all history into a single state vector. Nonetheless, we set $L$ to a relatively small value as a regularization mechanism in our low-data regime and for computational efficiency. We note that transformers typically require larger amounts of data and that their time complexity scales as $\mathcal{O}(L^2)$ with input length, potentially limiting clinical feasibility (Section \ref{section: breathing motion pred. with RNNs and attention-based ANNs}). Implementing early stopping during hyperparameter tuning could facilitate search over more extensive grids, including larger values of $L$, without prohibitive cost.

\subsubsection{Notes on RNN accuracy and hyperparameter selection}\label{section: discussion - comments on RNN performance}
% Change to "Notes on RNN and LMS accuracy and hyperparameter selection"?... 

% General comparison with (Pohl, 2025): prediction of PCA cpts with ETH Zürich (3.18Hz) vs marker position prediction (3.33Hz)
Predicting ETH Zürich-derived \gls{PCA} weights was a task comparable to 3.33Hz-sampled marker-position forecasting in \cite{pohl2025real}. Indeed, the sampling rates and per-sequence data partitioning into training, validation, and test sets were similar (30s/30s/remainder, in that study). Excluding \gls{RTRL} and transformers, which were respectively trained with fewer hidden units ($d \leq 40$) and not examined, in \cite{pohl2025real}, the ordering of the remaining non-baseline methods by horizon-averaged \gls{nRMSE} matched in both works; \gls{SnAp-1} yielded the lowest \gls{nRMSE}, whereas that of linear regression was substantially higher (0.34 vs. 1.66 in \cite{pohl2025real}). Nonetheless, across adaptive predictors, errors in \cite{pohl2025real} were lower than those reported in this work (Table \ref{table:signal pred nRMSE 3 PCA cpts avg over horizon}). This is partly because more training data was available in that study, as each sequence comprised 584 samples on average. In addition, the time-dependent marker locations in \cite{pohl2025real} were less noisy and more correlated than our \gls{PCA}-weight data derived from global image-based motion fields, which captured more complex, interacting dynamics (Section \ref{section: PCA breathing motion modeling results}). In both studies, the \gls{nRMSE} tended to increase with $h$, and linear regression was particularly competitive for one-step-ahead prediction.
% Rk: when viewed as 1D vectors, the principal components are uncorrelated, but there is no reason for the time-varying weights not to be correlated. 

% Hyperparameter optimization comparison with \cite{pohl2025real}
Hyperparameter-optimization trends for 3.33Hz sampling in \cite{pohl2025real} also aligned with those observed for the ETH Zürich sequences in this research (Fig. \ref{fig:hyperpar influence on PCA cpts prediction}). For \gls{UORO}, the validation \gls{nRMSE} attained its minimum at similar hidden layer sizes in both works ($d=70$ here and $d=60$ in \cite{pohl2025real}). Concerning \gls{SnAp-1}, the \gls{nRMSE} tended to decrease as $d$ increased and was minimized at $d=110$ in this work and at $d=120$ in \cite{pohl2025real}. In both studies, the optimal learning rate was $\eta = 0.01$ for \gls{UORO} and \gls{DNI}, whereas the validation error for \gls{SnAp-1} decreased with $\eta$ and reached its minimum at $\eta=0.02$. By contrast, prior work on respiratory motion forecasting using higher-frequency data, with sampling rates close to 30Hz, typically set $\eta$ between 0.001 and 0.005 \cite{lin2019towards, yu2020rapid, samadi2023respiratory}. In our setting with short sequences and low sampling rates, larger values of $\eta$ yielded better performance, plausibly because per-step changes were more pronounced and fewer time steps were available for online adaptation. This is consistent with \cite{pohl2025real}, in which the optimal $\eta$ decreased as the acquisition frequency increased.

% Paragraph already checked
% Why LMS does better at medium h on the ETH Zürich dataset, and comparison of RNNs with linear filters
\Gls{LMS} generally ranked higher for intensity-based metrics than for \gls{DVF} accuracy (Table \ref{table:frame pred perf}). Indeed, this algorithm has limited capability to forecast noisy signals; hence, the validation procedure tended to retain fewer components than for \glspl{RNN} (Fig. \ref{fig:nb PCA cpts optim}). Selecting a smaller value of $n_{\text{cp}}$ suppressed the influence of minor higher-order deformation modes, often dominated by noise, on predicted displacements. This effectively denoised the Lucas--Kanade optical-flow field, since the first one or two components mainly captured respiratory motion (Figs. \ref{fig:DIR and principal components sequence 1 ETH}--\ref{fig:1st cpt prediction (SnAp-1 vs vs UORO vs transformer) Magdeburg}). This observation %reinforces the idea 
suggests that imposing a regularization constraint on locally divergent \glspl{DVF} during deformable registration may further enhance frame-prediction performance. Regardless, \glspl{RNN} still achieved higher \glspl{SSIM} and $r$ values at long horizons than \gls{LMS} on the ETH Zürich data (Fig. \ref{fig:next frame pred perf vs hrz on ETH}). Furthermore, they surpassed \gls{LMS} at medium-to-long horizons across all metrics on the more irregular \gls{OvGU} sequences (Figs. \ref{fig:next frame pred perf vs hrz on Magdeburg ROI} and \ref{fig:next frame pred perf vs hrz on Magdeburg full image}). They also generally outperformed linear regression for intermediate-to-high values of $h$. This is attributable to their internal state, which acts as a memory and provides greater learning capacity, and the online training regime. Notably, linear regression outperformed \glspl{RNN} in \gls{3D} lung-tumor position forecasting in cine \gls{MRI} in \cite{li2023online}, but this may reflect the relatively short horizon in that study. We argue that while linear filters can predict relatively steady breathing well at shorter response times, online-trained \glspl{RNN} can better demonstrate their modeling potential when processing more irregular sequences or operating at longer horizons, consistent with the literature \cite{verma2010survey}. As \gls{MR-guided} radiotherapy systems evolve and acquisition rates increase, typical horizon-to-frame-rate ratios may also rise, further increasing the appeal of online-trained \acs{RNN}-based predictors even for relatively short look-ahead times.
% Note: \glspl{LSTM} were found to significantly outperform linear regression in \cite{lombardo2022offline} 

\subsection{Limitations and future work}

% Limitations regarding our datasets
Our \gls{MRI} datasets are relatively limited, as only ten subjects were involved; therefore, additional anatomies will be required to validate our findings. Data scarcity was a central limitation of this study and likely constrained forecasting performance, particularly for transformers (Section \ref{section: factors influencing transformer performance}). %, but also for dynamically trained \glspl{RNN} (due to short duration of the sequences). 
Nevertheless, our performance evaluation remains practically meaningful, given the variety of our records. Indeed, our acquisitions span thoracic and abdominal regions, exhibit markedly different noise and contrast levels, with several particularly noisy \gls{OvGU} sequences, and encompass a diverse range of breathing patterns. To our knowledge, this study is the first to investigate frame forecasting in \gls{2D} thoraco-abdominal cine-\gls{MRI} using publicly available data. The accompanying code is publicly available, enabling full reproducibility of our experiments \cite{pohl2024MRforecastingcode}. To better leverage scarce data and combine the advantages of offline and online learning, future work could explore pretraining temporal models on larger multi-subject datasets, followed by patient-specific parameter updates during inference. Beyond data limitations, one-shot single-sequence prediction was practically constrained by the computational resources required for fast, per-patient hyperparameter tuning; future research could examine more efficient tuning schemes.

% Limitations imposed by registration - reviewed
Out-of-plane motion and local variations in brightness or contrast can hamper deformable registration, thereby reducing image-forecasting accuracy (Section \ref{section: frame forecasting qualitative eval}). The moderate upper bound on accuracy corresponding to the oracle baseline on the \gls{OvGU} dataset (Table \ref{table:frame pred perf}) highlights that \gls{DIR} quality can substantially cap performance, as discussed in Section \ref{section: comparison with previous works MRI}. In particular, the Lucas--Kanade optical-flow algorithm can produce noisy \glspl{DVF} because its basic formulation does not explicitly involve spatial smoothness regularization. Moreover, it relies solely on the brightness-constancy assumption and does not learn or exploit representations of anatomical structures, unlike modern deep learning--based registration methods \cite{balakrishnan2019voxelmorph}. Future directions include the integration of more advanced registration techniques into our modular pipeline.

% Limitations regarding out-of-plane motion and PCA linearity - reviewed
Our framework assumes that future \gls{2D} frames can be approximated as the result of deforming an initial \gls{2D} reference frame. This hypothesis is restrictive because organ motion is \glsentrylong{3D} and tissue-brightness constancy does not reliably hold in \gls{MRI}. One possible remedy is to periodically update the motion model and reference frame when errors become large. Another option is to combine future \gls{DVF} estimation "with pure synthesis layers to better predict pixels that cannot be copied from other video frames," as suggested in \cite{liu2017video}. Although \gls{PCA} provides an interpretable representation of respiratory motion, the underlying deformation manifold is not strictly linear. Future work could investigate more expressive dimensionality-reduction methods, such as \gls{kPCA} or \glspl{AE}, to better capture complex motion patterns and mitigate domain shift in cross-dataset training and evaluation of population models.

% Challenges concerning the end-of-inspiration phase and diaphragm region, and discussion regarding the 2mm or 3mm clinically acceptable margin.
The \gls{EI} phase was subject to higher inter-cycle variability and was therefore harder to predict. This manifested as occasional misalignments of the diaphragm, which exhibited relatively high motion amplitudes, between the predicted and ground-truth frames (Figs. \ref{fig:next frame pred sq 1 RTRL vs pop transformer}, \ref{fig:sq 2 mean test intensity}--\ref{fig:sq 2 mean test deformation error}, and \ref{fig:visual comparison of ROI prediction with SnAp-1 and transformer at h=10 on OvGU dataset}). Similar \gls{EI}-phase errors, occurring particularly around the diaphragm, have been reported in related work (Section \ref{section:intro future frame forecasting in chest and liver image sequences}). Given those challenges, a 2mm upper bound on the \gls{TE} was proposed as a desirable target for motion compensation during radiotherapy \cite{murphy2004tracking}. Nonetheless, margins of up to 3mm have also been considered acceptable in specific contexts, such as ultrasound-based tracking \cite{preiswerk2014model}. On the ETH Zürich dataset, mean geometrical errors over the full frame and across horizons remained below 2mm for all non-baseline methods (Table \ref{table:frame pred perf}). \Gls{RNN} errors even stayed below 1.5mm across all horizons (Fig. \ref{fig:next frame pred perf vs hrz on ETH}). On the \gls{OvGU} dataset, however, the lowest horizon-averaged \gls{ROI}-based geometrical error, reached by both \gls{RTRL} and \gls{SnAp-1}, was 2.7mm, likely reflecting lower contrast and higher noise in the acquired frames. Furthermore, high-frequency oscillations in predicted trajectories near breathing irregularities (Figs. \ref{fig:PCA weights pred RTRL vs pop transformer}--\ref{fig:1st cpt prediction (SnAp-1 vs vs UORO vs transformer) Magdeburg}) may pose a challenge for radiotherapy systems with mechanical response limits, which cannot follow rapidly varying commands. We anticipate that the algorithmic improvements outlined above---including enhanced registration, richer motion-representation learning, and online adaptation of pretrained models---could help dampen this oscillatory behavior and reduce the mean error below the 2mm target. Finally, we note that errors measured on well-defined landmarks or clinical points of interest, rather than over the full \gls{ROI} or \gls{FOV}, may be closer to, or below, the 2mm margin guideline.

\section{Conclusion}
% Updated conclusion

% General ontext and novelty
In this research, we propose a new approach to forecast respiratory motion in cine \gls{MRI} to compensate for the latency of \gls{MR-guided} radiotherapy systems and, in turn, reduce healthy-tissue exposure to radiation. We apply standard \glspl{RNN} trained with online learning algorithms for the first time to one-shot future frame estimation in dynamic chest and liver \gls{MRI} sequences and compare them with lightweight encoder-only transformers in a low-data regime typical of medical imaging. To our knowledge, this is the first study to examine the impact of domain shift on cross-subject model training and evaluation in cine-\gls{MR} image forecasting. We combine two public datasets covering thoracic and abdominal imaging (from ETH Zürich and \gls{OvGU}, respectively), enabling evaluation across diverse breathing patterns and levels of noise and contrast. Our modular framework, which projects deformations between the incoming and reference frames onto the \gls{PCA} linear subspace and then forecasts the projection coordinates, contrasts with recent end-to-end deep-learning architectures \cite{nabavi2020respiratory, romaguera2020prediction}. It can decompose complex motion into simpler modes (e.g., cardiac and breathing components), offers high interpretability, is privacy-friendly, and requires limited data for model fitting.

% Quantitative assessment for each algorithm
% Our study investigates a wide range of horizons $h$ (up to 2.2s) and multiple complementary metrics, providing a comprehensive assessment of algorithm performance. 
% Is that first sentence worth citing in a conclusion I wish to shorten
\gls{RTRL} and \gls{SnAp-1} generally showed the highest cross-dataset and intra-dataset stability (Table \ref{table:stability performance}). They also attained the highest mean performance across horizons $h \leq 2.2\text{s}$, with geometrical errors of 1.4mm and 2.7mm over the full ETH Zürich images and the \gls{OvGU} frame \glspl{ROI}, respectively (Table \ref{table:frame pred perf}). Errors on the former dataset were on par with the literature, satisfying the clinical 2mm threshold guideline \cite{murphy2004tracking}. Nevertheless, errors on the latter dataset were higher, mostly due to challenges in deformable registration between high-noise, low-contrast images. %, especially in areas with high-amplitude motion. 
For example, the horizon-averaged \gls{ROI}-based \gls{SSIM} of \gls{SnAp-1} on the \gls{OvGU} frames (0.48) equaled approximately 75\% of the corresponding upper bound \gls{SSIM} attained by an oracle having access to the exact \gls{DVF} (0.63). Overall, performance decreased across algorithms as $h$ increased. \Gls{RNN} accuracy remained relatively high for medium-to-long horizons, primarily because online algorithms learn from recent respiratory patterns and can adapt to non-stationary changes on-the-fly with few training samples.
% tended to decrease -> decreased  (is that fine, as I wrote "overall")?
The sequence-specific transformer was competitive for low-to-medium horizons, but generalization remained limited for the population transformer, except at the shortest horizons on the \gls{OvGU} sequences. Transformers were particularly constrained by data scarcity, given their weak inductive biases, the small \glspl{SHL} considered ($L \leq 10\text{s}$, i.e., roughly two breathing cycles), limiting their ability to model long-term dependencies despite self-attention, and the distribution gap between the two datasets.
Linear regression was the best predictor at low horizons among non-baseline methods in nearly all settings.

% Qualitative evaluation
Overall, the generated frames were visually plausible, with sharpness and texture largely preserved even for noisy image sequences and long horizons.
% despite the high noise in the original \gls{OvGU} images and the long horizons considered. 
Organ contours were broadly faithful to the ground truth, although occasional diaphragm misalignments were observed at the \gls{EI} phase, characterized by higher motion variability.
% Liver vessels could be challenging to predict due to out-of-plane motion (e.g., blood flow-induced flicker), which our general strategy, based on modeling deformations as strictly planar and estimating future frames via vector-based resampling of the reference frame, did not account for. 
Out-of-plane motion, including vessel flickering induced by transverse blood flow, was challenging to predict, as we modeled deformations as strictly planar. Indeed, future frames were estimated via vector-based resampling of the initial frame, as all relevant content was assumed to appear in that reference image and remain visible over time. 
Moreover, pronounced oscillations of the predicted trajectories sometimes occurred around breathing irregularities, which may affect the mechanical guidance of the radiation beam.
% which may affect radiation-beam guidance

% Clinical implications and future works
Key algorithmic improvements include exploring more accurate registration methods, richer motion representations (e.g., \gls{kPCA} or \glspl{VAE}), faster, more efficient hyperparameter-tuning approaches, online adaptation of pretrained population models, and transformer variants specialized for time-series forecasting. Furthermore, training transformer-based models with more extensive data would help provide a fairer assessment of their capabilities. Subsequent studies should further assess robustness to irregular breathing, anatomical variability, and differences in acquisition protocols. While \gls{2D} cine-\gls{MRI} forecasting provides useful information about target motion, fully realizing its clinical benefits requires downstream modules, such as tumor and \gls{OAR} segmentation and \gls{3D} image generation from partial \gls{2D} views to compensate for the imaging limitations of \gls{MR-LINAC} systems. The implementation and evaluation of these components, integrated with video prediction, are left as future work.
% I may want to emphasize stochastic models, following the review from Opreah: e.g., stochastic RNNs, diffusion models... This will help keep my paper up to date while I would be able to showcase it better to a computer science audience..

\section*{Conflicts of interest statement}
The authors declare no conflict of interest.

\section*{Funding}

This work was supported by the Epson International Scholarship Foundation and the Japan Student Services Organization.

\section*{Acknowledgments}

We are thankful to Prof. Masaki Sekino, Prof. Ichiro Sakuma, Prof. Hitoshi Tabata (The University of Tokyo, Graduate School of Engineering) and Dr. Kiwoo Lee (Edogawa Hospital, Department of Radiology) for their thoughtful feedback and recommendations that helped enhance the quality of this research. We also express gratitude to Dr. Cristian Le Minh (Max Planck Institute) and Mr. Suryanarayanan N.A.V. (The University of Tokyo, Graduate School of Engineering) for their assistance regarding software and coding.

\section*{Data and code availability}

The data and code used in this research are publicly available online \cite{ETHdataset, gulamhussene__2d_2021, pohl2024MRforecastingcode}.

\bibliographystyle{model1-num-names}
%\bibliographystyle{cas-model2-names}

\bibliography{Future_frame_prediction_CMIG}

\FloatBarrier % Prevents appendix figures to go inside the references - require placeins package
\appendix
%\clearpage
\section{Appendix: Dataset characteristics}
\label{appendix: dataset characteristics}

The characteristics of the datasets used in our study\footnotemark~are provided in Table \ref{table:MRI general characteristics}.

\begin{table}[htb!]
\footnotesize
\setlength{\tabcolsep}{1.1pt}
\begin{center}
\begin{tabular}{lllllllll}
\hline
                           & ETH Zürich dataset              & \gls{OvGU} dataset    \\
\hline 
Body part                  & Thorax                          & Abdomen          \rule{0pt}{2.6ex}\\
Acquisition type           & Cross-sections from             & \acs{2D} navigator \\
                           & reconstructed \acs{4D}-\acs{MRI} & slices \\
Number of sequences        & 4                               & 8 \\
Whole-image \acs{FOV} (mm) & 270 $\times$ 270 (seq. 1--2)    & 255 $\times$ 320         \\
                           & 290 $\times$ 290 (seq. 3--4)    &          \\
In-plane resolution (mm)   & 1.0 $\times$ 1.0                & 1.82 $\times$ 1.82        \\
Through-plane resolution   & 1.0mm                           & 4.0mm \\
Sampling rate              & 3.18Hz                          & 6.0Hz         \\        
Frames per sequence        & 200                             & 498        \\        
Breathing regularity       & Regular                         & Rather irregular        \\        
Noise level                & Low                             & High           \\
Organ contrast             & High, stable                    & Low, variable        \\                                                                          
Maximum \acs{ROI} width    & n/a                             & 158mm \\
Maximum \acs{ROI} height   & n/a                             & 262mm \\  
Breathing cycles per seq. &                                 & \\
 - minimum--maximum         & 14--28                         & 13--26 \\
 - average                 & 16                              & 19.75 \\
\hline               
\end{tabular}
\end{center}
\caption{General characteristics of the two cine-\gls{MRI} datasets.}
\label{table:MRI general characteristics}
\end{table}

\footnotetext{For reference, the \gls{OvGU} sequences numbered from 1 to 8 in our study correspond to the following identifiers in \cite{gulamhussene__2d_2021}, in order: \datasetid{2020-11-10_KS81_Nav_Pur_1}, \datasetid{2020-11-12_QN76_Nav_Pur_1}, \datasetid{2020-11-17_CS31_Nav_Pur_2}, \datasetid{2020-11-17_JY02_Nav_Pur_2}, \datasetid{2020-11-23_ON65_Nav_Pur_2}, \datasetid{2020-11-23_PS11_Nav_Pur_1}, \datasetid{2020-11-25_II29_Nav_Pur_1}, \datasetid{2020-11-26_NE38_Nav_Pur_1}.}

\section{Appendix: Optimization of MR image registration parameters}%
\label{appendix: chest MR image registration optimization}

In this appendix, we describe the optimization of the pyramidal, iterative Lucas--Kanade optical-flow parameters using % per-sequence
grid search, illustrated with the ETH Zürich dataset (step 1.1 in Fig. \ref{fig:overall experimental setting}; see Section \ref{Breathing motion modeling with PCA}). The same methodology was applied to the \gls{OvGU} images for consistency. For each sequence, we minimize the registration error $E_{\text{ref}}(\vec{u})$ using the first $M_{\text{DIR}}$ frames, defined as follows:
\begin{equation} \label{eq:gt_OF_error_def}
E_{\text{ref}}(\vec{u}) = \sqrt{\frac{1}{(M_{\text{DIR}}-1)|I|}\sum_{k=2}^{M_{\text{DIR}}} \sum_{\vec{x}} \delta(\vec{u}, \vec{x}, t_k)^2} 
\end{equation}%
In this equation, $|I|$ and $\delta(\vec{u}, \vec{x}, t_k)$ refer, respectively, to the number of pixels in each frame and the instantaneous registration error at pixel $\vec{x}$ and time $t_k$ using the vector field $\vec{u}$ (Eq. \ref{eq:instant registration error}). The value of $\vec{u}$ that minimizes $E_{\text{ref}}(\vec{u})$ is considered the \emph{reference} (i.e., proxy ground-truth) \gls{DVF}, representing the motion to forecast. In contrast, $E_{\text{pred}}(n_{\text{cp}})$, defined in Eq. \ref{eq:val registration error}, is minimized with respect to $n_{\text{cp}}$ using the validation frames to accurately predict the future \gls{DVF} (Sections \ref{section: methods - optimization of n_cp} and \ref{section: optimization of nb of PCA cpts}). $M_{\text{DIR}}$ is set such that $t_{M_{\text{DIR}}} = 28.3\text{s}$; for online forecasting algorithms, $M_{\text{DIR}} = M_{\text{train}}$ (Table \ref{table:general experimental setup}). Table \ref{table:image registration parameter optimization} summarizes the ranges and selected values for the tuned parameters (definitions in \cite{pohl2021prediction}\footnotemark).%
\footnotetext{Variable notations were adapted from \cite{pohl2021prediction} for clarity. For instance, here, $n_{\text{layer}}^{\text{LK}}$ denotes the number of pyramid levels in the Lucas–Kanade optical-flow algorithm, distinct from the number of transformer-encoder layers, $n_{\text{layer}}$. $E_{\text{ref}}$ refers to the same quantity as $e_{\text{DVF}}$ in \cite{pohl2021prediction}.}%

\begin{table}[htb!]
\footnotesize
\setlength{\tabcolsep}{0.8pt}
\begin{tabular}{lll}
\hline
                                                       & Parameter               & Best  \\
                                                       & range                   & value(s) \\
\hline 
Standard deviation of the Gaussian filter                  & $\{0.1, 0.5, 1.0\} $    & 0.1 \\
applied to the initial image, $\sigma_{\text{init}}$          &                         &  \\ 
Standard deviation of the Gaussian filter             & $\{0.1, 0.5, 1.0\}$     & -  \\
used for downsampling at each layer, $\sigma_{\text{sub}}$         &                         &  \\ 
Standard deviation of the Gaussian kernel                  & $\{1.0, 2.0, 3.0, 4.0\}$& 2.0  \\
weighting the moment matrix, $\sigma_{\text{support}}$ &                         & or 3.0 \\
Number of pyramid layers, $n_{\text{layer}}^{\text{LK}}$      & $\{1, 2, 3\}$           & 2 or 3 \\
Number of iterations, $n_{\text{iter}}$                       & $\{1, 2, 3\}$           & 1 \\      
\hline
\end{tabular}
\caption{Parameter ranges for optimization of the pyramidal, iterative Lucas--Kanade optical-flow algorithm, along with the parameter values that experimentally yielded the lowest registration error $E_{\text{ref}}$ on the ETH Zürich sequences.} %
\label{table:image registration parameter optimization}
\end{table} 

\begin{figure*}[pos=htbp,align=\centering]
    \centering
    \includegraphics[width=0.195\textwidth]{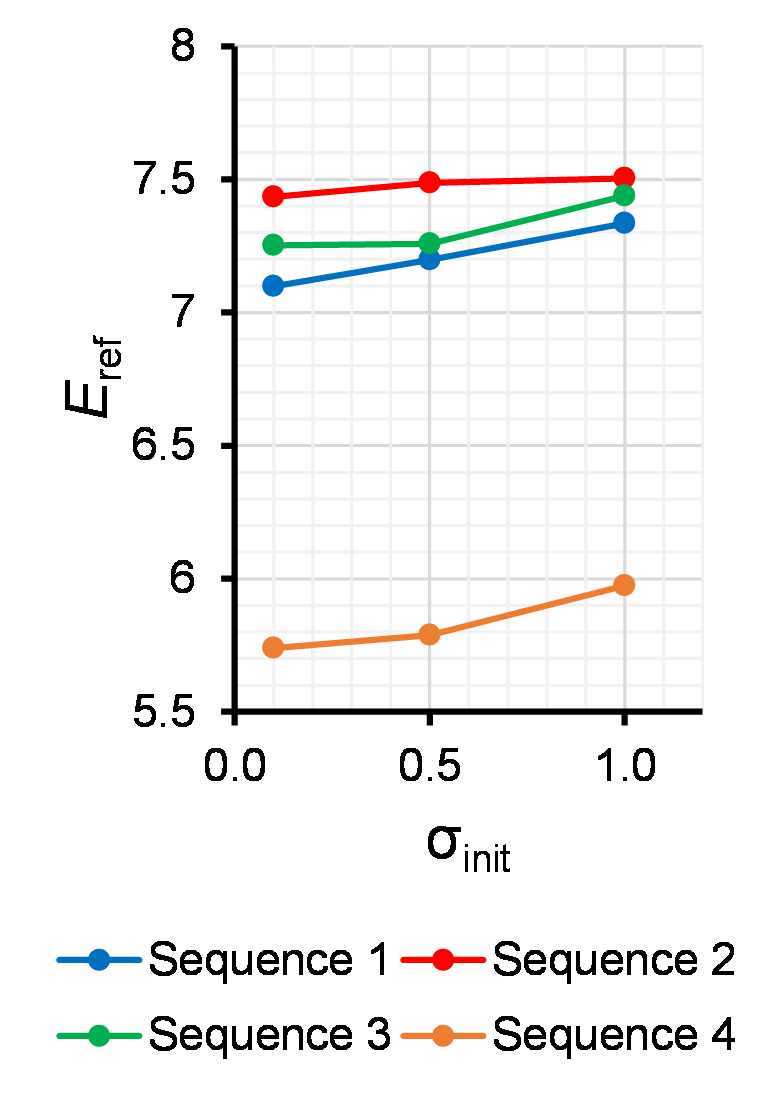} \label{subfig:DVF_err_sg_init_MRI}%
    \includegraphics[width=0.195\textwidth]{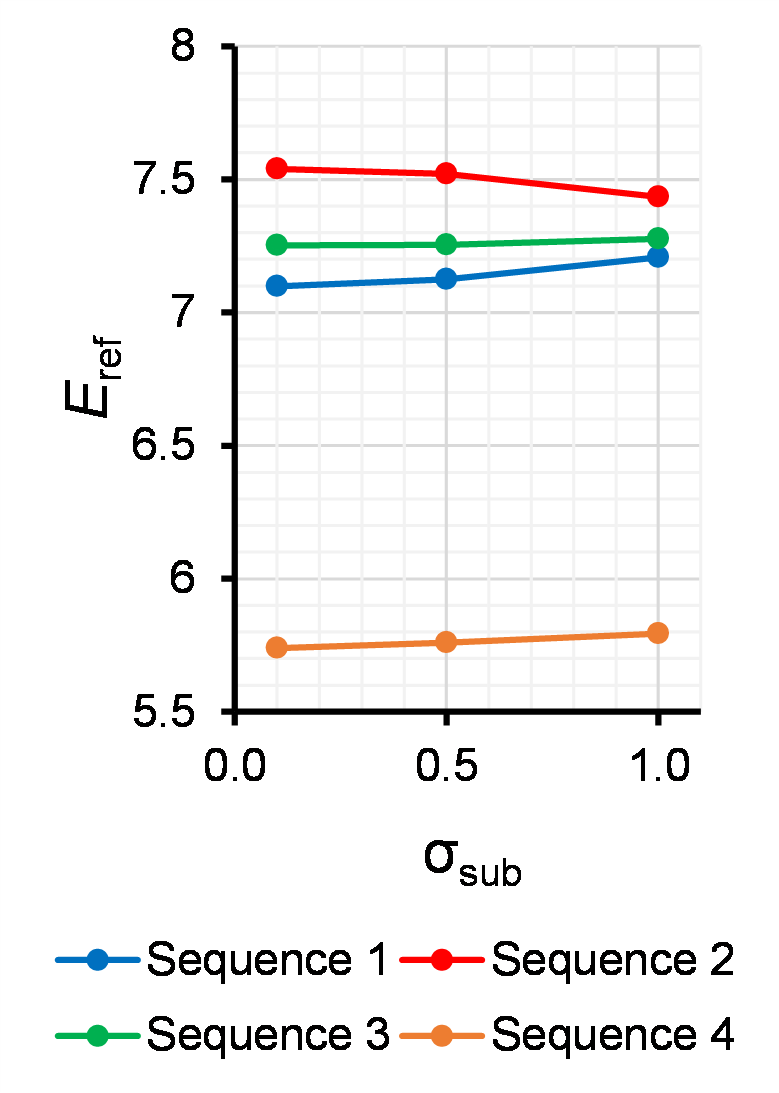} \label{subfig:DVF_err_sg_sub_MRI}%
    \includegraphics[width=0.195\textwidth]{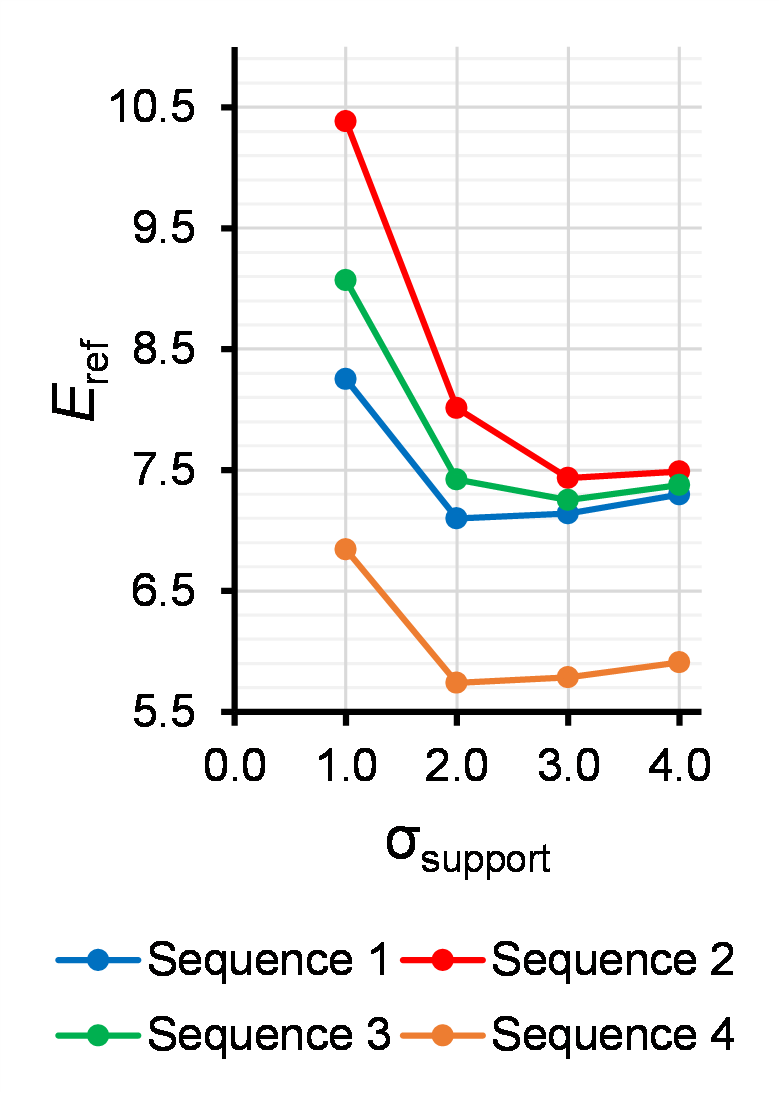} \label{subfig:DVF_err_sg_LK_MRI}%
    \includegraphics[width=0.195\textwidth]{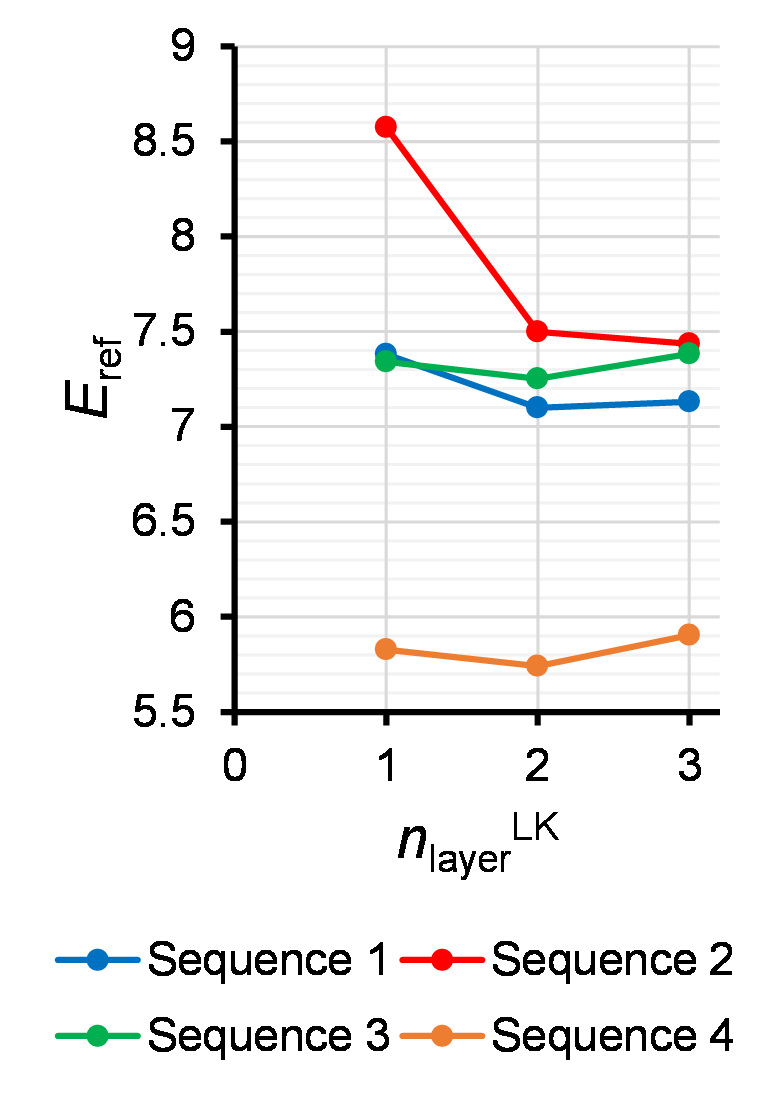} \label{subfig:DVF_err_nlayers_MRI}%
    \includegraphics[width=0.195\textwidth]{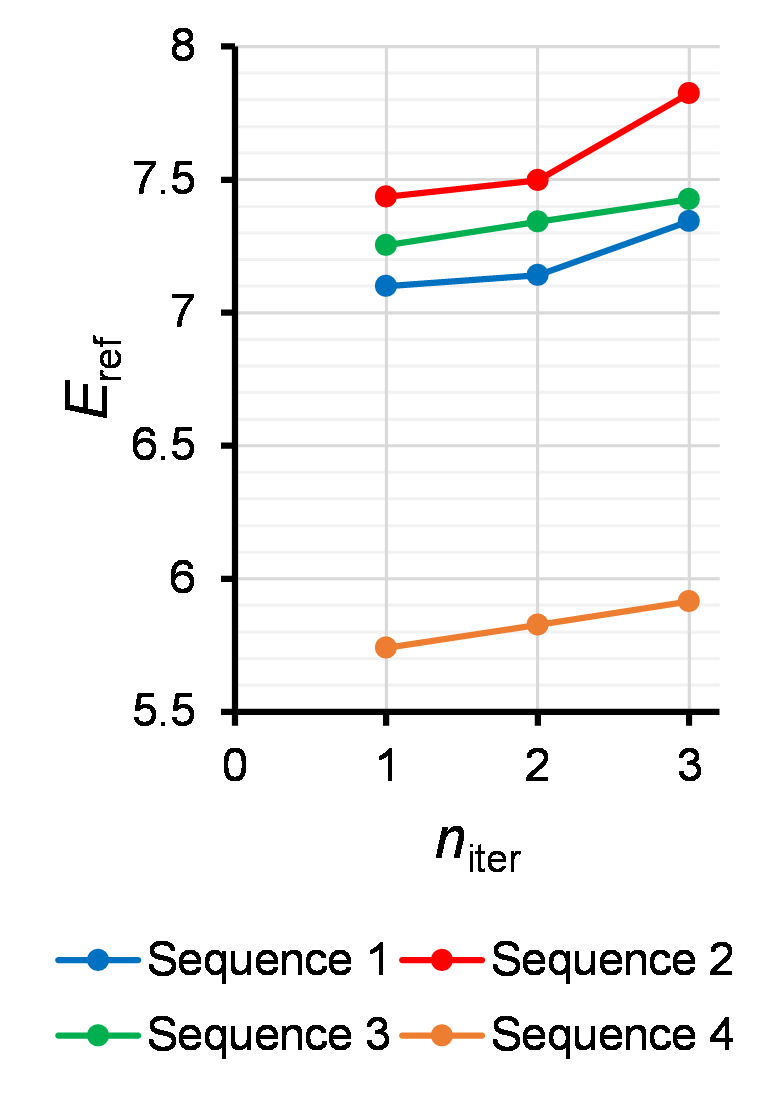} \label{subfig:DVF_err_niter_MRI}%
    \caption{\acs{RMS} registration error, $E_{\text{ref}}$, as a function of the parameters of the pyramidal, iterative Lucas--Kanade optical-flow algorithm, for each sequence in the ETH Zürich dataset (Eq. \ref{eq:gt_OF_error_def}). Given one parameter, each point in the corresponding graph represents the minimum of $E_{\text{ref}}$ over all combinations of the other parameters in the grid (Table \ref{table:image registration parameter optimization}).}%
    \label{fig:OF_grid_search_eval MRI}%
\end{figure*}

On the ETH Zürich sequences, $E_{\text{ref}}$ generally increased with $\sigma_{\text{init}}$ and $\sigma_{\text{sub}}$, suggesting that extra smoothing may be detrimental for the registration of moderately noisy images (Fig. \ref{fig:OF_grid_search_eval MRI}). $E_{\text{ref}}$ was a convex function of $\sigma_{\text{support}}$, with $\sigma_{\text{support}} = 1.0$ producing the highest errors. Using a single pyramid level yielded poor registration accuracy, consistent with the large amplitude of respiratory motion relative to the in-plane resolution (1mm isotropic pixel spacing after resampling). This aligns with prior recommendations to use a multiresolution scheme for optical-flow estimation in thoracic imaging \cite{xu2008lung, zhang2008use}. In medical imaging, multiscale designs are also commonly used in learning-based models that estimate spatial transformations (e.g., U-Net encoder--decoder component in \cite{balakrishnan2019voxelmorph} or multiscale residual blocks in \cite{romaguera2020prediction}).
% Likewise, the neural network for video forecasting in \cite{romaguera2020prediction} leveraged multi-scale residual blocks to predict spatial transformations. 
By contrast, % considering an iterative approach 
using several iterations reduced performance, as $E_{\text{ref}}$ increased with $n_{\text{iter}}$. $\sigma_{\text{support}}$ and $n_{\text{layer}}^{\text{LK}}$ had the largest impact on registration accuracy; their optimization reduced the marginal minimum of $E_{\text{ref}}$ (over all other parameters in the grid) by 19.6\% and 5.4\%, respectively. Overall, these trends are consistent with prior observations in thoracic \gls{CT} registration \cite{pohl2021prediction}.

% Putting the tables slightly earlier in the text to avoid ending with a floating table
\begin{table}[thb!]
%\normalsize
%\small
\footnotesize
\setlength{\tabcolsep}{0.8pt}
\begin{tabular}{ll}
%\hline
%\textbf{Parameters}                                                  &  \\[0.05cm]
\hline
\textbf{\gls{RNN} configuration} & \\%\rule{0pt}{2.6ex}\\ %[0.05cm]
\hline
Training loss                                                        & Instantaneous squared error \rule{0pt}{2.6ex}\\
Number and size of hidden layers                                     & Number: 1, size$^\dag$: $d$\\
Activation function $\Phi$                                           & Hyperbolic tangent \\
Optimization method                                                  & \acs{SGD} (learning rate$^\dag$ $\eta$)\\ 
Gradient clipping threshold                                          & 100.0 \\
Learnable parameter initialization                                   & Gaussian $\mathcal{N}(0, \sigma = 0.02)$\\[0.05cm]
\hline
\textbf{Transformer configuration} & \\%\rule{0pt}{2.6ex}\\ %[0.05cm]
\hline
Training loss                                                        & \acs{MSE} loss \rule{0pt}{2.6ex}\\
Embedding dimension$^\dag$                                           & $d_{\text{emb}}$ \\
Number of encoder layers$^\dag$                                      & $n_{\text{layer}}$ \\
Number of heads                                                      & 2 \\
Dimension of the \gls{FFN}                                           & Sequence-specific model: $2 d_{\text{emb}}$,\\
in each encoder layer                                                & population model: 16\\[0.05cm]
%Output \gls{FFN}                                                     & 1-layer network of dimension \\
%                                                                     & $\left \lfloor{\sqrt{L d_{\text{emb}} n_{\text{cp}}}}\right \rfloor $\\[0.05cm]
Output-head \gls{FFN} hidden width                                   & $\left \lfloor{\sqrt{L d_{\text{emb}} n_{\text{cp}}}}\right \rfloor $\\[0.05cm]
Activation function                                                  & \acs{ReLU} \\
Optimization method                                                  & \Gls{ADAM} (learning rate$^\dag$ $\eta$)\\
Dropout probability                                                  & 0.5 \\
Batch size                                                           & 32 \\
Number of epochs                                                     & Sequence-specific model: 50,\\
                                                                     & population model: max. 300\\[0.05cm]
Early stopping patience                                              & 30 (population model only)\\
Learnable parameter initialization                                   & Fan-in/fan-out–aware\\
                                                                     & (Kaiming/Xavier-style)\\[0.05cm]
Resampling method used for                                           & Bandlimited FIR interpolation\\
cross-dataset training                                               & (via polyphase filtering)\\[0.05cm]
Data augmentation (pop. only):                                       & \\
 - amplitude-scaling probability                                     & 0.8 \\
 - amplitude-scaling range                                           & [0.8, 1.2] \\
 - component-permutation prob.                                       & 0.5 \\ 
 - probability to add bias or drift\protect\footnotemark             & 0.3 \\ % drift
 - maximum additive slope allowed                                    & 0.05 $\times$ amplitude \\
 - maximum additive offset allowed                                   & 0.2 $\times$ amplitude \\ 
\hline
\end{tabular}
\caption{Configuration of \glspl{RNN} and transformers used in our experiments. Early stopping is implemented only for population transformers; the \enquote{patience} parameter refers to the number of epochs to wait for a loss decrease before terminating training. Quantities marked with $^\dag$, and $L$ (the \gls{SHL}), are tuned via grid search on the validation set (cf. Table \ref{table:models comparison} for the allowed parameter ranges).}
\label{table:RNNs_configuration}
\end{table} 

\footnotetext{In our training implementation for the population transformer, drift augmentation adjusts future targets based on the terminal input-window offset without accounting for the horizon; this likely had a minor effect on performance as we set the maximum random linear slope to a small value (0.05). Horizon-aligned augmentation is left for future work as an improvement.}

\begin{table}[thb!]
%\normalsize
%\small
\footnotesize
\setlength{\tabcolsep}{2.0pt}
\begin{tabular}{ll}
\hline
%\textbf{Parameters}                                                  &  \\[0.05cm]
%\hline
\textbf{General validation settings} & \\ %\rule{0pt}{2.6ex}\\ %[0.05cm]
\hline
Metric for selecting                                  & \acs{nRMSE} (Eq. \ref{eq:predicted weights nRMSE}) \rule{0pt}{2.6ex}\\
hyperparameters (except $n_{\text{cp}}$)                                 & \\[0.05cm]
%algorithm hyperparameters                                        & \\ [0.05cm]
Metric for selecting $n_{\text{cp}}$                                        & $E_{\text{pred}}(n_{\text{cp}})$ (Eq. \ref{eq:val registration error})\\[0.05cm]
Validation range for $n_{\text{cp}}$                                  & $n_{\text{cp}} \in \{ 1, 2, 3, 4\} $\\ [0.05cm]
\hline
\multicolumn{2}{l}{\textbf{Neural-network stochasticity mitigation}} \\% \rule{0pt}{2.6ex}\\ %[0.05cm]
\hline
Number of runs for                                                   & \acs{RTRL} and sequence- \rule{0pt}{2.6ex}\\
hyperparameter selection, $n_{\text{val}}$                           & specific transformer: 10,\\
                                                                     & population transformer: 5,\\
                                                                     & other algorithms: 250\\ [0.05cm]
Number of runs to evaluate test-set                                  & Same value as $n_{\text{val}}$ \\
\acs{PCA}-score prediction accuracy, $n_{\text{test}}^{\text{PCA}}$  & (Section \ref{section: prediction of weights on ETH Zurich data}) \\[0.05cm]
Number of runs to evaluate test-set                                  & \acs{RTRL} and transformers: 5,\\ 
image-prediction accuracy, $n_{\text{test}}$                         & other algorithms: 25 \\ [0.05cm]     
Number of runs   & Same value as $n_{\text{test}}$\\
for selecting $n_{\text{cp}}$, $n_{\text{val}}^{\text{dim}}$                                        & (Section \ref{section: methods - optimization of n_cp})\\ [0.05cm]
\hline
\textbf{Sequence temporal split} & \\ %\rule{0pt}{2.6ex}\\ %[0.05cm]
\hline
Number of images for \gls{DIR}                                       & ETH Zürich: 90,  \rule{0pt}{2.6ex}\\
parameter optimization, $M_{\text{DIR}}$                             & \acs{OvGU}: 170 ($t_{M_{\text{DIR}}}=28.3 \text{ s}$)\\[0.05cm]
Last training time-step index, $M_{\text{train}}$                           &  \\
 - \textit{Online predictors}                                        & ETH Zürich: 90, \\
                                                                     & \acs{OvGU}: 170 ($t_{M_{\text{train}}}\!=\!28.3 \text{s}$) \\ [0.05cm]
 - \textit{Offline predictors}                                       & ETH Zürich: 160, \\
                                                                     & \acs{OvGU}: 303  ($t_{M_{\text{train}}}\!=\!50.4 \text{s}$) \\ [0.05cm]
Last validation time-step                            & ETH Zürich: 180, \\ 
index, $M_{\text{val}}$                                                                     & \acs{OvGU}: 340 ($t_{M_{\text{val}}} = 56.6\text{s}$) \\ [0.05cm]
Proportion of each sequence for                                      & first 80\% of the sequence\\
training the population transformer                                  & (validation: last 20\%)\\[0.05cm]
Test time interval (all models):                                     &  \\ [0.05cm]
 - \emph{ETH Zürich}                                                 & last 6.3s of the sequence\\ 
                                                                     & (20 time steps)\\ 
 - \emph{\acs{OvGU}}                                                 & last 26.3s of the sequence\\
                                                                     & (158 time steps)\\[0.05cm] 
\hline
\end{tabular}
\caption{Parameters and settings related to the general experimental setup.}
\label{table:general experimental setup}
\end{table}

\section{Appendix: Notes on the PCA respiratory motion model}%
\label{appendix: PCA respiratory motion model}

In this section, we derive the equations of the motion model introduced in Section \ref{Breathing motion modeling with PCA}, starting from the conventional formulation of \gls{PCA}. This motion model is sequence-specific and similar to the original one proposed in \cite{zhang2007patient}, although here we apply \gls{PCA} only to internal motion and do not rely on surrogate data. We form the matrix $X \in \mathbb{R}^{M \times 2|I|}$ that contains all motion information for a given sequence up to time step $M$:%
% that stacks the vectorized deformation fields 
% We first define the matrix $X$ containing motion information for a given sequence up to time step $M \in \mathbb{N}$:%
%https://tex.stackexchange.com/questions/45551/missing-inserted-while-adding-resizebox
\begin{equation} \label{eq:data matrix}
\resizebox{1.0\hsize}{!}{$
X = 
\begin{bmatrix}
u_x(\vec{x_1}, t_1) & u_y(\vec{x_1}, t_1) & u_x(\vec{x_2}, t_1) & \dots & u_y(\vec{x}_{|I|}, t_1) \\
  &   & \dots &   &  \\
u_x(\vec{x_1}, t_M) & u_y(\vec{x_1}, t_M) & u_x(\vec{x_2}, t_M) & \dots & u_y(\vec{x}_{|I|}, t_M) \\
\end{bmatrix}
$}
\end{equation}%
Here, $u_x(\vec{x}, t)$ and $u_y(\vec{x}, t)$ refer, respectively, to the $x$- and $y$-components of $\vec{u}(\vec{x}, t)$, the deformation vector at pixel $\vec{x}$ and time $t$. To simplify notation in this appendix, we set $M$ to the number of training images ($M_{\text{train}}$). We denote by $\mu_X \in \mathbb{R}^{1 \times 2|I|}$ the row vector of column-wise means of $X$, that is, the mean deformation field over time:% 
% We define the mean deformation $\mu_X \in \mathbb{R}^{1 \times 2|I|}$ as the line vector containing the mean of each column of $X$:%
\begin{equation} \label{eq:mean vector}
\mu_X = [\mu_x(\vec{x_1}), \mu_y(\vec{x_1}), \mu_x(\vec{x_2}),  \dots, \mu_y(\vec{x}_{|I|})]
\end{equation}
In this equality, $\mu_x(\vec{x_i})$ and $\mu_y(\vec{x_i})$ refer to the averages of $u_x(\vec{x_i}, t_k)$ and $u_y(\vec{x_i}, t_k)$, respectively, over time steps $k \in \{1, \dots, M\}$. We define the centered data matrix $X_{\text{c}}$, whose columns each have zero mean, as:%
\begin{equation}\label{eq:centered data matrix}
X_{\text{c}} = X - \mathbb{1}_{M \times 1} \mu_X
\end{equation}

Given an arbitrary integer $n_{\text{cp}}$, \gls{PCA} seeks two matrices, $W \in \mathbb{R}^{M \times n_{\text{cp}}}$ and $U \in \mathbb{R}^{2|I| \times n_{\text{cp}}}$, that minimize the quantity $\|X_{\text{c}} - W U^T \|_F$, where $\|\cdot\|_F$ denotes the Frobenius norm, subject to the following two conditions:%
\begin{itemize}
	\item $W^T W$ is a diagonal matrix, 
	\item $U^T U = I_{n_{\text{cp}}}$ (identity matrix of size $n_{\text{cp}}$).
\end{itemize}%
Specifically, we first perform the spectral decomposition of $Y = X_{\text{c}} X_{\text{c}}^T$, i.e., we compute the matrix of its eigenvectors $V$ and the diagonal matrix $\Lambda$ that contains the square roots of its eigenvalues, such that:%
\begin{equation}
Y = V \Lambda^2 V^T    
\end{equation}%
We introduce the following notations:%
\begin{equation}
\Lambda =
\begin{bmatrix}
    \lambda_1 & & \\
    & \ddots & \\
    & & \lambda_M
\end{bmatrix}
\quad \text{with} \, \lambda_1 \geq \dots \geq \lambda_M \geq 0
\end{equation}
\begin{equation}
V = \big[V_1, \dots, V_M\big]
\end{equation}
We multiply each column $V_i$ by the sign of its first non-zero entry (which thereby becomes positive). We include this additional normalization step to ensure that the \gls{PCA} output remains consistent regardless of the eigendecomposition algorithm used. $W$ and $U$ are computed using the first $n_{\text{cp}}$ eigenpairs, as follows:%
\begin{equation}
W = [V_1, \dots, V_{n_{\text{cp}}}] 
  \begin{bmatrix}
    \lambda_1 & & \\
    & \ddots & \\
    & & \lambda_{n_{\text{cp}}}
  \end{bmatrix}    
\end{equation}
\begin{equation}
  U = X_{\text{c}}^T [V_1, \dots, V_{n_{\text{cp}}}]
  \begin{bmatrix}
    1/\lambda_1 & & \\
    & \ddots & \\
    & & 1/\lambda_{n_{\text{cp}}}
  \end{bmatrix}  
\end{equation}%
%$U$ is the principal components matrix, and its columns are the principal components. $W$ is referred to as the weight matrix. 
$X$, $W$, and $U$ approximately satisfy the following relationship:%
\begin{equation} \label{eq: PCA matrix relationship}
X - \mathbb{1}_{M \times 1} \mu_X = W U^T
\end{equation}

The rows of $W$ contain the time-dependent \gls{PCA} weights, and the columns of $U$ represent the spatial deformation modes. Accordingly, we denote the entries of $W$ and $U$ as follows:%
\begin{equation} \label{eq:W U entries}
W = 
\begin{bmatrix}
w_1(t_1) & \dots & w_{n_{\text{cp}}}(t_1) \\
\dots & \dots & \dots \\
w_1(t_M) & \dots & w_{n_{\text{cp}}}(t_M) \\
\end{bmatrix}
\end{equation}
\begin{equation}
U = 
\begin{bmatrix}
u_1^x(\vec{x_1}) & \dots & u_{n_{\text{cp}}}^x(\vec{x_1}) \\
u_1^y(\vec{x_1}) & \dots & u_{n_{\text{cp}}}^y(\vec{x_1}) \\
u_1^x(\vec{x_2}) & \dots & u_{n_{\text{cp}}}^x(\vec{x_2}) \\
\dots & \dots & \dots \\
u_1^y(\vec{x}_{|I|}) & \dots & u_{n_{\text{cp}}}^y(\vec{x}_{|I|}) \\
\end{bmatrix}
\end{equation}%
%Using the latter notations, we infer from Eq. \ref{eq: PCA matrix relationship} that for all pixel index $i \in \{1, \dots, |I|\}$ and time step $k \in \{1, \dots, M\}$:%
%\begin{equation}
%\left\{
%\begin{aligned}
%u_x(\vec{x_i}, t_k) & = \mu_x(\vec{x_i}) + \sum_{j=1}^{n_{\text{cp}}} w_j(t_k) u_j^x(\vec{x_i})\\
%u_y(\vec{x_i}, t_k) & = \mu_y(\vec{x_i}) + \sum_{j=1}^{n_{\text{cp}}} w_j(t_k) u_j^y(\vec{x_i})
%\end{aligned} \right.
%\end{equation}
Using those notations and defining $\vec{\mu}(\vec{x_i}) = [\mu_x(\vec{x_i}),\allowbreak \mu_y(\vec{x_i})]^T$ and $\vec{u_j}(\vec{x_i}) = [u_j^x(\vec{x_i}),\allowbreak u_j^y(\vec{x_i})]^T$,
%\footnotemark, 
Eq. \ref{eq: PCA matrix relationship} can be rewritten as Eq. \ref{eq:PCA respiratory model},
%We also define $\vec{\mu}(\vec{x_i}) = [\mu_x(\vec{x_i}), \mu_y(\vec{x_i})]^T$ and $\vec{u_j} = [u_j^x(\vec{x_i}),\allowbreak u_j^y(\vec{x_i})]^T$. Calculating each entry in $X$ using Eq. \ref{eq: PCA matrix relationship} and the notations above yields Eq. \ref{eq:PCA respiratory model} component-wise.
which holds for the time steps used to build $X$, and is extended to $k > M$ via projection (as discussed below).
%These two latter equations can be combined and rewritten as Eq. \ref{eq:PCA respiratory model} by defining $\vec{\mu}(\vec{x_i}) = [\mu_x(\vec{x_i}), \mu_y(\vec{x_i})]^T$ and $\vec{u_j} = [u_j^x(\vec{x_i}), u_j^y(\vec{x_i})]^T$. 
%Eq. \ref{eq:PCA respiratory model} is the explicit geometrical form of Eq. \ref{eq: PCA matrix relationship}, the latter being expressed using a more abstract linear algebra framework. Both describe the \gls{PCA} motion model, which approximates high-dimensional time-dependent \glspl{DVF}, $\vec{u}(\vec{x}, t)$, by a linear combination of a few static independent vector fields, $\vec{u_j}(\vec{x})$, generating a low-dimensional linear subspace, weighted by the coefficients $w_j(t)$. 
Notably, with our convention for eigenvector signs (positive first-row entry), 
%the sign of each weight trajectory $w_j(t)$ is fixed consistently across decompositions. Noticeably, 
we have $w_j(t_1) \geq 0$ for all $j \in \{1, \dots, n_{\text{cp}}\}$ (Figs. \ref{fig:DVF principal components and weights sequence 4} and \ref{fig:1st cpt prediction (SnAp-1 vs vs UORO vs transformer) Magdeburg}).
%Noticeably, the normalization step for $V$ described earlier (making all coefficients on its first row positive) implies that the first non-zero (scalar) time-dependent weight is always positive (Figs. \ref{fig:DVF principal components and weights sequence 4}, \ref{fig:1st PCA component prediction in Sq 6 of OgVU dataset}), that is:
%\begin{equation}
%\forall j \in \{1, \dots, n_{\text{cp}}\}, w_j(t_1) \geq 0
%\end{equation}

%\footnotetext{We define $\vec{\mu}(\vec{x_i}) = [\mu_x(\vec{x_i}), \mu_y(\vec{x_i})]^T$ and $\vec{u_j}(\vec{x_i}) = [u_j^x(\vec{x_i}),\allowbreak u_j^y(\vec{x_i})]^T$.}

Moreover, the formula $U^T U = I_{n_{\text{cp}}}$, expressing the orthonormality of the principal components, is equivalent to Eq. \ref{eq:principal components orthonormality}. Using this relationship, Eq. \ref{eq: PCA matrix relationship} can be rewritten as follows:%
\begin{equation} \label{eq: PCA matrix relationship rewritten}
W = X_{\text{c}} U
\end{equation}
For $k \geq 1$, we denote by $X_{\text{c}}(t_k)$ the row vector obtained by flattening the \gls{DVF} at time $t_k$ and centering it using $\mu_X$, the mean \gls{DVF} over the training time steps (Eq. \ref{eq:mean vector}):%
% For $k \in \mathbb{N}$, we define the following row vector, corresponding to the \gls{DVF} at time $t_k$, centered using the mean \gls{DVF} of the training set:%
\begin{equation} \label{eq:instantaneous centered DVF}
\resizebox{1.0\hsize}{!}{$
X_{\text{c}}(t_k) = [u_x(\vec{x_1}, t_k), u_y(\vec{x_1}, t_k), u_x(\vec{x_2}, t_k), \dots, u_y(\vec{x}_{|I|}, t_k)] - \mu_X
$}
\end{equation}
For $k \leq M$, $X_{\text{c}}(t_k)$ is the $k^{\text{th}}$ row of $X_{\text{c}}$; thus, Eq. \ref{eq: PCA matrix relationship rewritten} is equivalent to:%
%$X_{\text{c}}(t_k)$ is the $k^{\text{th}}$ row of the centered data matrix $X_{\text{c}} \in \mathbb{R}^{M_{\text{train}} \times 2|I|}$ when $k \leq M_{\text{train}}$. Eq. \ref{eq: PCA matrix relationship rewritten} can be rewritten as:%
\begin{equation} \label{eq:instantaneous PCA weights calculation}
[w_1(t_k), \dots, w_{n_{\text{cp}}}(t_k)] = X_{\text{c}}(t_k) U
\end{equation}

This equation expresses the weights at time $t_k$ as the projection of $X_{\text{c}}(t_k)$ onto the $n_{\text{cp}}$-dimensional linear subspace of $\mathbb{R}^{2 |I|}$ spanned by the orthonormal columns of $U$.
%The latter equation expresses the fact that the time-dependent weights at time $t_k$ can be computed by projecting the flattened centered \gls{DVF} vector, $X_{\text{c}}(t_k)$, onto the linear subspace of $\mathbb{R}^{2 |I|}$ spanned by the (orthogonal) columns of $U$.
We assume that spatiotemporal dynamics remain relatively stable throughout the image sequence. Accordingly, during inference, we define the weights at time $t_k$ as the projection of $X_{\text{c}}(t_k)$, estimated from the incoming image, onto the same subspace.
% During inference, we keep the principal components from the training data and define the weights as the projection of the centered \gls{DVF} $X_{\text{c}}(t_k)$ computed with the incoming data onto the same subspace. 
In other words, Eq. \ref{eq:instantaneous PCA weights calculation} defines the weights for $k > M$, by projection onto the fixed basis $U$
% In other words, Eq. \ref{eq:instantaneous PCA weights calculation} is also valid for $k > M_{\text{train}}$. 
(Fig. \ref{fig:Geometrical viewpoint}). Eq. \ref{eq:PCA weight calculation} follows directly from Eq. \ref{eq:instantaneous PCA weights calculation}.   

%\FloatBarrier % Prevents tables to go before the section heading? Does not work
\section{Appendix: Experimental setup}
\label{appendix: RNN experimental setup}

Tables \ref{table:RNNs_configuration} and \ref{table:general experimental setup} summarize the \gls{RNN} and transformer configuration and the parameters and settings for the general experimental setup, respectively. We clip the loss-gradient norm to 100.0 for \glspl{RNN} and 2.0 for \gls{LMS}, respectively \cite{pascanu2013difficulty}. The forecasting pipeline and performance evaluation step are implemented in MATLAB for consistent evaluation across methods. Transformer training and inference are implemented in Python, interfaced with the MATLAB evaluation code, and conducted on consumer-grade \glspl{GPU} (NVIDIA GeForce RTX 3060 and RTX 4060).

%\flushcolsend % why do I get undefined control sequence error
%\onecolumn

\section{Appendix: Further results on performance variation with the horizon}
\label{appendix: performance variation with h on Magdeburg}

\begin{figure*}[p] % or [t], or [p] to dedicate a float page
  \centering

  % --- your figure content ---
    \includegraphics[width=.33\textwidth]{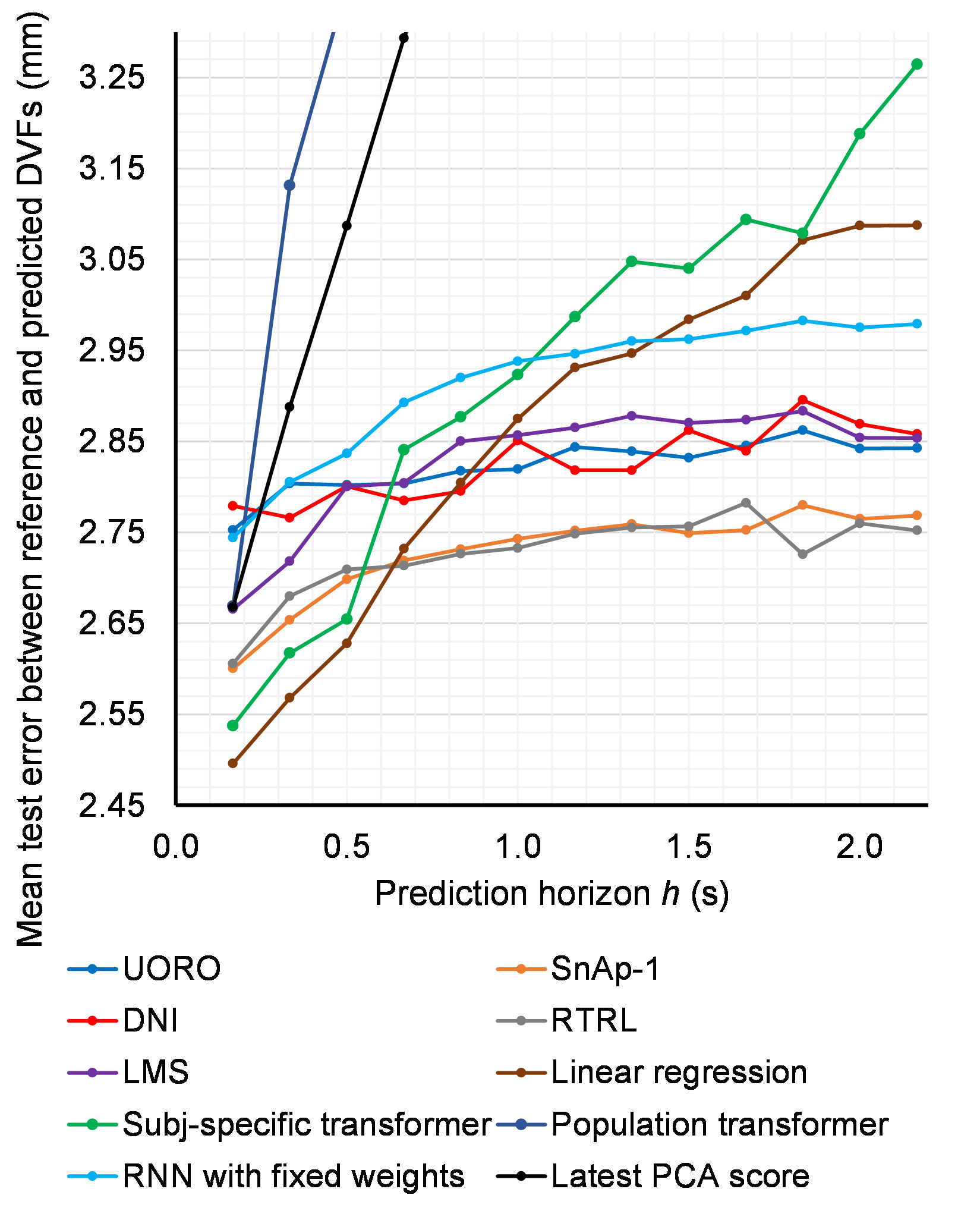}% 
    \includegraphics[width=.33\textwidth]{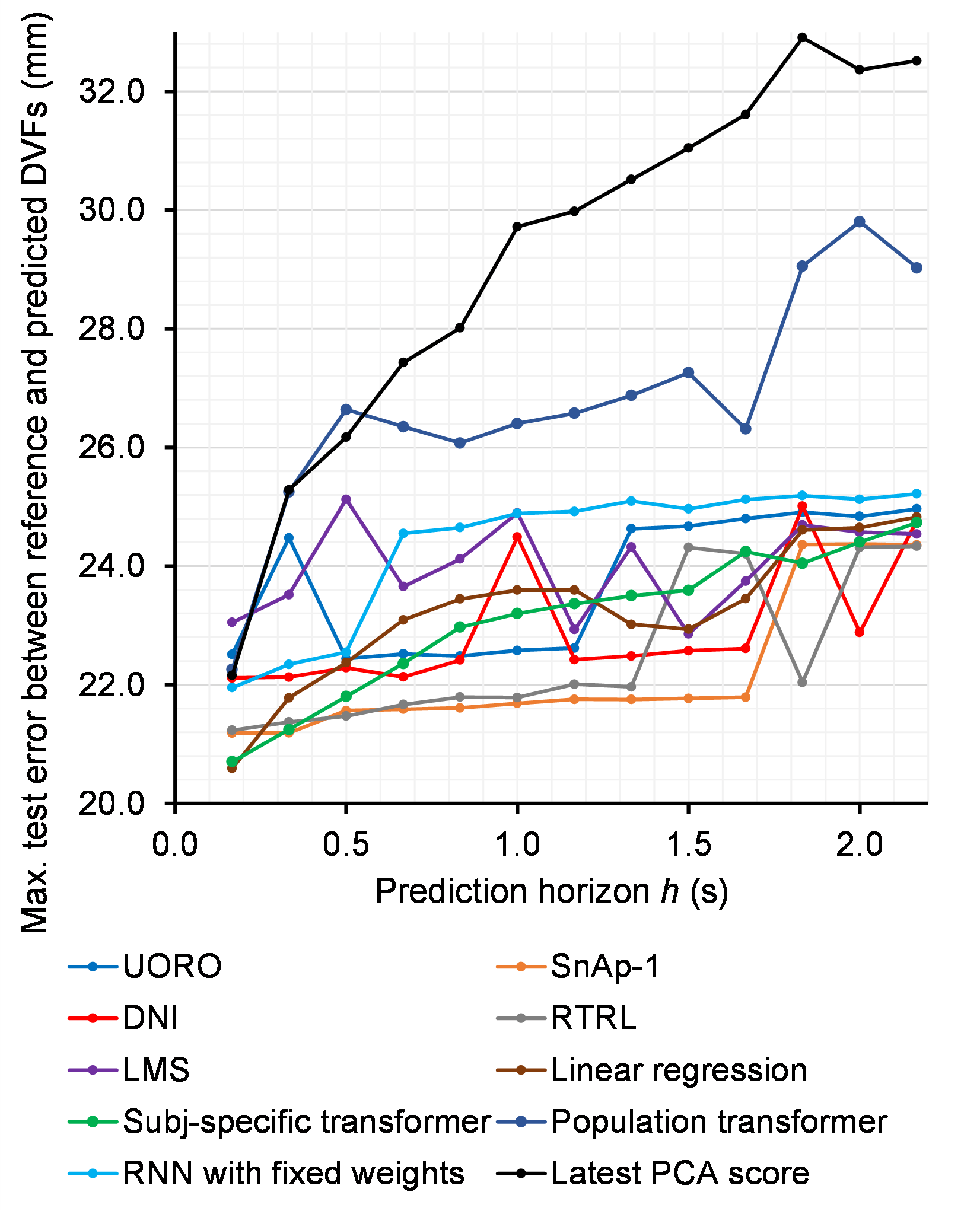} %     
    \includegraphics[width=.33\textwidth]{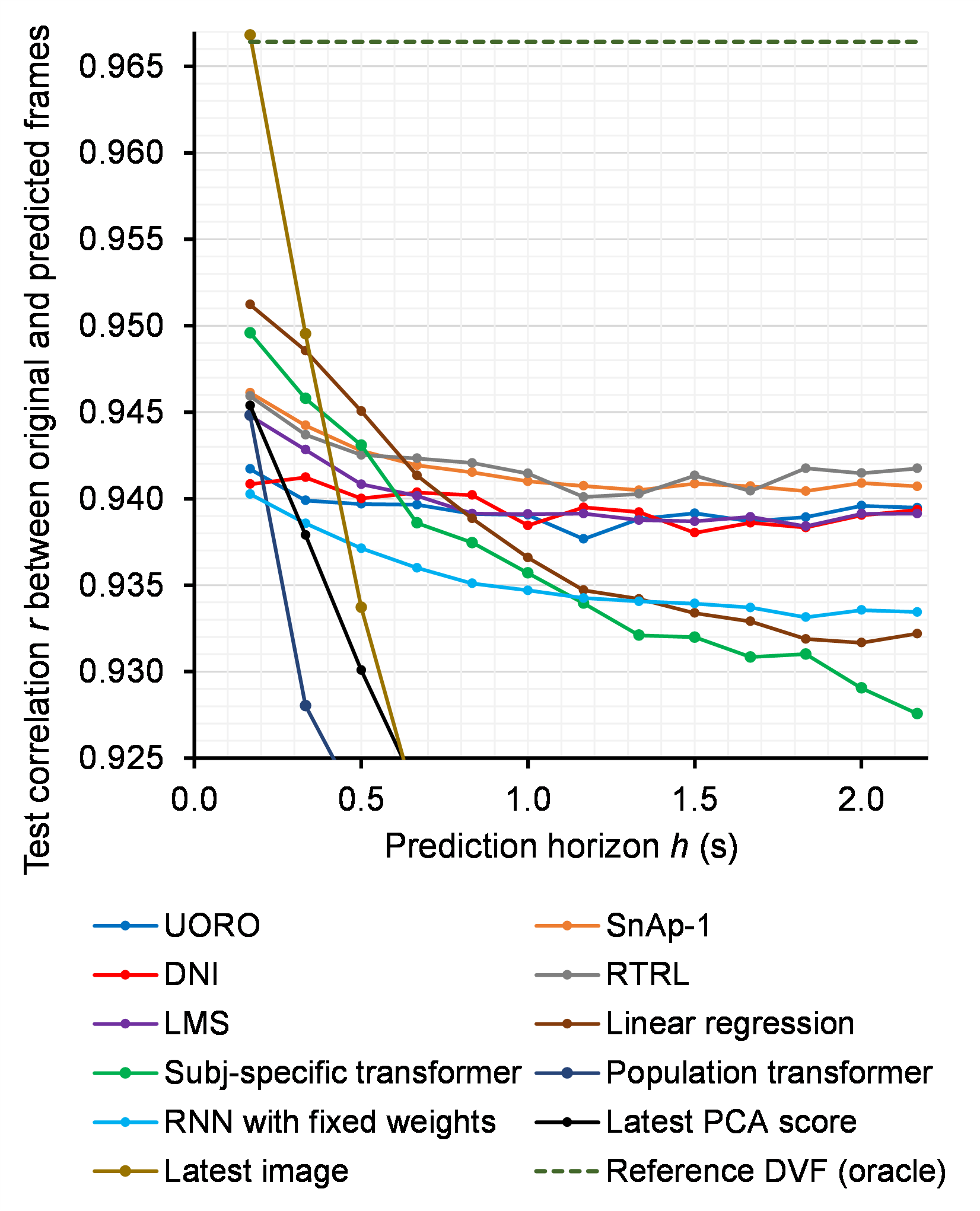}%  
    \includegraphics[width=.33\textwidth]{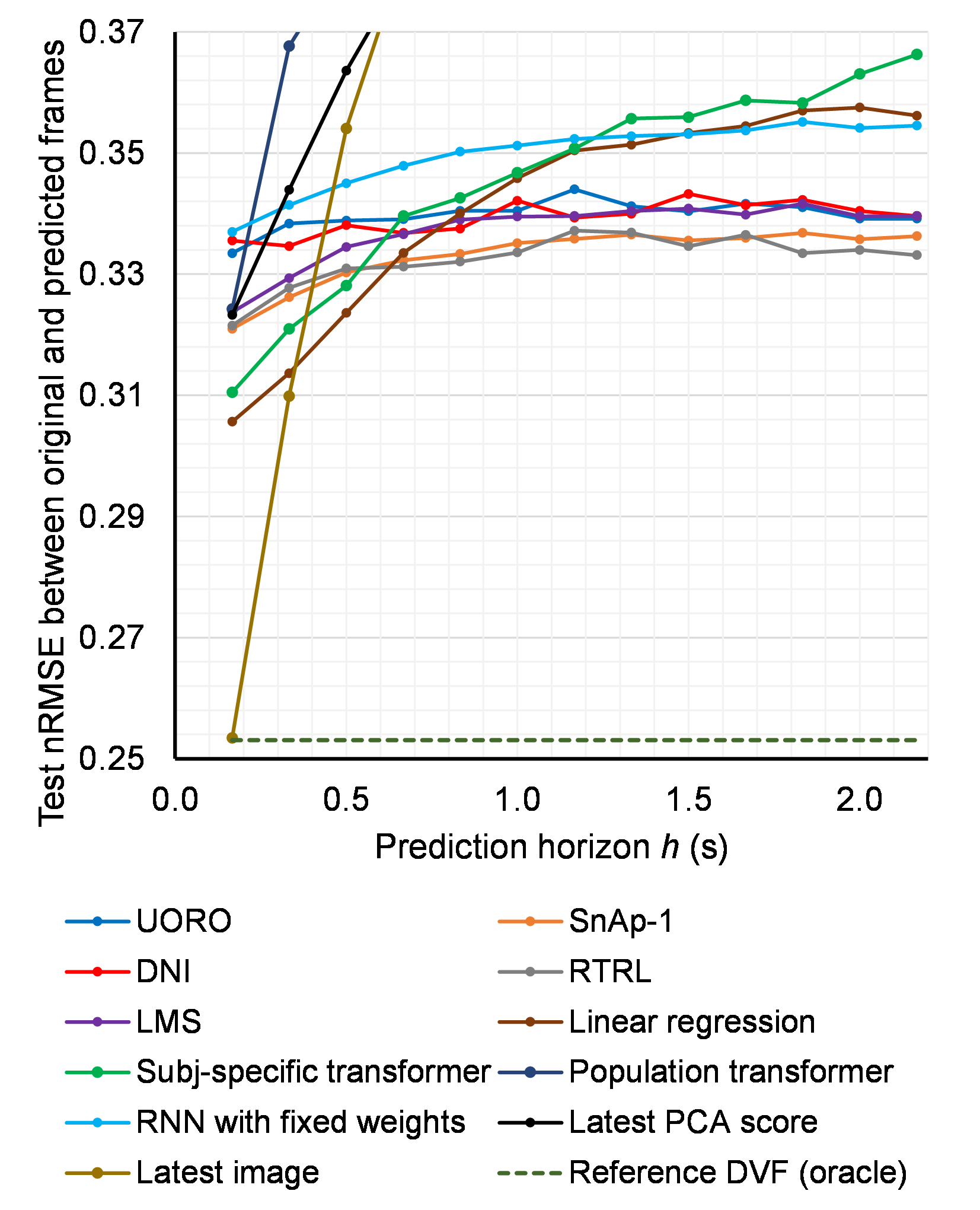}%     
    \includegraphics[width=.33\textwidth]{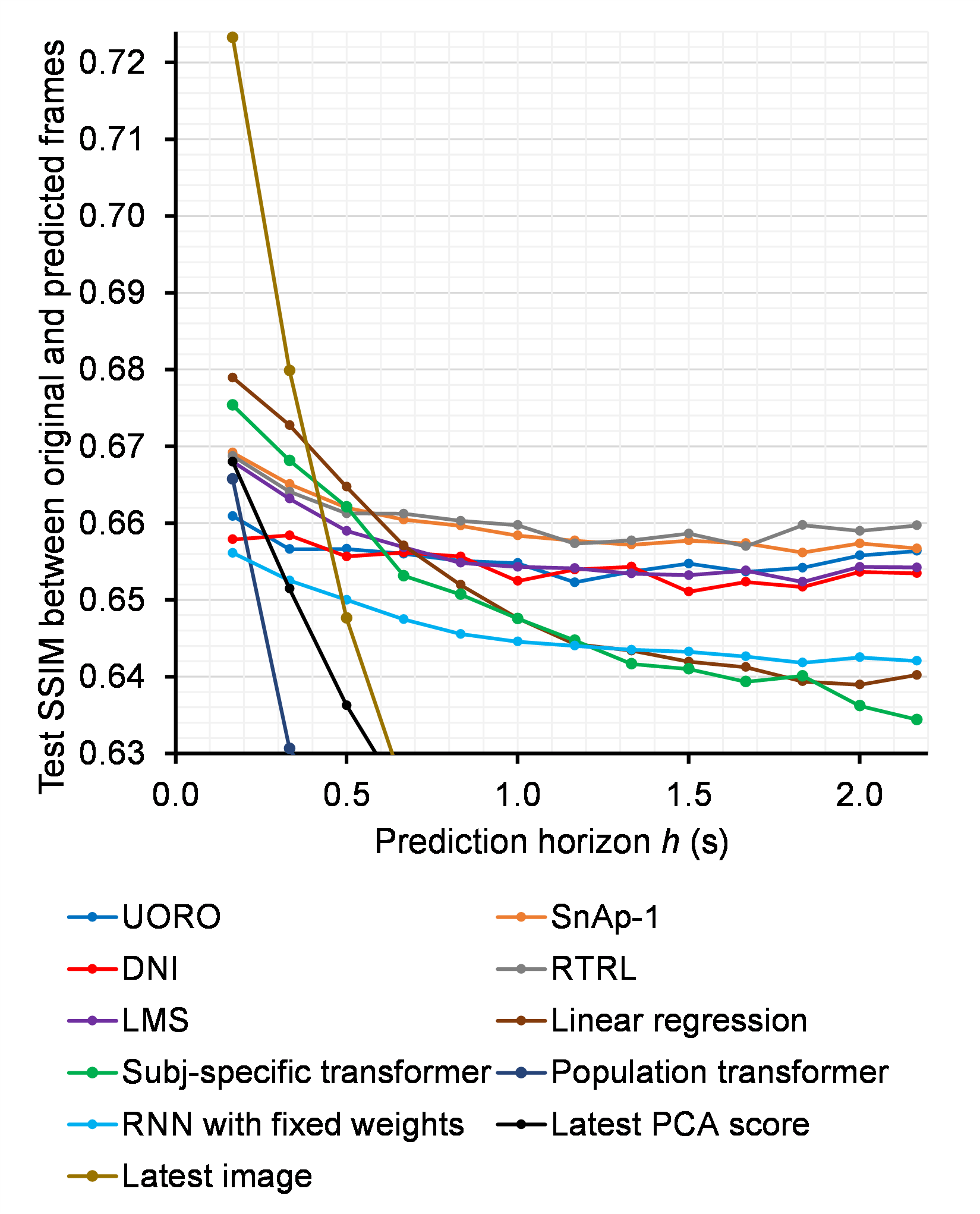} %             
    \caption{Test-set frame-forecasting performance for each algorithm as a function of the horizon $h$, for the \gls{OvGU} dataset (evaluation on the entire images). Each point represents the average of a given metric over the eight image sequences and $n_{\text{test}}$ runs. Hyperparameters were optimized for each sequence (except the population transformer) and each value of $h$ via grid search on the validation set. The mean of each curve (i.e., the performance averaged over $h$ for each method) is reported in the \enquote{\gls{OvGU} (whole image)} columns of Table \ref{table:frame pred perf}. The \gls{SSIM} of the oracle baseline is not shown, as it exceeded that of the other methods by a large margin.} %protect\footnotemark[\value{figfn}]%\protect\footnotemark.}
	\label{fig:next frame pred perf vs hrz on Magdeburg full image} 

  \smallskip
  %\medskip
  %\bigskip

  % --- your table content ---
  \scriptsize
  \setlength{\tabcolsep}{4.5pt}
  %\footnotesize  
  %\setlength{\tabcolsep}{2pt}

\begin{tabular}{llllllll}
\hline
                                    && \multicolumn{2}{l}{ETH Zürich}       & \multicolumn{2}{l}{\gls{OvGU} (whole image)} & \multicolumn{2}{l}{\gls{OvGU} (\acs{ROI})} \\[0.05cm]
\cline{3-8}
                                    && Test \acs{SSIM}& Mean test \gls{DVF} & Test \acs{SSIM} & Mean test \gls{DVF}        & Test \acs{SSIM}& Mean test \gls{DVF} \rule{0pt}{2.6ex}\\
                                    && decrease (\%)  & error increase (\%) & decrease (\%)   & error increase (\%)        & decrease (\%)  & error increase (\%)          \\[0.05cm]
\hline
%\textbf{Prediction methods} &&&&&&\rule{0pt}{2.6ex}\\
\textbf{Prediction}         & \acs{UORO}                            & \textbf{0.2}  & \textbf{3.4}   & \textbf{0.03} & \textbf{1.4}                  & \textbf{0.1}  & \textbf{2.2} \\
\textbf{algorithms}            & \acs{SnAp-1}                          & 0.6           & 7.0            & 1.3           & 4.3                           & 4.1           & 8.6 \\
                            & \acs{DNI}                             & \textbf{0.4}  & \textbf{5.5}   & 0.8           & 3.3                           & \textbf{2.5}  & 6.1 \\
                            & \acs{RTRL}                            & 0.6           & 6.8            & \textbf{0.7}  & \textbf{2.7}                  & \textbf{2.5}  & \textbf{5.0} \\
                            & \acs{LMS}                             & 0.9           & 7.0            & 1.4           & 5.0                           & 4.6           & 11.0 \\
                            & Linear regression                     & 1.3           & 19.4           & 4.8           & 20.2                          & 14.7          & 42.9 \\
                            & Sequence-specific transformer         & 1.6           & 17.3           & 5.1           & 24.7                          & 14.9          & 48.4 \\
                            & Population transformer                & 1.3           & 9.6            & 3.1           & 14.7                          & 10.2          & 25.2 \\
% \textbf{Baselines} &&&&&&\rule{0pt}{2.6ex}\\
\textbf{Baselines}          & \acs{RNN} with a frozen hidden layer  & 0.8           & 7.5            & 1.6           & 6.2                           & 5.7           & 12.4 \\
                            & Latest \acs{PCA} weight as prediction & 2.1           & 5.5            & 9.7           & 41.6                          & 30.4          & 79.8 \\
                            & Latest image as prediction            & 3.7           & n/a            & 15.6          & n/a                           & 40.4          & n/a \\[0.05cm]
\hline
\multicolumn{2}{l}{Average over non-baseline \acs{RNN} algorithms}      & 0.4           & 5.3            & 0.7           & 3.0                           & 2.2           & 5.6 \rule{0pt}{2.6ex}\\
\multicolumn{2}{l}{Average over all non-baseline algorithms}   & 0.9           & 9.5            & 2.1           & 9.5                           & 6.7           & 18.7 \\
\hline
\end{tabular}
  \captionof{table}{Test-set performance stability with respect to the horizon $h$ for each method, expressed as the relative decrease in the \acs{SSIM} and relative increase in the mean \gls{DVF} error averaged over each dataset (and evaluation region---\gls{ROI} or whole image---for the \gls{OvGU} data) between $h=0.3\text{s}$ and $h=2.2\text{s}$.
  %$h=1$ (0.31s) and $h=7$ (2.20s) for the ETH Zürich acquisitions and between $h=2$ (0.33s) and $h=13$ (2.17s) for the \gls{OvGU} acquisitions.
  These relative differences are computed from the data points in Figs. \ref{fig:next frame pred perf vs hrz on ETH}, \ref{fig:next frame pred perf vs hrz on Magdeburg ROI}, and \ref{fig:next frame pred perf vs hrz on Magdeburg full image}. %\protect\footnotemark[\value{tabfn}]. 
  In each column, the values corresponding to the two most stable non-baseline algorithms are bolded (ties included).}
  \label{table:stability_horizon}
\end{figure*}

Fig. \ref{fig:next frame pred perf vs hrz on Magdeburg full image} shows the performance of each algorithm as a function of the horizon $h$ on the \gls{OvGU} dataset (full-frame evaluation). Stability with respect to $h$ is detailed in Table \ref{table:stability_horizon}.

% Immediately before the figure* float (in normal text flow)
%\refstepcounter{footnote}\setcounter{figfn}{\value{footnote}}% reserve N
%\refstepcounter{footnote}\setcounter{tabfn}{\value{footnote}}% reserve N+1

% in the document

% Put the footnote texts exactly where you want them to print
%\footnotetext[\value{figfn}]{The \gls{SSIM} associated with the oracle (warping the initial frame with the Lucas--Kanade optical flow) was not represented because it was much higher than that of the other methods, for readability.}
%\footnotetext[\value{tabfn}]{The values used to compute the relative differences correspond to the data points in Figs. \ref{fig:next frame pred perf vs hrz on ETH}, \ref{fig:next frame pred perf vs hrz on Magdeburg ROI}, and \ref{fig:next frame pred perf vs hrz on Magdeburg full image}.}

%\printglossary[type=acronym] % does not work at the moment on TexMaker but works on Overleaf

% after your last content, just before \end{document}
\clearpage  % ensure all floats have been shipped out

\end{document}